\RequirePackage{fix-cm}
\documentclass[smallextended]{svjour3}       
\smartqed  
\usepackage[bookmarksdepth=2]{hyperref}
\usepackage[dvipsnames]{xcolor}
\usepackage{amssymb}
\usepackage{pifont}
\newcommand{\cmark}{\ding{51}}%

\usepackage{colortbl} 
\usepackage{graphicx}
\usepackage{makecell}

\usepackage{longtable}
\usepackage{xltabular}
\usepackage{afterpage}
\usepackage{subcaption}
\usepackage{xspace}
\usepackage{listings}
\usepackage{tabularx}
\usepackage{tabulary}
\usepackage{rotating}
\usepackage{tablefootnote}
\usepackage{enumitem}
\usepackage[most]{tcolorbox}
\usepackage{adjustbox}
\usepackage{booktabs}
\usepackage{fullpage}
\usepackage{diagbox}
\usepackage{float}
\usepackage{amsmath}
\usepackage{array}
\usepackage{subcaption}
\usepackage[strict]{changepage}
\usepackage{threeparttable}
\usepackage{wasysym}
\usepackage[title,page]{appendix}
\usepackage{multirow} 
\definecolor{summary_box}{RGB}{128,128,128}
\usepackage{xcolor} 
\usepackage{csquotes}

\newtcbtheorem{Summary}{\bfseries Summary of Research Question}{enhanced,drop shadow={black!5!white},
  coltitle=white,
  top=0.3in,
  attach boxed title to top left=
  {xshift=1.5em,yshift=-\tcboxedtitleheight/2},
  boxed title style={size=small,colback=summary_box}
}{summary}
\usepackage{algpseudocode}
\usepackage{algorithm}
\begin{document}

\title{On the Impact of Black-box Deployment Strategies for Edge AI on Latency and Model Performance
}
\subtitle{Empirical study on ONNX}


\author{Jaskirat Singh         \and
    Emad Fallahzadeh           \and
     Bram Adams                \and
     Ahmed E. Hassan
}


\institute{Jaskirat Singh \at 
    School of Computing, Queen's University, Kingston, Ontario, Canada\\
    \email{jaskirat.s@queensu.ca}\\
    \url{https://orcid.org/0009-0001-0112-7674}
    \and
    Emad Fallahzadeh \at
    School of Computing, 
    Queen's University, Kingston, Ontario, Canada. \\              \email{emad.fallahzadeh@queensu.ca}\\           %
              \url{https://orcid.org/0009-0005-5024-4868}
    \and
    Bram Adams \at
    School of Computing, Queen's University, Kingston, Ontario, Canada. \\
    \email{bram.adams@queensu.ca}\\
    \url{https://orcid.org/0000-0001-7213-4006}
    \and
    Ahmed E. Hassan \at School of Computing, Queen's University, Kingston, Ontario, Canada\\
    \email{hassan@queensu.ca}\\
                \url{https://orcid.org/0000-0001-7749-5513}
}

\date{Received: date / Accepted: date}

\maketitle
~\textit{Context:} Cloud-based black-box model deployment faces challenges related to latency and privacy due to data transmission across Wide Area Networks. On the other hand, Mobile-based black-box deployment prioritizes privacy at the expense of higher latency due to limited computational resources. To address these issues, Edge AI enables the deployment of black-box models across Mobile, Edge, and Cloud devices using a wide range of operators able to distribute a model's components, terminate inference early, or even quantize a model's computations, offering latency and privacy benefits. Existing surveys classify Edge AI model inference techniques into eight families, including Quantization, Early Exiting, and Partitioning, but they often treat these operators in isolation, overlooking their potential synergies and practical integration in real-world scenarios. Deciding what combination of operators to use across the Edge AI tiers to achieve specific latency and model performance requirements is still an open question for MLOps Engineers. ~\textit{Objective:} This study aims to empirically assess the accuracy vs inference time trade-off of different black-box Edge AI deployment strategies, i.e., combinations of deployment operators and deployment tiers.~\textit{Method:} In this paper, we conduct inference experiments involving three deployment operators (i.e., Partitioning, Quantization, Early Exit), three deployment tiers (i.e., Mobile, Edge, Cloud) and their combinations on four widely-used Computer-Vision models to investigate the optimal strategies from the point of view of MLOps developers. The analysis is conducted in a containerized environment using CUDA for Cloud GPU acceleration and ONNX for model interoperability, covering a wide range of network bandwidths.~\textit{Results:} Our findings suggest that Edge deployment using the hybrid Quantization + Early Exit operator could be preferred over Non-Hybrid operators (Quantization/Early Exit on Edge, Partition on Mobile-Edge) when faster latency is a concern at medium accuracy loss. However, when minimizing accuracy loss is a concern, MLOps Engineers should prefer using only a Quantization operator on Edge at a latency reduction or increase, respectively over the Early Exit/Partition (on Edge/Mobile-Edge) and Quantized Early Exit (on Edge) operators. In scenarios constrained by Mobile CPU/RAM resources, a preference for Partitioning across Mobile and Edge tiers is observed over Mobile deployment. For models with smaller input data samples (such as FCN), a network-constrained Cloud deployment can also be a better alternative than Mobile/Edge deployment and Partitioning strategies. For models with large input data samples (ResNet, ResNext, DUC), an Edge tier having higher network/computational capabilities than the Cloud/Mobile tier can be a more viable option than Partitioning and Mobile/Cloud deployment strategies. Smaller input data-sized models like FCN fit well in the Cloud, even with low bandwidth ($\leq$10 Mbps). Larger input data-sized models, like ResNe(x)t and DUC, need more bandwidth ($\geq$50 Mbps) for Cloud latency convergence. Partitioned-based strategies for large intermediate-sized models like FCN and DUC also need at least 50 Mbps for latency convergence. Overall, the Cloud tier performs better than the Edge and Mobile tiers for Non-Partitioning operators when the MEC bandwidth is at least 50 Mbps. However, its latency performance declines in lower bandwidth scenarios. Furthermore, Mobile-Edge Partitioning-based strategies are a better alternative compared to Mobile-Cloud and Edge-Cloud alternatives.


\section{Introduction}
\label{sec:introduction}
Artificial Intelligence (AI) on the Edge (also ~\enquote{Edge Intelligence} or ~\enquote{Edge AI})~\cite{zhou2019edge}, an interdisciplinary field derived from Edge computing and AI, receives a tremendous amount of interest from both the industry and academia. This is primarily due to its low latency, privacy preservation, and potential independence from network connectivity. Edge AI leverages widespread Edge resources instead of relying solely on Cloud or Mobile, leading to more efficient AI insights for inference and training tasks. For example, in our experiments, we consider inference tasks in a typical Edge AI environment involving an Edge device (tier) near a resource-scarce Mobile device (tier) and a resource-abundant Cloud device (tier) far from the Edge device (tier).

Traditional monolithic deployments such as deploying large AI models entirely on a Cloud or a Mobile tier affect the overall performance in terms of Key Performance Indicators (KPIs). For example, deploying entire AI models on the Cloud provides faster computation in model inference due to the available GPU resources. However, it leads to high transmission latency, monetary cost, and privacy leakage when transmitting large amounts of input data across the Wide-Area Network (WAN) to a centralized data center for AI applications (e.g., real-time video analytics). On-device inference, running entire AI applications on the Mobile tier to process the input data locally, provides data privacy protection but suffers from high computation latency because many AI applications require high computational power that significantly outweighs the capacity of resource-constrained Mobile tiers~\cite{DBLP:journals/corr/abs-2105-02613}.

Edge computing essentially pushes Cloud-like services to network Edge servers that are in closer proximity to Mobile tiers and data sources~\cite{wang2020convergence}. This offers several benefits compared to the traditional Cloud-based paradigm (i.e., low transmission latency, data privacy protection, and low monetary cost) and the Mobile-based paradigm (i.e., faster computational latency). However, this comes at the expense of increased computational latency compared to the Cloud and a higher data privacy threat compared to the Mobile.

Various operators for Edge AI model inference are proposed to address the above challenges faced by monolithic Mobile, Edge, and Cloud deployments. Zhou et al.~\cite{zhou2019edge} provide a detailed survey on seven major families of deployment operators. Among the three model optimization operator families, Model Compression includes operators like Weight Pruning, Knowledge Distillation, and Quantization to reduce computation and storage; Model Partition provides computational offloading across the tiers and latency/energy-oriented optimization; and Model Early-Exit performs partial DL model inference at Early Exit points, trading accuracy for speed. Among the other families, Edge Caching focuses on reusing previous results of the same inference task for faster response, Input Filtering detects differences between inputs to avoid redundant computation, Multi-Tenancy supports scheduling multiple DL-based tasks in a resource-efficient manner, and Model Selection uses Input-oriented, accuracy-aware optimization.

Currently, MLOps Engineers continuously experiment with different combinations of operators to find an optimal balance in latency and model prediction performance. White-box operators require substantial time for re-training or fine-tuning a model by changing weights and structure. This demands a deep understanding of the internal workings of the model, including its architecture, parameters, and training process. Furthermore, the resulting model may behave unpredictably compared to previously tested versions. In contrast, black-box operators allow quicker adaptation to models without requiring an in-depth understanding of their internal architecture or parameters. These operators apply transformations to pre-trained models and are often favored in scenarios where model transparency is limited, especially in DNN models.

Given black-box Edge AI operators, the challenge becomes: 1) Where (on which tiers) to deploy models in an Edge AI setting; and 2) How to post-process models using operators to make them compatible with those tiers. The combination of a choice of tier and operator forms an Edge AI deployment strategy. While there are many deployment operators and tiers, MLOps Engineers currently rely on trial and error to find the best configuration. Hence, this study initiates a catalog of empirical data to eventually enable recommendation systems that assist MLOps Engineers in deciding the most appropriate deployment strategy for their context.

The main contribution of this study is an in-depth empirical comparison between competing Edge AI deployment strategies to suggest recommendations for deploying DNN models for MLOps Engineers. Specifically, we compare strategies mapping common black-box deployment operators, including the Identity operator (which serves as a baseline with no model transformation), Partitioned, Early Exit, Quantized, and their combinations (Quantized Early Exit [QE] and Quantized Early Exit Partitioned [QEP]) to three common deployment tiers (i.e., Mobile, Edge, and Cloud) and their combinations in an Edge AI environment. Second, for each of the Edge AI deployment strategies, we evaluate the end-to-end (round-trip) latency in an Edge AI setup (Mobile, Edge, and Cloud tiers). Third, we focus on measuring the latency of deployment strategies across a wide range of varying input (i.e., image) sizes using sequential inference requests. Our study analyzes the optimal trade-off in terms of inference latency and accuracy among competing Edge AI deployment strategies. We address the following research questions: \begin{itemize} \item {RQ1: What is the impact of monolithic deployment in terms of inference latency and accuracy across the considered tiers?} \item {RQ2: What is the impact of the Quantized operator in terms of inference latency and accuracy within and across the considered tiers?} \item {RQ3: What is the impact of the Early Exit operator in terms of inference latency and accuracy within and across the considered tiers?} \item {RQ4: What is the impact of the Partitioned operator in terms of inference latency and accuracy across the considered tiers?} \item {RQ5: What is the impact of hybrid operators in terms of inference latency and accuracy within and across the considered tiers?} \item {RQ6: What is the impact of network bandwidth variations on the deployment strategies in terms of inference latency?} \end{itemize} Answering these research questions guides MLOps Engineers and researchers in the AI field to better understand and assess the impact of how and where black-box models are deployed in an Edge AI environment. The results of this paper provide valuable insights for MLOps Engineers debating the most feasible choices for performing inferences in Edge AI contexts.

The assessment in this study simulates an Edge AI deployment architecture for interconnected Mobile, Edge, and Cloud tiers using Docker containers. These containers provide a lightweight and consistent environment for realistic hardware specifications and network conditions. Network conditions, including bandwidths between tiers, are emulated using Linux Traffic Control. AI inference experiments utilize ONNX runtime executors tailored to the hardware limitations of the tiers, processing .onnx models after applying specific operators. Input data consists of larger-sized image samples to evaluate system scalability under computationally demanding scenarios. Accuracy measurements are also performed within each tier using the full validation dataset. This setup enables a comprehensive analysis of inference latency and accuracy across various deployment strategies under realistic conditions.

Our empirical evaluation reveals that black-box deployment strategies significantly impact inference latency and accuracy across Edge AI tiers. We find that hybrid strategies, particularly Quantization + Early Exit on Edge, offer the best latency-accuracy trade-off when a medium accuracy loss is acceptable. When accuracy preservation is paramount, Quantization alone on Edge outperforms other configurations. In resource-constrained mobile environments, Mobile-Edge Partitioning provides preferable latency over full Mobile deployments. Moreover, Cloud deployment becomes effective for small input models even at lower bandwidths ($\leq$ 10 Mbps), whereas larger input models require $\geq$ 50 Mbps for performance parity. Moreover, network bandwidth plays a critical role in shaping optimal deployment strategies.

The rest of this paper is structured as follows. Section~\ref{sec:background} discusses the background of the study. Section~\ref{sec:related_work} presents prior works in this field. Section~\ref{sec:approach} explains the approach, including subjects, experimental setup, metrics for evaluating model performance, motivation and approach for each research question, and data analysis. Section~\ref{sec:results} describes the results for the five research questions. Section~\ref{sec:discussion} discusses the results and compares the results of the RQs. Section~\ref{sec:threats} proposes the threats to the validity of the paper, followed by the conclusion in Section~\ref{sec:conclusion_and_futurework}.

\begin{figure} \includegraphics[width=1\textwidth]{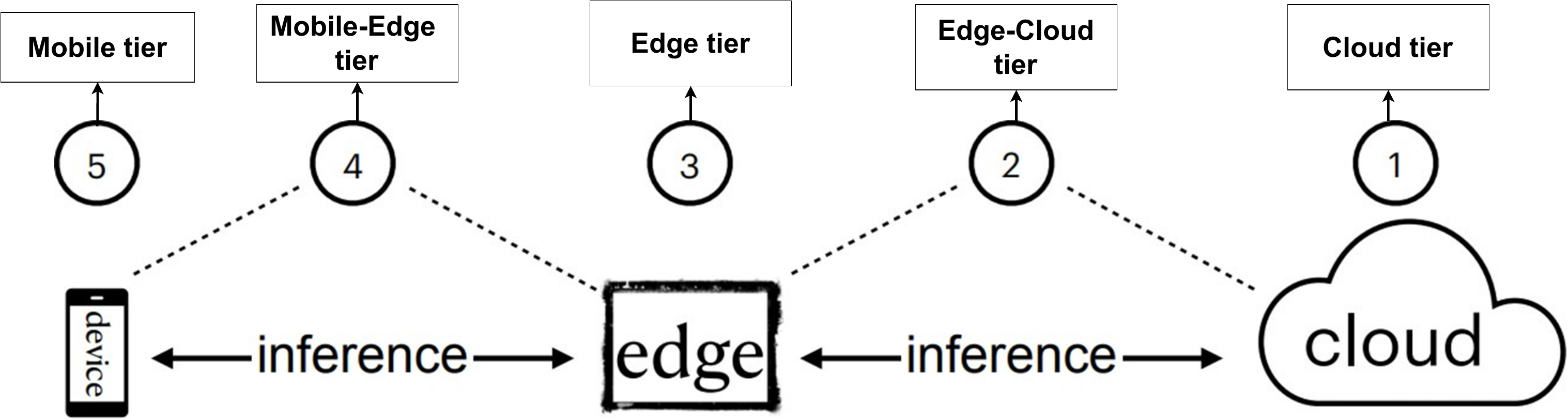} \caption{Graphical overview of Single-tier (Mobile, Edge, Cloud) and Multi-tier (Mobile-Edge, Edge-Cloud) Edge AI deployment Strategies} \label{edge_environment} \end{figure}

\section{Background}
\label{sec:background}
\subsection{Deep Learning Architecture}
\label{sec:deep_learning_architecture}
The architecture of a Deep Learning (DL) model is composed of layers that transform the input data using mathematical operations to produce an output. Layers are organized sequentially to form a Deep Neural Network (DNN) architecture~\cite{lecun2015deep}. The resulting computational graph represents the flow of data and computations through a neural network, i.e., a representation of how the layers are connected and how data moves from one layer to another. Each layer is a node in the graph, and the connections between nodes (Edges) show the data flow. A node encapsulates the entire computation performed by that layer, including all its individual elements. However, the individual elements within a layer, such as neurons, weights, biases, activation functions, and other internal components, are not explicitly represented as separate nodes in the graph. This graph structure allows frameworks to efficiently execute forward and backward passes (propagation) during training and inference processes.

During the training process, the forward pass computes predictions and loss, while the backward pass computes gradients for weight updates to minimize the loss. During the inference process, only the forward pass is used to make predictions or generate output based on input data. Weights are the learnable parameters associated with the layers in the model. These parameters are learned during training to optimize the model's performance. Activations are the intermediate outputs produced by layers as data flows through the computational graph~\cite{alzubaidi2021review}.

\subsection{Monolithic Edge AI deployment}
\label{sec:Monolithic Edge AI deployment}
The traditional deployment of an entire model on Mobile or Cloud (as shown in Figure~\ref{edge_environment}) for inference has some important limitations. For example, deploying the entire AI model on a Mobile device leads to slower computation, whereas on the Cloud it leads to higher transmission latency and potential data privacy threats. Various researchers experiment with ways to deploy models on the Edge (as shown in Figure~\ref{edge_environment}) due to its closer proximity to the Mobile device and faster computational capabilities than a Mobile device~\cite{10.1145/3409977,liu2022bringing}. However, Edge devices/servers are often smaller and less powerful than centralized servers or Cloud resources, which limits their processing power. To address these limitations, there is a need for alternative deployment strategies (i.e., Model Partitioning, Model Compression, Model Early Exiting) that aim to mitigate these challenges.

\subsection{Multi-tier Edge AI Partitioning}
\label{sec:Model Partitioning}
The Multi-tier Edge AI deployment strategy partitions (splits) an AI model between Edge-Cloud and Edge-Mobile tiers, respectively, as shown in Figure~\ref{edge_environment}). This overcomes the limitations of traditional deployment strategies by enhancing data privacy, increasing computation efficiency, reducing memory requirements, improving scalability, and providing a flexible architecture that can be easily adapted to different use cases and deployment scenarios.

To illustrate the applicability of this approach, consider a real-time traffic monitoring and incident detection system within a Mobile-Edge-Cloud infrastructure. Cameras mounted on traffic lights, vehicles, and drones continuously capture high-definition video feeds of a busy intersection. These video streams are transmitted to an Edge server located nearby, which processes the footage in real-time using lightweight Computer Vision models. The Edge server detects key objects such as vehicles, pedestrians, and potential road hazards, enabling immediate actions like adjusting traffic signals or issuing warnings for approaching vehicles when pedestrians are detected.

Simultaneously, the Cloud aggregates data from multiple Edge servers across the city, providing a centralized platform for large-scale analysis. By applying Deep Learning (DL) models, the Cloud identifies traffic patterns, predicts congestion zones, and optimizes traffic management strategies. These insights are then used for long-term planning, such as predicting rush hour congestion or adjusting traffic light timings across multiple intersections. This scenario exemplifies how Multi-tier Edge AI Partitioning can optimize resource usage and enhance system performance by distributing computational tasks appropriately across Mobile, Edge, and Cloud tiers.

The key insight is the observation that most of the strategies used to enable Partitioned model inference across, for example, Mobile and Edge devices, operate on the computation graph underlying modern DL architectures~\cite{Goodfellow-et-al-2016}. Such graphs express the different elements of neural networks, as well as the calculations performed while traversing edges in the graph. By finding the right edges to split, one can implement model Partitioning, effectively obtaining two or more sub-graphs.

Overall, model Partitioning is a powerful strategy for optimizing the deployment of DNN models on resource-constrained devices. This family of Edge AI operators partitions models across the tiers of an Edge AI Environment:

\begin{itemize}
\item Mobile-Edge Partition: The Mobile device receives the input data, performs inference on the first half of the model, and then sends the intermediate output to the Edge device. Afterward, the Edge device performs inference on the second half of the model using the received intermediate output from the Mobile device. Finally, the Edge device sends the final output back to the Mobile device.

\item Mobile-Cloud Partition: The Mobile device receives the input data, performs inference on the first half of the model, and sends the intermediate output to the Cloud device via the Edge device. Then, the Cloud device performs inference on the second half of the model using the received intermediate output from the Edge device. Finally, the Cloud device sends the final output back to the Mobile device via the Edge device.

\item Edge-Cloud Partition: The Mobile device first receives the input data and sends it to the Edge device as it is. Then, the Edge device performs inference on the first half of the model using the received input data and sends the intermediate output to the Cloud device. The Cloud device further performs inference on the second half of the model using the received intermediate output from the Edge device. Finally, the Cloud device sends the final output back to the Edge device, which in turn sends it back to the Mobile device.
\end{itemize}

\subsection{ONNX Run-time for inference}
\label{sec:ONNX and ONNX Run-time}
ONNX (Open Neural Network Exchange)\footnote{https://onnx.ai/} is a standard format built to represent inter-operable AI models that run on a variety of hardware platforms and devices. The core of the ONNX model is the computational graph, which represents the structure and computations of the model. The computational graph consists of nodes that represent individual operations or layers in the model. The nodes take input tensor(s), perform the specified operation, and produce output tensor(s). The nodes have attributes that define the type of operation, input and output tensors, and any parameters or weights associated with the operation.

The ONNX Runtime\footnote{https://github.com/microsoft/onnxruntime} is a high-performance inference Engine for deploying ONNX models to production for real-time AI applications. It is optimized for deployment on various devices (i.e., Mobile, Edge, and Cloud) and hardware platforms (i.e., CPUs, GPUs, and specialized accelerators). We consider ONNX as the subject models' format and the ONNX Run-time as an inference Engine for this study.

ONNX Runtime Execution Providers are a set of plug-ins that enable the execution of ONNX models on a wide range of hardware and software platforms. The ONNX Run-time supported Execution Providers studied and considered as a back-end for hardware acceleration while performing model inference are the CUDA Execution Provider, which uses GPU for computations, and the default CPU Execution Provider, which uses CPU cores for computation.

\subsection{Intel Neural Compressor}
\label{sec:intel_neural_compressor}
Intel Neural Compressor (INC) is an open-source toolkit designed to optimize DL models for better performance on Intel hardware. It provides a comprehensive set of features to enable Quantization, Pruning, and other optimizations, making models run faster and consume less power without significant loss in accuracy. INC tool is specifically designed to optimize neural networks for deployment on Intel hardware, such as CPUs or FPGAs.

\section{Related Work}
\label{sec:related_work}
\begin{table}[htbp]
\begin{adjustwidth}{-1.2cm}{}
    \centering
    \caption{Distinctive Features: Prior Work vs. Our Approaches}
    \begin{tabular}{|l|c|c|c|c|c|c|c|c|c|c|c|c|c|c|c|}
    \toprule
        \textbf{Ref.} & \textbf
        {P} & \textbf{E} & \textbf{Q} & \textbf{OO} & \textbf{BT} & \textbf{SS} & \textbf{M} & \textbf{E} & \textbf{C} & \textbf{ILA} & \textbf{ID} & \textbf{TD} & \textbf{RoI} & \textbf{AoA} &  \textbf{CaO} \\
        \midrule
        
        Our work & \cmark V & \cmark XV & \cmark XV& ~ & \cmark & \cmark & \cmark & \cmark & \cmark & S & \cmark & ~ & \cmark & \cmark & \cmark \\
        
        ~\cite{yang2023adaptive,dong2021joint,mohammed2020distributed} & \cmark & ~ & ~ & ~ & \cmark & \cmark & \cmark & \cmark & ~ & A & \cmark & ~ & \cmark & ~ & ~ \\
                                
        ~\cite{kang2017neurosurgeon} & \cmark & ~ & ~ & ~ & \cmark & ~ & \cmark & ~ & \cmark & S & \cmark & \cmark & ~ & ~ & ~ \\
       
         ~\cite{hu2019dynamic,liang2023dnn} & \cmark & ~ & ~ & ~ & \cmark & ~ & ~ & \cmark & \cmark & A & \cmark & ~ &  \cmark & ~ & ~ \\
              
         ~\cite{jeong2018computation} & \cmark & ~ & ~ & ~ & \cmark & ~ & \cmark & \cmark & ~ & S & \cmark & ~ & ~ & ~ & ~ \\
        
         ~\cite{pagliari2020crime} & \cmark & ~ & ~ & ~ & \cmark & ~ & \cmark & \cmark & \cmark & A & ~ & \cmark & \cmark & ~ & ~ \\
        
        ~\cite{li2018auto}  &   L & ~ &   L & ~ & \cmark & ~ & ~ & \cmark & \cmark & S & \cmark & ~ & ~ & ~ & \cmark \\

        ~\cite{zhou2019distributing} & \cmark & ~ & ~ & ~ & \cmark & ~ & ~ & \cmark & ~ & A & \cmark & ~ & \cmark & ~ & ~ \\ 

        ~\cite{choi2018deep,eshratifar2019bottlenet} & \cmark & ~ & ~ & BI & ~ & ~ & \cmark & ~ & \cmark & S & \cmark & ~ & ~ & \cmark & \cmark  \\

        ~\cite{eshratifar2019jointdnn} & \cmark & ~ & ~ & BI & ~ & ~ & \cmark & ~ & \cmark & S & \cmark & ~ & ~ & ~ & \cmark \\

        ~\cite{cohen2020lightweight} & \cmark & ~ & ~ & BI & \cmark & ~ & ~ & \cmark & \cmark & ~ & \cmark & ~ & ~ & \cmark & \cmark  \\
        
        ~\cite{hu2020fast}  & \cmark & ~ & ~ & BI & ~ & ~ & ~ & \cmark & ~ & A & \cmark & ~ & \cmark & \cmark & \cmark \\
        
        ~\cite{choi2020back,assine2021single} & \cmark & ~ & ~ & BI & ~ & ~ & \cmark & ~ & \cmark & ~ & \cmark & ~ & ~ & \cmark & \cmark  \\
        
        ~\cite{shao2020bottlenet++,jankowski2020joint} & \cmark & ~ & ~ & BI & ~ & ~ & \cmark & \cmark & ~ & ~ & \cmark & ~ & ~ & \cmark & \cmark  \\
        
        ~\cite{lee2021splittable} & \cmark & ~ & ~ & BI & ~ & ~ & ~ & \cmark & \cmark & A & \cmark & ~ & \cmark & \cmark & \cmark \\
        ~\cite{matsubara2022sc2,matsubara2022supervised,matsubara2019distilled,matsubara2020split,yao2020deep} & \cmark & ~ & ~ & BI & ~ & ~ & \cmark & \cmark & ~ & S & \cmark & ~ & ~ & \cmark & \cmark \\
                        
        ~\cite{sbai2021cut} & \cmark & ~ & ~ & BI & ~ & ~ & ~ & \cmark & \cmark & ~ & \cmark & ~ & ~ & \cmark & \cmark \\

         ~\cite{matsubara2022bottlefit} & \cmark & ~ & ~ & BI & ~ & ~ & \cmark & \cmark & ~ & A & \cmark & ~ & \cmark & \cmark & \cmark \\
        
        ~\cite{gormez20222,gormez2022class} & ~ & \cmark & ~ & ~ & \cmark & ~ & ~ & ~ & ~ & ~ & \cmark & ~ & ~ & \cmark & ~ \\
        
        ~\cite{lo2017dynamic} & ~ & \cmark & ~ & ~ & ~ & ~ & ~ & \cmark & \cmark & ~ & \cmark & ~ & ~ & \cmark & ~ \\
        ~\cite{pomponi2021probabilistic,wang2020dual,yang2020resolution,xing2020early,wang2019dynexit} & ~ & \cmark & ~ & ~ & ~ & ~ & ~ & ~ & ~ & ~ & \cmark & ~ & ~ & \cmark & ~\\
                
        ~\cite{soldaini2020cascade,xin2020deebert,elbayad2019depth} & ~ & \cmark & ~ & ~ & ~ & ~ & ~ & ~ & ~ & ~ & ~ & \cmark & ~ & \cmark & ~ \\
                        
        ~\cite{xin2020early}  & ~ & \cmark & ~ & ~ & ~ & ~ & ~ & ~ & ~ & S & ~ & \cmark & \cmark & \cmark & ~\\

        ~\cite{zhou2020bert}  & ~ & \cmark & ~ & ~ & ~ & ~ & ~ & ~ & ~ & S & ~ & \cmark & ~ & \cmark & ~\\
        
        ~\cite{wolczyk2021zero,chiang2021optimal,teerapittayanon2016branchynet} & ~ & \cmark & ~ & ~ & ~ & ~ & ~ & ~ & ~ & S & \cmark & ~ & \cmark & \cmark & ~ \\
                
        ~\cite{neshatpour2019icnn} & ~ & \cmark & ~ & PR & ~ & ~ & ~ & ~ & ~ & S & \cmark & ~ & \cmark & \cmark & \cmark \\
        
        ~\cite{na2023genetic} &  I &  I & ~ & ~ & ~ & \cmark & ~ & \cmark & ~ & A & \cmark & ~ & \cmark & ~ & \cmark \\ 

        ~\cite{teerapittayanon2017distributed} &  I &  I & ~ & ~ & ~ & ~ & \cmark & \cmark & \cmark & ~ & \cmark & ~ & ~ & \cmark & \cmark \\
        
        ~\cite{zeng2019boomerang} &  I &  I & ~ & ~ & ~ & ~ & \cmark & \cmark & ~ & S & \cmark & ~ & ~ & \cmark & \cmark \\
        
         ~\cite{laskaridis2020spinn} &  I &  I & ~ & ~ & ~ & ~ & \cmark & ~ & \cmark & S & \cmark & ~ & \cmark & \cmark & \cmark \\ 

        ~\cite{matsubara2021neural} &  I &  I & ~ & BI & ~ & ~ & \cmark & \cmark & ~ & S & \cmark & ~ & ~ & \cmark & \cmark \\
                
        ~\cite{phuong2019distillation,li2019improved} & ~ & \cmark & ~ & KD & ~ & ~ & ~ & ~ & ~ & ~ & \cmark & ~ & ~ & \cmark & \cmark \\
                
        ~\cite{liu2020fastbert} & ~ & \cmark & ~ & KD & ~ & ~ & ~ & ~ & ~ & ~ & ~ & \cmark & ~ & \cmark & \cmark \\     
        ~\cite{banner2019post,cai2020zeroq,choukroun2019low,fang2020post,garg2021confounding,he2018learning,lee2018quantization,meller2019same,nagel2019data,shomron2021post} & ~ & ~ & \cmark & ~ & \cmark & ~ & ~ & ~ & ~ & ~ & \cmark & ~ & ~ & \cmark & ~ \\   
        
        ~\cite{garg2022dynamic,hubara2020improving,zhao2019improving} & ~ & ~ & \cmark & ~ & \cmark & ~ & ~ & ~ & ~ & ~ & \cmark & \cmark & ~ & \cmark & ~ \\

        ~\cite{li2021brecq} & ~ & ~ & \cmark & ~ & \cmark & ~ & ~ & ~ & ~ & NM & \cmark & ~ & \cmark & \cmark & ~ \\
        
        ~\cite{fan2020training,shen2021once,sakr2022optimal} & ~ & ~ & \cmark & ~ & ~ & ~ & ~ & ~ & ~ & ~ & \cmark & \cmark & ~ & \cmark & ~ \\
        
        ~\cite{tailor2020degree} & ~ & ~ & \cmark & ~ & ~ & ~ & ~ & ~ & ~ & A & \cmark & ~ & \cmark & \cmark & ~ \\

        ~\cite{hawks2021ps} & ~ & ~ & \cmark & PR & ~ & ~ & ~ & ~ & ~ & ~ & \cmark & ~ & ~ & \cmark & \cmark \\

        \bottomrule
\end{tabular}
\begin{tablenotes}
\item[]\textsuperscript{1} P: Partitioning, Q: Quantization, E: Early Exiting, X: Quantized Early Exit, V: Quantized Early Exit Partitioned, L: Combination of Partitioning and Quantization, I: Combination of Partitioning and Early Exiting, OO: Other Operators (BI: Bottleneck Injection, PR: Pruning, KD: Knowledge Distillation), BT: Black-Box Transformations, SS: Simulated Setup (empty cell represents Real or Emulated Hardware Setup), M: Mobile, E: Edge, C: Cloud, ILA: inference Latency Approach (S: sequential, A: Asynchronous, NM: Not Mentioned), ID: Image Data, TD: Textual Data, RoI: Range of Input Sizes (for latency evaluation), AoA: Analysis of accuracy, CaO: Comparison across Operators, Empty cells of Mobile, Edge, and Cloud mean they are not an Edge AI setup
\end{tablenotes}
\label{relatedwork}
\end{adjustwidth}
\end{table}

In previous surveys~\cite{zhou2019edge,wang2020convergence,deng2020edge,murshed2021machine}, eight families of Edge AI model inference operators are discussed: Model Compression (Quantization, Weight Pruning, Knowledge Distillation), Model Partition, Model Early-Exit, Edge Caching, Input Filter, and Model Selection. Moreover, Matsubara et al.~\cite{matsubara2022split} conducted a comprehensive survey of the various approaches for Partitioning, Early Exiting, and their combinations with each other and with other operators (such as Bottleneck Injection, Pruning, and/or Knowledge distillation). As summarized in Table~\ref{relatedwork}, many studies focus on in-depth analysis of individual operators or comparing two different operators. Only one study~\cite{matsubara2021neural} considers three operators, suggesting that there is limited research comparing a larger set of operators within the context of Edge AI. Evaluating three (or more) operators allows for a more comprehensive analysis of their relative performances and trade-offs, which is a focus point of our study.

Among the various kinds of operators, we focus on operators that correspond to model transformations, i.e., modifications in the structure, parameters, or behavior of ML models. This subset of operators can be further categorized into white-box and black-box operators. In previous studies, the white-box operators discussed are Model Pruning and Knowledge Distillation, and the black-box operator discussed is Model Partitioning. Some of the operators like Model Quantization, Model Early Exiting, and Bottleneck Injection can be performed in both black-box and white-box manner. In real-world scenarios, applying white-box operators to ML models requires thorough domain expertise about their internal workings to fine-tune or retrain them, due to which there is a practical need to focus on black-box operators that can optimize models without fine-tuning or retraining. Therefore, we narrowed down our study to the black-box operators (i.e., Model Partition, Model Early Exit, Model Quantization) and their combinations to provide empirical data aimed at understanding how to optimize models robustly and tackle the challenges posed when deploying to heterogeneous Edge AI environments. Among the black-box operators, we excluded the Bottleneck Injection operator as it focuses on intermediate data compression techniques (such as lossless/lossy compression, clipping, etc) and does not inherently perform transformations on the DNN models themselves, which is the goal of our study.

Table~\ref{relatedwork} compares the distinctive features of our study and previous studies. Among the ten studies focusing on Non-Hybrid Partitioning operators, all of them were performed by black-box transformations, which suggests that this operator requires no retraining or fine-tuning. Among the 25 Early Exit studies, two studies were performed in a black-box manner, which shows that it is feasible to perform Early Exiting without retraining or fine-tuning. Among 21 Quantization studies, the majority of them (16) were performed in a black-box manner, indicating that this approach is more commonly used. In terms of hardware setup, four of the previous studies consider simulated setup, indicating that it is feasible to consider this kind of setup for testing the operator's performance. In terms of Mobile, Edge, and Cloud tiers, only two studies consider all three tiers, suggesting that this area of research is relatively less explored. In terms of Mobile, Edge, and Cloud tiers, ILA, and RoI, no study considers sequential inference of a range of inputs (with varying sizes) in an Edge AI setup (Mobile, Edge, and Cloud tiers), which was explored in our study. In terms of input data, the majority of studies (64) consider image data, indicating that this type of data is more commonly used for the mentioned operators.

In Table~\ref{relatedwork}, there are limited studies (only two) that consider all three tiers (i.e., Mobile, Edge, and Cloud) for Edge AI setup in their experiments. Considering all three tiers collectively provides a more holistic view of real-world deployment scenarios with varying computational and network conditions. Therefore, we considered all three tiers to ensure a comprehensive and versatile approach in our Edge AI setup. Four prior studies~\cite{yang2023adaptive,na2023genetic,dong2021joint,mohammed2020distributed} employ a simulated Edge AI setup instead of real hardware, the former is more cost-effective and more accessible than real hardware, while also providing a controlled environment, making it easier to isolate and micro-benchmark the latency performance of individual operators. On the other hand, while simulations can closely approximate the behavior of real hardware, they may not replicate all the nuances and complexities of a real-world environment such as hardware/network variability, power consumption, and real-time constraints. 

Among the studies considering interconnected multi-tier networks, the majority (20) consider sequential inference in comparison to asynchronous inference (i.e., 9) for latency evaluation across the tiers. In sequential inference, the inference tasks proceed in a step-by-step manner across the tiers of the network and are dependent on each other (i.e., the next inference task waits for the completion of the previous inference task). In asynchronous inference, the inference tasks across the tiers are performed concurrently or independently from each other. We considered sequential inference as it allows us to isolate the performance characteristics of individual operators in a controlled environment. In other words, it allows for uncovering a more deterministic impact of input data sizes and network/computational resources of heterogeneous Mobile, Edge, and Cloud tiers on the operators's latency performance, similar to how micro-benchmarks operate. 

The majority of the prior studies focus on image data instead of textual or speech data as input for inference of these black-box operators due to its prevalence in real-time deployment scenarios. Image data is often more complex and less interpretable than text or speech data, requiring more bandwidth for transmission and more storage space than text or speech data typically due to its larger size. It requires more computational resources to process possibly due to their usage in computationally intensive tasks like object detection, image classification, and image segmentation. As a result, this is why CV models are commonly studied in prior work, requiring image data as input for inference. As such, we narrowed down the scope of the data and model in our study to the CV domain.

To conclude, our paper provides a novel empirical study of Edge AI deployment strategies, which are mappings of the black-box operators (i.e., Partitioning, Early Exiting, Quantization), and their combinations, onto Edge AI tiers (i.e., Mobile, Edge, and Cloud), and their combinations, to analyze the optimal trade-off in terms of latency and accuracy in real-time deployment scenarios. The previous studies, as mentioned in Table~\ref{relatedwork}, combine Partitioning with either the Early Exiting~\cite{na2023genetic,teerapittayanon2017distributed,zeng2019boomerang,laskaridis2020spinn,matsubara2021neural} or Quantization operator~\cite{li2018auto}. In our study, we went one step further and analyzed unexplored combinations among these three operators, like Quantized Early Exit and Quantized Early Exit Partitioned. To our knowledge, there is no comprehensive study in previous work (Table~\ref{relatedwork}) on the comparative analysis of these three black-box operators and their specific combinations in the context of Edge AI to decide which operator is optimal in which deployment scenario. Secondly, our study in comparison to previous studies (Table~\ref{relatedwork}), evaluates the end-to-end (round trip) latency of the deployment strategies in an Edge AI setup (Mobile, Edge, and Cloud tiers). The third contribution is our focus on measuring the latency of deployment strategies across a wide range of varying input (i.e., image) sizes using sequential inference requests (which have not been explored in previous studies). This contribution helps in analyzing the impact of input data on the proposed deployment strategies.

Below, we discuss existing work related to the three operator families which we considered in our study.
\subsection{Partitioning}
\label{partioning}

As explained in Section~\ref{sec:Model Partitioning}, the Model Partitioning operator performs black-box transformations splitting a given model into head (1st half Partition) and tail (2nd half Partition) sub-models at a Partition point such that the two sub-models, when feeding the output of the head into the input of the tail, produces the same output as the original model. While in some studies~\cite{choi2018deep,cohen2020lightweight}, the Partitioning point is chosen heuristically, in the majority of studies performing Model Partitioning~\cite{eshratifar2019jointdnn,jeong2018computation,kang2017neurosurgeon,li2018auto,zeng2019boomerang,pagliari2020crime,dong2021joint,yang2023adaptive,mohammed2020distributed,liang2023dnn,hu2019dynamic}, various factors like computational load, network cost, energy consumption, input data sizes and/or privacy risk are evaluated for each of the Partitioning points of the DNN models during deployment across the Edge AI environment to inform the selection. There is no generalized optimal Partition point, as it varies for models with different architectures~\cite{matsubara2022split}. Therefore, in our study, we simplified our approach by considering equal-size (MB) Partitioning to do a fair evaluation of each of the subjects considered in our study.

Many of the CV models (i.e., AlexNet, VGG 11/16/19, DenseNet, ViT, NiN, ResNet 18/34/50/56, GoogLeNet, AgeNet, GenderNet, Inception-v3, BNNs, eBNNs) considered in previous studies~\cite{yang2023adaptive,dong2021joint,mohammed2020distributed,kang2017neurosurgeon,liang2023dnn,hu2019dynamic,jeong2018computation,pagliari2020crime,li2018auto,zhou2019distributing,choi2018deep,eshratifar2019jointdnn,teerapittayanon2017distributed,zeng2019boomerang,laskaridis2020spinn} for model Partitioning based on black-box transformations have weak accuracy performance and model complexity within reach of resource-constrained Mobile tiers. However, in our study, more accurate and complex state-of-the-art CV models are considered (such as Wide ResNet-101, ResNext-101, FCN, and DUC) to analyze the latency vs accuracy trade-off in an Edge AI environment with heterogeneous Mobile, Edge, and Cloud tiers.

The Bottleneck Injection (BI) operators have also been previously studied in combination with Model Partitioning to reduce the transmission, computation, and energy costs across the Mobile, Edge, and Cloud tiers. These introduce artificial bottlenecks to DNN models by compressing the intermediate data, modifying the DNN architecture, and/or both. The Bottleneck Injection techniques that do not involve re-training of DNN models include Intermediate Data Compression using Quantization, Tiling, Clipping, Binarization, Entropy Coding, and Lossy/Lossless Compression, which are analyzed in a few previous studies~\cite{choi2018deep,eshratifar2019jointdnn,cohen2020lightweight}. However, most of the BI operators require extensive re-training of models as they modify the DNN architectures with Auto Encoders~\cite{eshratifar2019bottlenet,hu2020fast,choi2020back,shao2020bottlenet++,lee2021splittable,jankowski2020joint,yao2020deep,assine2021single,sbai2021cut,matsubara2022bottlefit,matsubara2022sc2,matsubara2022supervised}, Head Network Pruning~\cite{jankowski2020joint}, and Head Network Distillation~\cite{matsubara2019distilled,matsubara2020split,matsubara2021neural,assine2021single,sbai2021cut,matsubara2022bottlefit,matsubara2022sc2,matsubara2022supervised}. The mentioned black-box and white-box BI operators may affect the accuracy performance due to intermediate data compression and architectural modifications, respectively.

In our study, we treat the Partitioning operator as a black-box transformation of a model into two sub-models (this number is commonly used in previous studies) that do not modify the input or intermediate representations in the models (i.e., the final output will not change), hence, preserving the accuracy. This is important, because the Bottleneck Injection operators can be costly and time-consuming (especially the ones involving architectural modifications), and would change the known, possibly certified behavior of an existing model. 

\subsection{Partitioning Approach}
\label{sec:rq4_motivation_approach}
We considered the simple, but effective aspect of equal-size (MB) Partitioning to perform a fair evaluation of the subjects. Equal-sized Partitioning allows each Partitioned model to have a similar level of complexity and workload, which can help to balance the computational load across the tiers of the Edge AI Environment, ignoring the heterogeneity of tiers. This fairness can be crucial for assessing the models objectively and avoiding biases introduced by variations in computational capabilities across tiers or certain tiers being underutilized or overloaded, promoting efficient resource utilization. Moreover, due to the variation in the structural properties of the subjects considered for CV tasks, this straightforward approach might lead to a larger amount of intermediate data being transferred to the Mobile, Edge, and Cloud network.

For each of the subjects (i.e., ResNet, ResNext, FCN, DUC) considered in our study, we observed that the size of the sub-graphs of the computational graph gradually increases while traversing from the input to the output node. As we progress through the sub-graphs, the receptive field (sensitive to the region of the input image) of nodes expands, incorporating information from a larger context, which often increases the size of the sub-graphs. This observation is consistent with the typical architecture of DNNs, where lower layers capture low-level features, and higher layers combine these features to form more complex representations. Therefore, choosing the Partition point(s) closer to the end might balance the size between the two sub-models, as shown in the Partitioning examples in the Appendix.

So, with that in mind, Algorithm~\ref{alg:heuristic_Partitioning} inspects the ONNX computational graph of subjects traversing from the end and heuristically selects the Partition point, i.e., node connection(s), that splits the model into two nearly equal-sized sub-models. This algorithm is different from the Early Exit algorithm~\ref{alg:early_exit} because Partitioning focuses on finding the node connection(s) that can Partition the model into equal sizes, which requires manually checking the sizes of Partitioned models in our study (proof-of-concept algorithm). It deviates from the Early Exit approach, in which we find and skip identically structured sub-graphs by analyzing their structures and input/output node dimension in ONNX computational graphs.

The models were partitioned so that the connection(s) used for doing that connect the outputs of the 1st half-Partitioned model to the input of the 2nd Half-Partitioned model. For ResNet and ResNext, the connection as shown in Figure~\ref{approach_partitioning_resnet_resnext}, available at the start of the 9th sub-graph (from the end), was used as the Partitioning point, as it showed the lowest difference in size between the sub-models (i.e, 7-8 MB) (based on line 5 to 16 in algorithm~\ref{alg:heuristic_Partitioning}). These sub-graphs are structured in a way such that one of its branches contains the interconnected nodes and the other branch is a connection (without any interconnected nodes) that eventually merges at the end of the sub-graph. It is not feasible to use the connections within the branches of these sub-graphs for Partitioning as it requires each of these branches to have inter-connected nodes to cut the sub-graph.

The architecture of FCN consists of two main branches that extend from the input node and merge at the last sub-graph located at the end. One of the branches consists of all the heavy-weight sub-graphs and the other branch consists mainly of an elongated connection, i.e., a connection that spans across the network without containing complex computations or heavy-weight sub-graphs (line 22 to line 23 in Algorithm~\ref{alg:heuristic_Partitioning}). These two side branches merge into a sub-graph at the end, which computes the final output. Therefore, the Partitioning of FCN requires cutting the connection within these two branches to maintain the information flow, as shown in Figure~\ref{approach_partitioning_fcn}. The first connection is situated at the start of the forth sub-graph from the end within the heavy-weight side branch, and the other connection (within the lightweight side branch) is situated just before the merge (line 24 in Algorithm~\ref{alg:heuristic_Partitioning}). The variation in FCN's Partitioned model sizes is limited to 3MB (based on line 5 to 16 in Algorithm~\ref{alg:heuristic_Partitioning}). As explained earlier (for ResNet/ResNext), these sub-graphs lack interconnected nodes in one of their branches, making it unfeasible to use their branches' connections for Partitioning.

For DUC, the connection situated at the start of the forth and third sub-graph from the end shows sub-models with 37 MB and 9MB variation, respectively (line 29 in Algorithm~\ref{alg:heuristic_Partitioning}). The forth sub-graph from the end contains inter-connected nodes within both its branches and, therefore, cutting the connection within each of its branches can lead to an even more effective balance in size for DUC's sub-models. Therefore, we explored the connections within the branches of this sub-graph and Partitioned the DUC model at the connections shown in Figure~\ref{approach_partitioning_duc}, which resulted in sub-models with around 1MB size variation (based on line 30 to 40 in Algorithm~\ref{alg:heuristic_Partitioning}). This shows that the challenges associated with equal-size Partitioning might vary for models with different architectures and therefore manual analysis of their computational graphs and Partitioned sub-model sizes is beneficial in such cases.

\begin{algorithm}
\caption{Equal-Size Partitioning of Subject models}
\label{alg:heuristic_Partitioning}
\begin{algorithmic}[1]
\State \textbf{Input:} ONNX computational graph $G_{\text{subject}}$ for each subject model
\State \textbf{Output:} Equal-size Partitioned sub-models
\State Initialize Min Size Difference $\Delta_{\text{min}} \leftarrow \infty$
\State Initialize Partition Point $P \leftarrow $ null

\Function{Partition}{$G_{\text{graph}}$, $\Delta_{\text{min}}$, $P$}
    \State Reverse traverse $G_{\text{graph}}$ starting from the end
     \While{not reached the beginning of $G_{\text{graph}}$}
        \State Use connection at the start of the current sub-graph $S_{\text{current}}$ for Partitioning $G_{\text{graph}}$
        \State Calculate size difference $\Delta$ of Partitioned sub-models
        \If{$\Delta < \Delta_{\text{min}}$}
            \State $\Delta_{\text{min}} \leftarrow \Delta$
            \State $P \leftarrow$ connection at the start of $S_{\text{current}}$
        \EndIf
    \EndWhile
    \State \textbf{return} $P, \Delta_{\text{min}}$
\EndFunction

\Function{PartitionResNe(x)t}{$G_{\text{subject}}$, $\Delta_{\text{min}}$, $P$}
    \State $P, \Delta_{\text{min}} \leftarrow$ Partition($G_{\text{subject}}$, $\Delta_{\text{min}}$, $P$)
    \State \textbf{return} Partitioned sub-models of $G_{\text{subject}}$ using $P$ based on $\Delta_{\text{min}}$
\EndFunction

\Function{PartitionFCN}{$G_{\text{subject}}$, $\Delta_{\text{min}}$, $P$}
    \State Select heavy-weight side branch $G_{\text{heavy-weight}}$
    \State Select light-weight side branch $G_{\text{light-weight}}$
    \State $P \leftarrow$ Use one connection within the $G_{\text{light-weight}}$ just before the merge for Partitioning $G_{\text{subject}}$
    \State $P, \Delta_{\text{min}} \leftarrow$ Partition($G_{\text{heavy-weight}}$, $\Delta_{\text{min}}$, $P$)
    \State \textbf{return} Partitioned sub-models of $G_{\text{subject}}$ using $P$ based on $\Delta_{\text{min}}$
\EndFunction

\Function{PartitionDUC}{$G_{\text{subject}}$, $\Delta_{\text{min}}$, $P$}
    \State $P, \Delta_{\text{min}} \leftarrow$ Partition($G_{\text{subject}}$, $\Delta_{\text{min}}$, $P$)
    \State Reverse traverse $G_{\text{subject}}$ starting from the end
     \While{not reached the beginning of $G_{\text{subject}}$}
        \If{the side branches of the current $S_{\text{current}}$ contain inter-connected nodes}
        \State Select connections within the branches of the $S_{\text{current}}$ for Partitioning
        \State Calculate size difference $\Delta$ of Partitioned sub-models
        \If{$\Delta < \Delta_{\text{min}}$}
        \State $\Delta_{\text{min}} \leftarrow \Delta$
         \State $P \leftarrow$ connections within the $S_{\text{current}}$
        \EndIf
        \EndIf
        \EndWhile
    \State \textbf{return} Partitioned sub-models of $G_{\text{subject}}$ using $P$ based on $\Delta_{\text{min}}$
\EndFunction
\end{algorithmic}
\end{algorithm}

The manual splitting of the subject's Identity models into two equally sized (in MB) sub-models is performed using ONNX Python APIs. We use the extract\_model() function on each of the Identity models to perform the following tasks: 1) Extract 1st half of the Partitioned model by traversing from the input node to the Partition point; 2) Extract 2nd half of the Partitioned model by traversing from the Partition point to the output node. At the Partition point, the output tensor name(s) of the 1st half-Partitioned model is identical to the input tensor name(s) of the 2nd half-Partitioned model. The 1st half-Partitioned sub-model is used for inference in either the Mobile or Edge tier, while the 2nd half-Partitioned sub-model is used for inference in either the Edge or Cloud tier.

\subsection{Early Exiting}
\label{earlyexiting}
Early Exiting is another family of Edge AI deployment operators allowing DNN models to make early predictions without having to wait until the entire computation process (full forward pass) is completed by terminating the execution at Early Exits/Classifiers/Sub-branches~\cite{scardapane2020should}. The benefits of model Early Exiting include faster inference speed, reduced energy consumption, and increased model efficiency at the cost of lower accuracy performance. Implementing Early Exiting on a trained model improves the model's runtime performance, especially if the model was initially designed for high accuracy and not optimized for efficiency.

In the majority of the previous studies, the Early Exits require re-training of the base models either by joint training or separate training. In Joint training, the (Early) Exits are trained simultaneously with a model~\cite{elbayad2019depth,laskaridis2020spinn,lo2017dynamic,pomponi2021probabilistic,soldaini2020cascade,teerapittayanon2016branchynet,teerapittayanon2017distributed,wang2019dynexit,wang2020dual,xin2020early,xing2020early,yang2020resolution,zeng2019boomerang,zhou2020bert} by defining a loss function for each of the classifiers and minimizing the weighted sum of cross-entropy losses per sample. In contrast, in separate training~\cite{liu2020fastbert,matsubara2021neural,xin2020deebert,wolczyk2021zero}, model training is performed in the first stage then the training of the Early Exits is performed, such that the pre-trained model parameters are frozen. There are some studies~\cite{gormez20222,gormez2022class} that perform Early Exiting by comparing the output of an Early Exit with the corresponding class means using Euclidean distance. If the output of an Early Exit is not close enough to a class mean, the execution continues and the same process is performed for the next Early Exit in the DNN model. In other words, the Early Exiting is performed dynamically at inference in a black-box manner.

In our study, we achieve a similar effect of Early Exiting by performing manual and static modifications on the computational graphs of the subject models to short-circuit (skip) the similarly structured graph computations. The motivation behind this approach lies in the flexibility and customization it offers in the ONNX framework to MLOps Engineers. 

\subsection{Early Exit Approach}
\label{sec:rq3_motivation_approach}
In our study, the Early Exit process involves modifying the architecture of the pre-trained subjects to include intermediate outputs and adding the necessary logic to allow the models to Exit Early at an intermediate stage in the neural network, where this stage includes intermediate outputs that can be used for prediction. We create Early Exit models by manually terminating the model early using ONNX python APIs~\footnote{https://github.com/onnx/onnx/blob/main/docs/PythonAPIOverview.md}. Since the Early Exit mechanism damages the accuracy of inference, a relatively slower Early Exit near the end of the DNN will gain better accuracy performance~\cite{teerapittayanon2016branchynet,zeng2019boomerang}. For this reason, we traverse the ONNX computational graphs for all subjects in reverse order (line 4 to 6 in Algorithm~\ref{alg:early_exit}), i.e., starting from the end (output node) to check the sub-graphs having identical structures, then short-circuiting them to create an Early Exit. Here, the sub-graph denotes a branch network of graphical nodes. 

This process takes into account specific considerations related to the model architecture of subjects and the desired trade-off between accuracy and inference speed. When dealing with DNNs, each sub-graph may have specific requirements for the dimensionality of its input and output nodes. If the subsequent sub-graphs have different structures and dimensions, skipping them could lead to incompatible input/output configurations, disrupting the overall flow of the computational process. In our case, the skipping of sub-graphs on each of the subject architectures is based on their identical structure and input/output node dimension. For each of the subjects, we skipped two identical consecutive sub-graphs while traversing from the end (line 10 to 13 in Algorithm~\ref{alg:early_exit}). Skipping more than two was not feasible due to the variation in structure as well as the input/output node dimension of the sub-graphs preceding them (line 14 to 15 in Algorithm~\ref{alg:early_exit}). Skipping only one sub-graph might not yield a significant speedup, as the reduction in model size would be limited to 1.07x to 1.12x of the Identity models (line 15 to 17). However, by skipping two consecutive sub-graphs, the model size can be reduced by 1.15x to 1.27x relative to the Identity models (line 15 to 17), resulting in a more substantial speedup during inference. In the architectures of these subjects (refer to example Figure~\ref{approach_earlyexiting_resnet_resnext} in appendix), there are other sub-graphs (more than two) having identical structures. Skipping them could result in higher latency as they are placed at the early stages of the graph, which are lighter in weight compared to the ones we selected. For them, skipping a higher number of sub-graphs would be required to yield significant model size reduction and faster latency performance. This would result in significant accuracy loss as these sub-graphs are positioned at the earlier stages of the graph, which capture fundamental features or representations of the input data, which are essential for accurate predictions.

Concretely, on the Identity model of each subject, Algorithm~\ref{alg:early_exit} first uses the extract\_model() function to perform the following tasks: 1) Extract an Early Exit sub-graph having an input node at the start and an Early Exit point at the end; 2) Extract a decision sub-graph having the Early Exit point at the start and the output node at the end. The purpose of the decision sub-graph is to make the final prediction based on the information available up to the Early Exit point. Then, we use the merge\_models() function to merge the Early Exit sub-graph with the decision sub-graph. This effectively allows us to either execute the entire model (i.e., the Identity model) or to skip the last two consecutive and identical sub-graphs (in terms of structure and input/output node dimension), essentially Exit the model halfway. We provided graphical illustrations of the Early Exit operation on the ONNX computational graphs of the subjects in Figure~\ref{approach_earlyexiting_resnet_resnext}~\ref{approach_earlyexiting_fcn}~\ref{approach_earlyexiting_duc}.

\begin{algorithm}
\caption{Early Exit based on Sub-Graph Similarity}
\label{alg:early_exit}
\begin{algorithmic}[1]
\State \textbf{Input:} ONNX computational graph for each subject model
\State \textbf{Output:} Modified computational graph with Early Exit
\State Initialize Max Size Difference $\Delta_{\text{max}} \leftarrow 0$
\For{each subject model}
    \State Extract the ONNX computational graph $G_{\text{subject}}$
    \State Reverse traverse $G_{\text{subject}}$ starting from the end
    \State Initialize skip count $count \gets 0$ 
    \While{not reached the beginning of $G_{\text{subject}}$}
        \State Extract current sub-graph $S_{\text{current}}$
        \State Extract preceding sub-graph $S_{\text{preceding}}$
        \If{$S_{\text{current}}$ has identical structure and input/output dimensions as $S_{\text{preceding}}$}
            \State Remove $S_{\text{current}}$ from $G_{\text{subject}}$
            \State Increment $count$ by 1
            \State Store the updated graph as $G_{\text{early\_exit}}$
            \State Compute size difference $\Delta \gets \text{Size}(G_{\text{subject}}) - \text{Size}(G_{\text{early\_exit}})$
            \If{$\Delta > \Delta_{\text{max}}$}
                \State Update $\Delta_{\text{max}} \gets \Delta$
            \EndIf
        \ElsIf{$S_{\text{current}}$ does not have identical structure and input/output dimensions as $S_{\text{previous}}$}
            \State \textbf{Break}     
        \EndIf
    \EndWhile
\EndFor
\end{algorithmic}
\end{algorithm}

\subsection{Quantization}
\label{quantization}

Quantization~\cite{gray1998quantization} is a member of the Model Compression family of Edge AI deployment operators, where the neural network's calculation reduces from full precision (i.e., 32-bit floating point format) to reduced precision (e.g., 16-bit, 8-bit integer point format) to decrease both the computational cost and memory footprint, making inference more scalable on resource-restricted devices~\cite{krishnamoorthi2018quantizing}. As explained in previous literature surveys~\cite{gholami2022survey,wu2020integer}, there are two popular Quantization approaches used in machine learning to optimize DNNs for deployment on hardware with limited numerical precision, such as CPUs, GPUs, and custom accelerators, i.e., QAT, and PTQ.

QAT incorporates Quantization into the training process itself. This is done using techniques such as fake Quantization or simulated Quantization, which simulate the effects of Quantization on the weights and activations during the forward and backward passes of training. The QAT Quantization method involves training data and back-propagation for its fine-tuning process, which requires a full training pipeline, which takes significant extra training time and can be computationally intensive when dealing with large and complex neural networks~\cite{fan2020training,shen2021once,sakr2022optimal,tailor2020degree,hawks2021ps}. In particular, the standard forward/backward passes are executed on a model that uses floating-point precision, and the model parameters are Quantized after each gradient update. By training the model to be more robust to Quantization, QAT results in models that are more accurate after Quantization than PTQ. QAT typically involves two stages: calibration, where the appropriate range of values for the weights and activations is determined, and fine-tuning, where the model is trained with the Quantized weights and activations.

An alternative to the more resource-intensive QAT method is PTQ. PTQ involves reducing the weights and activations of a pre-trained model to lower integer bits, all without the need for fine-tuning (i.e., in a black-box manner)~\cite{banner2019post,cai2020zeroq,choukroun2019low,fang2020post,garg2021confounding,he2018learning,lee2018quantization,meller2019same,nagel2019data,shomron2021post,garg2022dynamic,hubara2020improving,zhao2019improving,li2021brecq}. This can be done using techniques such as static or dynamic PTQ. In dynamic PTQ, the Quantization parameters are dynamically calculated for the weights and activations of a model during runtime and are specific for each inference, while for static Quantization, the Quantization parameters are pre-calculated using a calibration data set and remain static during each inference. The advantage of PTQ lies in its low and often negligible overhead. Unlike QAT, which relies on a substantial amount of labeled training data for retraining, PTQ is advantageous in scenarios where data is limited or unlabeled. Moreover, we observed that 16 studies considered the black-box Quantization (PTQ) and only 5 studies focused on white-box Quantization (QAT) as shown in Table~\ref{relatedwork}. This suggests that black-box Quantization is more common and is therefore considered for evaluation in our study. We keep the pipeline more straightforward by performing Static PTQ Quantization, which just requires representative data (i.e., validation set in our study) to compute statistics such as mean and standard deviation of weights and activations.

\subsection{Quantization Approach}
\label{sec:rq2_motivation_approach}

In our study, we used an INC with ONNX Runtime (CPU) backend to perform static PTQ on the subjects. Static PTQ uses a calibration dataset to determine the Quantization parameters, such as scaling factors and zero points for the model. These parameters are essential for representing the floating-point weights and activations of a model in lower-precision fixed-point formats, which are required for Quantization. The calibration dataset is used to represent a representative subset of the input data that the model is likely to encounter during inference. For each subject, its validation set is passed as the calibration data to capture the data distribution and help identify appropriate Quantization parameters for the model to maintain the desired level of accuracy.

The main advantage of using this technique is that it can lead to a significant reduction in memory requirements and computation time while still maintaining model accuracy. This is especially important in scenarios where the model needs to be deployed on resource-constrained devices, such as Mobile or Edge devices. In static PTQ, the weights and activations of a pre-trained model are Quantized to a fixed precision (i.e., 8-bit integers) by the INC.

\subsection{Hybrid Approach}
\label{sec:rq5_motivation_approach}
The RQ5 evaluates the impact of combining the three deployment operators (i.e., Quantization, Early Exit, and Partitioning). We perform Early Exit on the Quantized models by skipping identically structured sub-graphs from the end of the ONNX computational graphs, which is identical to the approach of Early Exit in Identity models, as explained in RQ3 (Section~\ref{sec:rq3_motivation_approach}). We provide graphical illustrations of the Quantized Early Exit operation on the ONNX computational graphs of the subjects as shown in Figures~\ref{approach_quantized_earlyexit_resnet_resnext}~\ref{approach_quantized_earlyexit_fcn}~\ref{approach_quantized_earlyexit_duc}.

We manually Partition the Quantized Early Exit models into two nearly equal-sized sub-models to generate the Quantized Early Exit Partitioned operator using a similar procedure as for the Partitioning of Identity models in RQ4 (Section~\ref{sec:rq4_motivation_approach}). Here, the second half-Partitioned model contains the Early Exit operation, allowing it to make early predictions. The first-half and second-half Partitioned sub-models are used for inference in Mobile/Edge and Edge/Cloud tier, respectively. We provided graphical illustrations of the Quantized Early Exit Partitioned operator on the ONNX computational graphs of the subjects as shown in Figures~\ref{approach_quantized_earlyexit_partition_resnet_resnext},~\ref{approach_quantized_earlyexit_partition_fcn}, and~\ref{approach_quantized_earlyexit_partition_duc}.

\section{Methodology}
\label{sec:approach}

This section presents the methodology adopted to address the research questions (RQs) introduced earlier. Our approach is grounded in the Goal/Question/Metric (GQM) paradigm~\cite{basili1993applying}, which provides a structured framework for defining measurement goals, formulating relevant research questions, and identifying appropriate metrics to assess the outcomes. By aligning our methodology with the GQM model, we ensure that our evaluation is systematic, goal-driven, and traceable from high-level objectives to concrete measurements.

\subsection{Goal, Research Questions, and Metrics}
The goal of the experiment is to analyze Edge AI deployment strategies to evaluate their impact on latency and accuracy performance from the perspective of MLOps engineers. In this context, a deployment strategy refers to a combination of operators and tiers. The operators include Identity, Quantization, Early Exit, Quantized Early Exit, and Quantized Early Exit Partition. The tiers include Mobile, Edge, Cloud, Mobile-Edge, Edge-Cloud, and Mobile-Cloud.

Table~\ref{tab:gqm-mapping} summarizes how our overarching goal maps to each research question and the associated evaluation metrics.

\begin{table}[h]
\centering
\caption{GQM Mapping of Goals, Research Questions, and Metrics}
\label{tab:gqm-mapping}
\begin{tabular}{|p{3.8cm}|p{5.3cm}|p{5.3cm}|}
\hline
\textbf{Goal} & \textbf{Research Question} & \textbf{Metrics} \\
\hline

Assess the baseline performance of Identity models deployed monolithically across Mobile, Edge, and Cloud tiers. 
& \textbf{RQ1}: What is the impact of Monolithic deployment (Identity models) in terms of latency and accuracy across tiers? 
& Inference latency (ms), Top-1 accuracy (baseline) \\
\hline

Evaluate how Quantization affects inference latency and accuracy within and across deployment tiers. 
& \textbf{RQ2}: What is the impact of Quantization on latency and accuracy within and across tiers? 
& Inference latency (ms), Top-1 accuracy, relative latency reduction (\%) \\
\hline

Analyze the effect of Early Exit on latency and accuracy in different deployment environments. 
& \textbf{RQ3}: What is the impact of Early Exit on latency and accuracy within and across tiers?
& Inference latency, Top-1 accuracy, number of early exits taken \\
\hline

Examine the effectiveness of Partitioning in reducing latency across multi-tier setups. 
& \textbf{RQ4}: What is the impact of Partitioning across tiers?
& Inference latency, Latency difference vs. Monolithic deployments \\
\hline

Assess the impact of hybrid operators combining Quantization, Early Exit, and Partitioning on latency and accuracy. 
& \textbf{RQ5}: What is the impact of hybrid operators (Quantized Early Exit, Quantized Early Exit Partitioning)?
& Inference latency, Accuracy, Trade-off analysis (latency vs. accuracy) \\
\hline

Determine how network bandwidth variation influences the latency of deployment strategies. 
& \textbf{RQ6}: What is the impact of bandwidth variation on deployment strategies?
& Inference latency under 1, 10, 50, 100, 150, and 200 Mbps \\
\hline

\end{tabular}
\end{table}

\subsubsection{RQ1: What is the impact of Monolithic deployment in terms of inference latency and accuracy across the considered tiers?}
\label{sec:rq1_motivation_approach}
\paragraph{Motivation}
This question aims to empirically assess the possible differences in terms of inference performance between the three tiers (i.e., Mobile, Edge, and Cloud) during the Monolithic deployment of Identity models (i.e., models to which the Identity operator is applied, meaning no modification or optimization is performed on the original model). In our study, Monolithic deployment on each tier involves deploying an entire model, along with any necessary pre-processing, post-processing, and inference logic, as a single unit. The goal of this research question is to analyze the impact of factors like computational resources, network bandwidth, and input data on inference latency for the three Monolithic deployment scenarios. The inference accuracy of Identity models is also computed as the baseline for analyzing the performance in later RQs.

\subsubsection{RQ2: What is the impact of the Quantization operator in terms of inference latency and accuracy within and across the considered tiers?}
\paragraph{Motivation}
This question evaluates the impact of the Quantization operator through two key comparisons. First, we compare the inference latency and accuracy of Quantized models against Identity models within the same deployment tier (i.e., Mobile, Edge, and Cloud). Second, we analyze the effect of Quantization across the three Monolithic tiers to examine its behavior in different deployment environments. These comparisons are designed to empirically explore the trade-offs introduced by Quantization, particularly its potential to reduce latency while maintaining acceptable accuracy, thereby informing its suitability for deployment in resource-constrained versus more capable environments.

\subsubsection{RQ3: What is the impact of the Early Exit operator in terms of inference latency and accuracy within and across the considered tiers?}
\paragraph{Motivation}
To evaluate the impact of the Early Exit operator, we conduct two types of comparisons. First, we compare Early Exit models with their Identity counterparts within the same deployment tier (i.e., Mobile, Edge, and Cloud) to assess how introducing early exits affects latency and accuracy under similar resource constraints. Second, we analyze how the Early Exit operator performs across the three Monolithic tiers to understand its behavior in varying deployment environments. These comparisons are motivated by the need to understand whether Early Exit can effectively reduce inference latency while maintaining acceptable accuracy across different system configurations.

\subsubsection{RQ4: What is the impact of the Partitioning operator in terms of inference latency and accuracy across the considered tiers?}
\paragraph{Motivation}
This research question compares the inference latency of multi-tier Partitioning strategies (i.e., Identity models partitioned across Mobile-Edge, Edge-Cloud, and Mobile-Cloud) with that of Monolithic Identity deployments (i.e., complete deployment of Identity models on Mobile, Edge, or Cloud). Additionally, the Partitioned operator in the Mobile-Edge tier is compared with other operators—Identity, Quantization, and Early Exit—in the Edge tier. This tier selection is based on their superior latency performance relative to other combinations (e.g., Edge-Cloud or Mobile-Cloud for Partitioning, and Mobile or Cloud for other operators).

These comparisons aim to empirically assess the effectiveness of the Partitioning operator in reducing inference latency within Edge AI environments and to understand its relative benefits over monolithic and alternative deployment strategies.

\subsubsection{RQ5: What is the impact of hybrid Operators in terms of inference latency and accuracy within and across the considered tiers?}
\paragraph{Motivation}
This question compares the inference latency and accuracy performance of combined optimization strategies involving Quantization, Early Exit, and Partitioning. First, the Quantized Early Exit (QE) hybrid operator is evaluated against Non-Partitioned operators (i.e., Identity, Quantization, and Early Exit) within each Monolithic tier (Mobile, Edge, and Cloud). Second, the Quantized Early Exit Partitioned (QEP) operator is analyzed across Multi-tier setups (i.e., Mobile-Edge, Edge-Cloud, and Mobile-Cloud) and compared with the QE operator deployed monolithically. Additionally, the QEP operator in the Mobile-Edge tier is compared with Non-Hybrid operators (i.e., Identity, Quantization, and Early Exit) in the Edge tier. These tier choices reflect configurations that show the most promising latency benefits based on prior observations (e.g., Mobile-Edge for QEP and Edge for non-hybrid operators).

To assess performance degradation, the accuracy of QE is also compared against Non-Hybrid operators. Identity and QE were selected for Partitioning in this and the previous RQ (RQ4) due to their representing the extremes of model size, enabling clearer latency contrasts across tiers. The decision also aligns with prior studies that explored combining Partitioning with Quantization and Early Exit, as discussed in Section~\ref{sec:related_work}.

These comparisons aim to empirically evaluate how integrating multiple model optimization operators affects inference latency and accuracy in diverse deployment environments.

\subsubsection{RQ6: What is the impact of bandwidth variations on the deployment strategies in terms of inference latency?}
\label{sec:rq6_motivation_approach}
\paragraph{Motivation}
This question aims to empirically assess the impact of network bandwidth variations on the inference latency performance of various deployment strategies, including Identity, Quantization, Early Exit, Partitioning, and hybrid operators, across the Mobile, Edge, and Cloud tiers.

To study this, we focus on experiments involving a single input data sample, where bandwidth is treated as a key independent variable. We evaluate six commonly observed bandwidth levels—1 Mbps, 10 Mbps, 50 Mbps, 100 Mbps, 150 Mbps, and 200 Mbps—across both Mobile-Edge and Edge-Cloud connections. These values were selected to reflect a range of realistic network conditions, from constrained (1 Mbps) to ideal (200 Mbps) scenarios, as typically encountered in mobile and edge computing environments.

This setup allows us to isolate and measure the direct effect of bandwidth on inference latency under each deployment strategy. By varying only the bandwidth while holding other factors constant, we can better understand how network limitations influence latency and which deployment strategies are more resilient to such variations.

\subsection{Subjects Selection}
\label{sec:subjects}
A set of four pre-trained, state-of-the-art models from the ONNX Model Zoo~\footnote{https://github.com/onnx/models} and Pytorch Imagenet models store~\footnote{https://pytorch.org/vision/main/models.html} is used as a suitable and representative sample of subjects for the experiment. Testing too many models can be computationally expensive and time-consuming, especially when involving techniques like Quantization, Early Exit, Partitioning, and their combinations. Previous benchmarking studies~\cite{hampau2022empirical} use a limited set of widely recognized models to evaluate methods. Therefore, using four models aligns with this convention, providing sufficient statistical insights without overburdening the study and strike a balance between variety and manageability. In this study, we focus on computer-vision tasks as they often demand significant computational and network bandwidth resources during deployment in real-world scenarios. To support the generality of the results, we ensure that the models are heterogeneous in terms of architecture, size, scope, and data set. Furthermore, to obtain findings on realistic models, we selected four large and complex image classification and segmentation models, as shown in Table~\ref{Subjectmodels}.

Image classification is a type of machine learning task where the goal is to assign a label or category to an input image. This is achieved by training a model on a dataset of labeled images, then using that model to predict the labels of new, unseen images~\cite{lu2007survey}. Image segmentation is the process of dividing an image into multiple segments or regions. The goal of image segmentation is to simplify and/or change the representation of an image into something more meaningful and easier to analyze. The output of image segmentation is a set of segments that collectively cover the entire image or a set of contours extracted from the image~\cite{minaee2021image}.

The ILSVRC (ImageNet Large Scale Visual Recognition Challenge) dataset is used for evaluating the performance metrics of both the ResNet and ResNext subjects as it is widely used for training and evaluating image classification models. As the network architecture of the ResNet and ResNext subjects is similar, we use different versions of pre-trained weights for ResNet (i.e., IMAGENET1K\_V2) and ResNext (i.e., IMAGENET1K\_V1) from the torchvision package to obtain a better generalization of our results. We use the COCO (Common Objects in Context) and Cityscapes datasets for evaluating the performance metrics of the FCN and DUC subjects, respectively. For the Image Classification subjects, we exported the ResNet and ResNext models from torchvision.models subpackage to ONNX using the torch.onnx.export() function. For the Image Segmentation subjects, we use the models from ONNX Model Zoo.

\begin{table}[htbp]
\caption{Subjects of the experiment}
\resizebox{1.0\textwidth}{!}{
\begin{tabular}{lllll}
 \toprule
\textbf{Model Name} & \textbf{Model Size} & \textbf{Parameters}  & \textbf{Scope}  & \textbf{Dataset} \\ 
\midrule
ResNet~\cite{zagoruyko2017wide} & 484MB & 126.81M & Image Classification & ILSVRC 2012~\cite{ILSVRC15}\\ 
ResNext~\cite{xie2017aggregated} & 319MB & 83.35M & Image Classification & ILSVRC 2012~\cite{ILSVRC15}\\
FCN~\cite{long2015fully} & 199MB & 	51.89M & Image Segmentation & COCO 2017~\cite{lin2014microsoft}\\ 
DUC~\cite{wang2018understanding} & 249MB & 65.14M & Image Segmentation & CityScapes~\cite{Cordts2016Cityscapes}\\
\bottomrule
\end{tabular}
}
\label{Subjectmodels}
\end{table}

\subsection{Experimental Variables}

The experiment is structured using a factorial design comprising multiple independent and controlled factors. Following guidelines from Wohlin et al.~\cite{wohlin2012experimentation}, we distinguish between \textit{independent factors}, which are actively varied during the experiment, and \textit{controlled factors}, which are held constant to isolate the effects of the independent ones. The responses (dependent variables) are latency and accuracy.

\paragraph{Independent Factors} 
The primary independent factors include:
\begin{itemize}
    \item \textbf{Operator Configuration (5 levels)}: Quantization, Early Exit, Partitioning, Quantized Early Exit, and Quantized Early Exit Partitioned.
    \item \textbf{Deployment Tier (6 levels)}: Mobile, Edge, Cloud, Mobile-Edge, Edge-Cloud, and Mobile-Cloud.
    \item \textbf{Network Bandwidth (6 levels)}: 1, 10, 50, 100, 150, and 200 Mbps (varied in experiments with single input samples; fixed in others as described below).
\end{itemize}

In experiments involving multiple input samples, bandwidth is treated as a controlled factor and fixed at 200 Mbps for Mobile-Edge and 1 Mbps for Edge-Cloud tiers to avoid confounding effects.

\paragraph{Controlled Factors (Fixed During the Experiment)}  
Controlled factors are variables that are deliberately kept constant across all experimental conditions to maintain internal validity and ensure that variations in the dependent variables are attributable only to the independent factors. These include:
\begin{itemize}
    \item \textbf{Model Architecture}: The same subject models (e.g., ResNet/ResNeXt, FCN, DUC) are used across all configurations.
    \item \textbf{Input Data and Preprocessing}: Identical datasets and input transformations are applied across all experimental runs.
    \item \textbf{Deployment Tools}: The ONNX runtime and Intel Neural Compressor are consistently used across all experiments.
    \item \textbf{Hardware Configuration}: Hardware specifications (CPU, GPU, RAM) for Mobile, Edge, and Cloud tiers are fixed. CPU for Mobile/Edge: Intel(R) Xeon(R) E7-4870 2.40GHz, CPU for Cloud: Intel(R) Xeon(R) Platinum 8268, GPU: NVIDIA A100 GPU, RAM for Mobile: 4GB, RAM for Edge: 16GB, RAM for Cloud: 64GB RAM.
\end{itemize}

These controlled factors are considered constant \textit{design parameters} in the experiment and not varied across treatment conditions.

\paragraph{Dependent Variables (Responses)}  
The measured outcomes are:
\begin{itemize}
    \item \textbf{Inference Latency}: Time in milliseconds to complete an end-to-end inference request, including preprocessing, model computation, post-processing, and data transmission.
    \item \textbf{Inference Accuracy}: Evaluated using Top-1/Top-5 accuracy for classification models, and mIoU for segmentation models.
\end{itemize}

\subsubsection{Inference Latency} 
\label{sec:inference latency}
We define the inference latency as being based on the sum of pre-processing latency, model computational latency, post-processing latency, and transmission latency. The pre-processing latency refers to the time spent transforming input data to align with the requirements of the model. The model computational latency refers to the time it takes to perform the forward pass of a neural network, which involves feeding an input through the network, applying various mathematical operations, and producing an output. The post-processing latency refers to the time spent refining and interpreting the model's output after the model's forward pass. The transmission latency refers to the time it takes for data to travel from one tier to another in an Edge AI network.

The inference latency is collected via a timer that is started right before the launch of an inference test run and gets stopped when the model returns the output after successful execution. To that extent, we employ the default\_timer from the timeit Python package for measuring inference latency. In the results of the five research questions, we employed the term~\enquote{speedup} to signify the extent by which the median inference latency of a particular operator or tier is faster compared to the median inference latency of another operator or tier. The median inference latency here is the median value among the five repetitions, where each repetition involves running the inference test over 100 input data samples.

\subsubsection{Accuracy} 
\label{sec:accuracy}
For different domain-specific models, the default accuracy metric varies. In our study, the employed metrics for evaluating the accuracy of image classification subjects like ResNet and ResNext are Top1\% and Top5\% accuracy, while the metric used for image segmentation subjects, like FCN and DUC, is mIoU (Mean Intersection Over Union). The definition of the metrics used is as follows:
\begin{itemize}
\item Top5\% and Top1\% accuracy: Top5\% accuracy measures the proportion of validation samples where the true label is among the top 5 predictions with the highest confidence score. Top1\% accuracy is a more strict evaluation metric, as it measures the proportion of validation samples where the model's highest confidence prediction matches the true label. Both Top1\% and Top5\% accuracy are useful metrics in image classification tasks and are often reported together to provide a more comprehensive evaluation of the model's performance. Therefore, in our study, we measure both metrics to gain a better understanding of how well image classification models are performing. 

\item mIoU\%: This is a commonly used metric for evaluating the performance of image segmentation models. It measures the degree of overlap between the predicted segmentation masks and the ground truth masks and provides a measure of how well the model can accurately segment the objects in the image. The mIoU is calculated by first computing the Intersection over Union (IoU) for each class between the predicted mask and the ground truth mask, which is defined as the ratio of the intersection between the predicted and ground truth masks to their union. The IoU ranges from 0 to 1, with higher values indicating better overlap between the predicted and ground truth masks. The mIoU is then calculated as the average of the IoU scores across all classes in the dataset. 

The reason for using mIoU as an evaluation metric for image segmentation models is that it is sensitive to both false positives (areas predicted as belonging to a class when they do not) and false negatives (areas not predicted as belonging to a class when they should). This makes it a valuable metric for evaluating the overall accuracy of a segmentation model and can help identify areas where the model is performing poorly.

In general, the accuracy metrics were calculated by validating each subject model on its specific validation data set, having varying sizes of images to get a more accurate and comprehensive picture of their accuracy performance, as shown in Table~\ref{Subjectmodels}. The ResNet/ResNext subject models have been validated on the ILSVRC 2012 dataset (50k validation samples), while the FCN and DUC subject models were validated on the COCO 2017 dataset (5k validation samples) and CityScapes leftImg8bit dataset (500 validation samples), respectively. We used~\enquote{accuracy} as a common term in the RQ results for the 4 subjects' respective accuracy metrics.

\end{itemize}

\paragraph{Controlled Variables}  
The following variables are held constant across experiments to ensure internal validity and isolate the effects of the independent variables:
\begin{itemize}
    \item \textbf{Model Architecture}: ResNe(x)t, FCN, and DUC.
    \item \textbf{Input Data}: Identical datasets and preprocessing are applied across all configurations.
    \item \textbf{Framework/Tools}: ONNX and Intel Neural Compressor.
    \item \textbf{Hardware Configuration}: CPU, RAM, and/or GPU of the respective Mobile, Edge, and Cloud tiers.
\end{itemize}

All five operator configurations are applied uniformly to the subject models listed in Table~\ref{Subjectmodels}, ensuring a balanced experiment design in which each configuration is tested under equivalent conditions.

To ensure clarity regarding the experimental design across all research questions (RQs), we summarize the independent, dependent, and controlled variables in Table~\ref{summarization_of_the_variables}, titled \textit{Summarization of Variables per Research Question}. This table outlines the factor levels and fixed conditions for each RQ, following guidelines from Wohlin et al.~\cite{wohlin2012experimentation}, and serves as a reference for understanding how the experimental variables are handled throughout the study.

\begin{table}[htbp]
\caption{Summarization of Variables per Research Question}
\label{summarization_of_the_variables}
\resizebox{1.0\textwidth}{!}{
\begin{tabular}{llll}
 \toprule
\textbf{RQ} & \textbf{Independent Variables} & \textbf{Factor Levels~\cite{wohlin2012experimentation}} & \textbf{Controlled Variables} \\ 
\midrule
RQ1 & Operator, Tier, Bandwidth & 1 op × 3 tier & Model Architecture, Input Data, Hardware Configs \\ 
RQ2 & Operator, Tier, Bandwidth & 1 op × 3 tier & Model Architecture, Input Data, Hardware Configs \\
RQ3 & Operator, Tier, Bandwidth & 1 op × 3 tier & Model Architecture, Input Data, Hardware Configs \\ 
RQ4 & Operator, Tier, Bandwidth & 1 op × 3 tier & Model Architecture, Input Data, Hardware Configs \\
RQ5 & Operator, Tier, Bandwidth & 2 op × 3 tier & Model Architecture, Input Data, Hardware Configs \\
RQ6 & Operator, Tier, Bandwidth & 5 op × 3 tier & Model Architecture, Input Data, Hardware Configs \\
\bottomrule
\end{tabular}
}
\end{table}

\subsection{Hypotheses}
To formally test our research questions, we define statistical hypotheses involving the mean inference latency and accuracy across various deployment strategies. Below, we clarify the notation used in the hypothesis formulations:

\begin{itemize}
    \item $\mu_{\text{latency}}(\text{Op}, \text{Tier})$ denotes the \textbf{mean inference latency} when operator $\text{Op}$ is applied in deployment tier $\text{Tier}$.
    \item $\text{accuracy}(\text{Op}, \text{Tier})$ denotes the \textbf{inference accuracy} under the same conditions.
    \item Operator symbols ($\text{Op}$) refer to:
    \begin{itemize}
        \item $\text{Op}_{id}$: identity (baseline operator)
        \item $\text{Op}_q$: quantization
        \item $\text{Op}_e$: early exit
        \item $\text{Op}_p$: partitioning
        \item $\text{Op}_{qe}$: quantized early exit (hybrid)
    \end{itemize}
    \item Deployment tiers ($\text{Tier}$) include:
    \begin{itemize}
        \item $m$, $e$, $c$: monolithic deployments on mobile, edge, and cloud
        \item $me$, $ec$, $mc$: partitioned deployments across tiers
    \end{itemize}
    \item $\text{BW}_i$ represents the available network bandwidth (in Mbps), with $i \in \{1, 10, 50, 100, 150, 200\}$.
\end{itemize}

Each null hypothesis ($H^0$) assumes no statistically significant difference between the compared groups, while the alternative hypothesis ($H^A$) posits that at least one difference exists. The statistical hypotheses corresponding to our research questions are listed below.

\paragraph{RQ1:} Is there a significant difference among the three Monolithic deployments in terms of inference latency?

\begin{align*}
\text{H}_{11}^0 &:\quad \mu_{\text{latency}}(\text{Op}_{id}, \text{Tier}_m) = \mu_{\text{latency}}(\text{Op}_{id}, \text{Tier}_e) = \mu_{\text{latency}}(\text{Op}_{id}, \text{Tier}_c) \\
\text{H}_{11}^A &:\quad \exists\, i,j \in \{m, e, c\} \text{ such that } \mu_{\text{latency}}(\text{Op}_{id}, \text{Tier}_i) \neq \mu_{\text{latency}}(\text{Op}_{id}, \text{Tier}_j)
\end{align*}

\paragraph{RQ1 (Monolithic Latency Comparison):}  
\begin{itemize}
    \item $H_{11}^0$: There is no significant difference in latency across mobile, edge, and cloud for the identity operator.
    \item $H_{11}^A$: At least one pair of tiers has significantly different latency.
\end{itemize}

\paragraph{RQ2:} Does the Quantization operator affect inference latency and accuracy within and across tiers?

\begin{align*}
\text{H}_{21}^0 &:\quad \mu_{\text{latency}}(\text{Op}_q, \text{Tier}_m) = \mu_{\text{latency}}(\text{Op}_q, \text{Tier}_e) = \mu_{\text{latency}}(\text{Op}_q, \text{Tier}_c) \\
\text{H}_{21}^A &:\quad \exists\, i,j \in \{m, e, c\} \text{ such that } \mu_{\text{latency}}(\text{Op}_q, \text{Tier}_i) \neq \mu_{\text{latency}}(\text{Op}_q, \text{Tier}_j) \\[1em]
\\
\text{H}_{22}^0 &:\quad \mu_{\text{latency}}(\text{Op}_q, \text{Tier}_k) = \mu_{\text{latency}}(\text{Op}_{id}, \text{Tier}_k) \quad \forall k \in \{m, e, c\} \\
\text{H}_{22}^A &:\quad \exists\, k \in \{m, e, c\} \text{ such that } \mu_{\text{latency}}(\text{Op}_q, \text{Tier}_k) \neq \mu_{\text{latency}}(\text{Op}_{id}, \text{Tier}_k) \\[1em]
\\
\text{H}_{23}^0 &:\quad \text{accuracy}(\text{Op}_q, \text{Tier}_k) = \text{accuracy}(\text{Op}_{id}, \text{Tier}_k) \quad \forall k \in \{m, e, c\} \\
\text{H}_{23}^A &:\quad \exists\, k \in \{m, e, c\} \text{ such that } \text{accuracy}(\text{Op}_q, \text{Tier}_k) \neq \text{accuracy}(\text{Op}_{id}, \text{Tier}_k)
\end{align*}

\paragraph{RQ2 (Quantization Effects):}  
\begin{itemize}
    \item $H_{21}^0$: Quantization latency is equal across monolithic tiers.
    \item $H_{21}^A$: Latency differs across tiers for quantization.
    \item $H_{22}^0$: Quantization does not change latency compared to identity within each tier.
    \item $H_{22}^A$: Quantization changes latency compared to identity in at least one tier.
    \item $H_{23}^0$: Quantization does not change accuracy compared to identity in any tier.
    \item $H_{23}^A$: Accuracy is affected by quantization in at least one tier.
\end{itemize}

\paragraph{RQ3:} Does the Early Exit operator impact inference latency and accuracy within and across tiers?

\begin{align*}
\text{H}_{31}^0 &:\quad \mu_{\text{latency}}(\text{Op}_e, \text{Tier}_m) = \mu_{\text{latency}}(\text{Op}_e, \text{Tier}_e) = \mu_{\text{latency}}(\text{Op}_e, \text{Tier}_c) \\
\text{H}_{31}^A &:\quad \exists\, i,j \in \{m, e, c\} \text{ such that } \mu_{\text{latency}}(\text{Op}_e, \text{Tier}_i) \neq \mu_{\text{latency}}(\text{Op}_e, \text{Tier}_j) \\[1em]
\\
\text{H}_{32}^0 &:\quad \mu_{\text{latency}}(\text{Op}_e, \text{Tier}_k) = \mu_{\text{latency}}(\text{Op}_{id}, \text{Tier}_k) \quad \forall k \in \{m, e, c\} \\
\text{H}_{32}^A &:\quad \exists\, k \in \{m, e, c\} \text{ such that } \mu_{\text{latency}}(\text{Op}_e, \text{Tier}_k) \neq \mu_{\text{latency}}(\text{Op}_{id}, \text{Tier}_k) \\[1em]
\\
\text{H}_{33}^0 &:\quad \text{accuracy}(\text{Op}_e, \text{Tier}_k) = \text{accuracy}(\text{Op}_{id}, \text{Tier}_k) \quad \forall k \in \{m, e, c\} \\
\text{H}_{33}^A &:\quad \exists\, k \in \{m, e, c\} \text{ such that } \text{accuracy}(\text{Op}_e, \text{Tier}_k) \neq \text{accuracy}(\text{Op}_{id}, \text{Tier}_k)
\end{align*}

\paragraph{RQ3 (Early Exit Effects):}  
\begin{itemize}
    \item $H_{31}^0$: Early exit latency is equal across monolithic tiers.
    \item $H_{31}^A$: Latency differs across tiers for early exit.
    \item $H_{32}^0$: Early exit does not change latency compared to identity in any tier.
    \item $H_{32}^A$: Early exit affects latency in at least one tier.
    \item $H_{33}^0$: Early exit does not affect accuracy compared to identity.
    \item $H_{33}^A$: Accuracy changes due to early exit in at least one tier.
\end{itemize}

\paragraph{RQ4:} Does the Partitioning operator affect inference latency across different deployment tiers?

\begin{align*}
\text{H}_{41}^0 &:\quad \mu_{\text{latency}}(\text{Op}_{id}, \text{Tier}_i) = \mu_{\text{latency}}(\text{Op}_{id}, \text{Tier}_j) \quad \forall i,j \in \{me, ec, mc, m, e, c\} \\
\text{H}_{41}^A &:\quad \exists\, i,j \in \{me, ec, mc, m, e, c\} \text{ such that } \mu_{\text{latency}}(\text{Op}_{id}, \text{Tier}_i) \neq \mu_{\text{latency}}(\text{Op}_{id}, \text{Tier}_j)
\end{align*}

\paragraph{RQ4 (Partitioning Effects):}  
\begin{itemize}
    \item $H_{41}^0$: Latency is equal across all deployment types (monolithic and partitioned).
    \item $H_{41}^A$: At least one pair of deployment types shows a difference in latency.
\end{itemize}

\paragraph{RQ5:} Do hybrid operators (e.g., Quantized Early Exit) impact inference latency and accuracy?

\begin{align*}
\text{H}_{51}^0 &:\quad \mu_{\text{latency}}(\text{Op}_{qe}, \text{Tier}_m) = \mu_{\text{latency}}(\text{Op}_{qe}, \text{Tier}_e) = \mu_{\text{latency}}(\text{Op}_{qe}, \text{Tier}_c) \\
\text{H}_{51}^A &:\quad \exists\, i,j \in \{m, e, c\} \text{ such that } \mu_{\text{latency}}(\text{Op}_{qe}, \text{Tier}_i) \neq \mu_{\text{latency}}(\text{Op}_{qe}, \text{Tier}_j) \\[1em]
\\
\text{H}_{52}^0 &:\quad \mu_{\text{latency}}(\text{Op}_{qe}, \text{Tier}_k) = \mu_{\text{latency}}(\text{Op}_q, \text{Tier}_k) \quad \forall k \in \{m, e, c\} \\
\text{H}_{52}^A &:\quad \exists\, k \in \{m, e, c\} \text{ such that } \mu_{\text{latency}}(\text{Op}_{qe}, \text{Tier}_k) \neq \mu_{\text{latency}}(\text{Op}_q, \text{Tier}_k) \\[1em]
\\
\text{H}_{53}^0 &:\quad \text{accuracy}(\text{Op}_{qe}, \text{Tier}_k) = \text{accuracy}(\text{Op}_q, \text{Tier}_k) \quad \forall k \in \{m, e, c\} \\
\text{H}_{53}^A &:\quad \exists\, k \in \{m, e, c\} \text{ such that } \text{accuracy}(\text{Op}_{qe}, \text{Tier}_k) \neq \text{accuracy}(\text{Op}_q, \text{Tier}_k)
\end{align*}

\paragraph{RQ5 (Hybrid Operator Effects):}  
\begin{itemize}
    \item $H_{51}^0$: Latency for the hybrid operator (quantized early exit) is equal across monolithic tiers.
    \item $H_{51}^A$: Latency differs across tiers for the hybrid operator.
    \item $H_{52}^0$: Latency of the hybrid operator matches that of quantization in each tier.
    \item $H_{52}^A$: Hybrid operator latency differs from quantization in at least one tier.
    \item $H_{53}^0$: Accuracy remains unchanged between hybrid and quantization operator.
    \item $H_{53}^A$: Accuracy differs between hybrid and quantization in at least one tier.
\end{itemize}

\paragraph{RQ6:} What is the impact of bandwidth variations on the deployment strategies in terms of inference latency?

\begin{align*}
\text{H}_{61}^0 &:\quad \mu_{\text{latency}}(\text{Op}_x, \text{Tier}_k, \text{BW}_{i}) = \mu_{\text{latency}}(\text{Op}_x, \text{Tier}_k, \text{BW}_{j}) \quad \forall i,j \in \{1, 10, 50, 100, 150, 200\} \\
\text{H}_{61}^A &:\quad \exists\, i,j \in \{1, 10, 50, 100, 150, 200\} \text{ such that } \mu_{\text{latency}}(\text{Op}_x, \text{Tier}_k, \text{BW}_{i}) \neq \mu_{\text{latency}}(\text{Op}_x, \text{Tier}_k, \text{BW}_{j}) \\[1em]
\\
\text{H}_{62}^0 &:\quad \mu_{\text{latency}}(\text{Op}_x, \text{Tier}_k, \text{BW}_{i}) = \mu_{\text{latency}}(\text{Op}_y, \text{Tier}_k, \text{BW}_{i}) \quad \forall x, y \in \{\text{id}, q, e, p, qe\} \\
\text{H}_{62}^A &:\quad \exists\, x, y \in \{\text{id}, q, e, p, qe\} \text{ such that } \mu_{\text{latency}}(\text{Op}_x, \text{Tier}_k, \text{BW}_{i}) \neq \mu_{\text{latency}}(\text{Op}_y, \text{Tier}_k, \text{BW}_{i}) \\[1em]
\\
\text{H}_{63}^0 &:\quad \mu_{\text{latency}}(\text{Op}_x, \text{Tier}_k, \text{BW}_{i}) \neq \mu_{\text{latency}}(\text{Op}_x, \text{Tier}_l, \text{BW}_{i}) \\
\text{H}_{63}^A &:\quad \exists\, k,l \in \{m, e, c, me, mc, ec\} \text{ such that } \mu_{\text{latency}}(\text{Op}_x, \text{Tier}_k, \text{BW}_{i}) \neq \mu_{\text{latency}}(\text{Op}_x, \text{Tier}_l, \text{BW}_{i})
\end{align*}

\paragraph{RQ6 (Bandwidth Sensitivity):}
\begin{itemize}
    \item $H_{61}^0$: Latency for a given operator and tier is invariant across bandwidth levels.
    \item $H_{61}^A$: Latency changes with bandwidth for at least one operator-tier combination.
    \item $H_{62}^0$: Different operators have similar latency under fixed bandwidth and tier.
    \item $H_{62}^A$: Operator choice affects latency under a given bandwidth-tier setting.
    \item $H_{63}^0$: Latency varies across tiers for a fixed operator and bandwidth.
    \item $H_{63}^A$: There is a tier-level effect on latency at fixed bandwidth and operator.
\end{itemize}

\subsection{Study Design}
\label{sec:study_design}

The experiment follows a nested factorial design~\cite{wohlin2012experimentation} in which some levels of one factor (e.g., deployment operator) are valid only for certain levels of another factor (e.g., deployment tiers). In our case: Partitioning and Quantized Early Exit Partitioned are nested within Multi-tiers (Mobile-Edge, Edge-Cloud, Mobile-Cloud) and Quantization, Early Exit, and Quantized Early Exit are nested within single tiers (Mobile, Edge, Cloud). Thus, specific operator-tier combinations are selected based on feasibility and/or practical relevance. This reduced testing scope makes the design a nested factorial design.

Table~\ref{summarization_of_the_treatment_combinations} summarizes the treatment combinations used in our study. For each research question (RQ), it outlines the selected deployment operators, tiers, and corresponding bandwidth settings, along with the sample size per treatment. This table provides a comprehensive view of how the nested factorial design was instantiated across different deployment strategies and experimental conditions.

\begin{table}[htbp]
\caption{Summarization of the treatment combinations. Legend - M: Mobile, C: Cloud, E: Edge, I: Identity, Q: Quantized, E: Early Exit, P: Partition, QE: Quantized Early Exit, QEP: Quantized Early Exit Partition}
\resizebox{1.0\textwidth}{!}{
\begin{tabular}{lll}
 \toprule
 \textbf{RQ} & \textbf{Treatments (operators * tier * bandwidth)} & \textbf{Sample size per treatment}\\ 
\midrule
RQ1 & Identity (I) * [M,E,C] * [ME: 200 Mbps, EC: 1 Mbps] & 100 range of inputs \\ 
RQ2 & Quantized (Q) * [M,E,C] * [ME: 200 Mbps, EC: 1 Mbps] & 100 range of inputs \\
RQ3 & Early Exit (E) * [M,E,C] * [ME: 200 Mbps, EC: 1 Mbps] & 100 range of inputs \\ 
RQ4 & Partition (P) * [ME,EC,MC] * [ME: 200 Mbps, EC: 1 Mbps] & 100 range of inputs \\
RQ5 & \makecell[l]{Quantized Early Exit (QE) * [M, E, C, ME, EC, MC] * [ME: 200 Mbps, EC: 1 Mbps] \\ Quantized Early Exit Partition (QEP) * [M, E, C, ME, EC, MC] * [ME: 200 Mbps, EC: 1 Mbps]} & 100 range of inputs \\
RQ6 & [I, Q, E, P, QE, QEP] * [M, E, C, ME, EC, MC] * [ME, EC: 1,10,50,100,150,200 Mbps] & 1 single input \\
\bottomrule
\end{tabular}
}
\label{summarization_of_the_treatment_combinations}
\end{table}

From the overall set of eight operator families for Edge AI inference discussed in Section~\ref{sec:introduction}, we narrow down the scope of our study to delve deeper into strategies that can optimize models in a black-box manner for deployment on resource-constrained and network-constrained Edge AI deployment scenarios. In other words, the operators transform the models as-is instead of fine-tuning models (which would invalidate prior model validation results). Among the model optimization operators, we selected three operators based on their representativeness (i.e., capable of addressing different aspects of model optimization, such as improving inference speed, providing better data privacy, optimizing resource usage, and reducing model size) and feasibility (i.e., the implementability of operators in black-box models).

Eventually, we selected one representative operator (i.e., Quantization) out of the three Model Compression operators (i.e., Quantization, Weight Pruning, Knowledge Distillation) as they all focus on a common goal of reducing the size and complexity of black-box models while preserving their accuracy as much as possible. Using the Quantization operator in the ONNX Runtime framework offered by the Intel Neural Compressor tool, Quantization can be performed by using the three widely used techniques discussed earlier, i.e., Static PTQ, Dynamic PTQ, and QAT. Our study used static PTQ as dynamic PTQ requires higher computational overhead during inference than static PTQ. QAT was excluded from the selection as it involved re-training models~\cite{jacob2018quantization}, while we focused on post-processing black-box models.

Model Partitioning was selected as it provides computational load splitting across the tiers (i.e., Mobile, Edge, and Cloud) of an Edge AI environment during distributed inference, enabling more efficient utilization of resources and providing scalable deployment of models. It also aims to provide better data privacy than the Monolithic Edge and Cloud deployments by transmitting intermediate outputs rather than the raw input data across the tiers of the Edge AI Environment. 

The motivation for considering Early Exit as an operator is its aim to save computational resources and reduce the time required to predict by Exit early during the forward pass of the neural network. This would especially be valuable in scenarios where resources are constrained (i.e., Mobile and Edge tiers). For example, drones performing real-time object detection or navigation in constrained environments (e.g., disaster recovery or delivery scenarios) benefit from faster inferences through early Exit to make immediate decisions.

These three operators and their combinations are configured on black-box models for inference, depending on where the transformed (fragments of) black-box models will reside among the three tiers of the Edge AI Environment. This strategic configuration is vital for achieving optimal performance, minimizing latency, and improving the scalability of Edge AI deployment. By aligning these operators with the unique characteristics and constraints of each tier, a more effective and adaptable Edge AI ecosystem can be developed, catering to a wide range of use cases and scenarios.

We analyze the trade-off between two quantitative metrics, i.e., inference latency and inference accuracy. This analysis plays a crucial role in understanding the dynamic interplay between performance and latency within the Edge AI Environment, guiding the selection of optimal strategies based on a deployment Engineer's use cases and requirements. Some use cases might prioritize low latency at the expense of accuracy, while others could emphasize accuracy even if it leads to slightly higher latency~\cite{zhou2019edge}. The empirical data collected from deploying various operators on different tiers provides a quantitative basis for evaluating this trade-off and serves as a foundation for our long-term objective (outside the scope of this paper): the creation of recommendation systems to automatically suggest the most appropriate operators and deployment strategies for specific use cases, aligning with desired latency and accuracy requirements.

In our study, the deployment strategies are a mapping of deployment tiers to one or more deployment operators. The deployment tiers are the physical locations for model deployments, which include three Single-tier (i.e., Mobile, Edge, and Cloud) and three Multi-tier (i.e., Mobile-Edge, Edge-Cloud, Mobile-Cloud) environments. The Single-tiers refer to the deployment of entire models on single computing tiers to achieve Monolithic inference. The Multi-tiers refer to the deployment of Partitioned models across multiple computing tiers to achieve distributed inference. The deployment operators refer to the specific techniques that are applied to modify black-box models for efficient deployment and execution within the Edge AI Environment. They can be categorized into singular operators and hybrid operators. Singular operators are individual optimization techniques that are applied to models independently and hybrid operators are combinations of singular optimization techniques. 

We considered four singular deployment operators, i.e., Identity Operator (no modifications), Quantization Operator, Early Exit Operator, and Partition Operator, and two hybrid operators, i.e., Quantized Early Exit and Quantized Early Exit Partitioned. We limited the hybrid operators to two as developing and evaluating hybrid operators involves combining multiple singular operators, which can increase the complexity of the study. By including a smaller set of hybrid operators, we can perform a more detailed comparative analysis against singular operators.

We perform Partitioning and Early Exit manually to check their feasibility (implementation), since automation across all types of models does not exist thus far. This feasibility analysis and the results of our study will help MLOps Engineers determine whether it is worth investing time/resources for automating this in the future. For instance, the Early Exit criteria (i.e., skipping identically structured sub-graphs) require close analysis of ONNX computational graphs of the subject models, as explained in the Early Exit approach (Section~\ref{sec:rq3_motivation_approach}). The Partitioning criteria (i.e., two equal-sized sub-models), require manually checking the sizes of sub-models, while the connection(s) used for achieving this criterion vary for subjects with varying architectures like ResNe(x)t, FCN, and DUC, as explained in our Partitioning approach (Section~\ref{sec:rq4_motivation_approach}). Given the heterogeneity of the studied models, right now there is no automated tool that considers the mentioned criteria for these two operators across all models. To date, only one paper~\cite{luo2023split}, uses a genetic algorithm to evenly split models into diverse sub-models, but this was only evaluated on the ResNet model of our study. As a consequence, we obtained Partitioning of Identity models, Early Exit of Identity/Quantized models, and Partitioning of Quantized Early Exit models through manual modifications. By executing these modified models in various deployment scenarios, across Mobile, Edge, and Cloud tiers, we then collect empirical data to analyze how each operator, and their combinations, perform under real-world conditions.

Previous studies~\cite{li2018edge,matsubara2022split} indicated that model Partitioning does not affect the inference accuracy, it just sends the same intermediate results remotely instead of within the same tier. As the model's architecture inherently involves sequential processing, the Partitioning aligns well with this logic by ensuring that data flows through each Partitioned model in a manner consistent with the original model's design, hence preserving accuracy. By default, the dimensions (width, height) of the model's input node of three subjects are fixed, i.e., (224, 224) for ResNet/ResNext and (800, 800) for DUC, while for one subject (i.e., FCN), it is dynamic (dependent on the width and height of input data). Therefore, for FCN, the intermediate data dimension/size varies for input data samples having varying widths/heights, while for other subjects, it remains fixed.

\subsection{Experimental Execution}
\label{sec:experimental_setup}
\begin{figure}
  \includegraphics[width=1\textwidth]{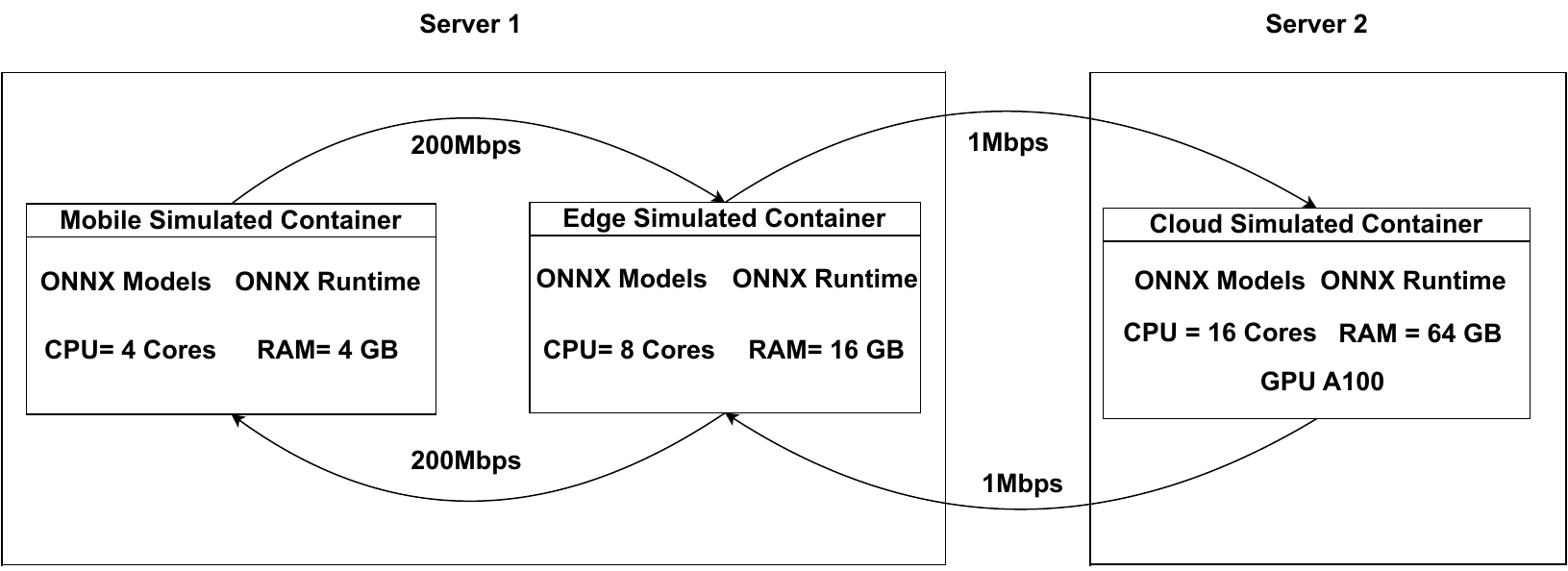}
\caption{Graphical illustration of Experimental Architecture for Edge AI}
\label{experimental_architecture}
\end{figure}

\subsubsection{Experimental Setup}
We simulated an Edge AI deployment architecture for Mobile, Edge, and Cloud tiers interconnected with each other as illustrated in Figure~\ref{experimental_architecture}. This architecture is designed to support AI inference tasks in both a Monolithic and distributed manner for various deployment operators. Based on previous studies~\cite{portabales2020dockemu,portabales2018dockemu}, we used Docker containers to simulate the hardware configurations of actual Mobile, Edge, and Cloud tiers. Docker is an open-source containerization technology ensuring a consistent and easily portable environment (or container)~\cite{docker}. A docker container is a lightweight and portable package that includes all the necessary dependencies, libraries, and configurations to run a software application~\cite{docker}. Docker provides a way to package and distribute applications in a standardized and portable format that can run on any platform, including Cloud, on-premise, and Edge~\cite{pahl2015containerization}.

This study utilized an experimental setup that may not fully generalize to all configuration scenarios. However, most configurations were selected based on prior studies to ensure comparability and validity. Additionally, we experimented with independent variables that we considered the most impactful, such as network bandwidth and the selected models. We further discuss the impact of Docker containers, models, operators, and GPU-specific execution providers in detail in the Discussion and Threats to Validity sections (Section~\ref{sec:discussion}, Section~\ref{sec:threats}). While both the Mobile and Edge tiers are simulated on a single server using Docker containers, resource constraints (e.g., CPU, RAM) and network conditions are carefully emulated to reflect the heterogeneous characteristics of these tiers. This setup enables a realistic evaluation of multi-tier deployment strategies in a controlled and repeatable environment. However, we acknowledge that certain hardware-specific characteristics, such as variations in physical device architectures, are not captured in this setup, which is common in similar simulation studies~\cite{portabales2020dockemu}. In particular, the Docker containers simulating Mobile and Edge tiers were configured with quad-core and octa-core CPUs (Intel(R) Xeon(R) E7-4870 2.40GHz) along with 4GB RAM and 16GB RAM, respectively. These configurations are based on previous studies~\cite{duan2021joint,dimolitsas2023multi}. We considered CPU-based Edge simulation to represent real-world scenarios where Edge devices often do not have dedicated GPUs due to power, size, or cost constraints. The Docker container simulating the Cloud was configured on a different server than the simulated Mobile and Edge containers. The simulated Cloud container contains 16-core CPUs (Intel(R) Xeon(R) Platinum 8268 CPU \@ 2.90GHz), 64GB RAM, and an NVIDIA A100 GPU, these configurations are based on previous studies~\cite{kunas2023optimizing,preuveneers2019towards}. The Cloud runs all inference experiments on its GPU using ONNX Runtime with CUDA Execution Provider. Our simulated setup of Mobile, Edge, and Cloud Docker containers closely mimics real-world hardware configurations as mentioned below:
\begin{itemize}
\item {The mobile Docker container mimics a lightweight Laptop (such as HP Chromebook x360) that has quad (4) core CPUs and 4 GB RAM~\footnote{https://www.hp.com/in-en/shop/hp-chromebook-x360-14a-ca0504tu-678m6pa.html}.}
\item {The edge container mimics a  mini server (resource configuration of a server within a K8S Edge Cluster), which has 8 CPU cores and 16 GB RAM~\cite{dimolitsas2023multi}.} 
\item {The cloud container mimics a virtual machine with 16 cores, 64 GB RAM, and a GPU (Nvidia-A100 as this is the only available GPU in our lab server)~\cite{kunas2023optimizing,preuveneers2019towards}.} 
\end{itemize}
For the three simulated Docker containers, we use the python:3.9-slim image as a base, on top of which we installed the necessary Python packages including a replica of the ONNX Run-time configuration (the out-of-the-box installation of the ONNX Run-time Python package). Here, the simulated Mobile/Edge/Cloud device is a virtual representation of a physical Mobile/Edge/Cloud device created and operated within a software-based simulation environment (i.e., Docker). The Docker simulations provide a flexible and convenient way to configure and customize virtual environments that mimic various hardware specifications, network conditions, and software configurations of real-time deployment scenarios. Moreover, the advantages of cost-effectiveness and controlled testing make Docker simulations an invaluable tool for conducting inference experiments.

The simulated Mobile and Edge containers are interconnected to a common network bridge in Docker to exchange API requests with each other. The Edge container further connects with the external, simulated Cloud container. We configured the Linux Traffic Control utility~\footnote{https://man7.org/linux/man-pages/man8/tc.8.html} inside each configured Docker container for simulating Mobile-Edge and Edge-Cloud network bandwidths. After applying a given combination of operators, we placed the resulting .onnx files for the subject models on the corresponding simulated devices. The Flask Framework handles incoming and outgoing requests across Mobile, Edge, and Cloud devices. Base64 encoding is used while transferring data across the devices as it allows the data to be transmitted in a more reliable and universally readable format. For ResNet/ResNext and FCN/DUC models, the final output that is transmitted across the Edge AI Environment is the predicted label or the segmented image, respectively, which both have smaller sizes than the network bandwidths of the Edge AI Environment.

Similar to earlier work,~\cite{nan2023large,zhang2023effect,suryavansh2019tango,fiandrino2019openleon,andres2018openleon}, an Edge-Cloud bandwidth of 1 Mbps was used for simulating the WAN transmission latency, and the Mobile-Edge bandwidth of 200 Mbps was used for simulating the WLAN (Wireless Local Area Network) transmission latency. The selected bandwidth values aim to represent typical network conditions found in WAN and WLAN environments. WAN connections are prevalent for communicating between Edge and Cloud over large geographical distances and often have lower bandwidth due to factors like network congestion and long-distance transmission. On the other hand, WLAN connections commonly used for Mobile and Edge devices placed in closer proximity to each other, tend to provide higher bandwidth. We also studied across a range of bandwidths, which is discussed in more detailed further below. 

\subsubsection{Operations done in preparation for each experiment}

The entire validation set for running inference experiments is computationally expensive and time-consuming, especially when dealing with resource-constrained and network-constrained scenarios. Therefore, for analyzing the impact of input data on inference latency, we conducted the inference experiments for the subjects using a representative subset of 100 image samples selected from their specific validation sets with a specific criterion: we ensured that these image samples had larger sizes compared to the remaining validation set. Larger-sized images often present greater computational challenges due to increased memory requirements and processing complexity~\cite{krizhevsky2012imagenet}. By selecting larger-sized image samples, the study can assess an upper bound for the inference latency performance and scalability of the models under investigation in resource-constrained and network-constrained scenarios. The recommendation for a minimum sample size of 100 is considered a typical number for the reliability of statistical analysis and to draw meaningful conclusions~\cite{guadagnoli1988relation}.

We conducted six inference latency trials to analyze the effects of varying deployment strategies, data sizes, and network bandwidths on inference latency across different scenarios. To ensure reliable and reproducible results, we divided the trials into two stages: an initial inference experiment and a final inference experiment. In the initial inference experiment, we performed 100 sequential runs on input test samples of varying sizes as a cache warm-up phase. This step stabilized the cache memory to mimic real-world continuous usage, where frequently accessed data populates the cache, as opposed to operating from a cold start. After the warm-up phase, we carried out the final inference experiment, which involved 500 sequential runs achieved through five repetitions on the same 100 input samples. These repetitions enhanced statistical significance and captured variability typical of real-world repetitive tasks. We logged the inference latency for each run in a text file and restarted the Mobile, Edge, and Cloud Docker containers after each experiment. A 20-second delay ensured consistent and isolated environments, further strengthening the reliability of our results.


To evaluate system performance under varying computational loads, we included models with different input sizes in our experiments. This consideration is critical for high-performance applications. Input size ranges across deployment strategies were as follows: ResNet/ResNeXt models ranged from 8 to 60 MB, DUC models from 19 to 22 MB, and FCN models from 2 to 5 MB.

Recognizing the correlation between model input size and network bandwidth and their impact on inference latency, we assessed the effect of Mobile-Edge and Edge-Cloud bandwidth variations. Based on commonly adopted practices in prior studies~\cite{alqarni2023odm,cui2024latency,nan2023large,zhang2023effect,zhuang2024decc}, we selected bandwidth values of 1, 10, 50, 100, 150, and 200 Mbps. To isolate the effect of network bandwidth and control for input size, we used the largest input sample in these experiments.

\subsubsection{Measurement Procedure and Tools}
\begin{figure}
  \includegraphics[width=1\textwidth]{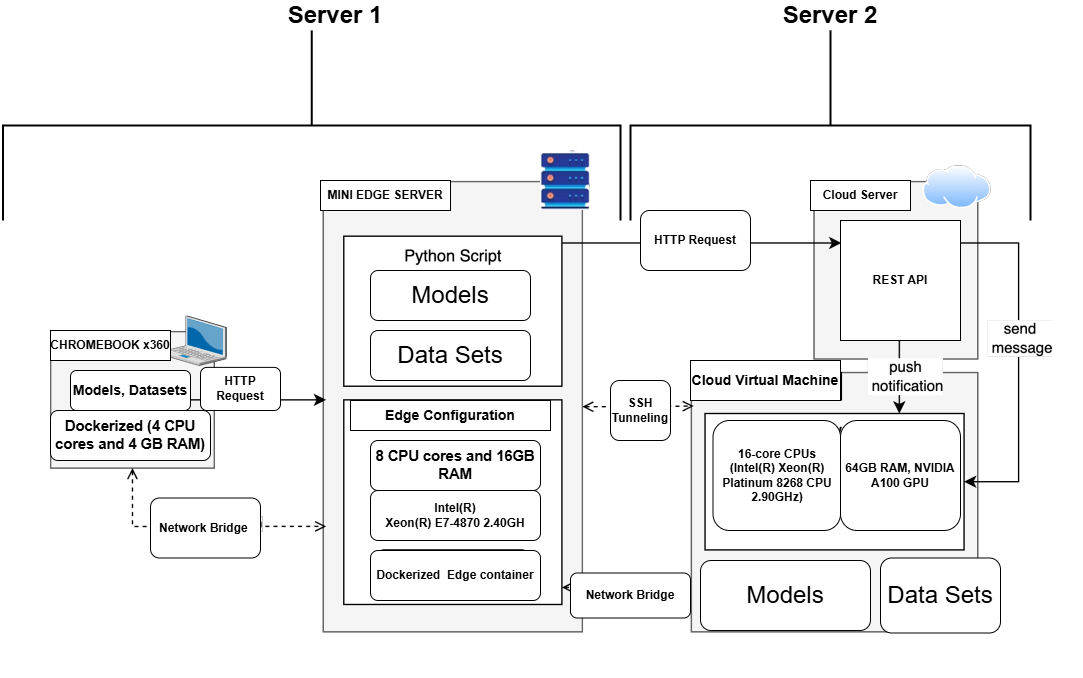}
\caption{Graphical illustration of Measurement Infrastructure}
\label{measurement_infrastructure}
\end{figure}

As shown in Figure~\ref{measurement_infrastructure}, our measurement infrastructure consists of 2 servers, i.e., We orchestrated the latency experiments as follows: Server 1 acted as the orchestrator, managing communication with the Cloud container on Server 2 (via the Mobile and Edge containers on Server 1), the Edge container on Server 1 (via the Mobile container on Server 2), and the Mobile container. Server 1 initiated the experiments and recorded raw latency data, storing the results in text files. We began each experiment with round-trip latency tests for different deployment strategies. To ensure consistency, we restarted the containers between subsequent deployment strategies and waited for 20 seconds to allow software reinitialization. During latency measurement, Server 1 sent input data to the Mobile container, which either processed the inference locally or offloaded computations to the Edge and Cloud containers. Once the containers completed the inference, they returned the output (predictions) to Server 1, completing the round trip.

We conducted inference latency experiments across a comprehensive set of configurations: twelve combinations of $<$Identity models(4), Monolithic tiers(3)$>$, twelve combinations of $<$Quantized models(4), Monolithic tiers(3)$>$, twelve combinations of $<$Early Exit models(4), Monolithic tiers(3)$>$, twelve combinations of $<$Partitioned models(4), Multi-tier setups(3)$>$, and twenty-four combinations of $<$Quantized Early Exit models(4), Deployment tiers(6)$>$. We evaluated model accuracy for Identity, Quantized, Early Exit, and Quantized Early Exit operators using their respective validation datasets, as shown in Table~\ref{accuracy}, and compared their accuracies.

We calculated the accuracy of the operators within each container using the complete validation dataset, as outlined in Figure~\ref{measurement_infrastructure}. For Identity, Quantization, Early Exit, and Quantized Early Exit operators, we independently evaluated accuracy within CPU-based Docker environments (Mobile, Edge) and a GPU-based Docker environment (Cloud) to assess the impact of hardware on performance. This setup provided a holistic evaluation of model generalizability across diverse platforms. Both Mobile and Edge environments utilized the same Runtime Execution Provider (CPU) for inference and shared an Intel(R) Xeon(R) E7-4870 processor (2.40GHz). However, they differed in CPU and memory configurations, allowing us to perform a nuanced analysis of performance across varying hardware setups.

\begin{table}[htbp]
\centering
\caption{Design of KW and Conover statistical tests for inference latency comparison across the tier(T) dimension, i.e., Mobile [M], Edge [E], Cloud [C], Mobile-Edge [ME], Edge-Cloud [EC], Mobile-Cloud [MC] for RQ1, RQ2, RQ3, RQ4, and RQ5.}
  \begin{minipage}[t]{0.45\textwidth}
  \centering
  \begin{tabular}{|c|c|c|c|c|c|c|}
\hline
\backslashbox{T}{O}    &     I     &      Q       &       E       &       P       &      QE  &  QEP\\
\hline
\midrule
M                                &     X    &              &             &          &      &   \\
\hline
E                                &     X    &              &             &          &     &    \\
\hline
C                                &     X    &              &             &          &      &   \\
\hline
ME                                &          &              &              &        &       &   \\
\hline
EC                                &          &              &             &         &       &    \\
\hline
MC                                &           &              &            &         &      &    \\
\hline
\end{tabular}
\captionsetup{justification=centering}
\subcaption{RQ1(X)}
\end{minipage}
\hfill
\begin{minipage}[t]{0.45\textwidth}
  \centering
  \begin{tabular}{|c|c|c|c|c|c|c|}
\hline
\backslashbox{T}{O}    &     I     &      Q       &       E       &       P      &      QE  &  QEP\\
\hline
\midrule
M                                &           &      X        &             &       &         &  \\
\hline
E                                &          &       X      &             &        &        &   \\
\hline
C                                &         &        X      &             &         &       &   \\
\hline
ME                                &          &               &             &        &      &   \\
\hline
EC                                &          &              &             &          &      &   \\
\hline
MC                                &           &              &            &           &      &   \\
\hline
\end{tabular}
\captionsetup{justification=centering}
\subcaption{RQ2(X)}
\end{minipage}
\vspace{0.5cm} 
\begin{minipage}[t]{0.45\textwidth}
  \centering
  \begin{tabular}{|c|c|c|c|c|c|c|}
\hline
\backslashbox{T}{O}    &     I     &      Q       &       E       &       P      &      QE  &  QEP\\
\hline
\midrule
M                                &           &              &      X       &       &         &  \\
\hline
E                                &          &              &       X      &        &        &   \\
\hline
C                                &         &              &        X     &         &       &   \\
\hline
ME                                &          &               &             &        &      &   \\
\hline
EC                                &          &              &             &          &      &   \\
\hline
MC                                &           &              &            &           &      &   \\
\hline
\end{tabular}
\captionsetup{justification=centering}
\subcaption{RQ3(X)}
\end{minipage}
\hfill
\begin{minipage}[t]{0.45\textwidth}
  \centering
  \begin{tabular}{|c|c|c|c|c|c|c|}
\hline
\backslashbox{T}{O}    &     I     &      Q       &       E       &       P       &      QE  &  QEP\\
\hline
\midrule
M                                &     X    &              &             &          &      &   \\
\hline
E                                &     X    &              &             &          &     &    \\
\hline
C                                &     X    &              &             &          &      &   \\
\hline
ME                                &          &              &              &      X  &       &   \\
\hline
EC                                &          &              &             &       X  &       &    \\
\hline
MC                                &           &              &            &       X  &      &    \\
\hline
\end{tabular}
\captionsetup{justification=centering}
\subcaption{RQ4(Y)}
\end{minipage}
\vspace{0.5cm}
\begin{minipage}[t]{0.45\textwidth}
  \centering
  \begin{tabular}{|c|c|c|c|c|c|c|}
\hline
\backslashbox{T}{O}    &     I     &      Q       &       E       &       P      &      QE  &  QEP\\
\hline
\midrule
M                                &           &              &             &       &     X,Y   &  \\
\hline
E                                &           &              &             &        &     X,Y  &   \\
\hline
C                                &           &              &             &         &    X,Y  &   \\
\hline
ME                                &          &               &             &        &      &  Y \\
\hline
EC                                &          &              &             &          &      &  Y \\
\hline
MC                                &           &              &            &           &      &  Y \\
\hline
\end{tabular}
\captionsetup{justification=centering}
\subcaption{RQ5(X) RQ5(Y)}
\end{minipage}
\label{approach_tier_dimention}
\end{table}

\begin{table}[htbp]
\centering
\caption{Design of KW and Conover test statistical tests for inference latency comparison across the Operator(O) dimension, i.e., Identity (I), Quantized (Q), Early Exit (E), Partitioned (P), Quantized Early Exit (QE), and Quantized Early Exit Partitioned (QEP) for comparison across 6 operators (RQ2, RQ3, RQ4, RQ5)}
  \begin{minipage}[t]{0.45\textwidth}
  \centering
  \begin{tabular}{|c|c|c|c|c|c|c|}
\hline
\backslashbox{T}{O}    &     I     &      Q       &       E       &       P       &       QE       &       QEP       \\
\hline
\midrule
M                       &     X     &       X      &     X       &               &      X        &                \\
\hline
E                       &           &              &             &                &               &                \\
\hline
C                       &           &              &             &                &               &                \\
\hline
ME                       &           &              &             &       X         &               &       X         \\
\hline
\end{tabular}
\captionsetup{justification=centering}
\subcaption{Comparison across 6 operators (RQ2, RQ3, RQ4, RQ5)}
\end{minipage}
\hfill
\begin{minipage}[t]{0.45\textwidth}
  \centering
  \begin{tabular}{|c|c|c|c|c|c|c|}
\hline
\backslashbox{T}{O}    &     I     &      Q       &       E       &       P       &       QE       &       QEP       \\
\hline
\midrule
M                       &          &             &            &               &              &                \\
\hline
E                       &     X    &     X        &      X       &                &     X          &                \\
\hline
C                       &           &              &             &                &               &                \\
\hline
ME                       &           &              &             &       X         &               &       X         \\
\hline
\end{tabular}
\captionsetup{justification=centering}
\subcaption{Comparison across 6 operators (RQ2, RQ3, RQ4, RQ5)}
\end{minipage}
\vspace{0.5cm} 

\begin{minipage}[t]{0.45\textwidth}
  \centering
   \begin{tabular}{|c|c|c|c|c|c|c|}
\hline
\backslashbox{T}{O}    &     I     &      Q       &       E       &       P       &       QE       &       QEP       \\
\hline
\midrule
M                       &          &             &            &               &              &                \\
\hline
E                       &         &             &             &                &               &                \\
\hline
C                       &     X      &       X      &      X       &                &     X          &                \\
\hline
ME                       &           &              &             &       X         &               &       X         \\
\hline
\end{tabular}
\captionsetup{justification=centering}
\subcaption{Comparison across 6 operators (RQ2, RQ3, RQ4, RQ5)}
\end{minipage}
\label{approach_operator_dimention}
\end{table}

\subsection {Data Analysis}
\label{sec:data analysis}
The inference latency experiments' results are analyzed using various statistical methods. Firstly, we use the Shapiro-Wilks test and Q-Q plot for each deployment strategy to assess the normality of the inference latency distribution and determine if parametric or non-parametric tests are appropriate for testing the hypotheses. After observing from the Shapiro-Wilks test and Q-Q plot that the data does not conform to a normal distribution, we employ the Kruskal-Wallis (KW) test~\cite{ostertagova2014methodology} to compare the inference latency distributions of independent groups (i.e., deployment strategies) based on two 
different dimensions (i.e., operator and tier dimension) and determine if there exists a significant difference among at least two of the independent groups (hypothesis testing). Further, the Conover test, a non-parametric post-hoc test, was used to perform pairwise comparisons of different deployment strategies across the tier and Operator Dimensions. This test was chosen to identify significant differences between strategies after a significant result was observed in the initial Kruskal-Wallis test. The Conover test's robustness to non-normal data ensures reliable comparisons of deployment strategies in scenarios where parametric assumptions may not hold. The design approach for the KW and Conover statistical tests across the tier and operator Dimension is illustrated in Table~\ref{approach_tier_dimention} and Table~\ref{approach_operator_dimention}, respectively. 

\subsubsection{Tier Dimension}
As shown in Table~\ref{approach_tier_dimention}, to investigate if there is a statistically significant difference between the three Monolithic deployment tiers (Single-tiers) in terms of inference latency of any of the Non-Partitioned operators, we perform a KW test ($\alpha$ = 0.05) for each of the four Identity models in RQ1, four Quantized models in RQ2, four Early Exit models in RQ3, and four Quantized Early Exit models in RQ5 by comparing their latency across the three Single-tiers (Mobile, Edge, Cloud). These four variants of Identity, Quantized, Early Exit, and Quantized Early Exit models indicate for each of the subjects (i.e., ResNet, ResNext, FCN, and DUC). Moreover, to investigate if there is a statistically significant difference between the Monolithic deployment strategies and Multi-tier Partitioning strategies, for each subject, we conduct a KW test ($\alpha$ = 0.05) across the Identity and Partitioned models in RQ4. We also perform this test across the Quantized Early Exit and Quantized Early Exit Partitioned models in RQ5, on their inference latency performance when deployed in the three Single-tier and three Multi-tier environments, respectively. 

After observing significant differences (KW Test: p-value $<$ 0.05), we further employ the Conover post-hoc test~\cite{conover1979multiple}. For the pairwise comparisons having significant differences (Conover test: adjusted p-value $<$ 0.05), we evaluate Cliff's delta effect size~\cite{cliff1993dominance} to assess the inference latency ranking of tiers based on the direction and magnitude of their difference in the corresponding RQs. 

\subsubsection{Operator Dimension}
For each subject, to investigate if there is a statistically significant difference between the six operators (Identity, Quantized, Early Exit, Partitioned, Quantized Early Exit, Quantized Early Exit Partitioned), we perform the KW tests shown in Table~\ref{approach_operator_dimention}. If there is a significant difference (KW Test: p-value $<$ 0.05), we further employ post-hoc Conover tests~\cite{conover1979multiple}. For the pairwise comparisons having significant differences (Conover test: adjusted p-value $<$ 0.05), we used Cliff's delta effect size~\cite{cliff1993dominance} to analyze how operators' inference latency ranks by looking at how much they differ (magnitude) and which way (sign) they differ in the corresponding RQs.

To evaluate the accuracy comparisons of the operators, we use the Wilcoxon Signed Rank Test~\cite{dwivedi2017analysis} to determine if there exists a significant difference for the Identity vs Quantized models in RQ2, Identity vs Early Exit models in RQ3, Quantized vs Quantized Early Exit models in RQ4, and Early Exit vs Quantized Early Exit models in RQ5. This test analyzes the differences between two paired groups, i.e., the accuracy measurements under two different operators for the same subjects (i.e., ResNet, ResNext, FCN, DUC) and environments (i.e., Mobile, Edge, Cloud). Each group has 18 samples of accuracy measurements, i.e., six accuracy metric values ([ResNet, ResNext] x [Top 1\%, Top 5\%] + [FCN, DUC] x [mIOU\%]) x 3 environments. In other words, in a particular operator's group, we are concatenating the accuracy metric(s), all of which are percentage values, of all four subjects, then comparing corresponding accuracy metrics using paired statistical tests.

To avoid false discoveries, Bonferroni Correction~\cite{holm1979simple} is applied to the p-value for each Wilcoxon test comparing two operators by considering a p-value less than or equal to 0.0125 as statistically significant. The adjusted significance level of 0.0125 is derived by dividing the conventional significance level ($\alpha$ = 0.05) by the number of multiple comparisons (four in our case, as each operator is compared four times). After this correction, if a significant difference is observed between the two operators, we utilize Cliff's Delta effect size~\cite{cliff1993dominance} to measure the magnitude/sign of their difference in corresponding RQs. 

On the other hand, for the Shapiro-Wilks, Kruskal-Wallis, and Posthoc Conover tests, we compare the obtained p-value with the significance level of alpha = 0.05. According to~\cite{hess2004robust}, we interpret effect size as negligible (d $<$ 0.147), small (0.147 $\le$ d $<$ 0.33), medium (0.33 $\le$ d $<$ 0.474), or large (d $\ge$ 0.474). Negative values for d imply that, in general, samples from the distribution on the left member of the pair had lower values. These findings, in combination with the box plots displaying the distributions, will provide us with an additional understanding of the extent to which one deployment strategy differs from another. In particular, we analyze the median in the box plots to evaluate how each deployment strategy's performance compares to the others.

\section{Results}
\label{sec:results}

\subsection{What is the impact of monolithic deployment in terms of inference latency and accuracy across the considered tiers? (RQ1)}
\label{sec:rq1}
\subsubsection{Data Exploration}
\textbf{For the Identity models having large input data sizes (ResNet, ResNext, and DUC), the Edge tier shows the lowest median inference latency of 5.64, 3.96, and 23.64 seconds, respectively.}
\begin{figure}[t]
\centering
\includegraphics[width=1\textwidth]{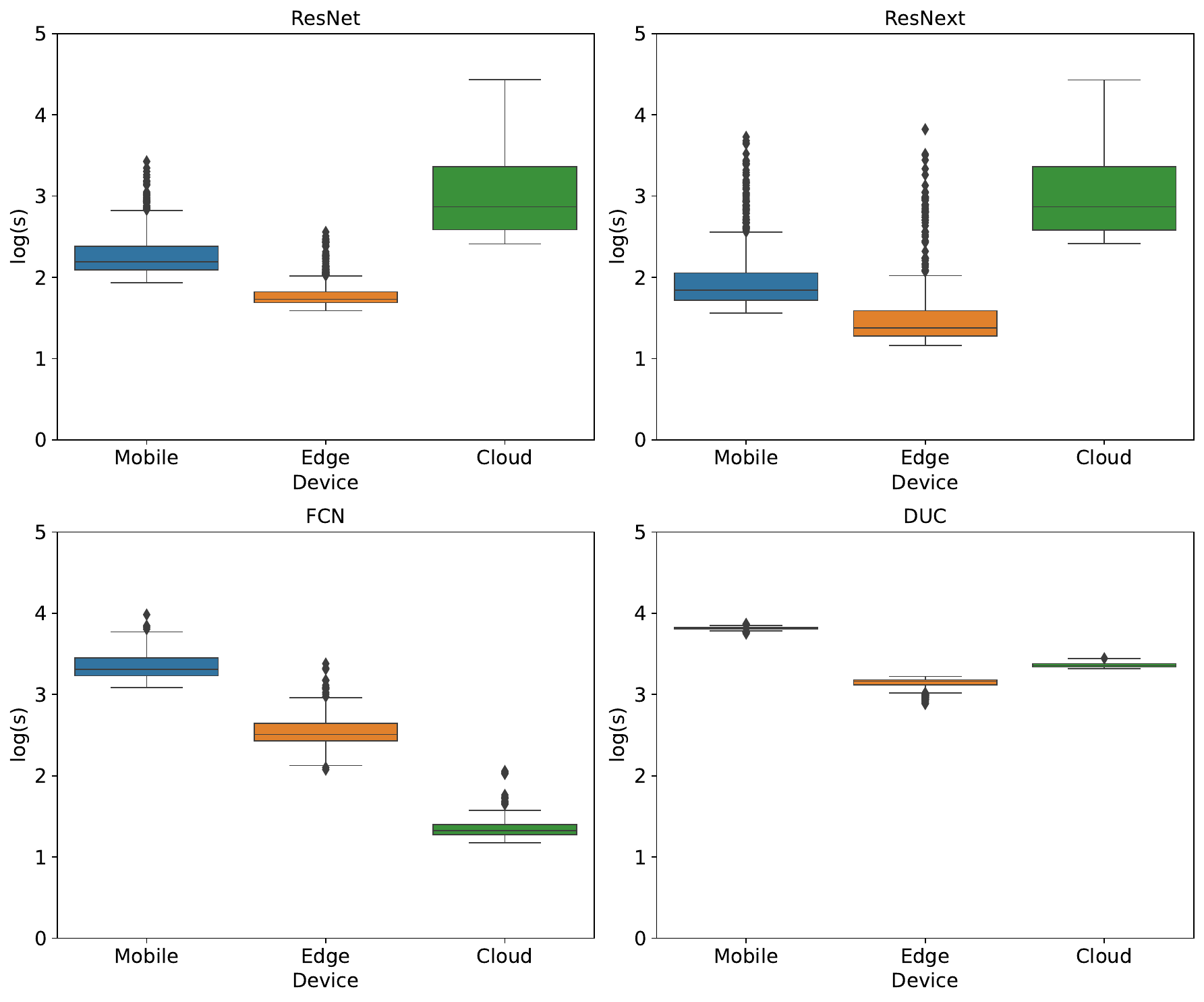}
\caption{Box plots of the measures collected for inference latency (seconds) from Mobile, Edge, and Cloud tiers for Identity versions of Subjects}
\label{rq1_graph}
\end{figure}

The box plots in Fig.~\ref{rq1_graph} represent the inference latency distribution in the logarithmic scale of the subject's Identity models across the Mobile, Edge, and Cloud tiers. The Edge tier shows a 3.12x, 4.43x, and 1.21x lower median inference latency (Table~\ref{rq1_descriptive_statistics}) compared to the Cloud tier for ResNet, ResNext, and DUC Identity models, respectively, with a large effect size (Table~\ref{rq1_Identity_statistical_results}). This is because of the Edge's higher network bandwidth capacity (200 Mbps) compared to Cloud (1 Mbps), which allows faster transmission of large input data samples for ResNet/ResNext (8 to 60 Mb) and DUC (19-22 Mb) during Edge deployment.

In contrast, the FCN Identity model shows the lowest median inference latency of 3.76 seconds in the Cloud tier, while the other 3 Identity models show the highest median inference latencies of 17.61, 17.61, and 28.61 seconds, respectively, in the Cloud tier (Table~\ref{rq1_descriptive_statistics}). This is because the input data samples for FCN are much smaller in size (2-5 Mb), which allows faster data transmission even in the network-constrained Cloud tier. Moreover, the Interquartile range (IQR) of the ResNet and ResNext is wider compared to FCN and DUC in the Cloud, indicating a more significant variability around the median. This is because the input data size range of ResNet and ResNext is around 17x higher compared to FCN and DUC, so a higher spread of inference latency values is observed for ResNet and ResNext. This behavior indicates that the input data sizes and network bandwidths play a critical role in the end-to-end inference latency during Edge and Cloud deployment.

The Edge tier shows 1.83x lower average median inference latency compared to the Mobile tier across the four Identity models. In particular, the Edge shows a drop of 3.30, 2.33, 15.10, and 21.81 seconds in median inference latency compared to Mobile for ResNet, ResNext, FCN, and DUC Identity models, respectively, along with a large effect size (Table~\ref{rq1_Identity_statistical_results}). These differences are possibly due to the larger computational resources (CPU/RAM) of the Edge tier compared to the Mobile tier, due to which the computations during model inference are faster in the Edge tier. In particular, Edge has 4x the RAM and twice as many CPU cores as Mobile. The Cloud tier shows an 8.52 and 23.62 seconds drop in median inference latency compared to both Mobile and Edge tiers, respectively only for the FCN Identity model as the abundant computational resources of the Cloud (16x/4x RAM and 4x/2x CPU relative to Mobile/Edge) and the lowest input data size range for the FCN subject (as mentioned earlier) accounts for both faster model inference and data transmission. 

The architectural complexity of models also plays a role in inference as among the 4 Identity models, the DUC Identity model shows the highest median inference latency of 45.46, 23.64, and 28.61 seconds on Mobile, Edge, and Cloud, respectively. The DUC Identity model has the highest number of graph nodes (i.e., 355), which contributes to the higher architectural complexity in comparison to the other Identity models (FCN: 260 graph nodes, ResNet/ResNext: 240 graph nodes). Here, the graph nodes represent the total number of operations in the ONNX computational graph of a model. 

\subsubsection{Normality Test}
Normality assessment was conducted using the Shapiro-Wilk test and visual inspection through QQ plots. For each Identity model across all deployment tiers, the Shapiro-Wilk test resulted in $p < 0.05$, rejecting the null hypothesis of normality, as shown in Table~\ref{rq1_descriptive_statistics}. This suggests that inference latency distributions significantly deviate from normality.

Additionally, QQ plots showed heavy skewness and long tails across all subjects and tiers, as shown in Table~\ref{normality_results_rq1}, further supporting the result of the Shapiro-Wilk test. As a result, non-parametric methods were employed for statistical comparisons in the subsequent analysis.

\subsubsection{Hypothesis Testing}

Given the non-normal latency distributions, the Conover test (a post-hoc non-parametric multiple comparisons method) was used to evaluate significant differences between deployment tiers. Table~\ref{rq1_Identity_statistical_results} presents $p$-values and Cliff’s Delta ($\delta$) effect sizes for each pairwise comparison. All pairwise comparisons across Mobile (M), Edge (E), and Cloud (C) showed statistically significant differences ($p < 0.05$). The Cliff’s Delta effect sizes were consistently large (L), indicating strong practical significance.

\begin{table}[htbp]
\centering
\caption{RQ1 results of the Cliff's Delta effect size and Conover test p-value between Mobile (M), Edge (E), and Cloud (C) deployment for $I_{\text{t}}$ (Identity version of ResNet), $I_{\text{x}}$ (Identity version of ResNext), $I_{\text{f}}$ (Identity version of FCN), and $I_{\text{d}}$ (Identity version of DUC).}
\begin{minipage}[t]{0.9\textwidth}
\centering
\begin{tabular}{|c|cc|cc|cc|}
\hline
\backslashbox{$I_{\text{x}}$}{$I_{\text{t}}$} & \multicolumn{2}{c|}{M} & \multicolumn{2}{c|}{E} & \multicolumn{2}{c|}{C} \\
\cline{2-7}
 & $p$ & $\delta$ & $p$ & $\delta$ & $p$ & $\delta$ \\
\hline
M & - & - & $9.6e^{-221}$ & L (0.9) & $7.6e^{-187}$ & -L (0.83) \\
\hline
E & $4.0e^{-99}$ & L (0.71) & - & - & 0.0 & -L (1.0) \\
\hline
C & $2.9e^{-152}$ & -L (0.86) & 0 & -L (0.92) & - & - \\
\hline
\end{tabular}
\centering
\vspace{0.5cm} 
\captionsetup{justification=centering}
\end{minipage}
\hfill
\begin{minipage}[t]{0.9\textwidth}
\centering
\begin{tabular}{|c|cc|cc|cc|}
\hline
\backslashbox{$I_{\text{d}}$}{$I_{\text{f}}$} & \multicolumn{2}{c|}{M} & \multicolumn{2}{c|}{E} & \multicolumn{2}{c|}{C} \\
\cline{2-7}
 & $p$ & $\delta$ & $p$ & $\delta$ & $p$ & $\delta$ \\
\hline
M & - & - & 0.0 & L (0.99) & 0.0 & L (1.0) \\
\hline
E & 0.0 & L (1.0) & - & - & 0.0 & L (1.0) \\
\hline
C & 0.0 & L (1.0) & 0.0 & -L (1.0) & - & - \\
\hline
\end{tabular}
\captionsetup{justification=centering}
\end{minipage}

\begin{threeparttable}
\begin{tablenotes}
\item[]\textsuperscript{1} In these, and later, tables, a Positive sign for the $i^{th}$ cell shows that the latency of column[i]$<$row[i], while a Negative sign for the $i^{th}$ cell shows that the latency of column[i]$>$row[i].
\item[]\textsuperscript{2} In these, and later, tables, the L, M, S, and N symbols mean Large, Medium, Small, and Negligible effect size, respectively.
\item[]\textsuperscript{3} In these, and later, tables, an empty cell means that the Cliff's Delta effect size was not considered because the pairwise comparison was not statistically significant based on the Conover test.
\end{tablenotes}
\end{threeparttable}
\label{rq1_Identity_statistical_results}
\end{table}

\begin{table}[htbp]
\caption{Accuracy performance of Identity, Quantized, Early Exit, and Quantized Early Exit versions of subjects within Mobile, Edge, and Cloud tiers.}
\resizebox{1.0\textwidth}{!}{
\begin{tabular}{cccccccccccc}
 \toprule
\multirow{2}{*}{Subject} & \multirow{2}{*}{Operator} & \multirow{2}{*}{Model Size} & \multicolumn{3}{c}{Top-1\%}  & \multicolumn{3}{c}{Top-5\%}  & \multicolumn{3}{c}{mIOU\%} \\
\cmidrule(lr){4-6}\cmidrule(lr){7-9}\cmidrule(lr){10-12}

 & & & Mobile & Edge & Cloud & Mobile & Edge & Cloud & Mobile & Edge & Cloud \\
\midrule
ResNet & Identity & 484 MB & 82.52 & 82.520 & 82.522 & 96.008 & 96.008 & 96.008 & - & - & - \\
ResNet & Quantized & 123 MB & 82.148 & 82.148 & 82.164 & 95.792 & 95.792 & 95.814 & - & - & - \\
ResNet & Early Exit & 380 MB & 76.586 & 76.586 & 76.59 & 93.442 & 93.442 & 93.446 & - & - & - \\
ResNet & Quantized Early Exit & 96 MB & 75.392 & 75.392 & 75.346 & 93.034 & 93.034 & 93.024 & - & - & - \\
\midrule
ResNext & Identity & 319 MB & 83.244 & 83.244 & 83.244 & 96.456 & 96.456 & 96.458 & - & - & -\\
ResNext & Quantized & 81 MB & 83.084 & 83.084 & 83.14 & 96.402 & 96.402 & 96.386 & - & - & -\\
ResNext & Early Exit & 250 MB & 77.276 & 77.276 & 77.284 & 93.92 & 93.92 & 93.924 & - & - & -\\
ResNext & Quantized Early Exit & 64 MB & 75.668 & 75.668 & 75.616 & 93.83 & 93.83 & 93.826 & - & - & -\\
\midrule
FCN & Identity & 199 MB & - & - & - & - & - & - & 66.7343 & 66.7343 & 66.7348 \\
FCN & Quantized & 50 MB & - & - & - & - & - & - & 66.38 & 66.38 & 66.35 \\
FCN & Early Exit & 164 MB & - & - & - & - & - & - & 55.16 & 55.16 & 55.16 \\
FCN & Quantized Early Exit & 42 MB & - & - & - & - & - & - & 54.36 & 54.36 & 54.32 \\
\midrule
DUC & Identity & 249 MB & - & - & - & - & - & - & 81.9220 & 81.9220 & 81.9223 \\
DUC & Quantized & 63 MB & - & - & - & - & - & - & 81.62 & 81.62 & 81.62 \\
DUC & Early Exit & 215 MB & - & - & - & - & - & - & 75.746 & 75.746 & 75.744 \\
DUC & Quantized Early Exit & 54 MB & - & - & - & - & - & - & 75.32 & 75.32 & 75.32 \\
\bottomrule
\end{tabular}
}
 \label{accuracy}
\end{table}

After rounding off the decimal digits (up to 4 places) in the accuracy metric values, we observed that for each subject, each of the 4 operators (i.e., Identity, Quantized, Early Exit, and Quantized Early Exit) exhibits identical performance between Mobile and Edge tiers, as shown in Table~\ref{accuracy}. The main reason seems to be the identical hardware (CPU processor) and software (packages) configuration of Mobile and Edge-simulated Docker containers. Table 9 shows a summary of results for RQ1. 

\begin{figure}[t]
\centering
\includegraphics[width=1\textwidth]{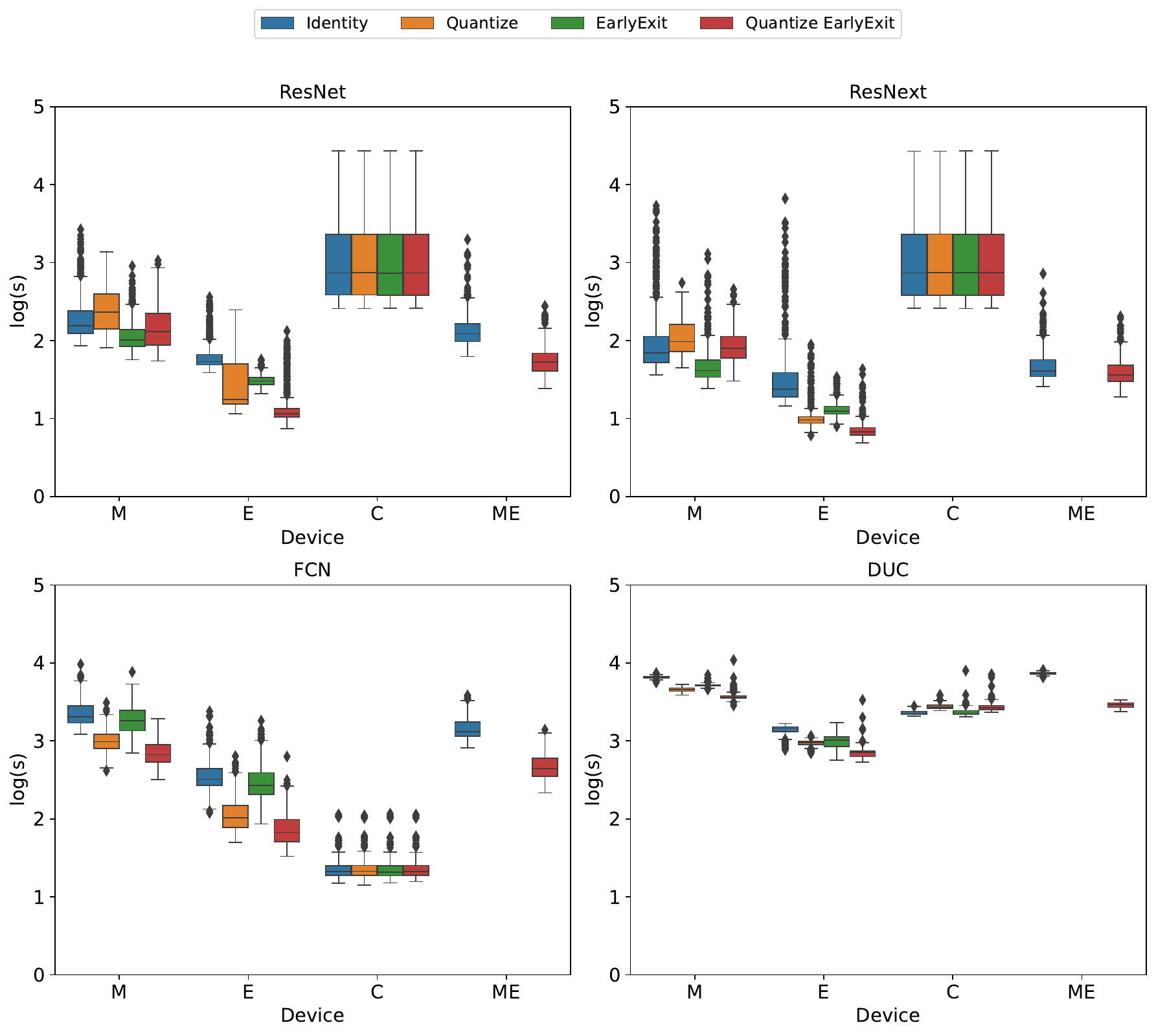}
\caption{Box plots of the measures collected for inference latency of Identity, Quantized, Early Exit, and Quantized Early Exit models in monolithic tiers (Mobile[M], Edge[E], Cloud[C]) and Partitioned Identity/Quantized Early Exit models in Mobile-Edge (ME) tier.}
\label{rq2_rq3_rq4_rq5_graph}
\end{figure}

\begin{Summary}{}{firstsummary}
Among the three monolithic deployment tiers, the Edge tier could be the preferred choice in terms of latency in scenarios where models (ResNet, ResNext, DUC) have large input data size requirements and the Mobile/Cloud tier has computational/bandwidth limitations. In contrast, for models having smaller input data size requirements (FCN), Cloud deployment could be the optimal choice over computationally constrained tiers (Mobile/Edge).
\end{Summary}

\begin{table}[h!]
\centering
\caption{Summary of results for Research Question 1}
\label{tab:rq1_result_summarization}
\resizebox{\textwidth}{!}{%
\begin{tabular}{|l|l|c|c|p{6.5cm}|}
\hline
\textbf{Model} & \textbf{Best Tier} & \textbf{Latency (s)} & \textbf{Gain over Next Best} & \textbf{Notes} \\ \hline
ResNet  & Edge  & 5.64  & 3.12x faster than Cloud & Large input size, bandwidth bottleneck in Cloud \\ \hline
ResNext & Edge  & 3.96  & 4.43x faster than Cloud & Similar to ResNet \\ \hline
DUC     & Edge  & 23.64 & 1.21x faster than Cloud & Highest graph complexity \\ \hline
FCN     & Cloud & 3.76  & 8.52s faster than Mobile & Small input size enables fast Cloud performance \\ \hline
\end{tabular}%
}
\end{table}

\subsection{What is the impact of the Quantization operator in terms of inference latency and accuracy within and across the considered tiers? (RQ2)}
\label{sec:rq2}
\subsubsection{Data Exploration}
\textbf{In Mobile, the Quantized models show 1.17x higher and 1.27x lower average median inference latency w.r.t Identity models for ResNet/Resnext and FCN/DUC, respectively. In Edge, the Quantized models show 1.48x lower average median inference latency than the Identity models across all subjects. In Cloud, no significant difference was shown among Quantized and Identity models across all subjects (except DUC).}

As shown in Figure~\ref{rq2_rq3_rq4_rq5_graph} (blue and orange box plots), the Quantized models show 1.15x to 1.19x higher and 1.17x to 1.37x lower median inference latency compared to the Identity models (Table~\ref{rq2_descriptive_statistics}, Table~\ref{rq1_descriptive_statistics}) in the Mobile tier, along with small to large effect sizes (Table~\ref{rq2_rq3_rq4_rq5_mobile_statistical_results}). The generated Quantized models are about 4x smaller than the Identity models of the subjects as shown in Table~\ref{accuracy}. Therefore, in an ideal situation, the Quantized models should show faster latency in comparison to Identity models due to model size reduction, as shown in FCN and DUC's latency results. But, for the ResNet and ResNext subjects, these models show slower latency, possibly due to their costly operations in the resource-constrained Mobile tier, leading to their higher CPU/Memory utilization than Identity models. In the Edge tier, the Quantized models show 1.20x to 1.64x lower median inference latency than the Identity models across the four subjects, along with large effect sizes (Table~\ref{rq2_rq3_rq4_rq5_edge_statistical_results}). This shows that the Quantized models perform faster than the Identity models in a high-resource environment (i.e., Edge).

During Cloud inference, the majority of graphical nodes (94.01\% to 100\%) are processed on the CUDA Execution provider for the Identity models, while for the Quantized models, 53.82\% to 86.52\% of the graphical nodes are processed on the CUDA Execution Provider and the remaining ones are processed on the CPU Execution Provider. Compared to Identity models, the percentage of graphical nodes that use CUDA Execution Provider is lower for Quantized models, which might imply that Quantized models are somewhat less optimized (or compatible) for GPU (i.e., CUDA) processing than Identity models. Among the four subjects, the Quantized model of DUC shows the lowest percentage of the graphical nodes (53.82\%) placed on the CUDA Execution Provider, resulting in its significantly slower inference latency than the DUC Identity model, for which all the graphical nodes are placed on the CUDA execution provider. For the remaining three subjects (i.e., ResNet, ResNext, FCN), the Quantized models show similar inference latency compared to the Identity models in the Cloud tier due to the higher percentage of graphical nodes placed on the CUDA execution provider, i.e., 85.62\% to 86.52\% and 94.01\% to 100\% for Quantized and Identity models, respectively, which is statistically significant (Wilcoxon Test: p-value = 0.03, $\alpha$ = 0.05) along with large effect size (1.0).

\textbf{The Quantized models show a small accuracy drop ($<$0.5\%) across the four subjects in comparison to Identity models.}

In terms of accuracy, the Quantized models demonstrate a marginal accuracy drop of 0.05\% down to 0.38\% compared to the Identity models, as shown in Table~\ref{accuracy}, While conducting the Wilcoxon test, the accuracy difference comes out to be statistically significant (p-value = 7.6e$^{-6}$, $\alpha$ = 0.0125). However, the effect size remains small (0.16). This trade-off suggests that Quantization is an effective operator for achieving faster inference in Edge (based on previous findings) without significantly compromising the accuracy performance.

\textbf{When comparing Quantization across the three monolithic deployment tiers, the Quantized models show 2.60x and 3.44x lower average median inference latency during Edge deployment than their deployment in Mobile and Cloud, respectively.}

The Quantized models across all subjects in the Edge exhibit 1.97x to 3.06x lower median inference latency compared to their inference in the Mobile, as shown in Figure~\ref{rq2_rq3_rq4_rq5_graph} (orange box plots), along with large effect sizes (Table~\ref{rq2_Quantized_statistical_results}) due to its higher computational resources. For ResNet, ResNext, and DUC subjects, the Quantized models in the Cloud demonstrate 1.58x to 6.61x higher median inference latency than their inference in the Edge tier due to the higher impact of the large input data sizes of these subjects on the transmission across the restricted Edge-Cloud network. However, the Quantized model for the FCN subject exhibits 1.97x lower median inference latency in the Cloud tier compared to the Edge tier with a large effect size (Table~\ref{rq2_Quantized_statistical_results}). This is due to the FCN's small input data sizes leading to a lower impact on data transmission across the restricted Edge-Cloud network. The reasoning behind these findings is similar and explained briefly in the RQ1 findings (Section~\ref{sec:rq1}). Table 14 shows a summary of results for RQ2.

\subsubsection{Normality Test}
The Shapiro-Wilk test and QQ plots were applied to assess the normality of latency distributions. For most model-subject combinations, the null hypothesis of normality was rejected ($p < 0.05$), indicating non-normal distributions of inference latency values(see Table~\ref{rq2_descriptive_statistics}). As a result, non-parametric methods like the Conover and Wilcoxon tests were selected for hypothesis testing.

\subsubsection{Hypothesis Testing}
According to the Conover test, the null hypothesis that there is no significant inference latency difference between the Quantized and Identity models was rejected for the DUC subject (2.0e$^{-247}$, $\alpha$ = 0.05) during Cloud deployment. The large effect size (Table~\ref{rq2_rq3_rq4_rq5_cloud_statistical_results}) for DUC shows that its Quantized model has a significantly higher distribution of inference latency magnitude compared to its Identity model. Conversely, for the remaining three subjects (i.e., ResNet, ResNext, FCN), the null hypothesis cannot be rejected, indicating that their Quantized models show similar or equivalent inference latency compared to their Identity models.

\begin{table}[htbp]
\centering
\caption{Cliff's Delta effect size and Conover Test p-value between $I_{\text{M}}$, $Q_{\text{M}}$, $E_{\text{M}}$, $P_{\text{ME}}$, $QE_{\text{M}}$, and $QEP_{\text{ME}}$ versions of subjects for RQ2, RQ3, RQ4, and RQ5}
\label{rq2_rq3_rq4_rq5_mobile_statistical_results}

\resizebox{\textwidth}{!}{%
\begin{tabular}{|c|cc|cc|cc|cc|cc|cc|}
\hline
\backslashbox{$R_{\text{x}}$}{$R_{\text{t}}$} & \multicolumn{2}{c|}{$I_{\text{M}}$} & \multicolumn{2}{c|}{$Q_{\text{M}}$} & \multicolumn{2}{c|}{$E_{\text{M}}$} & \multicolumn{2}{c|}{$P_{\text{ME}}$} & \multicolumn{2}{c|}{$QE_{\text{M}}$} & \multicolumn{2}{c|}{$QEP_{\text{ME}}$} \\
\cline{2-13}
 & $p$ & $\delta$ & $p$ & $\delta$ & $p$ & $\delta$ & $p$ & $\delta$ & $p$ & $\delta$ & $p$ & $\delta$ \\
\hline
$I_{\text{M}}$      & - & - & $4.0e^{-9}$ & -S (0.24) & $4.4e^{-67}$ & L (0.58) & $2.8e^{-26}$ & M (0.37) & $3.0e^{-20}$ & S (0.26) & $3.9e^{-270}$ & L (0.94) \\
\hline
$Q_{\text{M}}$      & $2.2e^{-23}$ & -M (0.34) & - & - & $1.7e^{-113}$ & L (0.69) & $2.6e^{-59}$ & L (0.54) & $3.2e^{-50}$ & M (0.44) & 0.0 & L (0.96) \\
\hline
$E_{\text{M}}$      & $9.1e^{-88}$ & L (0.6) & $5.8e^{-179}$ & L (0.79) & - & - & $2.3e^{-12}$ & -S (0.25) & $4.0e^{-17}$ & -S (0.24) & $5.3e^{-94}$ & L (0.80) \\
\hline
$P_{\text{ME}}$     & $7.0e^{-94}$ & L (0.63) & $8.3e^{-187}$ & L (0.83) &  &  & - & - &  &  & $6.5e^{-157}$ & L (0.87) \\
\hline
$QE_{\text{M}}$     & 0.001 & -N (0.12) & $6.6e^{-12}$ & S (0.29) & $7.8e^{-114}$ & -L (0.68) & $1.3e^{-120}$ & -L (0.71) & - & - & $6.2e^{-171}$ & L (0.85) \\
\hline
$QEP_{\text{ME}}$   & $2.8e^{-148}$ & L (0.75) & $9.8e^{-253}$ & L (0.91) & $5.2e^{-12}$ & S (0.24) & $8.3e^{-10}$ & S (0.23) & $2.5e^{-179}$ & L (0.82) & - & - \\
\hline
\end{tabular}%
}

\vspace{0.5cm}

\resizebox{\textwidth}{!}{%
\begin{tabular}{|c|cc|cc|cc|cc|cc|cc|}
\hline
\backslashbox{DUC}{FCN} & \multicolumn{2}{c|}{$I_{\text{M}}$} & \multicolumn{2}{c|}{$Q_{\text{M}}$} & \multicolumn{2}{c|}{$E_{\text{M}}$} & \multicolumn{2}{c|}{$P_{\text{ME}}$} & \multicolumn{2}{c|}{$QE_{\text{M}}$} & \multicolumn{2}{c|}{$QEP_{\text{ME}}$} \\
\cline{2-13}
 & $p$ & $\delta$ & $p$ & $\delta$ & $p$ & $\delta$ & $p$ & $\delta$ & $p$ & $\delta$ & $p$ & $\delta$ \\
\hline
$I_{\text{M}}$      & - & - & $8.1e^{-292}$ & L (0.92) & $6.8e^{-18}$ & S (0.25) & $6.2e^{-92}$ & L (0.65) & 0.0 & L (0.99) & 0.0 & L (1.0) \\
\hline
$Q_{\text{M}}$      & 0.0 & L (1.0) & - & - & $1.0e^{-196}$ & -L (0.77) & $5.3e^{-83}$ & -L (0.58) & $2.2e^{-64}$ & L (0.56) & $2.0e^{-191}$ & L (0.88) \\
\hline
$E_{\text{M}}$      & 0.0 & L (0.99) & $7.7e^{-322}$ & -L (0.93) & - & - & $2.0e^{-34}$ & M (0.38) & 0.0 & L (0.94) & 0.0 & L (0.99) \\
\hline
$P_{\text{ME}}$     & 0.0 & -L (0.98) & 0.0 & -L (1.0) & 0.0 & -L (1.0) & - & - & $3.7e^{-250}$ & L (0.88) & 0.0 & L (0.99) \\
\hline
$QE_{\text{M}}$     & 0.0 & L (0.99) & $1.2e^{-321}$ & L (0.94) & 0.0 & L (0.98) & 0.0 & L (1.0) & - & - & $9.5e^{-46}$ & L (0.57) \\
\hline
$QEP_{\text{ME}}$   & 0.0 & L (1.0) & 0.0 & L (1.0) & 0.0 & L (1.0) & 0.0 & L (1.0) & 0.0 & L (0.99) & - & - \\
\hline
\end{tabular}%
}

\begin{threeparttable}
\begin{tablenotes}
\item[]\textsuperscript{1} $I_{\text{M}}$, $Q_{\text{M}}$, $E_{\text{M}}$, $QE_{\text{M}}$ denote Identity, Quantized, Early Exit, and Quantized Early Exit models in the Mobile tier.
\item[]\textsuperscript{2} $P_{\text{ME}}$ and $QEP_{\text{ME}}$ denote Partitioned and Quantized Early Exit Partitioned models in the Mobile-Edge tier.
\item[]\textsuperscript{3} L, M, S, and N refer to Large, Medium, Small, and Negligible effect sizes, respectively.
\item[]\textsuperscript{4} An empty $p$ cell indicates the pairwise comparison was not statistically significant based on the Conover test.
\end{tablenotes}
\end{threeparttable}
\end{table}

\begin{table}[htbp]
\centering
\caption{Cliff's Delta effect size and Conover Test p-value between $I_{\text{E}}$, $Q_{\text{E}}$, $E_{\text{E}}$, $P_{\text{ME}}$, $QE_{\text{E}}$, and $QEP_{\text{ME}}$ versions of subjects for RQ2, RQ3, RQ4, and RQ5}
\label{rq2_rq3_rq4_rq5_edge_statistical_results}

\resizebox{\textwidth}{!}{%
\begin{tabular}{|c|cc|cc|cc|cc|cc|cc|}
\hline
\backslashbox{$R_{\text{x}}$}{$R_{\text{t}}$} & \multicolumn{2}{c|}{$I_{\text{E}}$} & \multicolumn{2}{c|}{$Q_{\text{E}}$} & \multicolumn{2}{c|}{$E_{\text{E}}$} & \multicolumn{2}{c|}{$P_{\text{ME}}$} & \multicolumn{2}{c|}{$QE_{\text{E}}$} & \multicolumn{2}{c|}{$QEP_{\text{ME}}$} \\
\cline{2-13}
 & $p$ & $\delta$ & $p$ & $\delta$ & $p$ & $\delta$ & $p$ & $\delta$ & $p$ & $\delta$ & $p$ & $\delta$ \\
\hline
$I_{\text{E}}$      & - & - & $3.8e^{-147}$ & L (0.58) & $9.3e^{-161}$ & L (0.99) & $2.9e^{-103}$ & -L (0.83) & 0.0 & L (0.91) & 0.0003 & N (0.13) \\
\hline
$Q_{\text{E}}$      & 0.0 & L (0.9) & - & - &  &  & 0.0 & -L (0.83) & $1.0e^{-105}$ & L (0.74) & $2.7e^{-114}$ & -L (0.55) \\
\hline
$E_{\text{E}}$      & $3.6e^{-194}$ & L (0.93) & $7.5e^{-44}$ & -L (0.69) & - & - & 0.0 & -L (1.0) & $4.5e^{-94}$ & L (0.81) & $8.8e^{-127}$ & -L (0.90) \\
\hline
$P_{\text{ME}}$     & $1.9e^{-74}$ & -L (0.53) & 0.0 & -L (0.95) & 0.0 & -L (1.0) & - & - & 0.0 & L (0.99) & $5.1e^{-135}$ & L (0.87) \\
\hline
$QE_{\text{E}}$     & 0.0 & L (0.98) & $5.3e^{-86}$ & L (0.78) & $4.4e^{-219}$ & L (0.93) & 0.0 & L (1.0) & - & - & 0.0 & -L (0.90) \\
\hline
$QEP_{\text{ME}}$   & $4.6e^{-36}$ & -M (0.42) & 0.0 & -L (0.93) & 0.0 & -L (0.99) & $1.4e^{-9}$ & S (0.23) & 0.0 & -L (0.99) & - & - \\
\hline
\end{tabular}%
}

\vspace{0.5cm}

\resizebox{\textwidth}{!}{%
\begin{tabular}{|c|cc|cc|cc|cc|cc|cc|}
\hline
\backslashbox{DUC}{FCN} & \multicolumn{2}{c|}{$I_{\text{E}}$} & \multicolumn{2}{c|}{$Q_{\text{E}}$} & \multicolumn{2}{c|}{$E_{\text{E}}$} & \multicolumn{2}{c|}{$P_{\text{ME}}$} & \multicolumn{2}{c|}{$QE_{\text{E}}$} & \multicolumn{2}{c|}{$QEP_{\text{ME}}$} \\
\cline{2-13}
 & $p$ & $\delta$ & $p$ & $\delta$ & $p$ & $\delta$ & $p$ & $\delta$ & $p$ & $\delta$ & $p$ & $\delta$ \\
\hline
$I_{\text{E}}$      & - & - & $1.3e^{-302}$ & L (0.92) & $7.7e^{-14}$ & S (0.25) & 0.0 & -L (0.98) & 0.0 & L (0.98) & $6.2e^{-46}$ & -L (0.48) \\
\hline
$Q_{\text{E}}$      & $7.9e^{-311}$ & L (0.90) & - & - & $5.9e^{-219}$ & -L (0.83) & 0.0 & -L (1.0) & $3.0e^{-36}$ & L (0.5) & 0.0 & -L (0.97) \\
\hline
$E_{\text{E}}$      & $3.9e^{-239}$ & L (0.83) & $2.2e^{-10}$ & -S (0.25) & - & - & 0.0 & -L (0.98) & 0.0 & L (0.95) & $2.1e^{-99}$ & -L (0.57) \\
\hline
$P_{\text{ME}}$     & 0.0 & -L (1.0) & 0.0 & -L (1.0) & 0.0 & -L (1.0) & - & - & 0.0 & L (1.0) & $1.4e^{-185}$ & L (0.99) \\
\hline
$QE_{\text{E}}$     & 0.0 & L (0.99) & $4.8e^{-249}$ & L (0.94) & $3.1e^{-321}$ & L (0.82) & 0.0 & L (1.0) & - & - & 0.0 & -L (1.0) \\
\hline
$QEP_{\text{ME}}$   & $6.4e^{-207}$ & -L (1.0) & 0.0 & -L (1.0) & 0.0 & -L (1.0) & $7.3e^{-166}$ & L (1.0) & 0.0 & -L (1.0) & - & - \\
\hline
\end{tabular}%
}

\begin{threeparttable}
\begin{tablenotes}
\item[]\textsuperscript{1} $I_{\text{E}}$, $Q_{\text{E}}$, $E_{\text{E}}$, $QE_{\text{E}}$ denote Identity, Quantized, Early Exit, and Quantized Early Exit models in the Edge tier.
\item[]\textsuperscript{2} $P_{\text{ME}}$ and $QEP_{\text{ME}}$ denote Partitioned and Quantized Early Exit Partitioned models in the Mobile-Edge tier.
\item[]\textsuperscript{3} L, M, S, and N refer to Large, Medium, Small, and Negligible effect sizes, respectively.
\item[]\textsuperscript{4} An empty $p$ cell indicates the pairwise comparison was not statistically significant based on the Conover test.
\end{tablenotes}
\end{threeparttable}
\end{table}

\begin{table}[htbp]
\centering
\caption{Cliff's Delta effect size and Conover Test p-value between $I_{\text{C}}$, $Q_{\text{C}}$, $E_{\text{C}}$, $P_{\text{ME}}$, $QE_{\text{C}}$, and $QEP_{\text{ME}}$ versions of subjects for RQ2, RQ3, RQ4, and RQ5}
\label{rq2_rq3_rq4_rq5_cloud_statistical_results}

\resizebox{\textwidth}{!}{%
\begin{tabular}{|c|cc|cc|cc|cc|cc|cc|}
\hline
\backslashbox{ResNext}{ResNet} & \multicolumn{2}{c|}{$I_{\text{C}}$} & \multicolumn{2}{c|}{$Q_{\text{C}}$} & \multicolumn{2}{c|}{$E_{\text{C}}$} & \multicolumn{2}{c|}{$P_{\text{ME}}$} & \multicolumn{2}{c|}{$QE_{\text{C}}$} & \multicolumn{2}{c|}{$QEP_{\text{ME}}$} \\
\cline{2-13}
 & $p$ & $\delta$ & $p$ & $\delta$ & $p$ & $\delta$ & $p$ & $\delta$ & $p$ & $\delta$ & $p$ & $\delta$ \\
\hline
$I_{\text{C}}$        & - & - &  &  &  &  & $4.6e^{-257}$ & L (0.95) & & &   0.0 & L (1.0) \\
\hline
$Q_{\text{C}}$        & & & - & - & & & $6.6e^{-257}$ & L (0.95) & & & 0.0 & L (1.0) \\
\hline
$E_{\text{C}}$        & & & & & - & - & $3.6e^{-257}$ & L (0.95) & & & 0.0 & L (1.0) \\
\hline
$P_{\text{ME}}$       & 0.0 & L (1.0) & 0.0 & L (1.0) & 0.0 & L (1.0) & - & - & $4.6e^{-257}$ & -L (0.95) & $1.3e^{-50}$ & L (0.87) \\
\hline
$QE_{\text{C}}$       & & & & & & & 0.0 & -L (1.0) & - & - & 0.0 & L (1.0) \\
\hline
$QEP_{\text{ME}}$     & 0.0 & L (1.0) & 0.0 & L (1.0) & 0.0 & L (1.0) & 0.0001 & S (0.23) & 0.0 & L (1.0) & - & - \\
\hline
\end{tabular}%
}

\vspace{0.5cm}

\resizebox{\textwidth}{!}{%
\begin{tabular}{|c|cc|cc|cc|cc|cc|cc|}
\hline
\backslashbox{DUC}{FCN} & \multicolumn{2}{c|}{$I_{\text{C}}$} & \multicolumn{2}{c|}{$Q_{\text{C}}$} & \multicolumn{2}{c|}{$E_{\text{C}}$} & \multicolumn{2}{c|}{$P_{\text{ME}}$} & \multicolumn{2}{c|}{$QE_{\text{C}}$} & \multicolumn{2}{c|}{$QEP_{\text{ME}}$} \\
\cline{2-13}
 & $p$ & $\delta$ & $p$ & $\delta$ & $p$ & $\delta$ & $p$ & $\delta$ & $p$ & $\delta$ & $p$ & $\delta$ \\
\hline
$I_{\text{C}}$        & - & - &  &  & &  & 0.0 & -L (1.0) & &  & $8.0e^{-298}$ & -L (1.0) \\
\hline
$Q_{\text{C}}$        & $1.9e^{-289}$ & -L (0.92) & - & - & &  & 0.0 & -L (1.0) & &  & $1.9e^{-287}$ & -L (1.0) \\
\hline
$E_{\text{C}}$        & 0.02 & -N (0.03) & $1.4e^{-263}$ & L (0.86) & - & - & 0.0 & -L (1.0) & &  & $5.5e^{-301}$ & -L (1.0) \\
\hline
$P_{\text{ME}}$       & 0.0 & -L (1.0) & $3.1e^{-276}$ & -L (1.0) & 0.0 & -L (1.0) & - & - & 0.0 & L (1.0) & $2.6e^{-57}$ & L (0.99) \\
\hline
$QE_{\text{C}}$       & $8.9e^{-221}$ & -L (0.85) & $6.8e^{-10}$ & S (0.20) & $1.5e^{-196}$ & -L (0.78) & 0.0 & L (1.0) & - & - & $5.3e^{-296}$ & -L (1.0) \\
\hline
$QEP_{\text{ME}}$     & 0.0 & -L (0.94) & $5.3e^{-16}$ & -S (0.30) & 0.0 & -L (0.90) & $6.3e^{-188}$ & L (1.0) & $3.8e^{-45}$ & -M (0.42) & - & - \\
\hline
\end{tabular}%
}

\begin{threeparttable}
\begin{tablenotes}
\item[]\textsuperscript{1} $I_{\text{C}}$, $Q_{\text{C}}$, $E_{\text{C}}$, $QE_{\text{C}}$ denote Identity, Quantized, Early Exit, and Quantized Early Exit models in the Cloud tier.
\item[]\textsuperscript{2} $P_{\text{ME}}$ and $QEP_{\text{ME}}$ denote Partitioned and Quantized Early Exit Partitioned models in the Mobile-Edge tier.
\item[]\textsuperscript{3} L, M, S, and N refer to Large, Medium, Small, and Negligible effect sizes, respectively.
\item[]\textsuperscript{4} An empty $p$ cell indicates the pairwise comparison was not statistically significant based on the Conover test.
\end{tablenotes}
\end{threeparttable}
\end{table}

\begin{table}[htbp]
\centering
\caption{RQ2 results of the Cliff's Delta effect size and Conover test p-value between Mobile (M), Edge (E), and Cloud (C) deployment of Quantized models ($Q_{\text{t}}$, $Q_{\text{x}}$, $Q_{\text{f}}$, $Q_{\text{d}}$ denote Quantized versions of ResNet, ResNext, FCN, and DUC respectively).}
\begin{minipage}[t]{0.9\textwidth}
\centering
\begin{tabular}{|c|cc|cc|cc|}
\hline
\backslashbox{$Q_{\text{x}}$}{$Q_{\text{t}}$} & \multicolumn{2}{c|}{M} & \multicolumn{2}{c|}{E} & \multicolumn{2}{c|}{C} \\
\cline{2-7}
 & $p$ & $\delta$ & $p$ & $\delta$ & $p$ & $\delta$ \\
\hline
M & - & - & $3.3e^{-230}$ & L (0.93) & $6.5e^{-142}$ & -L (0.74) \\
\hline
E & 0.0 & L (0.99) & - & - & 0.0 & -L (1.0) \\
\hline
C & 0.0 & -L (0.98) & 0.0 & -L (1.0) & - & - \\
\hline
\end{tabular}
\centering
\vspace{0.5cm}
\captionsetup{justification=centering}
\end{minipage}
\hfill
\begin{minipage}[t]{0.9\textwidth}
\centering
\begin{tabular}{|c|cc|cc|cc|}
\hline
\backslashbox{$Q_{\text{d}}$}{$Q_{\text{f}}$} & \multicolumn{2}{c|}{M} & \multicolumn{2}{c|}{E} & \multicolumn{2}{c|}{C} \\
\cline{2-7}
 & $p$ & $\delta$ & $p$ & $\delta$ & $p$ & $\delta$ \\
\hline
M & - & - & 0.0 & L (1.0) & 0.0 & L (1.0) \\
\hline
E & 0.0 & L (1.0) & - & - & 0.0 & L (0.99) \\
\hline
C & 0.0 & L (1.0) & 0.0 & -L (1.0) & - & - \\
\hline
\end{tabular}
\captionsetup{justification=centering}
\end{minipage}

\begin{threeparttable}
\begin{tablenotes}
\item[]\textsuperscript{1} A Positive sign for the $i^{th}$ cell shows that the latency of column[i] $<$ row[i], while a Negative sign means column[i] $>$ row[i].
\item[]\textsuperscript{2} L, M, S, and N symbols denote Large, Medium, Small, and Negligible effect sizes, respectively.
\item[]\textsuperscript{3} Empty cells indicate non-significant comparisons based on the Conover test.
\end{tablenotes}
\end{threeparttable}
\label{rq2_Quantized_statistical_results}
\end{table}

\begin{Summary}{}{secondsummary}
The Quantization operator could be the preferred choice over the Identity operator across the four subjects when faster latency (1.48x) is a concern in the Edge tier, at a small accuracy loss (\textless0.4\%). Among the three monolithic deployment tiers, the Edge again is the most suitable deployment tier for the Quantization operator when factors like large input data size and constrained computational (Mobile)/ network (Cloud) environment play a crucial role. In contrast, Cloud deployment again is a better option for this operator when factors like small input data sizes and constrained computational environments (Mobile/Edge) are important.
\end{Summary}

\begin{table}[h!]
\centering
\begin{tabular}{|l|c|c|c|l|}
\hline
\textbf{Model} & \textbf{Best Tier} & \textbf{Latency Change} & \textbf{Accuracy Drop} & \textbf{Notes} \\
\hline
ResNet & Edge & 1.64x faster than Identity & 0.37\% & Costly on Mobile; Cloud uses more CPU than GPU \\
ResNext & Edge & 1.59x faster than Identity & 0.16\% & Same as ResNet; best in Edge, neutral in Cloud \\
FCN & Cloud & 1.97x faster than Edge & 0.35\% & Small input size favors Cloud; ideal for low-bandwidth \\
DUC & Edge & 1.58x faster than Cloud & 0.30\% & Cloud performance bottlenecked by GPU compatibility \\
\hline
\end{tabular}
\caption{Summary of results for Research Question 2}
\end{table}

\subsection{What is the impact of the Early Exit operator in terms of inference latency and accuracy within and across the considered tiers? (RQ3)}
\label{sec:rq3}
\subsubsection{Data Exploration}
\textbf{In Mobile and Edge, the Early Exit models show a 1.15x and 1.21x lower average median inference latency than the Identity models, respectively. In the Cloud, the Early Exit models show no practically significant difference in inference latency compared to the Identity models.}

The Early Exit models show 1.05x to 1.25x and 1.08x to 1.32x lower median inference latency than the Identity models (Table~\ref{rq3_descriptive_statistics}, Table~\ref{rq1_descriptive_statistics}) in the Mobile and Edge tiers, respectively, as shown in Figure~\ref{rq2_rq3_rq4_rq5_graph} (blue and green box plots), along with small or large effect sizes (Table~\ref{rq2_rq3_rq4_rq5_mobile_statistical_results}, Table~\ref{rq2_rq3_rq4_rq5_edge_statistical_results}). The utilization of intermediate predictions in the Early Exit models allows it to leverage information from earlier stages of the neural network, leading to model size reduction (1.15x to 1.27x) and faster inference compared to the Identity models in restricted-constrained tiers (i.e., Mobile and Edge).

\textbf{In Mobile, the Quantized models show 1.44x higher and 1.18x lower average median inference latency w.r.t Early Exit models for ResNet/ResNext and FCN/DUC, respectively. In Edge, the Quantized models show 1.23x lower average median inference latency than the Early Exit models across the four subjects. In the Cloud, no significant inference latency difference was shown among Quantized and Early Exit models across all subjects (except DUC).}

During Mobile deployment, the Quantized models for ResNet and ResNext subjects show 1.42x to 1.45x higher median inference latency than the Early Exit models, as shown in Figure~\ref{rq2_rq3_rq4_rq5_graph} (orange and green box-plots), along with large effect sizes (Table~\ref{rq2_rq3_rq4_rq5_mobile_statistical_results}). In contrast, for FCN and DUC subjects, the Quantized models show 1.05x to 1.31x lower median inference latency than the Early Exit models, along with large effect sizes (Table~\ref{rq2_rq3_rq4_rq5_mobile_statistical_results}). Even though the Quantized model sizes of the ResNet and ResNext subjects are 3.08x lower than the Early Exit models, they still show slower latency results, possibly due to the costly Quantization operations of these two subjects in a low-resource environment (i.e., Mobile) similar to the reasoning discussed in RQ1 (Section~\ref{sec:rq1}) when comparing Quantized models with Identity models in the Mobile tier. 

In the Edge tier, the Quantized models show 1.03x to 1.51x lower median inference latency than the Quantized models across the four subjects. In comparison to Early Exit models (Table~\ref{rq2_rq3_rq4_rq5_edge_statistical_results}), the inference latency of Quantized models is similar (ResNet) or significantly faster with small (DUC) or large (ResNext, FCN) effect sizes. This shows that in the Edge tier, the Quantized models are overall a better option than the Early Exit models in terms of faster latency. 

\textbf{The Early Exit models cost a medium accuracy drop of 2.53\% to 11.56\% and 2.35\% to 11.21\% in comparison to Identity and Quantized models, respectively.}

In terms of accuracy, the Early Exit models show a significant statistical difference (Wilcoxon Test: p-value = 7.6e$^{-6}$, $\alpha$ = 0.0125) with a medium effect size (0.38) in comparison to Identity and Quantized models. As presented in Table~\ref{accuracy}, the Early Exit models reveal an accuracy drop ranging from 2.53\% to 11.56\% and 2.35\% to 11.21\% when compared with Identity and Quantized models. This finding suggests that Early Exit models are less effective in accuracy performance relative to both Identity and Quantized models. Based on this and previous findings, the Quantization operator can achieve faster inference while maintaining a reasonably high level of accuracy, making it a preferred choice over the early exit operator. 

\textbf{When comparing Early Exiting across the three monolithic deployment tiers, the Early Exit models show 1.92x and 2.90x lower average median inference latency during Edge deployment than their deployment in Mobile and Cloud tiers, respectively.}

The Early Exit models in the Edge outperform the ones on the Mobile tier by 1.69x to 2.29x in terms of median inference latency, as shown in Figure~\ref{rq2_rq3_rq4_rq5_graph} (green box plots), with large effect sizes (Table~\ref{rq3_earlyexit_statistical_results}) due to the higher computational resources of Edge. For ResNet, ResNext, and DUC subjects, the Early Exit models in the Cloud tier exhibit 1.40x to 5.91x higher median inference latency compared to the Edge, with large effect sizes (Table~\ref{rq3_earlyexit_statistical_results}) due to the higher impact of the large input data sizes of these subjects on the transmission across the restricted Edge-Cloud network during Cloud deployment. However, for the FCN subject, the Early Exit model in the Cloud tier experiences a 3.03x lower median inference latency compared to the Edge tier, with a large effect size (Table~\ref{rq3_earlyexit_statistical_results}). This is due to the FCN's small input data sizes leading to a lower impact on data transmission across the restricted Edge-Cloud network during Cloud deployment. The reasoning behind these findings is similar and explained briefly in the RQ1 findings (Section~\ref{sec:rq1}). Table 16 shows a summary of results for RQ3.

\subsubsection{Normality Test}
The Shapiro-Wilk test and QQ plots reveal non-normality across all configurations and deployment tiers for latency data (Table~\ref{rq3_descriptive_statistics}), justifying the use of non-parametric tests such as the Conover test and Cliff’s Delta for pairwise comparisons.

\subsubsection{Hypothesis Testing}
According to the Conover test, for ResNet, ResNext, and FCN, the null hypothesis that there is no significant difference between the Early Exit and Identity model in the Cloud cannot be rejected, indicating that the Early Exit models during Cloud deployment show similar or equivalent inference latency compared to the Identity models. For the DUC subject, the null hypothesis was rejected, although with a negligible effect size (Table~\ref{rq2_rq3_rq4_rq5_cloud_statistical_results}), suggesting that the difference is likely not practically significant. The main reason for not having a significant difference is the ample availability of computational resources in the Cloud tier compared to resource-constrained Mobile and Edge tiers due to which the impact of Early Exiting on subjects is not significant in comparison to the Identity models in the Cloud tier. 

In the Cloud tier, the Quantized models show no statistically significant difference (according to the Conover test) in inference latency in comparison to Early Exit models for all subjects (except DUC). The reasoning for these findings is due to the lower compatibility of DUC's Quantization nodes with the CUDA Execution Provider during Cloud deployment, which is similar to those discussed briefly in the first finding of RQ1 (Section~\ref{sec:rq1}) when comparing Quantized and Identity models in the Cloud tier.

\begin{table}[htbp]
\centering
\caption{RQ3 results of the Cliff's Delta effect size and Conover test p-value between Mobile (M), Edge (E), and Cloud (C) deployment of Early Exit models ($E_{\text{t}}$, $E_{\text{x}}$, $E_{\text{f}}$, and $E_{\text{d}}$ denote Early Exit versions of ResNet, ResNext, FCN, and DUC respectively).}
\begin{minipage}[t]{0.9\textwidth}
\centering
\begin{tabular}{|c|cc|cc|cc|}
\hline
\backslashbox{$E_{\text{x}}$}{$E_{\text{t}}$} & \multicolumn{2}{c|}{M} & \multicolumn{2}{c|}{E} & \multicolumn{2}{c|}{C} \\
\cline{2-7}
 & $p$ & $\delta$ & $p$ & $\delta$ & $p$ & $\delta$ \\
\hline
M & - & - & 0.0 & L (1.0) & 0.0 & -L (0.98) \\
\hline
E & 0.0 & L (0.9) & - & - & 0.0 & -L (1.0) \\
\hline
C & 0.0 & -L (0.98) & 0.0 & -L (1.0) & - & - \\
\hline
\end{tabular}
\centering
\vspace{0.5cm}
\captionsetup{justification=centering}
\end{minipage}
\hfill
\begin{minipage}[t]{0.9\textwidth}
\centering
\begin{tabular}{|c|cc|cc|cc|}
\hline
\backslashbox{$E_{\text{d}}$}{$E_{\text{f}}$} & \multicolumn{2}{c|}{M} & \multicolumn{2}{c|}{E} & \multicolumn{2}{c|}{C} \\
\cline{2-7}
 & $p$ & $\delta$ & $p$ & $\delta$ & $p$ & $\delta$ \\
\hline
M & - & - & 0.0 & L (0.9) & 0.0 & L (1.0) \\
\hline
E & 0.0 & L (1.0) & - & - & 0.0 & L (1.0) \\
\hline
C & 0.0 & L (1.0) & 0.0 & -L (1.0) & - & - \\
\hline
\end{tabular}
\captionsetup{justification=centering}
\end{minipage}

\begin{threeparttable}
\begin{tablenotes}
\item[]\textsuperscript{1} A Positive sign for the $i^{th}$ cell shows that the latency of column[i] $<$ row[i], while a Negative sign means column[i] $>$ row[i].
\item[]\textsuperscript{2} L, M, S, and N symbols denote Large, Medium, Small, and Negligible effect sizes, respectively.
\item[]\textsuperscript{3} Empty cells indicate non-significant comparisons based on the Conover test.
\end{tablenotes}
\end{threeparttable}
\label{rq3_earlyexit_statistical_results}
\end{table}

\begin{Summary}{}{thirdsummary}
Similar to RQ2, the Quantized operator could be the preferred choice when faster latency (1.23x) is a concern in the Edge tier, at medium accuracy improvement (up to 11.21\%) than the Early Exit operator, which shows faster latency (1.21x) than the Identity models at medium accuracy drop (up to 11.56\%). Among the three monolithic deployment tiers, the Edge again is the most suitable deployment tier for the Early Exit operator when factors like large input data size and a constrained network(Cloud) environment play a crucial role. Cloud deployment is again a better option for this operator when factors like small input data size and constrained computational environments (Mobile/Edge) play a crucial role.
\end{Summary}

\begin{table}[h!]
\centering
\begin{tabular}{|l|c|c|c|p{6cm}|}
\hline
\textbf{Model} & \textbf{Best Tier} & \textbf{Latency Gain} & \textbf{Accuracy Drop} & \textbf{Notes} \\ \hline
ResNet  & Edge  & 1.21x faster than Identity & 11.56\% & No significant gain in Cloud; benefits from intermediate prediction on Edge \\
ResNext & Edge  & 1.21x faster than Identity & 10.77\% & Similar to ResNet; best used in constrained environments \\
FCN     & Cloud & No significant gain        & 2.53\%  & Cloud benefits due to small input size; no Edge advantage \\
DUC     & Edge  & 1.08x–1.32x faster         & 6.80\%  & Gains seen in Edge; Cloud gain negligible due to GPU usage limits \\
\hline
\end{tabular}
\caption{Summary of results for Research Question 3}
\label{tab:early_exit_rq3}
\end{table}

\subsection{What is the impact of the Partitioned operator in terms of inference latency and accuracy across the considered tiers? (RQ4)}
\label{sec:rq4}
\subsubsection{Data Exploration}
\begin{figure}[t]
\centering
\includegraphics[width=1\textwidth]{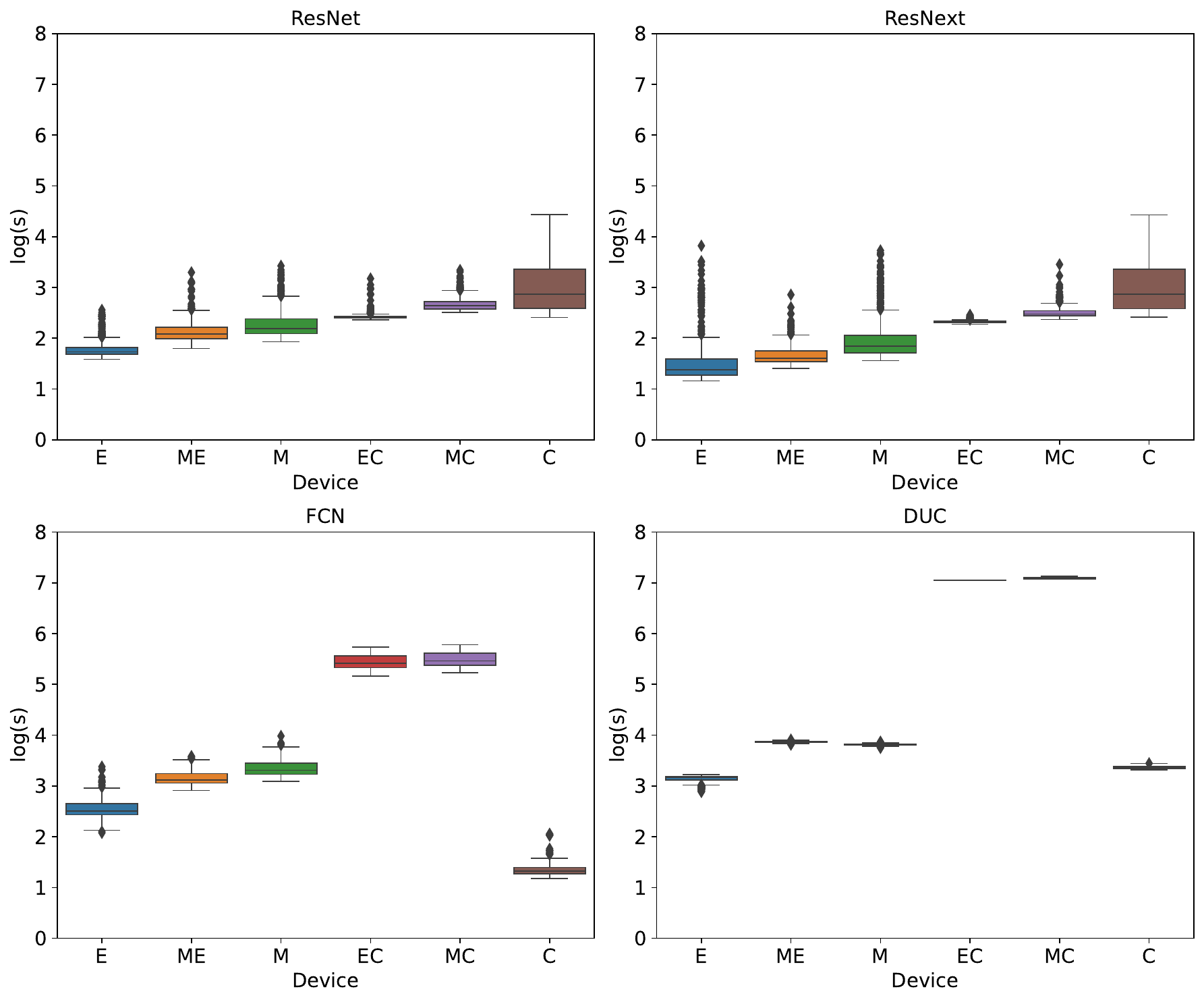}
\caption{Box plots of the inference latency measure for Multi-tier (Mobile-Edge [ME], Edge-Cloud [EC], Mobile-Cloud [MC]) Partitioned strategies and Single-tier (Mobile [M], Edge [E], Cloud [C]) monolithic deployment strategies}
\label{rq4_graph}
\end{figure}

\textbf{Among the three Multi-tier Partitioned strategies, the Mobile-Edge Partitioned strategy shows 9.37x and 9.86x lower average median inference latency compared to Edge-Cloud and Mobile-Cloud Partitioned strategies, respectively.}

The Mobile-Edge Partitioned strategy outperforms the Edge-Cloud and Mobile-Cloud Partitioned strategies in terms of median inference latency as depicted in Figure~\ref{rq4_graph} (blue, orange, and green box-plots). In particular, it achieves a significant speedup, ranging from 1.37x to 24.17x and 1.74x to 24.95x compared to the Edge-Cloud and Mobile-Cloud Partitioned strategies (Table~\ref{rq4_descriptive_statistics}), respectively, with large effect sizes (Table~\ref{rq4_partition_statistical_results}). 

The primary reason for this behavior is the impact of the size of intermediate data during transmission across the Edge-Cloud network. In the Edge-Cloud and Mobile-Cloud Partitioned strategies, the intermediate results generated after the first half of inference need to be transmitted from the Edge tier to the Cloud tier for the second half of inference. This transmission incurs additional latency, particularly when the intermediate data size of subjects (i.e., 6.12Mb for ResNet/ResNext, 118.12 to 200Mb for FCN, 781.25Mb for DUC) is larger than the limited Cloud network bandwidth (1 Mbps). Conversely, the Mobile-Edge Partitioned strategy is less influenced by the size of the intermediate data due to the higher Mobile-Edge network bandwidth (200 Mbps). Hence, the impact of the intermediate data size on inference latency is a crucial factor contributing to the observed performance advantage of the Mobile-Edge Partitioned strategy.

On the other hand, the Edge-Cloud Partitioned strategy achieves a speedup ranging from 1.03x to 1.26x when compared to the Mobile-Cloud Partitioned strategy, as shown in Figure~\ref{rq4_graph} (yellow and green box-plots), along with small or large effect sizes (Table~\ref{rq4_partition_statistical_results}). These differences can be attributed to the faster computational capabilities of the Edge-Cloud tier compared to the Mobile-Cloud tier.

\begin{table}[htbp]
\centering
\caption{RQ4 Results of the Cliff's Delta effect size and Conover test p-value between Mobile-Edge (ME), Edge-Cloud (EC), and Mobile-Cloud (MC) Partitioned strategies}
\begin{minipage}[t]{0.9\textwidth}
\centering
\begin{tabular}{|c|cc|cc|cc|}
\hline
\backslashbox{$P_{\text{x}}$}{$P_{\text{t}}$} & \multicolumn{2}{c|}{ME} & \multicolumn{2}{c|}{EC} & \multicolumn{2}{c|}{MC} \\
\cline{2-7}
 & $p$ & $\delta$ & $p$ & $\delta$ & $p$ & $\delta$ \\
\hline
ME & - & - & $1.8e^{-117}$ & L (0.77) & 0.0 & L (0.94) \\
\hline
EC & $8.1e^{-192}$ & L (0.98) & - & - & $1.3e^{-148}$ & L (0.96) \\
\hline
MC & 0.0 & L (0.99) & $1.0e^{-90}$ & L (0.99) & - & - \\
\hline
\end{tabular}
\centering
\vspace{0.5cm}
\captionsetup{justification=centering}
\end{minipage}
\hfill
\begin{minipage}[t]{0.9\textwidth}
\centering
\begin{tabular}{|c|cc|cc|cc|}
\hline
\backslashbox{$P_{\text{d}}$}{$P_{\text{f}}$} & \multicolumn{2}{c|}{ME} & \multicolumn{2}{c|}{EC} & \multicolumn{2}{c|}{MC} \\
\cline{2-7}
 & $p$ & $\delta$ & $p$ & $\delta$ & $p$ & $\delta$ \\
\hline
ME & - & - & 0.0 & L (1.0) & 0.0 & L (1.0) \\
\hline
EC & 0.0 & L (1.0) & - & - & $4.4e^{-20}$ & S (0.27) \\
\hline
MC & 0.0 & L (1.0) & 0.0 & L (1.0) & - & - \\
\hline
\end{tabular}
\captionsetup{justification=centering}
\end{minipage}

\begin{threeparttable}
\begin{tablenotes}
\item[]\textsuperscript{1} $P_{\text{t}}$, $P_{\text{x}}$, $P_{\text{f}}$, $P_{\text{d}}$ denotes Partitioned versions of ResNet, ResNext, FCN, and DUC models, respectively.
\item[]\textsuperscript{2} L, M, S, and N symbols denote Large, Medium, Small, and Negligible effect sizes, respectively.
\item[]\textsuperscript{3} Empty cells indicate non-significant comparisons based on the Conover test.
\end{tablenotes}
\end{threeparttable}
\label{rq4_partition_statistical_results}
\end{table}

\textbf{The Edge Identity deployment strategy shows a 1.63x lower average median latency than the Mobile-Edge Partitioned strategy, which shows a 1.13x lower average median inference latency compared to the Mobile Identity deployment strategy.}

The Edge Identity deployment strategy outperforms the Mobile-Edge Partitioned strategy, exhibiting a speedup ranging from 1.26x to 2.02x, as shown in Figure~\ref{rq4_graph} (pink and dark blue boxplots), along with large effect sizes (Table~\ref{rq4_Partition_vs_Identity_statistical_results}). One possible explanation for this behavior is tier heterogeneity. In the Mobile-Edge deployment scenario, where distributed inference of partitioned models takes place, the tiers involved possess varying computation capabilities. Consequently, the processing speeds may vary, with slower tiers (such as Mobile) impacting the overall inference latency. This suggests that the use of a Partitioned model may not be necessary for scenarios where a monolithic deployment tier, such as Edge, is sufficiently capable of handling the computational load of the entire model.

In turn, the Mobile-Edge Partitioned strategy for all subjects (except DUC) achieves a speedup ranging from 1.11x to 1.26x compared to the Mobile Identity deployment strategy, as shown in Figure~\ref{rq4_graph} (dark blue and red box plots), along with medium to large effect sizes (Table~\ref{rq4_Partition_vs_Identity_statistical_results}). Given that the Mobile tier in our study has the lowest computational resources compared to the more powerful Edge tier, offloading half of the computational load to the Edge tier alleviates the burden on the Mobile tier, resulting in reduced inference latency. These findings suggest that deploying partitioned models across resource-constrained tiers (i.e., Mobile and Edge) is more effective than deploying the entire model solely on the Mobile tier, due to the distribution of computational load during inference. For the DUC subject, the Mobile-Edge Partitioned strategy shows a 1.05x lower median inference latency than the Mobile Identity deployment strategy, with a large effect size. This is possible because of the large intermediate data size (781.25Mb) of the DUC model during distributed inference, which led to transmission overhead, even across the high Mobile-Edge network bandwidth of 200 Mbps, leading to a slower latency than the Mobile tier.

\textbf{For ResNet/ResNext subjects, the Mobile-Edge, Edge-Cloud, and Mobile-Cloud strategies show 2.85x, 1.65, and 1.36 lower average median latency, respectively compared to the Cloud Identity deployment strategy. In contrast, in the case of FCN/DUC subjects, they show a 3.84x, 49.98x, and 52.14x higher average median inference latency.}

For ResNet/ResNext subjects, the Mobile-Edge, Edge-Cloud, and Mobile-Cloud Partitioned strategies achieve a speedup of 2.18x to 3.52x, 1.58x to 1.73x, and 1.25x to 1.48x, respectively, compared to the Cloud Identity deployment strategy, as illustrated in Figure~\ref{rq4_graph}, along with medium to large effect sizes (Table~\ref{rq4_Partition_vs_Identity_statistical_results}). The lower intermediate data size (6.12 Mb) in comparison to the input data size (8 to 60 Mb) of these two subjects speeds up their transmission across the Mobile, Edge, and Cloud tiers for the 3 Multi-tier Partitioned strategies. It suggests that Partitioned strategies can be a better alternative than Cloud deployment for subjects having intermediate data sizes lower than the input data.

For FCN/DUC subjects, the Mobile-Edge, Edge-Cloud, and Mobile-Cloud Partitioned strategies exhibit 1.67x to 6.01x, 40.38x to 59.59x, and 41.68x to 62.60x higher median inference latency, respectively, than the Cloud Identity deployment strategy, with large effect sizes (Table~\ref{rq4_Partition_vs_Identity_statistical_results}). For DUC, the intermediate data size (781.25 Mb) is much higher in comparison to the input data size (19-22 Mb), which led to its transmission overhead across both the Mobile-Edge (200 Mbps) and Edge-Cloud (1 Mbps) networks for the 3 Multi-tier Partitioned strategies. Conversely, for FCN, the intermediate data size varies between 118.12 Mb to 200 Mb, which is still much higher than their input data size (2-5 Mb), leading to transmission overhead. This is especially the case for the Edge-Cloud and Mobile-Cloud Partitioned strategies, which require transmission across the constrained Edge-Cloud network (1 Mbps). For the same subject, the slower latency of Mobile-Edge Partitioned than of Cloud deployment is majorly due to the computational advantage of the Cloud compared to the Mobile-Edge tier. This suggests that Cloud deployment can be a better alternative than Partitioned strategies for subjects having input data sizes smaller than the intermediate data. Moreover, for the FCN subject, which has the smallest input data sizes among the four subjects, its Cloud Identity deployment strategy also shows faster latency than its Mobile/Edge Identity deployment strategies, as explained in RQ1 findings (Section~\ref{sec:rq1}).

\textbf{The Edge Early Exit/Quantized deployment strategy shows 1.96x/2.41x lower average median latency than the Mobile-Edge Partitioned strategy at a medium/small accuracy loss.}

The Edge Early Exit/Quantized deployment strategy shows lower median inference latency than the Mobile-Edge Partitioned strategy, ranging from 1.67x to 2.35x/ 1.87x to 2.44 across the four subjects, as shown in Figure~\ref{rq2_rq3_rq4_rq5_graph} (blue, orange, green box-plots), along with large effect sizes (Table~\ref{rq2_rq3_rq4_rq5_edge_statistical_results}). In terms of accuracy, the Early Exit and Quantized operators show medium (2.53\% to 11.56\%) and small (0.05\% to 0.38\%) accuracy loss relative to Identity (or Partitioned) models as stated in RQ2 and RQ3 findings. This indicates that the Quantized and Early Exit operators at the Edge tier are a better alternative than the Partitioned operator at the Mobile-Edge tier in scenarios where sacrificing a small to medium level of accuracy may be acceptable to achieve faster latency. Table 19 shows a summary of results for RQ4.

\subsubsection{Normality Test}
To assess the applicability of parametric statistical tests, the Shapiro-Wilk test was conducted for each subject-model pair across ME, EC, and MC strategies. In all cases, p-values were less than 0.05, indicating deviations from normality (Table~\ref{rq4_descriptive_statistics}). This conclusion was further supported by the Q-Q plots (Table~\ref{normality_results_rq4}), which displayed noticeable divergence from the expected linear pattern. Therefore, non-parametric tests were employed in subsequent analyses.

\subsubsection{Hypothesis Testing}
All partitioning strategies (ME, EC, MC) differ statistically and practically in how they affect model performance in terms of pairwise comparisons using the Conover test, as shown in Table~\ref{rq4_partition_statistical_results}.

\begin{table}[htbp]
\centering
\caption{RQ4 results of the Cliff's Delta effect size and Conover test p-value between Multi-tier (Mobile-Edge [ME], Edge-Cloud [EC], Mobile-Cloud [MC]) Partitioned Strategies and Single-tier (Mobile [M], Edge [E], Cloud [C]) Monolithic Strategies.}
\vspace{0.3cm}

\begin{minipage}[t]{0.9\textwidth}
\centering
\begin{tabular}{|c|cc|cc|cc|}
\hline
\backslashbox{$P_{\text{t}}$}{$I_{\text{t}}$} & \multicolumn{2}{c|}{M} & \multicolumn{2}{c|}{E} & \multicolumn{2}{c|}{C} \\
\cline{2-7}
 & $p$ & $\delta$ & $p$ & $\delta$ & $p$ & $\delta$ \\
\hline
ME & $2.0e^{-36}$ & -M (0.37) & $9.4e^{-112}$ & L (0.83) & 0.0 & -L (0.95) \\
\hline
EC & $3.5e^{-29}$ & L (0.54) & 0.0 & L (0.96) & $2.0e^{-221}$ & -L (0.95) \\
\hline
MC & $1.8e^{-267}$ & L (0.73) & 0.0 & L (1.0) & $1.1e^{-12}$ & -M (0.33) \\
\hline
\end{tabular}
\captionsetup{justification=centering}
\end{minipage}

\vspace{0.5cm}

\begin{minipage}[t]{0.9\textwidth}
\centering
\begin{tabular}{|c|cc|cc|cc|}
\hline
\backslashbox{$P_{\text{x}}$}{$I_{\text{x}}$} & \multicolumn{2}{c|}{M} & \multicolumn{2}{c|}{E} & \multicolumn{2}{c|}{C} \\
\cline{2-7}
 & $p$ & $\delta$ & $p$ & $\delta$ & $p$ & $\delta$ \\
\hline
ME & $9.0e^{-77}$ & -L (0.63) & $9.1e^{-12}$ & L (0.53) & 0.0 & -L (1.0) \\
\hline
EC & $2.9e^{-36}$ & L (0.68) & $7.9e^{-266}$ & L (0.82) & $4.2e^{-241}$ & -L (1.0) \\
\hline
MC & $8.3e^{-211}$ & L (0.74) & 0.0 & L (0.84) & $1.7e^{-52}$ & -L (0.76) \\
\hline
\end{tabular}
\captionsetup{justification=centering}
\end{minipage}

\vspace{0.5cm}

\begin{minipage}[t]{0.9\textwidth}
\centering
\begin{tabular}{|c|cc|cc|cc|}
\hline
\backslashbox{$P_{\text{f}}$}{$I_{\text{f}}$} & \multicolumn{2}{c|}{M} & \multicolumn{2}{c|}{E} & \multicolumn{2}{c|}{C} \\
\cline{2-7}
 & $p$ & $\delta$ & $p$ & $\delta$ & $p$ & $\delta$ \\
\hline
ME & $2.0e^{-103}$ & -L (0.65) & $1.0e^{-269}$ & L (0.98) & 0.0 & L (1.0) \\
\hline
EC & 0.0 & L (1.0) & 0.0 & L (1.0) & 0.0 & L (1.0) \\
\hline
MC & 0.0 & L (1.0) & 0.0 & L (1.0) & 0.0 & L (1.0) \\
\hline
\end{tabular}
\captionsetup{justification=centering}
\end{minipage}

\vspace{0.5cm}

\begin{minipage}[t]{0.9\textwidth}
\centering
\begin{tabular}{|c|cc|cc|cc|}
\hline
\backslashbox{$P_{\text{d}}$}{$I_{\text{d}}$} & \multicolumn{2}{c|}{M} & \multicolumn{2}{c|}{E} & \multicolumn{2}{c|}{C} \\
\cline{2-7}
 & $p$ & $\delta$ & $p$ & $\delta$ & $p$ & $\delta$ \\
\hline
ME & 0.0 & L (0.98) & 0.0 & L (1.0) & 0.0 & L (1.0) \\
\hline
EC & 0.0 & L (1.0) & 0.0 & L (1.0) & 0.0 & L (1.0) \\
\hline
MC & 0.0 & L (1.0) & 0.0 & L (1.0) & 0.0 & L (1.0) \\
\hline
\end{tabular}
\captionsetup{justification=centering}
\end{minipage}

\vspace{0.3cm}

\begin{threeparttable}
\begin{tablenotes}
\item[]\textsuperscript{1} $P_{\text{t}}$, $P_{\text{x}}$, $P_{\text{f}}$, $P_{\text{d}}$ denote Partitioned versions of ResNet, ResNext, FCN, and DUC respectively.
\item[]\textsuperscript{2} $I_{\text{t}}$, $I_{\text{x}}$, $I_{\text{f}}$, $I_{\text{d}}$ denote Identity versions of ResNet, ResNext, FCN, and DUC respectively.
\item[]\textsuperscript{3} A Positive sign for the $i^{th}$ cell means latency of column[i] $<$ row[i], and Negative means column[i] $>$ row[i].
\item[]\textsuperscript{4} L, M, S, and N symbols mean Large, Medium, Small, and Negligible effect sizes, respectively.
\item[]\textsuperscript{5} Empty cells (not shown) would indicate non-significant results based on the Conover test ($p > 0.05$).
\end{tablenotes}
\end{threeparttable}

\label{rq4_Partition_vs_Identity_statistical_results}
\end{table}

\begin{Summary}{}{fourthsummary}
The Edge Identity/Early Exit/Quantized deployment strategy shows faster latency (1.63x/1.96x/2.41x) at no/medium/small accuracy loss than the ME Partitioned strategy, which exhibits faster latency (1.13x) compared to the Mobile Identity deployment strategy (1.13x), EC/MC Partitioned strategy (9.37x/9.86x), and Cloud Identity deployment strategy (2.85x for ResNet/ResNext) in deployment scenarios where factors like input/intermediate data size and computational/network resources play a crucial role. 

In scenarios where the subjects have smaller input data sizes (i.e., FCN) such that their transmission across the bandwidth-constrained Cloud tier is not a major concern, their monolithic Cloud Identity deployment is much more effective than their Multi-tier Partitioned strategies and Edge/Mobile Identity deployment strategies. 

\end{Summary}

\begin{table}[h!]
\centering
\begin{tabular}{|l|c|c|p{6.5cm}|}
\hline
\textbf{Model} & \textbf{Best Partitioning Tier} & \textbf{Latency Gain} & \textbf{Notes} \\
\hline
ResNet  & Mobile-Edge & 1.13x faster than Mobile & 9.37x faster than EC, 9.86x faster than MC; Edge Identity is still faster (1.63x) \\
ResNext & Mobile-Edge & 1.13x faster than Mobile & Similar trend as ResNet; large intermediate data size limits EC/MC strategies \\
FCN     & Cloud       & Cloud outperforms ME & Small input size makes Cloud preferable over any partitioned strategy \\
DUC     & Edge        & Edge Identity faster (1.96x) & Intermediate data too large (781MB) to benefit from partitioning \\
\hline
\end{tabular}
\caption{Summary of results for Research Question 4}
\label{tab:partition_rq4}
\end{table}

\subsection{What is the impact of Hybrid operators in terms of inference latency and accuracy within and across the considered tiers? (RQ5)}
\subsubsection{Data Exploration}
\label{sec:rq5}
\subsubsection{Quantitative Analysis of Quantized Early Exit operator on monolithic deployment tiers}
\label{sec:rq5_1}
\textbf{In Mobile, for two subjects (FCN/DUC), the Quantized Early Exit models show 1.45x, 1.13x, and 1.35x lower average median inference latency than the Identity, Quantized, and Early Exit models, respectively. In Edge, for all subjects, the Quantized Early Exit models show 1.75x, 1.17x, and 1.45 lower average median inference latency than the Identity, Quantized, and Early Exit models, respectively. In Cloud, the Quantized Early Exit models show no significant difference from the Identity, Quantized, and Early Exit models for all subjects (except DUC).}

In the Mobile tier, for FCN and DUC, the Quantized Early Exit models show 1.29x to 1.62x and 1.16x to 1.54x lower median inference latency than the Identity and Early Exit models, respectively (Figure~\ref{rq2_rq3_rq4_rq5_graph}) along with large effect sizes (Table~\ref{rq2_rq3_rq4_rq5_mobile_statistical_results}, Figure~\ref{rq4_descriptive_statistics}, Figure~\ref{rq1_descriptive_statistics}, Figure~\ref{rq3_descriptive_statistics}). Conversely, for the ResNet/ResNext subject, the Quantized Early Exit models show 1.07x lower/1.05x higher (small effect sizes) and 1.11x (negligible effect size)/1.33x (large effect size) higher median inference latency than the Identity and Early Exit models, respectively. Even though the Quantized Early Exit models have 4.61x to 5.04x and 3.90x to 3.98x lower size than the Identity and Early Exit models respectively, still they show slower latency for ResNet, ResNext, or both during Mobile deployment, which indicates that these two subjects' Quantization Operations are costly in lower Memory/CPU environments, as mentioned in RQ2 findings (Section~\ref{sec:rq2}). Moreover, the Quantized Early Exit models show 1.09x to 1.28x lower median inference latency than the Quantized models across the four subjects, with small to large effect sizes. This is possibly due to the addition of early exiting, which reduces computations during inference in comparison to a Quantized variant without any early exit similar to RQ3 findings (Section~\ref{sec:rq3}). 

During Edge deployment, across the four subjects, the Quantized Early Exit models exhibit 1.36x to 1.97x, 1.13x to 1.20x, 1.17x to 1.82x lower median inference latency than the Identity, Quantized, and Early Exit models, respectively, as depicted in Figure~\ref{rq2_rq3_rq4_rq5_graph}, with large effect sizes (Table~\ref{rq2_rq3_rq4_rq5_edge_statistical_results}). This suggests that the Quantized Early Exit models exhibit greater robustness on the Edge tier in comparison to the Mobile tier. The higher computational resource on the Edge tier is likely a contributing factor to this outcome.

For the DUC subject, the Quantized Early Exit models show slower latency than the Identity and Early Exit models with large effect sizes (Table~\ref{rq2_rq3_rq4_rq5_cloud_statistical_results}). We believe that this is due to the lowest percentage of graphical nodes processed with the CUDA Execution Provider for DUC during Cloud deployment, similar to the reasoning explained briefly for the RQ2 results (see first finding in Section~\ref{sec:rq2}). For the same subject, the Quantized Early Exit model shows faster latency than the Quantized model, but with a small effect size, possibly due to the minor influence of Early Exiting.

\textbf{The Quantized Early Exit models show a medium drop in accuracy relative to Identity, Quantized, and Early Exit models.}

In Table~\ref{accuracy}, an accuracy drop of 2.62\% to 12.41\%, 2.56\% to 12.02\%, 0.09\% to 1.66\% is observed when comparing the performance of the Quantized Early Exit models with the Identity, Quantized, and Early Exit models, respectively. These accuracy differences are statistically significant (Wilcoxon Test: p-value = 7.6e$^{-6}$, $\alpha = 0.0125$), with medium effect sizes. For applications prioritizing real-time inference in the Edge tier and willing to accept a medium decrease in accuracy compared to the Identity, Quantized, and Early Exit models, the utilization of a Hybrid (Quantized Early Exit) model may be a suitable choice.

\textbf{When comparing Quantized Early Exit across the three monolithic deployment tiers, the Quantized Early Exit models during Edge deployment show 2.63x and 3.89x lower average median inference latency than their deployment in Mobile and Cloud tiers, respectively.}

The Quantized Early Exit models in the Edge tier demonstrate 2.03x to 2.90x lower median inference latency compared to the Mobile tier as shown in Figure~\ref{rq2_rq3_rq4_rq5_graph} (red box plots), with large effect sizes (Table~\ref{rq5_Quantized_earlyexit_statistical_results}) due to Edge's higher computational resources. Among the Edge and Cloud tiers, the Quantized Early Exit model's median inference latency under-performs by 1.77x to 7.69x during Cloud deployment for three subjects (i.e., ResNet, ResNext, DUC), along with a large effect size (Table~\ref{rq5_Quantized_earlyexit_statistical_results}). This is due to the higher impact of large input data sizes of these subjects on the transmission across the restricted Edge-Cloud network during Cloud deployment. Conversely, for the FCN subject, the Quantized Early Exit model exhibits 1.93x lower median inference latency in the Cloud tier compared to the Edge tier, with a large effect size (Table~\ref{rq5_Quantized_earlyexit_statistical_results}). Similarly, this is due to FCN's small input data sizes, leading to a lower impact on data transmission across the constrained Edge-Cloud network during Cloud deployment. The reasoning behind these findings is similar to and explained briefly in the RQ1 findings (Section~\ref{sec:rq1}).

\begin{table}[htbp]
\centering
\caption{RQ5 Results of the Cliff's Delta effect size and Conover test p-value between Mobile (M), Edge (E), and Cloud (C) deployment for Quantized Early Exit models}
\begin{minipage}[t]{0.9\textwidth}
\centering
\begin{tabular}{|c|cc|cc|cc|}
\hline
\backslashbox{$QE_{\text{x}}$}{$QE_{\text{t}}$} & \multicolumn{2}{c|}{M} & \multicolumn{2}{c|}{E} & \multicolumn{2}{c|}{C} \\
\cline{2-7}
 & $p$ & $\delta$ & $p$ & $\delta$ & $p$ & $\delta$ \\
\hline
M & - & - & $2.5e^{-304}$ & L (0.99) & $9.4e^{-260}$ & -L (0.91) \\
\hline
E & 0.0 & L (1.0) & - & - & 0.0 & -L (1.0) \\
\hline
C & 0.0 & -L (1.0) & 0.0 & -L (1.0) & - & - \\
\hline
\end{tabular}
\centering
\vspace{0.5cm}
\captionsetup{justification=centering}
\end{minipage}
\hfill
\begin{minipage}[t]{0.9\textwidth}
\centering
\begin{tabular}{|c|cc|cc|cc|}
\hline
\backslashbox{$QE_{\text{d}}$}{$QE_{\text{f}}$} & \multicolumn{2}{c|}{M} & \multicolumn{2}{c|}{E} & \multicolumn{2}{c|}{C} \\
\cline{2-7}
 & $p$ & $\delta$ & $p$ & $\delta$ & $p$ & $\delta$ \\
\hline
M & - & - & 0.0 & L (1.0) & 0.0 & L (1.0) \\
\hline
E & 0.0 & L (1.0) & - & - & 0.0 & L (0.97) \\
\hline
C & $8.7e^{-316}$ & L (0.97) & 0.0 & -L (1.0) & - & - \\
\hline
\end{tabular}
\captionsetup{justification=centering}
\end{minipage}

\begin{threeparttable}
\begin{tablenotes}
\item[]\textsuperscript{1} $QE_{\text{t}}$, $QE_{\text{x}}$, $QE_{\text{f}}$, and $QE_{\text{d}}$ denote Quantized Early Exit versions of ResNet, ResNext, FCN, and DUC, respectively.
\item[]\textsuperscript{2} L, M, S, and N symbols denote Large, Medium, Small, and Negligible effect sizes, respectively.
\item[]\textsuperscript{3} Empty cells indicate non-significant comparisons based on the Conover test.
\end{tablenotes}
\end{threeparttable}
\label{rq5_Quantized_earlyexit_statistical_results}
\end{table}

\subsubsection{Quantitative Analysis of Quantized Early Exit Partitioned Strategy and its Comparison with monolithic deployment Strategies}
\label{sec:rq5_2}

\begin{figure}[t]
\centering
\includegraphics[width=1\textwidth]{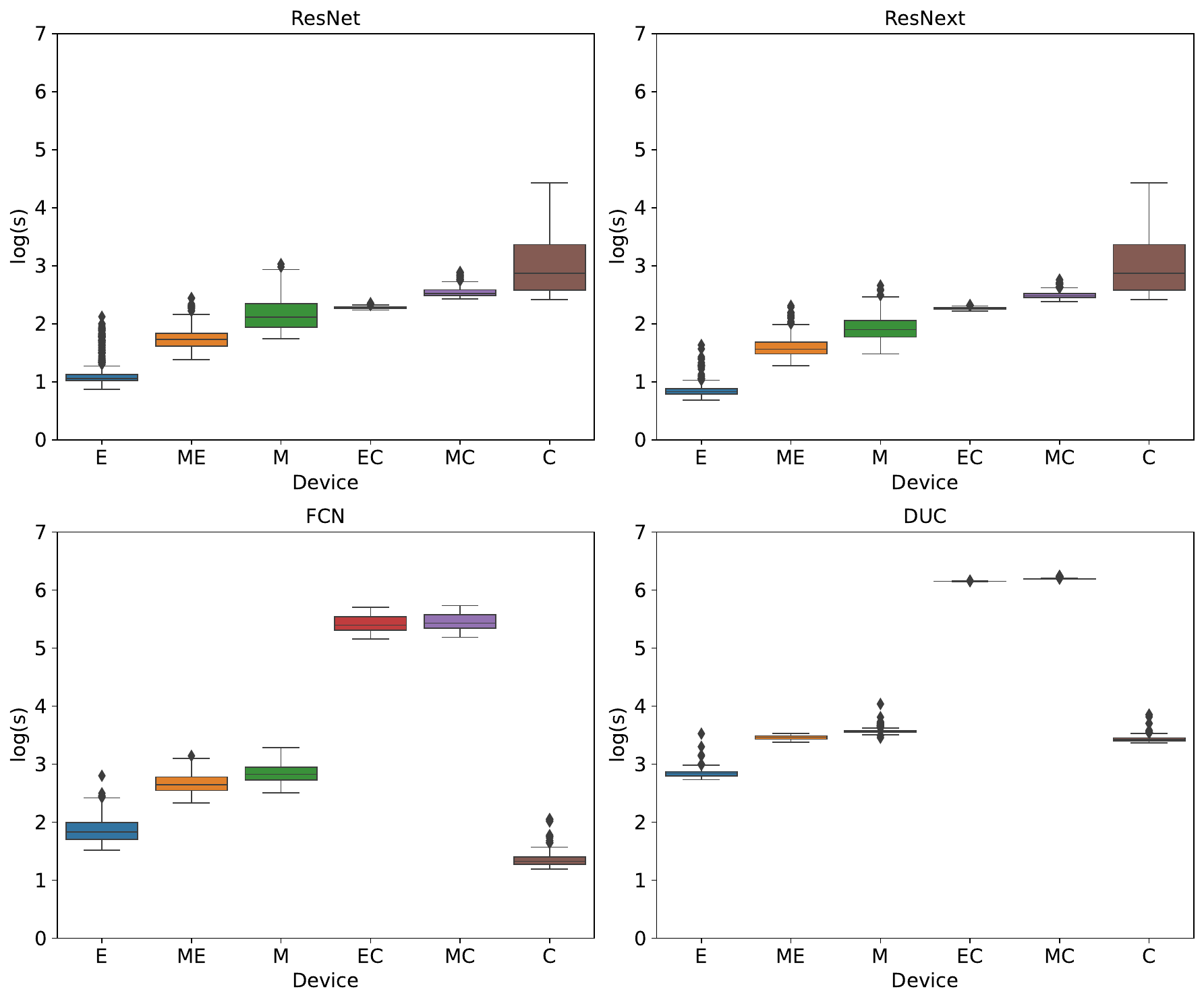}
\caption{Box plots of the inference latency measure for Multi-tier Quantized Early Exit Partitioned strategies and monolithic Quantized Early Exit deployment strategies involving 6 deployment tiers (i.e., Mobile-Edge [ME], Edge-Cloud [EC], Mobile-Cloud [MC], Mobile [M], Edge [E], Cloud [C])}
\label{rq5_graph}
\end{figure}

\textbf{Across the three Multi-tier Quantized Early Exit Partitioned strategies, the Mobile-Edge Quantized Early Exit Partitioned strategy shows an 8.51x and 9.04x lower average median inference latency than Edge-Cloud and Mobile-Cloud Quantized Early Exit Partitioned strategies, respectively.}

The Mobile-Edge Quantized Early Exit Partitioned strategy accounts for 1.73x to 15.65x and 2.21x to 16.20x lower median inference latency than the Edge-Cloud and Mobile-Cloud Quantized Early Exit Partitioned strategies, respectively (Figure~\ref{rq5_graph}), with large effect size (Table~\ref{rq5_partitioned_quantized_earlyexit_statistical_results}). On the other hand, the Edge-Cloud Quantized Early Exit Partitioned strategy shows 1.04x to 1.28x lower median inference latency than the Mobile-Cloud Quantized Early Exit Partitioned strategy as shown in Figure~\ref{rq5_graph}, along with small or large effect sizes (Table~\ref{rq5_partitioned_quantized_earlyexit_statistical_results}). This is due to similar reasons as explained in the RQ4 results (see first finding in Section~\ref{sec:rq4}).

\begin{table}[htbp]
\centering
\caption{RQ5 Results of the Cliff's Delta effect size and Conover test p-value between Mobile-Edge (ME), Edge-Cloud (EC), and Mobile-Cloud (MC) for Quantized Early Exit Partitioned Strategies}
\begin{minipage}[t]{0.9\textwidth}
\centering
\begin{tabular}{|c|cc|cc|cc|}
\hline
\backslashbox{$QEP_{\text{x}}$}{$QEP_{\text{t}}$} & \multicolumn{2}{c|}{ME} & \multicolumn{2}{c|}{EC} & \multicolumn{2}{c|}{MC} \\
\cline{2-7}
 & $p$ & $\delta$ & $p$ & $\delta$ & $p$ & $\delta$ \\
\hline
ME & - & - & $3.6e^{-305}$ & -L (-0.97) & 0.0 & -L (1.0) \\
\hline
EC & 0.0 & -L (0.99) & - & - & $7.0e^{-241}$ & -L (1.0) \\
\hline
MC & 0.0 & -L (1.0) & $1e^{-323}$ & -L (1.0) & - & - \\
\hline
\end{tabular}
\centering
\vspace{0.5cm}
\captionsetup{justification=centering}
\end{minipage}
\vspace{0.7cm}

\begin{minipage}[t]{0.9\textwidth}
\centering
\begin{tabular}{|c|cc|cc|cc|}
\hline
\backslashbox{$QEP_{\text{d}}$}{$QEP_{\text{f}}$} & \multicolumn{2}{c|}{ME} & \multicolumn{2}{c|}{EC} & \multicolumn{2}{c|}{MC} \\
\cline{2-7}
 & $p$ & $\delta$ & $p$ & $\delta$ & $p$ & $\delta$ \\
\hline
ME & - & - & 0.0 & -L (1.0) & 0.0 & -L (1.0) \\
\hline
EC & 0.0 & -L (1.0) & - & - & $3.8e^{-16}$ & -S (0.25) \\
\hline
MC & 0.0 & -L (1.0) & $1.5e^{-268}$ & -L (1.0) & - & - \\
\hline
\end{tabular}
\captionsetup{justification=centering}
\end{minipage}

\begin{threeparttable}
\begin{tablenotes}
\item[]\textsuperscript{1} $QEP_{\text{t}}$, $QEP_{\text{x}}$, $QEP_{\text{f}}$, and $QEP_{\text{d}}$ denote Quantized Early Exit Partitioned versions of ResNet, ResNext, FCN, and DUC, respectively.
\item[]\textsuperscript{2} L, M, S, and N symbols denote Large, Medium, Small, and Negligible effect sizes, respectively.
\item[]\textsuperscript{3} All p-values shown are statistically significant ($p < 0.05$); empty cells denote non-significant comparisons.
\end{tablenotes}
\end{threeparttable}
\label{rq5_partitioned_quantized_earlyexit_statistical_results}
\end{table}

\textbf{The Edge Quantized Early Exit deployment strategy shows 2.03x lower average median inference latency than the Mobile-Edge Quantized Early Exit Partitioned strategy, which shows 1.29x lower average median inference latency than the Mobile Quantized Early Exit deployment strategy.}

The Edge Quantized Early Exit strategy outperforms the Mobile-Edge Quantized Early Exit Partitioned strategy with large effect sizes (Table~\ref{rq5_Partition_vs_Identity_statistical_results}), exhibiting a speedup ranging from 1.85x to 1.94x, as shown in Figure~\ref{rq5_graph}. In turn, the Mobile-Edge Quantized Early Exit Partitioned strategy shows 1.09x to 1.47x lower median inference latency than the Mobile Quantized Early Exit strategy (large effect sizes) as shown in Figure~\ref{rq5_graph} and Table~\ref{rq5_Partition_vs_Identity_statistical_results}. In general, this suggests that distributed model deployment across Mobile and Edge tiers can be an optimal choice compared to monolithic model deployment in a resource-constrained environment (such as mobile) when faster inference is a concern at no accuracy loss. This is due to similar reasons provided in RQ4 results (see second finding in Section~\ref{sec:rq4}).

\textbf{For ResNet/ResNext, the Mobile-Edge, Edge-Cloud, and Mobile-Cloud Quantized Early Exit Partitioned strategies show 3.41x, 1.82x, and 1.44x lower average median latency, respectively than the Cloud Quantized Early Exit deployment strategy. In contrast, in the case of FCN/DUC, they show 2.39x, 36.92x, and 38.28x higher average median inference latency.}

For ResNet and ResNext, the Mobile-Edge, Edge-Cloud, and Mobile-Cloud Quantized Early Exit Partitioned strategies demonstrate a lower median inference latency of 3.12x to 3.70x, 1.80x to 1.84x, and 1.40x to 1.48x, respectively than the Cloud Quantized Early Exit strategy as shown in Figure~\ref{rq5_graph}, along with large effect sizes (Table~\ref{rq5_Partition_vs_Identity_statistical_results}). Conversely, for FCN and DUC, the same strategies exhibit a 1.04x to 3.74x, 15.28x to 58.57x, and 15.91x to 60.65x higher median inference latency, along with medium to large effect sizes (Table~\ref{rq5_Partition_vs_Identity_statistical_results}). A possible explanation for these results is similar to the RQ4 results (see third finding in Section~\ref{sec:rq4}). Moreover, for the FCN subject, which has the lowest input data sizes among the four subjects, its Cloud Quantized Early Exit deployment strategy also shows faster latency than its Mobile/Edge Quantized Early Exit deployment strategies, as explained previously in RQ5 findings (Section~\ref{sec:rq5}).

\textbf{The Edge Identity/Quantized/Early Exit deployment strategy shows 1.17x/1.72x/1.42x lower average median latency at medium accuracy gain when compared with Mobile-Edge Quantized Early Exit Partitioned strategy, which shows 1.39x lower average median latency than the Mobile-Edge Partitioned strategy at medium accuracy loss.}

The Edge Identity/Quantized/Early Exit deployment strategies show 1x to 1.35x/ 1.61x to 1.88x/ 1.24x to 1.59x lower median inference latency than the Mobile-Edge Quantized Early Exit Partitioned strategy across the four subjects, as shown in Figure~\ref{rq2_rq3_rq4_rq5_graph} (blue, orange, green, red box-plots), along with medium to large effect sizes (Table~\ref{rq2_rq3_rq4_rq5_edge_statistical_results}). In terms of accuracy, the Identity, Quantized, and Early Exit models show medium accuracy gain (i.e., 2.62\% to 12.41\%, 2.56\% to 12.02\%, 0.09\% to 1.66\%) relative to Quantized Early Exit (or Quantized Early Exit Partitioned) models as stated previously in the second finding of RQ5 (Section~\ref{sec:rq5_1}). This indicates that the Non-Hybrid operators (Identity, Quantized, and Early Exit) at the Edge tier are a better alternative than the Hybrid Quantized Early Exit Partitioned operator at the Mobile-Edge tier in scenarios where maximizing both latency and accuracy are of utmost importance. Moreover, the Mobile-Edge Quantized Early Exit Partitioned strategy shows 1.04x to 1.60x lower median inference latency with large effect sizes (Table~\ref{rq2_rq3_rq4_rq5_edge_statistical_results}) than the Mobile-Edge Partitioned strategy but at a cost of medium accuracy loss (2.62\% to 12.41\%). Table 23 shows a summary of results for RQ5.

\begin{table}[htbp]
  \centering
  \caption{RQ5 Results of the Cliff's Delta effect size and Conover test p-value between Multi-tier (Mobile-Edge [ME], Edge-Cloud [EC], Mobile-Cloud [MC]) Quantized Early Exit Partitioned Strategies and Single-tier (Mobile [M], Edge [E], Cloud [C]) Quantized Early Exit strategies}
  
  \begin{minipage}[t]{0.9\textwidth}
    \centering
    \begin{tabular}{|c|cc|cc|cc|}
      \hline
      \backslashbox{$QEP_{\text{t}}$}{$QE_{\text{t}}$} & \multicolumn{2}{c|}{M} & \multicolumn{2}{c|}{E} & \multicolumn{2}{c|}{C} \\
      \cline{2-7}
       & $p$ & $\delta$ & $p$ & $\delta$ & $p$ & $\delta$ \\
      \hline
      ME & $7.9e^{-228}$ & -L (0.85) & $3.7e^{-130}$ & L (0.90) & 0.0 & -L (1.0) \\
      \hline
      EC & $5.9e^{-12}$ & M (0.35) & 0.0 & L (1.0) & 0.0 & -L (1.0) \\
      \hline
      MC & $5.6e^{-319}$ & L (0.78) & 0.0 & L (1.0) & $6.3e^{-73}$ & -L (0.65) \\
      \hline
    \end{tabular}
  \end{minipage}

  \vspace{0.5cm}

  \begin{minipage}[t]{0.9\textwidth}
    \centering
    \begin{tabular}{|c|cc|cc|cc|}
      \hline
      \backslashbox{$QEP_{\text{x}}$}{$QE_{\text{x}}$} & \multicolumn{2}{c|}{M} & \multicolumn{2}{c|}{E} & \multicolumn{2}{c|}{C} \\
      \cline{2-7}
       & $p$ & $\delta$ & $p$ & $\delta$ & $p$ & $\delta$ \\
      \hline
      ME & $1.0e^{-213}$ & -L (0.82) & $5.4e^{-287}$ & L (0.99) & 0.0 & -L (1.0) \\
      \hline
      EC & $7.7e^{-218}$ & L (0.85) & 0.0 & L (1.0) & 0.0 & -L (1.0) \\
      \hline
      MC & 0.0 & L (0.98) & 0.0 & L (1.0) & $1.4e^{-172}$ & -L (0.80) \\
      \hline
    \end{tabular}
  \end{minipage}

  \vspace{0.5cm}

  \begin{minipage}[t]{0.9\textwidth}
    \centering
    \begin{tabular}{|c|cc|cc|cc|}
      \hline
      \backslashbox{$QEP_{\text{f}}$}{$QE_{\text{f}}$} & \multicolumn{2}{c|}{M} & \multicolumn{2}{c|}{E} & \multicolumn{2}{c|}{C} \\
      \cline{2-7}
       & $p$ & $\delta$ & $p$ & $\delta$ & $p$ & $\delta$ \\
      \hline
      ME & $1.1e^{-76}$ & -L (0.57) & $2.3e^{-290}$ & L (1.0) & 0.0 & L (1.0) \\
      \hline
      EC & 0.0 & L (1.0) & 0.0 & L (1.0) & 0.0 & L (1.0) \\
      \hline
      MC & 0.0 & L (1.0) & 0.0 & L (1.0) & 0.0 & L (1.0) \\
      \hline
    \end{tabular}
  \end{minipage}

  \vspace{0.5cm}

  \begin{minipage}[t]{0.9\textwidth}
    \centering
    \begin{tabular}{|c|cc|cc|cc|}
      \hline
      \backslashbox{$QEP_{\text{d}}$}{$QE_{\text{d}}$} & \multicolumn{2}{c|}{M} & \multicolumn{2}{c|}{E} & \multicolumn{2}{c|}{C} \\
      \cline{2-7}
       & $p$ & $\delta$ & $p$ & $\delta$ & $p$ & $\delta$ \\
      \hline
      ME & 0.0 & -L (0.99) & 0.0 & L (1.0) & $4.2e^{-55}$ & M (0.42) \\
      \hline
      EC & $2.1e^{-277}$ & L (1.0) & 0.0 & L (1.0) & 0.0 & L (1.0) \\
      \hline
      MC & 0.0 & L (1.0) & 0.0 & L (1.0) & 0.0 & L (1.0) \\
      \hline
    \end{tabular}
  \end{minipage}

  \begin{threeparttable}
    \begin{tablenotes}
      \item[]\textsuperscript{1} $QEP_{\text{t}}$, $QEP_{\text{x}}$, $QEP_{\text{f}}$, $QEP_{\text{d}}$ denote Quantized Early Exit Partitioned versions of ResNet, ResNext, FCN, and DUC, respectively.
      \item[]\textsuperscript{2} $QE_{\text{t}}$, $QE_{\text{x}}$, $QE_{\text{f}}$, $QE_{\text{d}}$ denote single-tier Quantized Early Exit versions of the same models.
      \item[]\textsuperscript{3} L, M, S, and N denote Large, Medium, Small, and Negligible Cliff’s Delta effect sizes.
      \item[]\textsuperscript{4} All p-values shown are significant ($p < 0.05$).
    \end{tablenotes}
  \end{threeparttable}

  \label{rq5_Partition_vs_Identity_statistical_results}
\end{table}

\subsubsection{Normality Test}
The Shapiro-Wilk test and QQ plots (Table~\ref{rq5_descriptive_statistics}, Figure~\ref{normality_results_rq5_1}, Figure~\ref{normality_results_rq5_2}) suggest that latency data for most deployment tiers and model variants deviate from normality. P-values are $<$ 0.05 across all cases. This non-normality validates the use of non-parametric tests (Conover and Cliff's Delta).

\subsubsection{Hypothesis Testing}
According to the Conover test, the null hypothesis that there is no significant difference between Quantized Early Exit and Identity, Quantized Early Exit and Quantized, Quantized Early Exit and Early Exit models in the Cloud cannot be rejected for three subjects (i.e., ResNet, ResNext, FCN), indicating their similar or equivalent latency performance in the Cloud. Due to the powerful computing resources of the Cloud, the impact of the Quantized Early Exit models is not significant compared to the Identity, Quantized, and Early Exit models. However, for the DUC subject, the null hypothesis was rejected with p-values of 8.9e$^{-221}$, 6.8e$^{-10}$ and 1.5e$^{-196}$ ($\alpha$ =0.05) for Quantized Early Exit vs Identity, Quantized Early Exit vs Quantized, and Quantized Early Exit vs Early Exit comparison, respectively, indicating significant inference latency difference between them.

\begin{Summary}{}{fifthsummary}
The Quantized Early Exit operator shows 1.75x/1.17x/1.45 faster latency than the Identity/Quantized/Early Exit operator in the Edge tier at medium accuracy loss (up to 12.41\%/12.02\%/1.66\%).

The Edge deployment of Hybrid (Quantized Early Exit) and Non-Hybrid (Identity/Quantized/Early Exit) operators show 2.03x and 1.17x/1.72x/1.42x faster latency at no accuracy loss and medium (up to 12.41\%/12.02\%/1.66\%) accuracy gain, respectively than the ME Quantized Early Exit Partitioned strategy, which shows faster latency than the ME Partitioned strategy (1.39x at medium accuracy loss), EC/MC Quantized Early Exit Partitioned (8.51x/9.04x), Mobile Quantized Early Exit (1.29x), \& Cloud Quantized Early Exit (3.41x for ResNet/ResNext) strategies for scenarios influenced by input/intermediate data size of the subjects and computational/bandwidth resources of the Mobile, Edge, and Cloud tiers.

In scenarios where the subjects have smaller input data sizes (i.e., FCN) such that their transmission across the bandwidth-constrained Cloud tier is not a major concern, their monolithic Cloud Quantized Early Exit deployment is much more effective than their Multi-tier Quantized Early Exit Partitioned strategies and Edge/Mobile Quantized Early Exit deployment strategies. 

\end{Summary}

\begin{table}[h!]
\centering
\caption{Summary of results for Research Question 5}
\label{tab:rq5_result_summarization}

\resizebox{\textwidth}{!}{%
\begin{tabular}{|l|c|c|c|p{6.5cm}|}
\hline
\textbf{Model} & \textbf{Best Tier} & \textbf{Latency Gain} & \textbf{Accuracy Drop} & \textbf{Notes} \\ \hline
ResNet  & Edge & 1.75x vs Identity; 2.03x vs ME-QEP & 12.41\% & Best performance on Edge. ME-QEP faster than ME-Part (1.39x) \\
ResNext & Edge & 1.17x–1.72x vs Non-Hybrids; 2.03x vs ME-QEP & 12.02\% & Similar trends to ResNet; large effect sizes in latency gains \\
FCN     & Cloud & 1.93x faster than Edge for QE & 2.62\% & Cloud deployment preferable due to small input size \\
DUC     & Edge & 1.45x faster vs Early Exit; 2.03x vs ME-QEP & 1.66\% & High input size favors Edge deployment; ME-QEP adds moderate accuracy loss\\
\hline
\end{tabular}%
}
\end{table}

\subsection{What is the impact of network bandwidth variations on the deployment strategies in terms of inference latency? (RQ6)}
\label{sec:rq6}
\subsubsection{Data Exploration}
\textbf{Deploying smaller input data-sized models (FCN), is well-suited for Cloud tier with lower bandwidth ($\leq$10 Mbps). Larger input data-sized models (ResNe(x)t and DUC), perform better in Cloud deployments having moderate to high bandwidth ($\geq$50 Mbps).}
\begin{figure}[htbp!]
\centering
\includegraphics[width=1\textwidth]{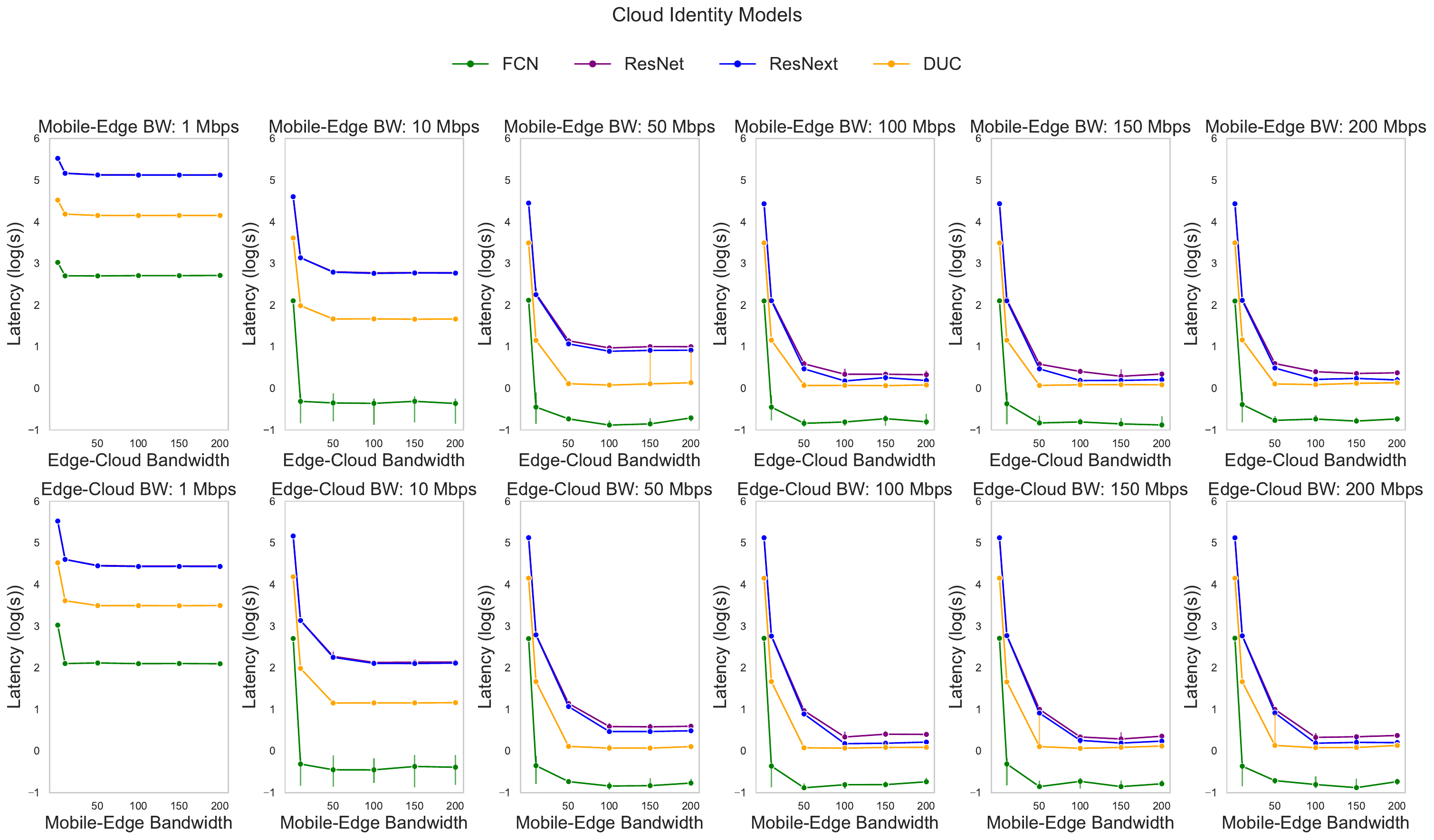}
\caption{Line graphs of the measures collected for inference latency vs network bandwidth of Cloud Identity models.}
\label{cloud_latency_Identity}
\end{figure}

Figure~\ref{cloud_latency_Identity} shows the line graphs of the measures collected for the latency of Identity models in the Cloud across the four subjects (i.e., FCN, ResNet, ResNext, DUC) for various network bandwidths. For the other models (Quantized, Early Exit, and Quantized Early Exit), similar line graphs were observed and therefore added to the replication package~\footnote{https://github.com/SAILResearch/wip-24-jaskirat-black-box-edge-operators}. As both Mobile-Edge and Edge-Cloud bandwidth increases, latency tends to drop until either 50 or 100 Mbps, then stays constant. The latency performance gap between the models narrows as the bandwidth increases, indicating the models perform more similarly at higher bandwidth levels ($\geq$100 Mbps). For all cases, a steep latency drop is observed initially when bandwidth increases from 1 to 50 Mbps. Beyond 100 Mbps, the latency changes become steady. 

FCN (green line) consistently shows the lowest latency across all scenarios in the Cloud, indicating its efficiency in handling workloads under varying bandwidth conditions when comparing the different plots in a row. 
This is because the FCN’s input data size (5 Mb) is the smallest one across all the datasets. 
This suggests that deploying models with small input sizes like FCN is ideal for Cloud deployment even in bandwidth-constrained settings ($\leq$10 Mbps), which complements the RQ1-5 findings for FCN-based Cloud operators. ResNext and ResNet (blue and purple lines) models tend to have the highest latency among the models, especially in low network bandwidth conditions (i.e., $\leq$10 Mbps), which again aligns with the RQ1-5 findings. As the input data size of ResNe(x)t (i.e., 60 Mb) is much higher than the network-constrained bandwidth (i.e.,$\leq$10 Mbps), its latency is higher compared to other models at $\leq$10 Mbps. The DUC (orange line) demonstrates moderate latency compared to other models, such as FCN and ResNe(x)t in network-constrained scenarios (i.e., 10 Mbps). The DUC and ResNe(x)t models' latency performance reduces sharply as the bandwidth increases from 1 Mbps to 50 Mbps, eventually reaching a plateau at 50 or 100 Mbps, indicating a steady state. The input data size of DUC (22 Mb) and ResNe(x)t (60 Mb) exceeds the 10 Mbps bandwidth but falls below the 50 Mbps bandwidth (for DUC) and is slightly above the 50 Mbps bandwidth (for ResNe(x)t). As a result, latency decreases progressively for ResNe(x)t and DUC models as the bandwidth rises from 10 Mbps to 50 Mbps. This suggests that the ResNe(x)t and DUC models are suitable for Cloud deployment that operates in scenarios with at least moderate bandwidth availability ($\geq$50 Mbps) across Mobile-Edge and Edge-Cloud networks, but might not be ideal for constrained bandwidth settings ($\leq$10 Mbps), which complements the RQ1-5 findings for ResNe(x)t/DUC-based cloud models.

\begin{figure}[htbp!]
\centering
\includegraphics[width=1\textwidth]{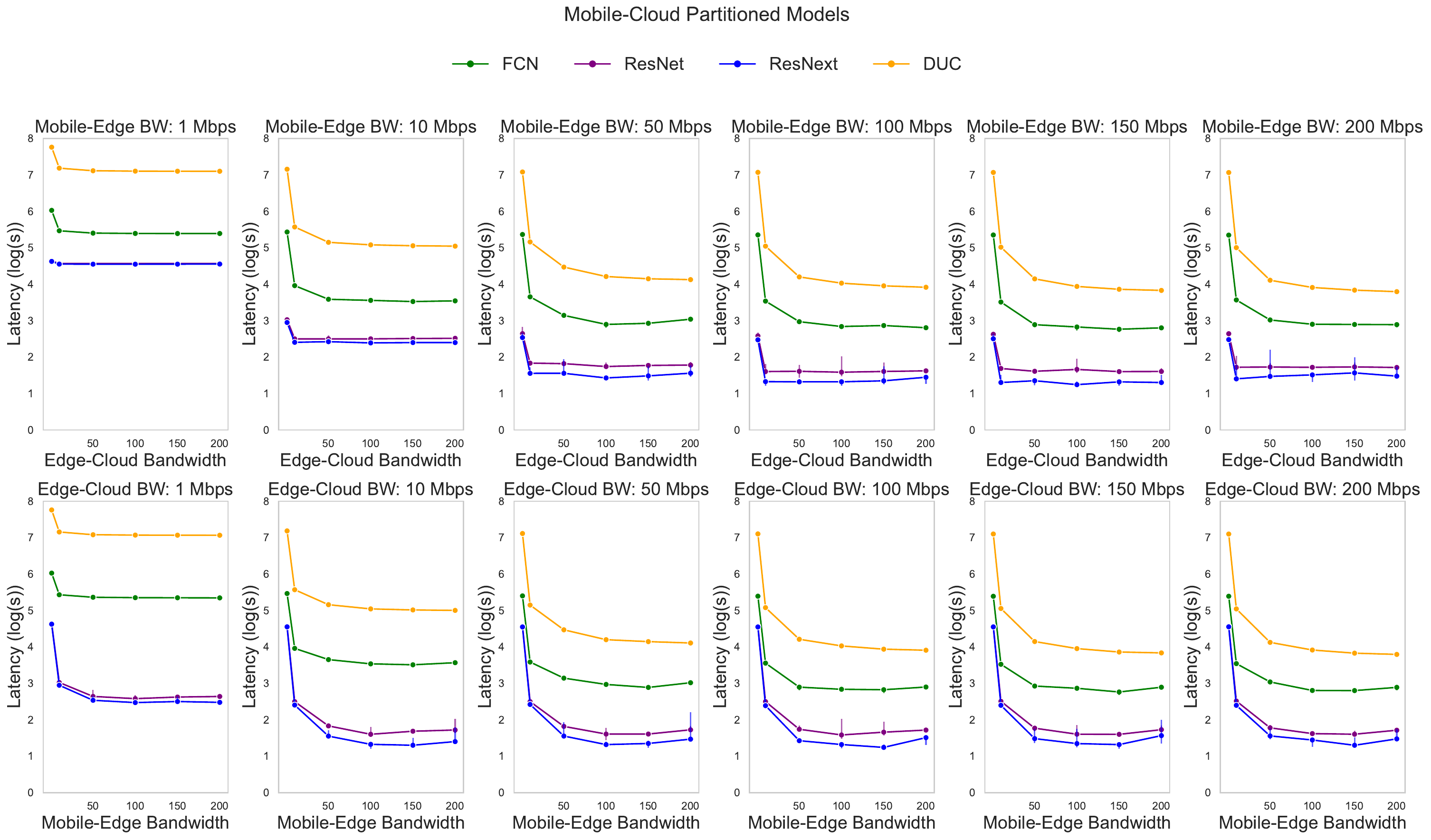}
\caption{Line graphs of the measures collected for inference latency vs network bandwidth of Mobile-Cloud Partitioned models.}
\label{mobile_cloud_partition_latency}
\end{figure}
\begin{figure}[htbp!]
\centering
\includegraphics[width=1\textwidth]{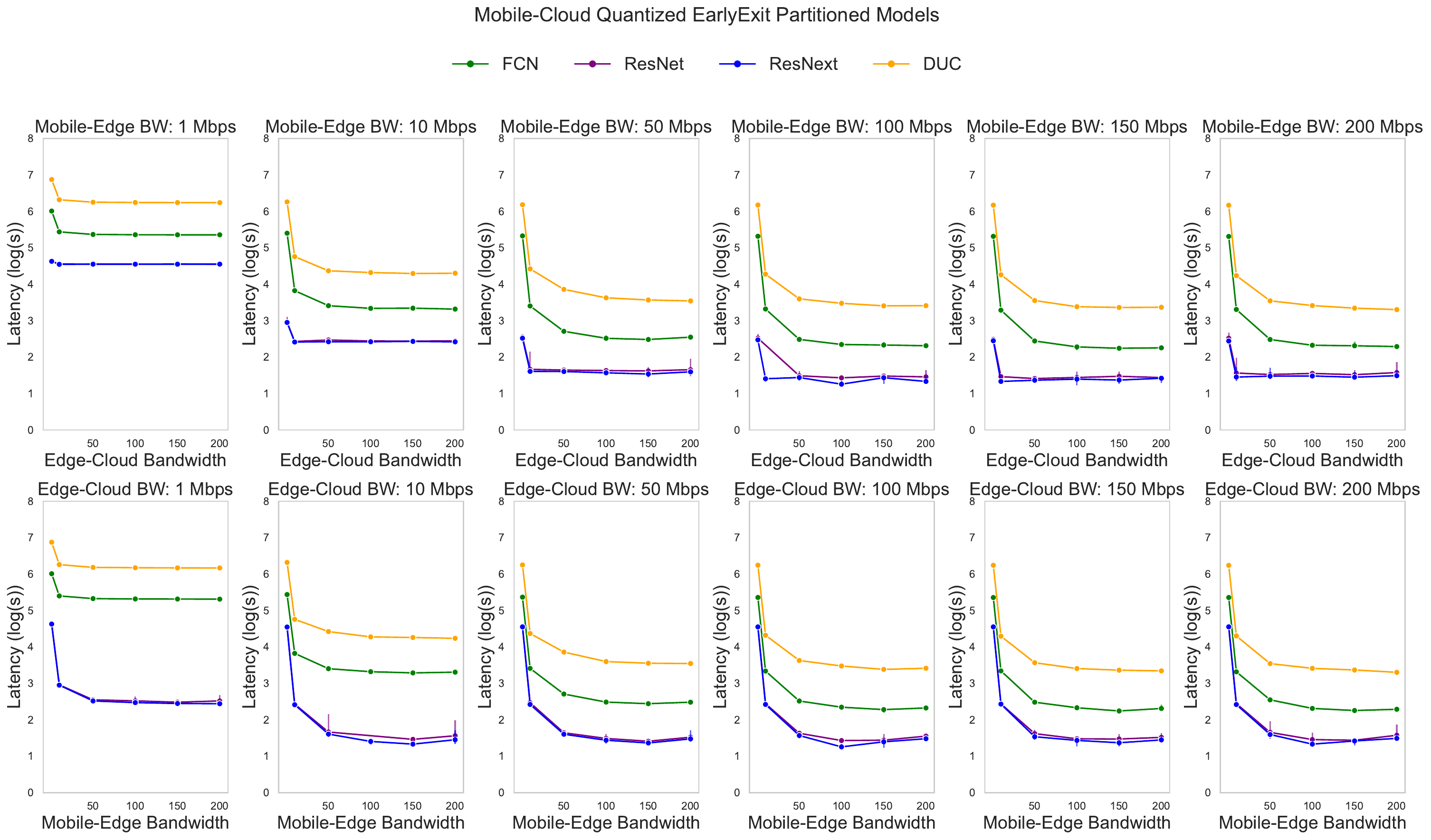}
\caption{Line graphs of the measures collected for inference latency vs network bandwidth of Mobile-Cloud Quantized Early Exit Partitioned models.}
\label{mobile_cloud_quantized_earlyexit_partition_latency}
\end{figure}
\begin{figure}[htbp!]
\centering
\includegraphics[width=1\textwidth]{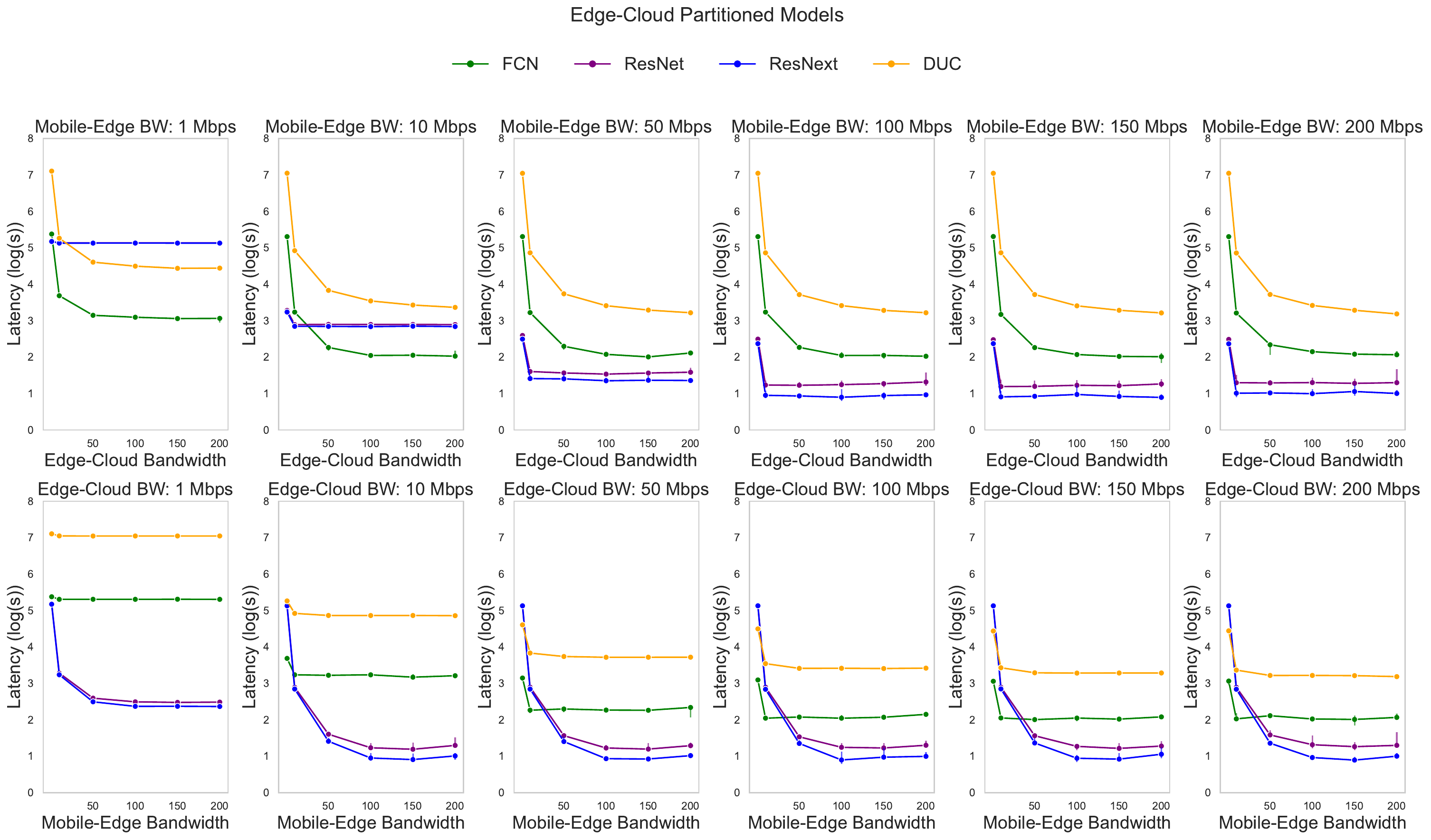}
\caption{Line graphs of the measures collected for inference latency vs network bandwidth of Edge-Cloud Partition models.}
\label{edge_cloud_partition_latency}
\end{figure}
\begin{figure}[htbp!]
\centering
\includegraphics[width=1\textwidth]{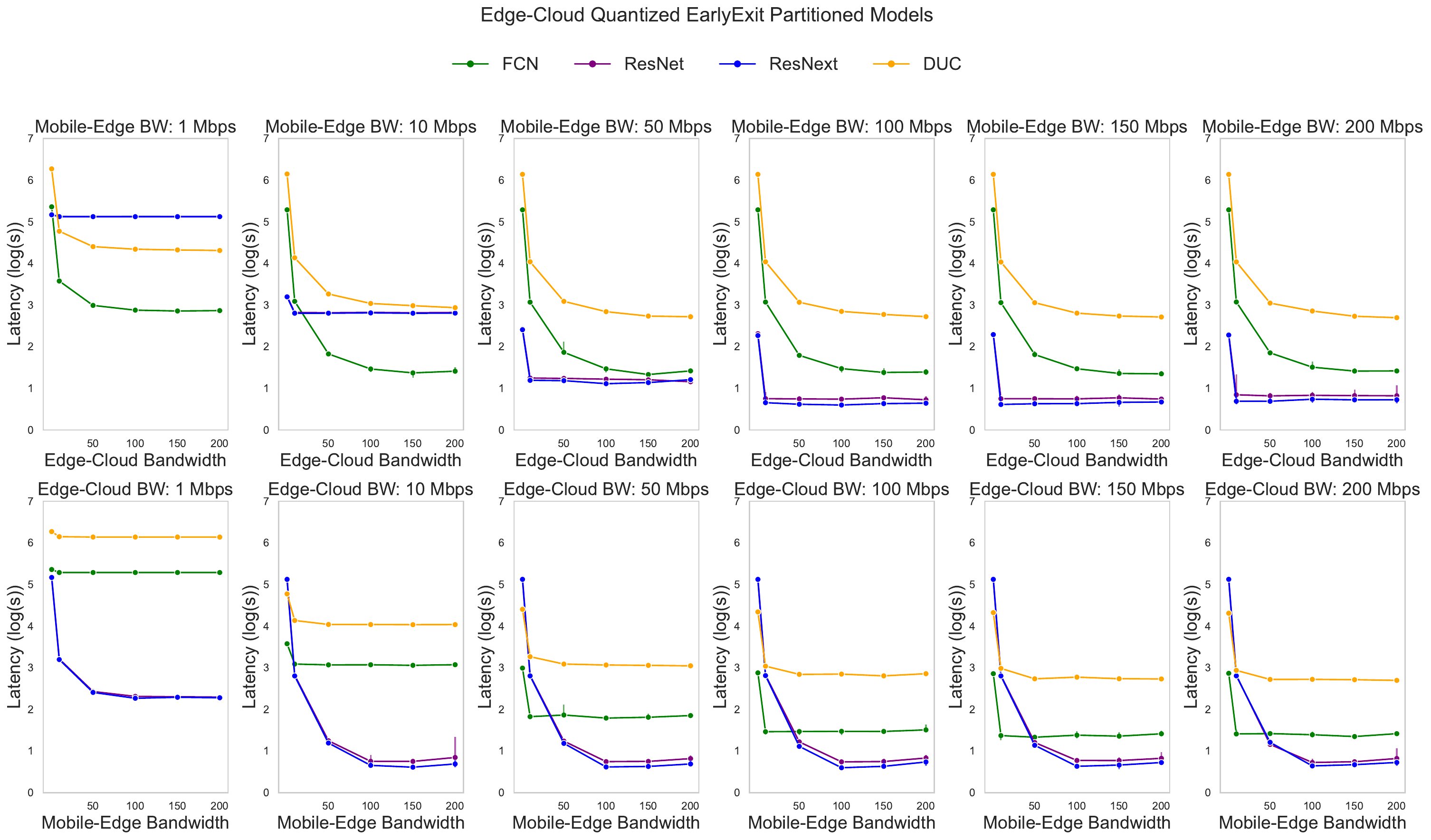}
\caption{Line graphs of the measures collected for inference latency vs network bandwidth of Edge-Cloud Quantized Early Exit Partitioned models.}
\label{edge_cloud_quantized_earlyexit_partition_latency}
\end{figure}

\textbf{For larger intermediate data-sized models (FCN, DUC), Partition-based strategies require $\geq$50 Mbps to achieve latency convergence. The Non-Partitioned models with large input data sizes (ResNe(x)t) are suitable for Mobile and Edge deployments at $\geq$50 Mbps.}

The Figure~\ref{mobile_cloud_partition_latency},~\ref{mobile_cloud_quantized_earlyexit_partition_latency},~\ref{edge_cloud_partition_latency},and~\ref{edge_cloud_quantized_earlyexit_partition_latency} show the line graphs of the measures collected for latency vs bandwidth of Mobile-Cloud Partitioned, Mobile-Cloud Quantized Early Exit Partitioned, Edge-Cloud Partitioned, and Edge-Cloud Quantized Early Exit Partitioned models across the four subjects (i.e., FCN, ResNet, ResNext, DUC). For these Partitioned-based models, the top row shows the steady state of ResNe(x)t latency when Edge-Cloud bandwidth is varied and the Mobile-Edge bandwidth is kept fixed, possibly due to the lower influence of intermediate data size (6.12 Mb) on Edge-Cloud bandwidths, similar to the observations in RQ4 and RQ5 findings (Section~\ref{sec:rq4},~\ref{sec:rq5_2}) when analyzing Partitioning-based strategies. Moreover, the steady line graph (top row) drops to lower latency as the Mobile-Edge bandwidth increases from 1 to 50 Mbps. A sudden drop in latency is observed when increasing the Mobile-Edge bandwidth from 1 to 50 Mbps at fixed Edge-Cloud bandwidths (bottom row) because the influence of input data size (60 Mb) transmission decreases on the latency. For Mobile-Cloud Partitioned-based FCN models, the latency gradually decreases with the increase in either Edge-Cloud bandwidth (top row) or Mobile-Edge bandwidth (bottom row) and converges at 50 or 100 Mbps. However, for Edge-Cloud Partitioning-based FCN models, the latency plateaus at 50 or 100 Mbps of Edge-Cloud bandwidth at fixed Mobile-Edge bandwidths (top row), and the steady line graphs (bottom row) drops to lower latency as the Edge-Cloud bandwidth increases from 1 to 100 Mbps. The reason is the low impact of intermediate data size (135 Mb) on latency at higher bandwidths ($\geq$50 Mbps). These findings are complement to the RQ4 and RQ5 findings. 

\begin{figure}[htbp!]
\centering
\includegraphics[width=0.8\textwidth]{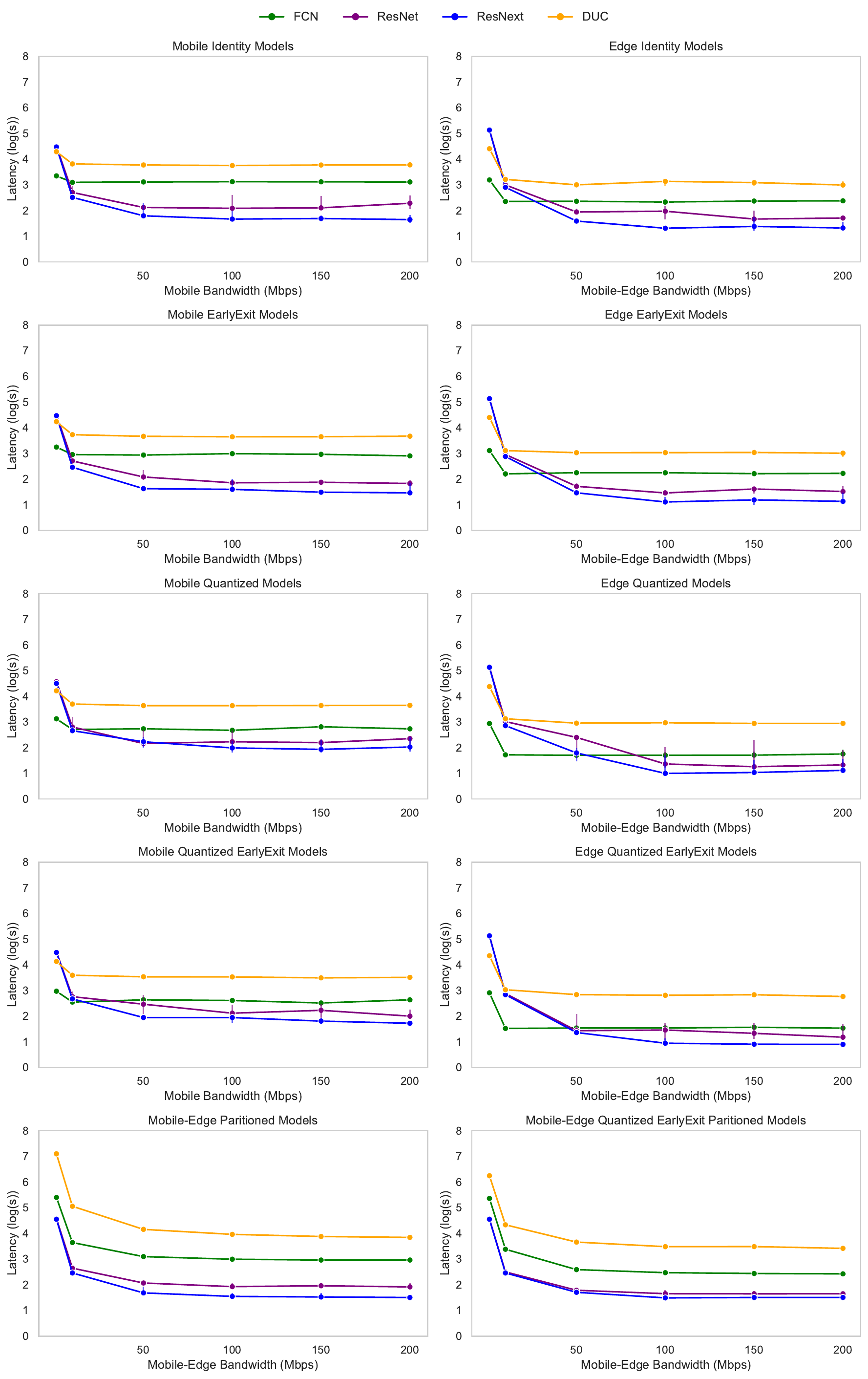}
\caption{Line graphs of the measures collected for inference latency vs network bandwidth of Mobile-Edge Partitioned, Mobile-Edge Quantized Early Exit Partitioned, Mobile (Identity, Early Exit, Quantized, Quantized Early Exit), and Edge (Identity, Early Exit, Quantized, Quantized Early Exit).}
\label{mobile_edge_latency}
\end{figure}

Figure~\ref{mobile_edge_latency} shows the line graphs of the measures collected for latency vs bandwidth of Mobile-Edge (Partitioned, Quantized Early Exit Partitioned), Mobile/Edge (Identity, Early Exit, Quantized, and Quantized Early Exit) deployment strategies. For DUC (orange line), the Mobile-Edge Partitioned-based models show a sharp decrease in latency when Mobile-Edge bandwidth increases from 1 Mbps to 50 Mbps. In the same case, the Mobile/Edge (Identity, Early Exit, Quantized, Quantized Early Exit) deployment strategies show a marginal drop in latency. The reason is the higher size of intermediate data (781.25Mb) compared to the input data (22 Mb) of DUC and that's why the latency of Partitioned-based models across Mobile-Edge shows a sharp decrease as the bandwidth increases from 1 to 50 Mbps. For the other subjects, i.e., FCN (green line), ResNet (purple line), and ResNext (blue line), the intermediate data is not that significantly large compared to DUC; therefore, a smaller decrease in latency is shown from 1 to 50 Mbps. For ResNe(x)t, the input data size is the highest (3x to 12x) compared to the other subjects (FCN, DUC), which is why it shows a larger decrease in latency for the same interval when models are deployed on Mobile and Edge tiers. For FCN, the trend is overall quite steady for the models during Mobile and Edge deployment as the input data size is quite small (5 Mb) to make an impact on the latency for various network bandwidths. These findings at low-bandwidth scenarios of Mobile-Edge ($\leq$50 Mbps), complement the RQ1-5 findings of high Mobile-Edge bandwidth (i.e., 200 Mbps). Table 24 shows a summary of results for RQ6.

\subsubsection{Normality Test}
We conducted the Shapiro-Wilk normality test and examined QQ plots for each model variant and bandwidth configuration. Contrary to typical expectations, the Shapiro-Wilk test results indicated that the latency distributions for the majority of configurations did not significantly deviate from normality (p $>$ 0.05), as shown in Table~\ref{rq6_cloud_identity_descriptive_statistics_part1}, Table~\ref{rq6_cloud_identity_descriptive_statistics_part2}, Table~\ref{rq6_cloud_quantization_descriptive_statistics_part1}, Table~\ref{rq6_cloud_quantization_descriptive_statistics_part2}, Table~\ref{rq6_cloud_earlyexit_descriptive_statistics_part1}, Table~\ref{rq6_cloud_earlyexit_descriptive_statistics_part2}, Table~\ref{rq6_cloud_quantized_earlyexit_descriptive_statistics_part1}, Table~\ref{rq6_cloud_quantized_earlyexit_descriptive_statistics_part2}, Table~\ref{rq6_cloud_models_descriptive_statistics}, Table~\ref{rq6_mobile_models_descriptive_statistics}, Table~\ref{rq6_edge_models_descriptive_statistics}, Table~\ref{rq6_mobile_cloud_partitioning_descriptive_statistics_part1}, Table~\ref{rq6_mobile_cloud_partitioning_descriptive_statistics_part2}, Table~\ref{rq6_mobile_cloud_quantized_earlyexit_partitioning_descriptive_statistics_part1}, Table~\ref{rq6_mobile_cloud_quantized_earlyexit_partitioning_descriptive_statistics_part2}, Table~\ref{rq6_edge_cloud_partitioning_descriptive_statistics_part1}, Table~\ref{rq6_edge_cloud_partitioning_descriptive_statistics_part2}, and Table~\ref{rq6_edge_cloud_quantized_earlyexit_partitioning_descriptive_statistics}, Figure~\ref{normality_results_rq6_mobile_identity}, Figure~\ref{normality_results_rq6_edge_identity}, Figure~\ref{normality_results_rq6_mobile_quantization}, Figure~\ref{normality_results_rq6_edge_quantization}, Figure~\ref{normality_results_rq6_mobile_earlyexit}, Figure~\ref{normality_results_rq6_edge_earlyexit}, Figure~\ref{normality_results_rq6_mobile_quantized_earlyexit}, Figure~\ref{normality_results_rq6_edge_quantized_earlyexit}, Figure~\ref{normality_results_rq6_mobile_edge_partition}, and Figure~\ref{normality_results_rq6_mobile_edge_quantized_earlyexit_partition}. This suggests that latency data maintains a roughly symmetric distribution with minimal skewness or outlier influence across diverse scenarios. These findings were visually corroborated by the QQ plots, which showed points aligning closely with the reference line.

\subsubsection{Hypothesis Testing}
As shown from Table~\ref{tab:conover_cloud_identity_me1} to Table~\ref{tab:conover_mobile_edge_quantized_earlyexit_partition}, for all model configurations, the Kruskal-Wallis p-value is extremely small (p $<$ 0.05), which indicates strong evidence against the null hypothesis. This suggests that there are significant differences in latency across the different bandwidth conditions for each model. After finding significance with the Kruskal-Wallis test, Conover's pairwise post-hoc tests were used to pinpoint which specific pairs of bandwidth values differ significantly when Mobile-Edge bandwidth is fixed and Edge-Cloud bandwidth is varied, and when Edge-Cloud bandwidth is fixed and Mobile-Edge bandwidth is varied. In both scenarios, Conover post hoc tests showed that models like ResNet, ResNeXt, FCN, and DUC exhibited strong sensitivity to changes in the Edge-Cloud and Mobile-Edge bandwidths, with most low vs. high bandwidth comparisons producing significant p-values (typically p $<$ 0.05). In general, the findings indicate that both the ME and the EC bandwidths significantly influence the latency of the model, especially at lower bandwidth levels, while the performance differences diminish at higher bandwidths (e.g., 100–200 Mbps), suggesting a saturation effect.

\begin{Summary}{}{sisthsummary}
Deploying models with small input data size (i.e., FCN) is ideal for Cloud deployment with bandwidth-constrained settings ($\leq$10 Mbps). ResNe(x)t and DUC models with large input data size are suitable for Cloud deployment that operate in scenarios with at least moderate bandwidth availability ($\geq$50 Mbps) but might not be ideal for constrained bandwidth settings ($\leq$10 Mbps). For models with lower intermediate data size (i.e. ResNe(x)t), the Edge-Cloud/Mobile-Cloud Partitioned-based strategies have no impact of intermediate data on latency across network bandwidth variations. 

For models with higher intermediate data size (i.e., FCN, DUC), the Edge-Cloud/Mobile-Cloud Partitioned-based strategies either converge at higher bandwidths ($\geq$100 Mbps for FCN) or doesn't converge (for DUC). For Mobile-Edge Partitioned based strategies, all the subjects converge at 50 Mbps (except DUC, which converges at 100 Mbps due to higher intermediate data size). For ResNe(x)t, the input data is the largest (3 to 12x) compared to the other subjects (FCN, DUC), which is why it shows a larger decrease in latency at 50 Mbps for Mobile and Edge deployments before plateauing.
\end{Summary}

\begin{table}[h!]
\centering
\resizebox{\textwidth}{!}{
\begin{tabular}{|l|c|c|c|p{6.5cm}|}
\hline
\textbf{Model} & \textbf{Ideal Bandwidth (Mbps)} & \textbf{Best Tier} & \textbf{Latency Behavior} & \textbf{Notes} \\ \hline
ResNet  & $\geq 50$  & Edge/Cloud & High drop from 1–50 Mbps & Input size (60MB) leads to latency reduction as bandwidth improves \\
ResNext & $\geq 50$  & Edge/Cloud & Large decrease, then plateaus & Cardinality increases memory usage; sensitive to bandwidth change\\
FCN     & $\leq$10  & Cloud      & Stable across bandwidths & Small input/intermediate size makes it bandwidth-resilient \\
DUC     & $\geq$100 & Edge       & Sharp drop, then plateau & Huge intermediate size (781MB); converges at 100 Mbps\\
\hline
\end{tabular}
}
\caption{Summary of results for Research Question 6}
\label{tab:rq6_result_summarization}
\end{table}

\section{Discussion}
\label{sec:discussion}
\subsection{Interpretation of Results}
Among the three Monolithic deployment tiers, the Edge tier consistently demonstrates significantly faster inference latency performance for each operator examined in the respective research questions (i.e., Identity operator in RQ1, Quantized operator in RQ2, Early Exit operator in RQ3, and Quantized Early Exit operator in RQ5). This outcome underscores the significance of deployment operators involved in the Edge tier compared to Mobile and Cloud tiers when facing computational limitations, network bandwidth constraints, and large input data transmission. The Edge tier's closer proximity to the Mobile and higher computational resources allows faster processing and reduced transmission latency, resulting in improved inference latency performance.

In the Cloud tier, the comparisons between the Quantized operator and the Identity operator (RQ2) as well as between the Quantized Early Exit operator and the Early Exit/Identity operator (RQ5) do not show any significant difference (Conover Test) for any subjects, except DUC. The main challenge is the inability of the CUDA execution provider in the ONNX Runtime inference Engine to fully support the CUDA kernels used for the quantized graphical nodes during Cloud deployment. The unavailability of the CUDA kernel for the nodes of the Quantized/Quantized Early Exit models leads to their execution being run on the CPU instead. This introduces some overhead due to the lower processing power of the CPU compared to the CUDA. Future research efforts can delve into solutions for optimizing the Quantization in a GPU-based Cloud environment. One solution might be to modify the model structure and its operations to avoid nodes that lack CUDA kernels. Techniques such as operator fusion, where multiple operations are combined into a single operation, could be beneficial~\cite{chen2018tvm}. Another solution could be to develop custom CUDA kernels for the nodes of the Quantized/Quantized Early Exit models that currently lack them. This would require a deep understanding of both CUDA programming and the specific operations performed by the incompatible nodes. Utilizing other GPU-specific execution providers like TensorRT instead of CUDA was not considered due to the reasoning provided in Section~\ref{sec:threats}.

On Cloud, the comparisons between the Early Exit and Identity operator (RQ3) as well as between the Quantized Early Exit and Quantized operator (RQ5) show no significant difference for the Resnet/ResNext/FCN subject and even if the difference is significant (in the case of DUC) based on the post-hoc test, the effect size remains negligible to small. This suggests that the advantages provided by these specialized operators in terms of speedup may be less pronounced in Cloud deployments, where computational resources are typically more abundant. In the above scenarios, the use of the Identity operator alone may be sufficient, and incorporating specialized operators like Quantization, Early Exit or their combinations may not yield significant benefits in terms of improving latency.

Among the three multi-deployment tiers (Mobile-Edge, Edge-Cloud, and Mobile-Cloud), the Mobile-Edge tier consistently exhibits faster latency performance for each operator examined in their respective research questions (i.e., Partition operator in RQ4 and Quantized Early Exit operator in RQ5). The key contributing factor to the Mobile-Edge tier's superiority is the higher network bandwidth it offers, which facilitates faster transmission of intermediate outputs during distributed inference. The Mobile-Edge distributed inference consistently shows faster latency performance than stand-alone Mobile inference for Identity models in RQ4 and Quantized Early Exit models in RQ5 due to the computational load distribution. The Partitioning of Identity/Quantized Early Exit models should ideally not influence the accuracy drops as it aims to divide the model into smaller components without altering the computations or operations performed, as stated in~\cite{matsubara2022split}. In other words, the computations within the Partitioned models are consistent with the original model's computations. The models considered in previous studies for CV tasks are relatively small (i.e, lower size) and less complex, which potentially resulted in findings that Mobile deployment is a better alternative than Mobile-Edge Partitioning (Kang et al~\cite{kang2017neurosurgeon}). However, in our study, where complex models are considered as the subjects, Partitioning across the Mobile and Edge is a better alternative than doing the local computing of the whole model on a resource-constrained Mobile tier, especially when faster latency is a concern at no accuracy loss. 

This study considers multiple dimensions, including operators, data models, network bandwidth, and tiers, requiring us to control one parameter at a time to analyze its impact on the others. In RQ1–RQ5, we fixed the network bandwidth at 1 Mbps for Mobile-Edge and 200 Mbps for Edge-Cloud to examine the effects of different operators and data models. This choice was deliberate, as these values are commonly used (as discussed in Section~\ref{sec:approach}), whereas RQ6 explored a range of bandwidths to assess how variations influence latency performance. The results from RQ1–RQ5 indicate that operators (Identity, Quantized, Early Exit, and Quantized Early Exit) exhibit better latency performance on Edge compared to Cloud, primarily due to the Edge tier’s higher bandwidth (200 Mbps vs. 1 Mbps) and the impact of input data transmission size. In RQ6, a similar relative performance was observed when Mobile-Edge and Edge-Cloud bandwidths were $\le$ 50 Mbps; however, when bandwidth exceeded 50 Mbps, the Cloud tier outperformed the Edge tier. Furthermore, RQ6 results show that Mobile-Edge Partitioning-based strategies consistently outperform both Mobile-Cloud and Edge-Cloud Partitioning-based strategies, reinforcing findings from RQ4 and RQ5. Additionally, Mobile-Edge Quantized Early Exit Partitioning-based models demonstrate superior latency performance compared to Mobile-Edge Partitioned models, further confirming that bandwidth constraints significantly impact partitioned strategies, as previously observed in RQ4 and RQ5. 

We examine how network bandwidth variations, explored in RQ6, influence the outcomes of RQ1–RQ5. Under the constrained 1 Mbps Cloud bandwidth assumed in RQ1–RQ5, the Edge tier consistently provides superior latency performance (RQ1–RQ3). However, when the Cloud bandwidth increases to 50 Mbps or more (RQ6), the Cloud tier becomes the optimal choice, emphasizing the significant role of network capacity. In RQ4, Mobile-Edge partitioning is most efficient under the 1 Mbps Cloud bandwidth, but at higher bandwidths, Edge-Cloud partitioning outperforms it by combining fast data transfer with Cloud computation. In RQ5, QE on Edge and QEP on Mobile-Edge perform best under low Cloud bandwidth, whereas at $>$50 Mbps, QE on Cloud emerges as the fastest monolithic strategy, and QEP on Edge-Cloud becomes the most effective hybrid strategy, surpassing Mobile-Edge setups. Mobile-Edge bandwidth also plays a critical role in the effectiveness of partitioned strategies such as Mobile-Edge Partitioned and Mobile-Edge QEP. While these strategies perform well under the 200 Mbps assumption used in RQ1–RQ5, their performance deteriorates when the Mobile-Edge bandwidth is reduced (e.g., 1–50 Mbps), as the cost of transmitting large intermediate data offsets the benefits of distributed inference. Likewise, models such as DUC, which generate large intermediate outputs, show poor performance under constrained Mobile and Mobile-Edge bandwidth conditions, but perform significantly better when bandwidth exceeds 100 Mbps.

\begin{figure}[t]
\centering
\includegraphics[width=1\textwidth]{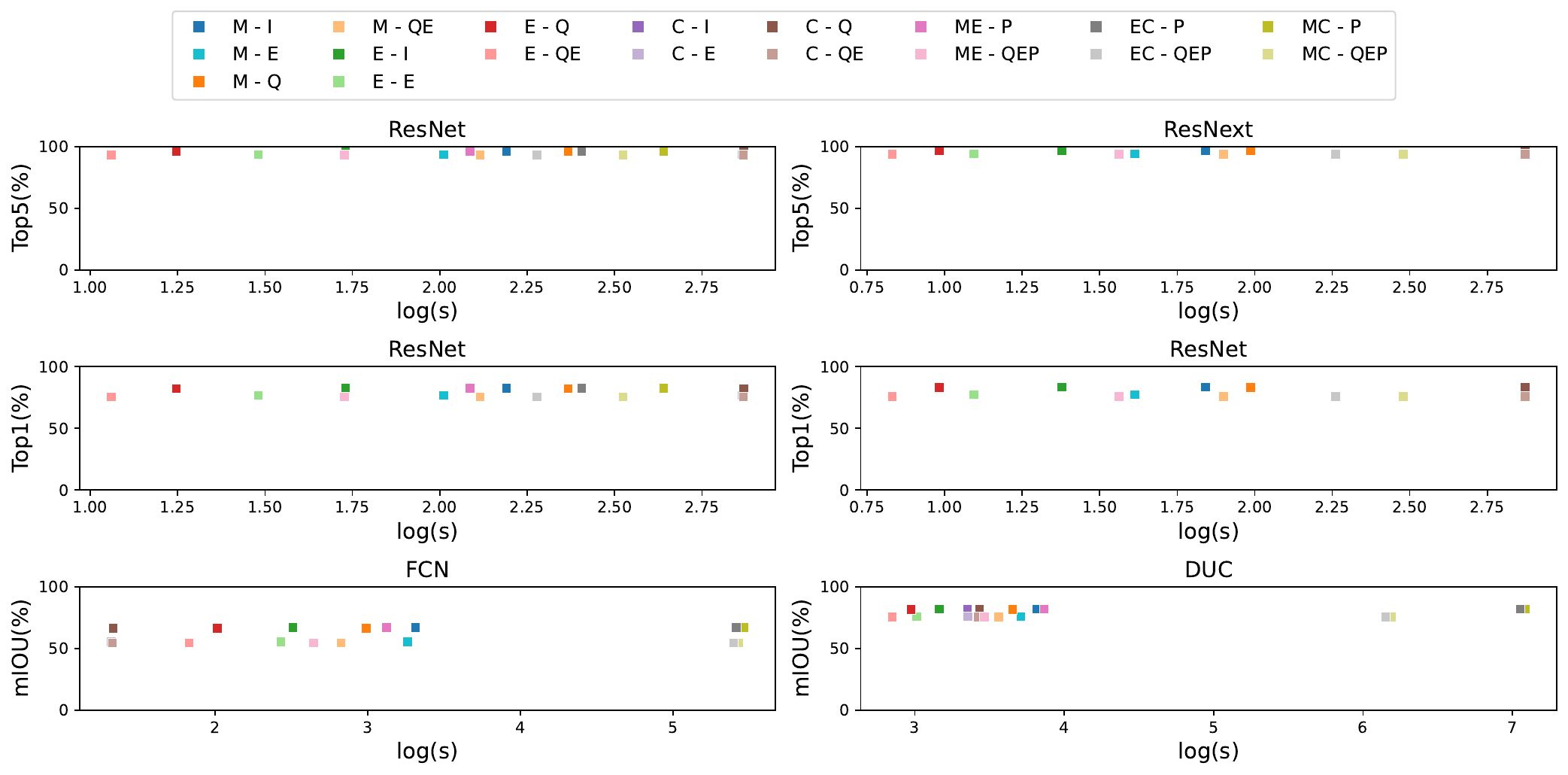}
\begin{minipage}{0.9\textwidth}
M-I: Mobile Identity, M-E: Mobile Early Exit, M-Q: Mobile Quantized, M-Quantized Early Exit: Mobile Quantized Early Exit, E - I: Edge Identity, E- E: Edge Early Exit, E-Q: Edge Quantized, E - Quantized Early Exit: Edge Quantized Early Exit, C - I: Cloud Identity, C - E: Cloud Early Exit, C - Q: Cloud Quantized, C - Quantized Early Exit: Cloud Quantized Early Exit, ME - P: Mobile Edge Partitioned, ME - Quantized Early Exit: Mobile-Edge Quantized Early Exit Partitioned, EC - P: Edge Cloud Partitioned, EC - Quantized Early Exit: Edge Cloud Quantized Early Exit Partitioned, MC - P: Mobile Cloud Partitioned, MC - Quantized Early Exit: Mobile Cloud Quantized Early Exit Partitioned.
\end{minipage}
\caption{Scattered plots of the measures collected for latency vs accuracy of the deployment strategies when evaluated on a range of input samples.}
\label{latency_vs_accuracy_scatterted_plot}
\end{figure}

In the accuracy vs latency scattered plots (Figure~\ref{latency_vs_accuracy_scatterted_plot}) of the deployment strategies when evaluated on a range of input samples, we can see that for ResNet, ResNext, and DUC, the Edge Quantized Early Exit (E - Quantized Early Exit) deployment strategy (light red scattered points) is Pareto-dominated by other strategies in terms of latency. Whereas, for FCN, the Cloud Quantized (C - Q) deployment strategy (dark brown scatter point) is Pareto-dominated by other strategies in terms of all objectives (i.e., latency and accuracy).

The significant presence of outliers in the box plots of ResNext on Edge (Figure~\ref{rq1_graph}) is possibly due to the higher memory usage while achieving inference latency measurements on Edge. This is primarily due to its cardinality-based architecture, which provides better accuracy than Resnet at the cost of increased computational and memory requirements~\footnote{https://www.ikomia.ai/blog/resnext-cnn-cardinality-efficiency-explained}. Here is a detailed breakdown of the outliers in the latency results of ResNext on Edge:
\begin{enumerate}
    \item Cardinality in ResNext: ResNeXt introduces the concept of cardinality, which refers to the number of parallel paths or branches in a block. Each ResNeXt block splits the input into multiple branches, processes them independently, and then aggregates the results. While this improves representational power and flexibility, it increases:
    \begin{itemize}
    \item Intermediate activations: Each branch produces its own intermediate feature maps, increasing the total memory required to store these activations during forward and backward passes.
    \item Parameter storage: Each branch has its own convolutional layers, leading to a greater number of weights to store in memory. For larger cardinalities, more parameters are distributed across branches. This increases the memory required to store weights and biases.
    \end{itemize}
\item Aggregation of Branch Outputs
\begin{itemize}
\item After processing through the branches, ResNeXt aggregates their outputs (usually by summation). This process temporarily requires additional memory to store the outputs of all branches before combining them. 

\item In ResNet, this step is simpler since there is only a single path per block, avoiding this additional memory overhead.
\end{itemize}
\item Wider Representations: ResNeXt achieves greater representational power by increasing cardinality rather than depth or width. While this improves accuracy, it also:
    \begin{itemize}
        \item Increases the size of intermediate tensors: Wider representations mean larger activation maps, which consume more memory.
        \item Requires storing gradients: During back-propagation, the gradients of these wider activations must also be kept in memory, further increasing the memory demand.
    \end{itemize}
\item Redundant Memory Usage in Backpropagation: During training, intermediate activations are stored for gradient computation. In ResNext: 
\begin{itemize}
    \item Each branch has its own set of activations that must be retained.
    \item The memory required for these intermediate results grows linearly with the cardinality.
\end{itemize}
\item Suboptimal Hardware Utilization: Many hardware accelerators are optimized for simpler, sequential architectures like ResNet. The parallel branch design in ResNext can lead to inefficiencies in memory allocation and access, indirectly contributing to higher memory usage.
\end{enumerate}

These outliers, which represent higher inference latencies, narrow the gap between the Edge and Cloud latency of ResNext, thereby reducing the perceived advantage of Edge in specific scenarios. In particular, these outliers on Edge indicate that under memory-constrained conditions caused by ResNext cardinality-based architecture, Edge may experience occasional latency spikes, which could undermine its advantage over Cloud. The majority of data points and the median on Edge tightly cluster below Cloud's latency median, emphasizing the general trend of lower inference on Edge. For latency comparison, as we considered the median, which itself is outlier-insensitive, the presence of outliers does not necessarily affect the overall conclusions. This suggests that Edge indeed is a preferable choice over Cloud for ResNext. Moreover, while comparing the distribution of Edge and Cloud latency of Resnext, we see a large effect size (Table~\ref{rq1_Identity_statistical_results}) indicating a statistically significant difference.

\subsection{Comparison with Existing Literature}
In previous studies (Table~\ref{relatedwork}), the resolution of input data size is relatively small as they are validated only on small-scale datasets such as MNIST and CIFAR datasets, and/or they did not explicitly analyze the impact of high-resolution images (from Image Net, COCO, and CityScapes datasets) on the end-to-end latency evaluation of operators in Mobile, Edge, and Cloud tiers. Eshratifar et al.~\cite{eshratifar2019jointdnn} suggest that using either local computing only or Cloud computing only is not an optimal solution in terms of inference latency in comparison to model Partitioning. However, the key issue is that they considered a single image for the end-to-end sequential inference across the Mobile-Cloud tier, without exploring the variation of input data sizes. In our study, the impact of multiple and varying image sizes on end-to-end latency was explored, which resulted in more generalized findings. Based on our results, the subjects having low-resolution images (such as FCN) may favor network-constrained Cloud deployment in comparison to multi-tier Partitioning strategies and Mobile/Edge deployment, as the impact on transmission overhead for smaller-sized images reduces, and the Cloud, as usual, has better computational capabilities. For subjects (i.e., ResNet, ResNext, DUC) having large-sized image samples, Edge deployment is a better alternative than multi-tier Partitioning strategies and Mobile/Cloud deployment.

Prior work~\cite{eshratifar2019jointdnn,jeong2018computation,kang2017neurosurgeon,li2018auto,zeng2019boomerang,pagliari2020crime,dong2021joint,yang2023adaptive,mohammed2020distributed,liang2023dnn,hu2019dynamic} explores factors such as the computational load, network cost, energy consumption, and/or privacy risk for each of the DNN Partitioning points in an Edge AI setup to dynamically decide the optimal Partition point, and stated that the model Partitioning operator achieves significant latency speedup i.e., latency reduction compared to traditional Mobile and/or Cloud deployment, similar to our findings, i.e., Mobile-Edge distributed inference of Partitioned/Quantized Early Exit models is a better alternative than resource-constrained Mobile deployment of Identity/Quantized Early Exit models. When comparing the multi-tier distributed strategies of Partitioned/Quantized Early Exit models with the Cloud deployment of Identity/Quantized Early Exit models in RQ4/RQ5, the intermediate data size and input data size play a crucial role. For FCN and DUC, the intermediate data size of their Partitioned/Quantized Early Exit variants is larger than their input data sizes, correlating with faster Cloud inference than distributed inference. Conversely, for ResNet and ResNext, the intermediate data size of their Partitioned/Quantized Early Exit variants is smaller than their input data sizes, correlating with faster-distributed inference than Cloud inference. Different Partition points (specific graphical node connections (s) within the neural network architecture where the model is divided or split into two sub-models) might have different intermediate data sizes, and their impact on the latency might vary. However, in our study, we limited our research to a single Partition point, as the goal was to create equal-size Partitioned models (which require a single Partition point). Although our Partitioning approach simply and fairly splits the models statically into two sub-models to have equal sizes, one running at the Mobile/Edge, and the other one in the Edge/Cloud, it requires manual analysis of the ONNX computational graphs of the subject models, which varies in terms of graph complexity and architecture design. In terms of subject models considered for model Partitioning, previous studies performed their experiments on lightweight CV models, which are less complex and less accurate than the heavy-weight state-of-the-art CV models considered in our study.

In previous studies performing Early Exit~\cite{elbayad2019depth,laskaridis2020spinn,lo2017dynamic,pomponi2021probabilistic,soldaini2020cascade,teerapittayanon2016branchynet,teerapittayanon2017distributed,wang2019dynexit,wang2020dual,xin2020early,xing2020early,yang2020resolution,zeng2019boomerang,zhou2020bert,liu2020fastbert,matsubara2021neural,xin2020deebert,wolczyk2021zero}, there is a trade-off between accuracy and latency. This trade-off in Early Exit comes from the fact that exiting earlier in the network can reduce latency but may also result in less accurate predictions. This is because the early layers in a DNN generally extract low-level features, while the later layers extract high-level features that are more task-specific. Therefore, an early exit at early layers might miss important high-level features, leading to a decrease in accuracy. On the other hand, waiting for the network to reach the later layers can increase the accuracy but also increase the latency. In our case, the Early Exit was performed in the later stage of the models, which showed faster latency than the original model but at a medium accuracy loss. One of the reasons for this significant accuracy loss is that the Early Exit approach in our study is based on the condition of manually short-circuiting identically structured sub-graphs on pre-trained models. This means that the early exits are added in a black-box manner on pre-trained models without retraining them, which contributes to this accuracy loss. Previous studies perform DNN Early Exit that requires training of the models, which results in better accuracy performance for Early Exit even in earlier stages of the model.

As stated in previous studies~\cite{li2018auto,banner2019post,cai2020zeroq,choukroun2019low,fang2020post,garg2021confounding,he2018learning,lee2018quantization,meller2019same,nagel2019data,shomron2021post,garg2022dynamic,hubara2020improving,zhao2019improving,li2021brecq}, the Quantization operator in our study also shows a small (not significant) accuracy drop in comparison to the original model. In addition to that, we also compared the Quantization operator's performance with other operators like Early Exit and Quantized Early Exit operators, concluding that during Edge deployment, the Quantization can be used in scenarios where the least accuracy drop is of utmost importance w.r.t the original model, at the benefit of faster latency than Early Exit and the cost of slower latency than Quantized Early Exit operators.

\subsection{Implications for the Practitioners}

In terms of effort, the use of an automated tool like Intel Neural Compressor for applying the Quantization operator suggests a streamlined and automated process for Edge AI model deployment. This implies that, right now, the Quantization process can be performed without manual effort and intervention, as a tool automates the necessary modifications to achieve model Quantization. Yet, while applying Quantization to models from the ONNX Model ZOO and torchvision.models subpackage, we observed that some of the models contain unsupported ONNX operations that are not listed in the supported ONNX schema\footnote{https://github.com/onnx/onnx/blob/main/docs/Operators.md}, due to which the model may not be feasible for Quantization without additional modifications or workarounds. For instance, while converting models from Pytorch to ONNX format, the converted ONNX models can include custom layers or operations that are specific to the Pytorch framework and hence lack Quantization support in ONNX.

On the other hand, applying the Early Exit operator on Identity/Quantized models and the Partitioning operator on Identity/Quantized Early Exit models currently requires manual analysis of the ONNX computational graphs using the Netron Visualizer tool~\footnote{https://github.com/lutzroeder/netron} and manual modifications of the neural network using ONNX Python APIs. We observed that performing these deployment operators on ONNX models from other classes (e.g., textual inference task), was not always feasible due to the complex architecture of the ONNX computational graph. The complex architecture of such models includes various layers, connections, and branching structures that can make it difficult to identify suitable Partition points or early exit points. In such cases, applying these operators might require significant manual analysis and modifications of the model's computational graph, which can be time-consuming and error-prone. Below are the crucial points that highlights why the results are helpful for MLOps and in which cases:

\textbf{Empirical Foundation for Deployment Choices}
The study delivers data-backed insights on how different black-box deployment operators (Quantization, Early Exit, Partitioning) and their combinations affect latency and accuracy across Mobile, Edge, and Cloud tiers. MLOps engineers often struggle with trial-and-error deployment tuning—this study systematically removes the guesswork by evaluating 20+ configurations in controlled, realistic environments.

Our study offers a concrete empirical foundation to assist MLOps engineers in selecting appropriate deployment strategies for black-box models across heterogeneous Edge AI environments. For instance, when aiming to reduce latency under moderate accuracy constraints, Quantized Early Exit (QE) on Edge emerges as a promising solution due to its effective balance between performance and computational cost. Conversely, when preserving accuracy is critical, using Quantization alone on Edge is preferable, as it delivers latency benefits with minimal accuracy degradation compared to Early Exit or QE. In resource-constrained Mobile environments, Mobile-Edge Partitioning outperforms pure Mobile deployment by offloading heavy computation to nearby Edge tiers, reducing latency without sacrificing output fidelity. Additionally, the results show that Cloud deployment remains viable for smaller input models even under low-bandwidth conditions ($\leq$10 Mbps), while larger models require at least 50 Mbps for latency convergence. These insights can help MLOps teams systematically evaluate the trade-offs between latency, accuracy, and resource constraints, avoiding ad hoc trial-and-error tuning and promoting informed, performance-aware deployment strategies.

\textbf{Black-Box Compatibility}
The operators analyzed require no model re-training, which is ideal for real-world, production-grade MLOps settings where models are often closed-source or externally sourced. This makes the findings highly actionable across industries.

\textbf{Towards a Foundational Dataset for Recommendation Systems}
The empirical results of the study provide valuable insights into the latency and accuracy performance of operators across and within tiers under various deployment scenarios.  As such, we believe that the results could inspire work related to the automation of these operators and can be used as a foundation for developing recommendation systems for Edge AI operators. In particular, the results serve as core training data for automated systems such as AGI-based recommendation engines that learn optimal AI deployment strategies. This includes input-output mappings for a wide range of models, operators, tiers, and network conditions—allowing autonomous adaptation in real-time environments. 

However, we recognize that our current insights are derived from a simulated and controlled testbed, which may not capture all real-world heterogeneities (e.g., hardware variation, dynamic workloads). As such, we do not position our results as directly prescriptive but rather as a knowledge base from which data-driven heuristics could eventually be extracted and validated in real environments. To ensure trust and transparency, any future recommendation system must be accompanied by interpretable explanations, and its performance must be benchmarked against real-world deployment scenarios. Until such validation is achieved, we encourage practitioners to rely on the detailed experimental conditions we provide in this study to assess contextual relevance to their own settings.

While this study focuses primarily on latency and accuracy trade-offs, our observations about memory-boundedness suggest the importance of tracking CPU utilization, memory footprint, network throughput, and energy consumption. These system-level metrics are highly relevant for environmentally-conscious deployments and for fine-grained performance optimization. Future work will expand the current benchmarking pipeline to include these metrics, which are already partially accessible via Docker container statistics and external energy profiling tools (e.g., Intel RAPL, NVIDIA SMI). Doing so will allow us to further align Edge AI deployment strategies with sustainability goals.

In the Future, we plan to extend our testbed with tools like cgroups, perf, and powerstat to collect metrics such as CPU usage, RAM consumption, and energy draw during inference. This will support a multi-objective evaluation of deployment strategies, including carbon footprint and energy cost, critical variables in real-world AI systems design.

\textbf{Versatile B2B Applicability}
These results are not just academic—they can be directly applied in any B2B use case where AI models are deployed in an Edge AI Infrastructure (e.g., manufacturing, healthcare, smart cities, autonomous vehicles, and more).

\section{Threats to Validity}
\label{sec:threats}
Below, we discuss threats to the study validity and the strategies we applied to mitigate these threats, based on literature guidelines~\cite{wohlin2012experimentation}.

\textit{Construct Validity}: One possible threat is the mono-operation bias caused by having only one factor of computational configuration (RAM/CPU), Early Exiting point, and the Partitioning point. The Mobile-Edge network bandwidth of 200 Mbps and Edge-Cloud network bandwidth of 1 Mbps were used to simulate the close and distant proximity of Mobile-Edge and Edge-Cloud environment, respectively, based on earlier work~\cite{nan2023large,zhang2023effect,suryavansh2019tango,fiandrino2019openleon,andres2018openleon} for testing deployment strategies on a range of samples in RQ1-5. We further expand our analysis to include additional experiments evaluating deployment strategies on a single largest input sample under a broader range of commonly used bandwidth settings (1, 10, 50, 100, 150, and 200 Mbps) across both Mobile-Edge and Edge-Cloud networks~\cite{alqarni2023odm,cui2024latency,zhuang2024decc,cui2024latency,nan2023large,zhang2023effect} in RQ6. The additional findings in RQ6 results (Section~\ref{sec:rq6}) allow us to provide a more comprehensive understanding of the latency performance trade-offs under varying bandwidth conditions and emphasize the impact of bandwidth on deployment strategies. The computational simulations of Mobile, Edge, and Cloud tiers are also based on previous studies~\cite{dimolitsas2023multi,duan2021joint,kunas2023optimizing,preuveneers2019towards}. Variations in the computational resources can impact the end-to-end inference latency for the deployment strategies.

One of the key limitations of our study is the use of Docker containers for simulating resource-constrained tiers (Mobile, Edge) on a common server and resource-abundant tier (Cloud) on a separate server in our experimental setup. We recognize that a distributed setup using multiple physical machines would provide additional realism. Our Docker-based simulated experimental setup was designed to ensure the reproducibility of our results. While we acknowledge that this approach may not fully replicate the complexities and challenges faced in real-world distributed MLOps deployment infrastructure, our primary objective was to provide a controlled and consistent environment for testing and validating our deployment strategies. This methodology aligns with prior research, as highlighted in Table~\ref{relatedwork} of the Related Work section (Section~\ref{sec:related_work}), where similar simulated setups have been employed. By adopting this approach, we establish a reproducible framework that can serve as a foundation for future studies. We further test a comprehensive range of network bandwidths ensuring our evaluation aligns with the diverse characteristics of real-world deployments.

However, our decision to use Docker containers on a single machine was intentional and driven by the following considerations:

\begin{itemize}
\item Prototyping a Unified Baseline for MLOps: MLOps engineers often start by testing their pipelines in controlled environments to ensure baseline functionality before scaling to distributed systems. Our choice mirrors this natural progression, ensuring that our findings are directly relevant to the foundational stages of MLOps pipeline development.

\item Reproducibility as a Scientific Priority: One of the key challenges in evaluating MLOps tools and strategies is achieving reproducibility across studies. A single-machine Docker-based setup provides a stable and standardized environment, reducing variability and enabling others to replicate our work with minimal dependencies or hardware constraints. This step is crucial in building trust and validating methods within the community.

\item Cost-Effective Innovation: MLOps research is inherently resource-intensive. By leveraging a Docker-based approach, we reduced the financial and logistical overhead of deploying experiments on distributed physical systems. This resource efficiency allowed us to focus on developing innovative insights into key MLOps challenges, with the understanding that future work can build on these foundations in more distributed contexts.

\item Alignment with MLOps Practices: Docker-based setups are widely used in MLOps workflows, particularly during the prototyping phase (Table 1 in Related Work Section). Our approach aligns with this practice, ensuring the immediate applicability of our results to real-world scenarios.
\end{itemize}

The limitations of the considered simulated setup:
\begin{itemize}
    \item Clock Speed and Thermal Factors: Docker provides a valuable feature for setting resource constraints, enabling the simulation of specific configurations like the number of CPU cores and available RAM. To further strengthen the study, it’s worth noting that certain hardware-level characteristics, such as clock speed variations and thermal management, may differ from real-world systems.
    
    \item GPU Utilization and Variability: The study’s cloud simulation utilizes an NVIDIA A100 GPU, which is a state-of-the-art piece of hardware. However, to enhance the study's applicability to real-world scenarios, it might be beneficial to consider the diversity of GPUs typically deployed in actual cloud environments, such as the T4, V100, or even older hardware. This could provide a more comprehensive understanding of performance across different cloud setups.
\end{itemize}

Influence of these limitations on the study findings:
\begin{itemize}
    \item Limited Real-World Representation: The results might not fully capture the nuances of actual hardware behaviors, such as clock speed variations or thermal management, which are significant in real-world settings.
    \item Performance Variability: Differences in computational and hardware characteristics may lead to outcomes that do not align perfectly with what would be observed on diverse real-time systems.
\end{itemize}

In a previous study~\cite{portabales2020dockemu}, Docker containers running on a server are used to simulate several resource-constrained tiers in a realistic IOT framework. This is similar to what we did for simulating the resource-constrained tiers (i.e., Mobile and Edge) in our study. In real-world scenarios, variations in hardware across different Edge/Mobile devices are possible and may impact the generalizability of our simulated setup. Simulating the impact of multiple factors is very costly as it involves running numerous experiments, each taking considerable time to complete and requiring significant computational resources. While we acknowledge the importance of diverse deployment scenarios having different computational/network configurations of Mobile, Edge, and Cloud tiers, the specific experimental setup was chosen strategically to provide a focused exploration of a typical Edge AI environment having a resource-scarce Mobile device, an Edge device's closer proximity, and higher computational capacity w.r.t Mobile device, together with a resource-abundant Cloud device with network constraints.

In our study, we focus on Early Exit at a single stage of the neural network. However, it is important to note that early exiting at multiple stages of the network can result in varying accuracy performance. Early exiting at a later stage may yield higher accuracy but slower inference due to more processing required before making predictions. On the other hand, the early exit at an earlier stage may provide faster inference but with lower accuracy since the predictions are made based on less processed information. The detailed explanation for selecting the early exit criteria is mentioned in the RQ3 approach (Section~\ref{sec:rq3_motivation_approach}), and requires manual inspection of the subjects' ONNX computational graphs. Similarly, for Partitioning, we considered the Partition point that leads to equal-sized sub-models for effective and fair load distribution across the tiers involved in distributed inference. Our study acknowledges the limitations associated with this simplified model Partitioning approach. We recognize that real-world scenarios often entail more complex Partitioning strategies as mentioned in related work (Section~\ref{relatedwork}), especially when dealing with intricate model architectures, such as Partitioning at each of the layers of DNN models and analyzing their impact on various factors like computational, transmission, and/or energy cost. 

In our study, the decision to adopt a simplified approach stems from the need for a fair evaluation of the subject models. By manually inspecting the ONNX computational graphs of the DNN models, we aim to establish a baseline understanding of the challenges and dynamics involved in equal-size Partitioning of the DNN models with varying and complex architectures, specifically in the ONNX framework. The manual insights of the computational graphs of the subjects will assist MLOps Engineers in understanding the factors involved in constructing an automated tool to dynamically decide the optimal Partition point to achieve equal-size sub-models for subjects with varying architectures. This approach might have an adverse impact on the transmission of intermediate data during distributed inference across the Mobile, Edge, and Cloud tiers, as shown for two of the subjects (FCN, DUC) in our study. Different Partition points may have different intermediate data sizes which can yield latency benefits during distributed inference across Mobile, Edge, and Cloud tiers in scenarios where the intermediate data size is lower than the input data size, as shown (ResNet and ResNext) in our study.

\textit{Conclusion Validity}
We considered the Static PTQ approach over the Dynamic PTQ approach due to its faster inference capabilities. Dynamic PTQ requires additional computational overhead during inference because of the dynamic recalibration process in which the model's weights and activations are recalibrated based on the input data's statistics during inference. On the other hand, Static PTQ involves quantizing the model's weights and activations only once during the model conversion phase, without the need for recalibration during inference. Since the quantization parameters are precomputed and do not change during inference, the quantization process is much simpler and requires fewer computations during inference. 

\textit{Internal Validity}: The risk of how history might affect the inference latency results of the deployment strategies is reduced by performing all measurements in the same Edge AI environment, using the same infrastructure. To maintain uniformity and minimize variations, we developed automated scripts to execute the inference experiments for the deployment strategies sequentially, one after another. Before starting an inference experiment, we took the necessary step of restarting the Docker containers to eliminate any potential residual effects from the previous inference experiment. In our experiments, we considered sequential inference (i.e., 1 request at a time) instead of parallel inference (i.e., multiple requests at a time). Sequential inference allows us to efficiently utilize the available resources for each deployment strategy, ensuring a more accurate representation of their true inference latency performance~\cite{zhang2021nn}, similar to how micro-benchmarks operate. The goal of our study was not load testing~\cite{jiang2015survey}, where the focus would be on measuring the system's ability to handle multiple concurrent inference requests. Performing parallel inference may lead to resource contention, which could obscure the true impact of deployment strategies. Similarly, the Scalability aspects, such as the impact of increasing the number or complexity of models deployed simultaneously in a real-world setting, are not explored, as they are outside the scope of this study, since our micro-benchmarks will not focus on system-level measurements.

It is worth mentioning that the ONNX Run-time inference Engine may perform worse on the first input received than on subsequent inputs during inference experiments of deployment strategies, mainly because of a required warm-up inference. To remove such bias, we used a trial inference experiment. For each deployment strategy, a trial experiment of 100 inference runs is performed sequentially to reach a steady state of the cache, then the final inference experiment of 500 runs = 100 (input samples) x 5 (repetitions) is performed sequentially (and repeatedly) without any cool-down period between subsequent runs to simulate the scalability of each deployment strategy. Note that the relatively high standard deviation in the inference latency measure for each deployment strategy might have been caused by having 5 repetitions per sample run; this potential source of bias can be mitigated by increasing the number of such repetitions. In our study, it is costly to do this due to the computational and transmission overhead caused by factors like large model size and input/intermediate data transmission. 

Input size is a potential threat to internal validity, as it can significantly affect both latency and resource usage across models. To address this, we carefully selected 100 large-size input samples from the validation dataset for the multi-input trial experiments. This selection aimed to evaluate the impact of varying input sizes on latency, as described in Section~\ref{sec:experimental_setup}. Additionally, for the single-input trial experiments, we used the largest input sample available to assess latency across different deployment strategies and network conditions. Testing with larger input sizes allowed us to evaluate how well the simulated environment can handle heavier computational workloads, simulating real-world scenarios such as medical imaging, video games, and high-resolution photography. This approach also helped assess the scalability of the system and identify potential bottlenecks, such as increased latency due to memory, processing power, or bandwidth limitations. By carefully considering and controlling for input size in our experimental setup, we aimed to minimize its potential confounding effects on the results.

\textit{External Validity}: The selection of the subjects might constitute another potential threat as it was performed manually. Thus, the selected set of subjects could not be regarded as an accurate representation of the whole population. The first iteration in this process consisted of choosing a set of subjects from the ONNX Zoo and PyTorch Models, initially aiming to have two representatives from each class (i.e., the inference task). However, we observed a lack of already trained models from some of the classes and what is more, most of the models were not feasible for Partitioning/Early Exiting due to the complex ONNX graph architectures, or for Quantization due to some operations or layers in the ONNX graph architectures that may not have Quantization support in the neural compressor tool. As a result of that, we ended up with a selection of four subject models, from the CV category. This threat is reduced by aiming to diversify the inference tasks they performed (i.e., Image Classification and Image Segmentation). 

The majority of the previous studies, as shown in Table~\ref{relatedwork}, focus on CV tasks for the operators due to their major impact on various factors like computational load and data transmission during deployment in an Edge AI environment. Therefore, the choice of CV domain for subject models and datasets is quite common. However, we do acknowledge that different types of models (e.g., Natural Language Processing and Speech Recognition) may exhibit different behaviors in response to deployment strategies. In our study, we limited the scope of our study to three black-box operators, i.e., Partition, Quantization, and Early Exiting, as these are most commonly used, which do limit the comprehensiveness of the studied operators. Overall, the selection of subjects (datasets/models) and deployment operators is one of the external threats concerning this study. This threat can be mitigated in the future by repeating the experiment on other domain-specific subjects (e.g. Natural Language Processing, Speech Recognition) and white-box operators (e.g., Quantization Aware Training, Weight Pruning, Knowledge distillation) as mentioned in Table~\ref{relatedwork}. 

The effectiveness and compatibility of the Intel Neural Compressor tool for Quantization might vary for different models or frameworks (i.e., ONNX), affecting the reproducibility of the study in different environments (i.e., CPU, GPU). In our results, for some subjects like ResNet and ResNext, their Quantized and Quantized Early Exit models show slower latency performance than Identity and Early Exit models in resource-constrained environments (i.e., Mobile). Yet, for other subjects (like FCN and DUC), their Quantized and Quantized Early Exit models show faster latency performance than Identity and Early Exit models in the same environment. Conversely, in high-resource environments (i.e., Edge), all subjects' Quantized and Quantized Early Exit models show faster latency than the Identity and Early Exit models. This shows that for different subjects the compatibility of this tool (i.e., Intel Neural Compressor) for Quantization might vary in terms of latency performance in computationally varying environments. Moreover, this tool might show different latency or accuracy behavior for models in different formats (such as Pytorch and Tensorflow). Moreover, the visual analysis of the computational graphs might vary with different models, frameworks, or visualization tools.


Among the GPU-specific Execution Providers (e.g., CUDA and TensorRT), we selected CUDA for our GPU-based environment (i.e., Cloud) due to its notable inference benefits\footnote{https://developer.nvidia.com/blog/end-to-end-ai-for-nvidia-based-pcs-cuda-and-tensorrt-execution-providers-in-onnx-runtime}\footnote{https://github.com/chaiNNer-org/chaiNNer/discussions/2437} over TensorRT. Nonetheless, relying on CUDA introduces certain limitations. Framework-specific optimizations, hardware-specific dependencies, and differences in precision or ecosystem support could affect the generalizability of our results. For example, TensorRT, while optimized for high-performance inference, may outperform CUDA in scenarios with static model execution graphs. Furthermore, CUDA’s dependency on NVIDIA GPUs limits our findings, potentially biasing performance outcomes compared to hardware like AMD or ARM-based systems.

Despite these constraints, there were several reasons for this choice. First, CUDA is faster to initialize as it evaluates only small network building blocks during its exhaustive search, leveraging the cuDNN inference library for granular neural network operations. This makes CUDA more versatile in handling inference tasks with varying sizes, easier to use, and less demanding in terms of setup. In contrast, TensorRT evaluates entire graphs and explores all execution paths, which can take several minutes for large ONNX models. Additionally, TensorRT’s allocation of workspace memory for intermediate buffers leads to higher memory usage, and its frequent engine recomputation for varying input sizes—such as in our study involving diverse image sizes—can result in slower performance. Thus, alternative providers like TensorRT were deemed suboptimal for our specific requirements.

Instead of using actual devices, we used Docker containers for simulating the hardware and network configurations of physical Mobile, Edge, and Cloud devices to create real-time deployment scenarios. The Docker simulations allow flexibility by easily configuring the network/hardware settings. Setting up and maintaining actual devices can be expensive and require more experience, especially for simulating different configurations and deployment scenarios. Docker containers provide a cost-effective way to create virtual environments that closely mimic the behavior of real hardware and network bandwidth configurations without the need for additional physical resources. The latest versions of all the tools and packages were employed on the simulated devices in the experimental setup (Section~\ref{sec:approach}). The generalization factor can be improved by replicating the experiment on different hardware and network configurations of the devices. In other words, real-world deployment considerations, such as network variability, security implications, or dynamic Edge environments, should be considered for generalization.

In our study, we computed the inference accuracy performance independently on multiple representative deployment tiers (i.e., Mobile, Edge, Cloud) for four different types of accuracy-sensitive operators (i.e., Identity, Quantized, Early Exiting, and Quantized Early Exiting) to provide valuable insights into the model's generalizability across different hardware targets, specifically CPUs and GPUs. The ONNX models are designed to be hardware-agnostic and can be deployed on various hardware devices without significant modifications. This allows the models to achieve consistent accuracy performance across different deployment environments, as long as the hardware supports the necessary operations and computational capabilities.

We applied the operators to the subjects in ONNX format and performed the inference of the transformed models using the ONNX Runtime Engine due to optimized deployment performance benefits, as suggested by previous studies~\cite{openja2022empirical,hampau2022empirical}. The feasibility of scripting black-box transformations using Python ONNX APIs was another reason for considering subject models in ONNX format, instead of other formats (like Pytorch and Tensorflow).

\section{Conclusions and Future Work}
\label{sec:conclusion_and_futurework}
Deploying black-box models (DNNs) efficiently in an Edge AI setting introduces unique challenges for MLOps Engineers and software practitioners. The black-box models require specific considerations for optimization in resource-constrained and network-constrained deployment scenarios. This paper aims to be an important stepping stone in the field of MLOps, in particular for the deployment of black-box models, to evaluate the benefits and trade-offs of Edge AI deployment strategies involving mappings of $<$operators, tiers$>$, by evaluating their performance in terms of quantitative metrics like latency and accuracy. While previous works focused on exploring and addressing individual operators (i.e., Partitioned, Early Exit, Quantization), our study has systematically compared the individual operators and their unexplored combinations in an Edge AI Environment using empirical data of four major CV subjects for testing the various deployment strategies. 

The MLOps Engineers could prefer Mobile-Edge distributed inference when faster latency is a concern in deployment scenarios where the mobile tier has strict resource (CPU/RAM) requirements. For models with smaller input data size requirements, their deployment at the Cloud tier with limited network bandwidth capacity can also be a better alternative than Model Partitioned across Mobile, Edge, and Cloud tiers and Mobile/Edge deployment. For models with large input data size requirements, Edge deployment can be a priority over Model Partitioned and Mobile/Cloud deployment in scenarios where the Edge has higher computational/network capabilities than the Mobile/Cloud. Among the studied operators, the Edge deployment of the Quantized Early Exit operator could be the preferred choice over the Edge deployment of the Early Exit/Quantized operator and Mobile-Edge deployment of the Partition operator when faster latency is a requirement at medium accuracy loss. In contrast, for MLOps Engineers having requirements of the minimal accuracy loss w.r.t the original model, the Edge deployment of the Quantized operator could be the preferred choice at the benefit of faster latency over the Edge/Mobile-Edge deployment of the Early Exit/Partition operator and the cost of slower latency over the Edge deployment of Quantized Early Exit operator. Deploying Non-Partitioned models with small input data size (i.e., FCN) is ideal for Cloud deployment even in bandwidth-constrained settings ($\leq$10 Mbps). Whereas, deploying the Non-Partitioned models with large input data size (ResNe(x)t, DUC) is suitable for Cloud deployment with moderate bandwidth availability ($\geq$50 Mbps). For models with higher intermediate data sizes (i.e., FCN, DUC), the Partition-based strategies need higher bandwidths ($\geq$50 Mbps) for latency convergence. For Non-Partitioned models with large input data sizes (ResNe(x)t), the Mobile and Edge deployment latencies converge at 50 Mbps. In general, the Cloud tier outperforms the Edge and Mobile tier for the Non-Partitioning operators when MEC bandwidth is at least 50 Mbps, but remains suboptimal under lower bandwidth conditions. Additionally, Mobile-Edge Partitioning-based strategies latency performance consistently exceeds Mobile-Cloud and Edge-Cloud alternatives.

Our study focuses on the impact of network bandwidth variations across Mobile-Edge and Edge-Cloud environments, as this is a critical factor for latency performance evaluation. However, system-level factors such as CPU/memory usage, network throughput, and energy consumption were not included in our current analysis. These factors may also influence overall latency performance and merit further investigation. While we scoped this study to specifically analyze the effects of network bandwidth, future work could incorporate these additional variables to provide a more holistic evaluation of system performance under varying resource constraints. The provided empirical results for these operators on the four Image Classification and Segmentation subjects give valuable insights into the speed and performance of operators across and within tiers under various deployment scenarios, which could inspire work related to the automation of these operators for future studies in the field of MLOps. Additionally, the empirical approach employed in our study, and the empirical results obtained, can be used as a foundation for developing recommendation systems for Edge AI operators.



\section{Compliance with Ethical Standards}
\noindent \textbf{Conflict of Interest:} All authors certify that they have no affiliations with or involvement in any organization or entity with any financial interest or non-financial interest in the subject matter or materials discussed in this manuscript.

\noindent \textbf{Funding:} This study was funded by NSERC.

\noindent \textbf{Ethical Approval:} This article does not contain any studies with human participants or animals performed by any of the authors.

\noindent \textbf{Informed Consent:} Not applicable.

\section{Data Availability Statement}
The models generated and analyzed during the current study are available at the following link: 

https://github.com/SAILResearch/wip-24-jaskirat-black-box-edge-operators.git.

\def\UrlBreaks{\do\/\do-\do.} 
\bibliographystyle{acm}
\label{bibliography}
\bibliography{citations.bib}
\clearpage
\begin{appendices}
\section{Graphical Illustrations of Manual Operators}
\label{Graphical Illustrations of Manual Operators}

\begin{figure}[!htbp]
\centering
\includegraphics[width=1\textwidth]{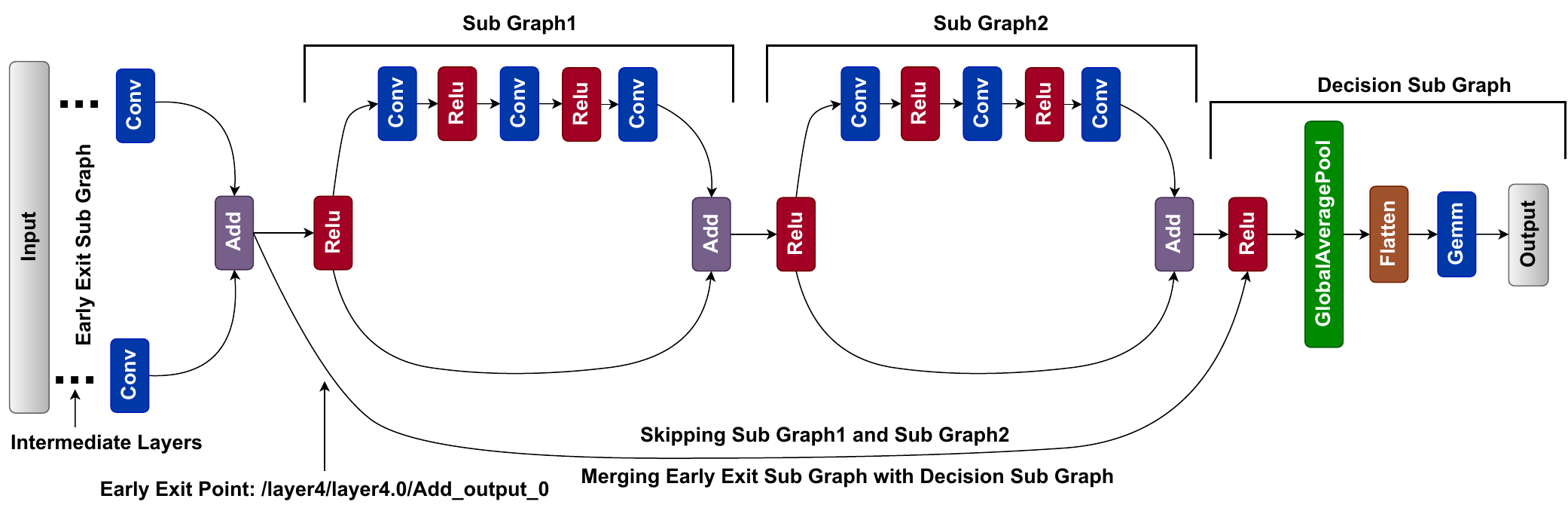}
\caption{Graphical Illustration of Early Exiting for ResNet and ResNext}
\label{approach_earlyexiting_resnet_resnext}
\end{figure}

\begin{figure}[!htbp]
\centering
\includegraphics[width=1\textwidth]{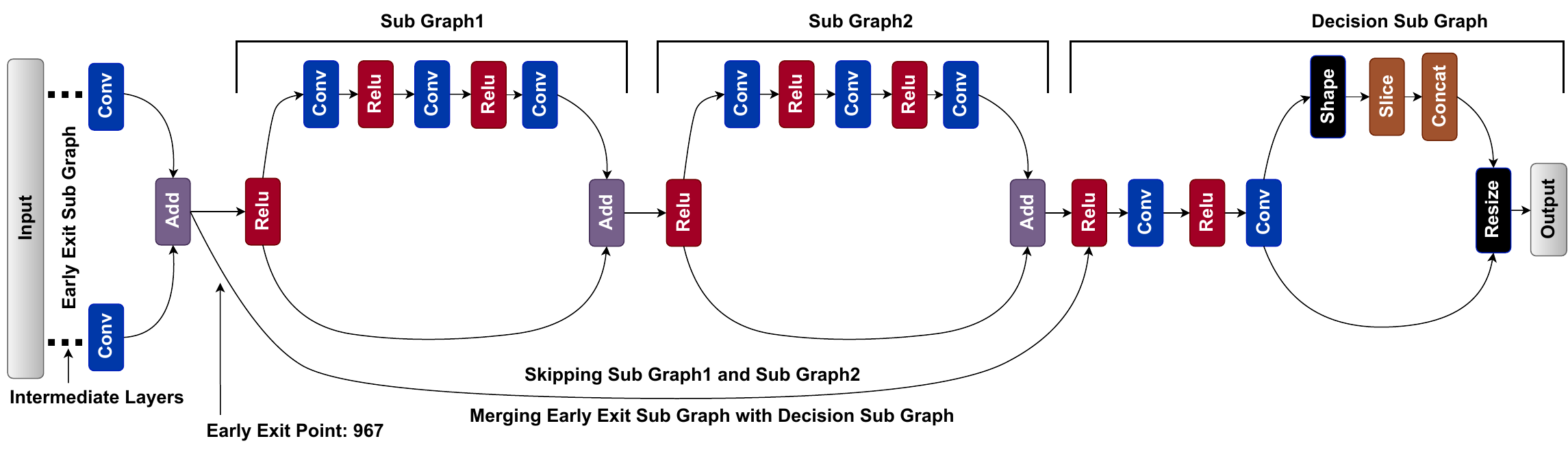}
\caption{Graphical Illustration of Early Exiting for FCN}
\label{approach_earlyexiting_fcn}
\end{figure}
\begin{figure}[!htbp]
\centering
\includegraphics[width=1\textwidth]{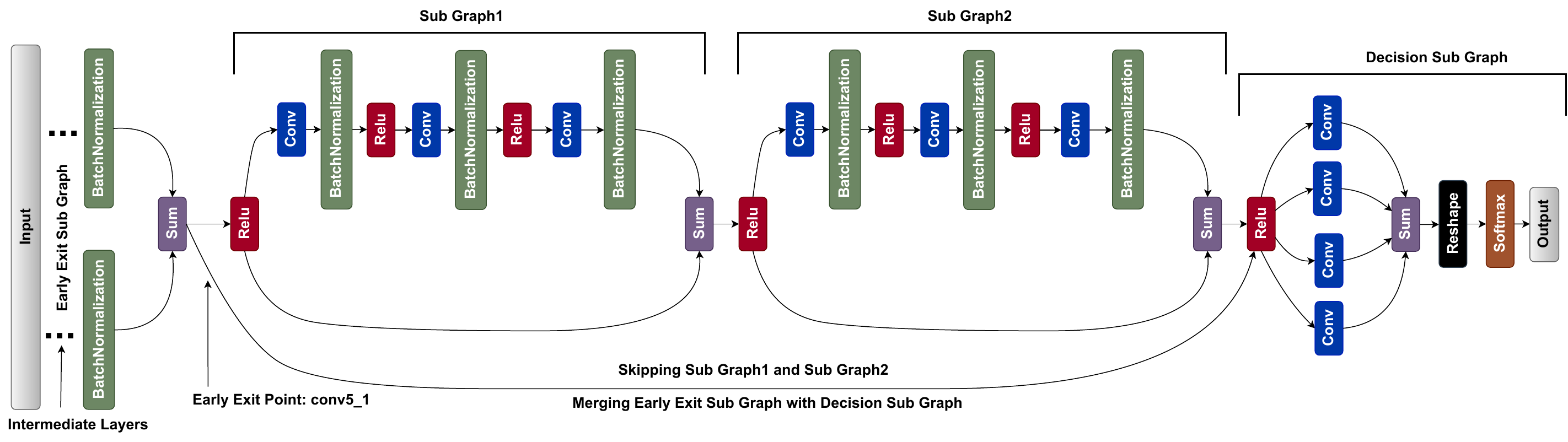}
\caption{Graphical Illustration of Early Exiting for DUC}
\label{approach_earlyexiting_duc}
\end{figure}
\begin{figure}[!htbp]
\centering
\includegraphics[width=1\textwidth]{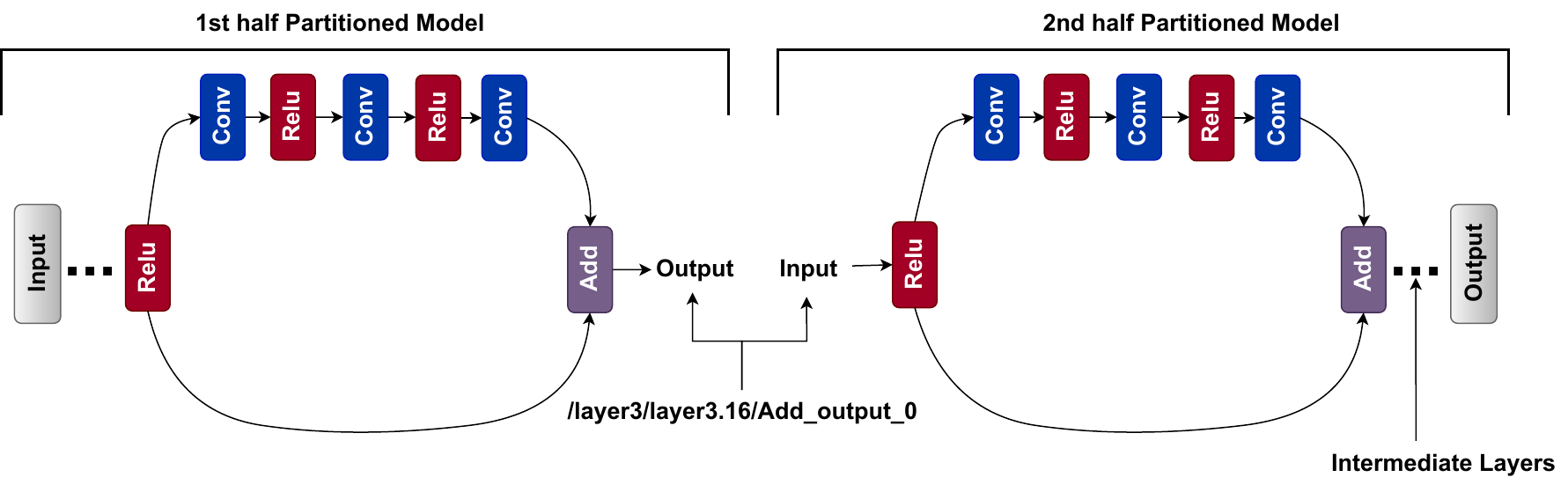}
\caption{Graphical Illustration of Model Partitioning for ResNet and ResNext}
\label{approach_partitioning_resnet_resnext}
\end{figure}
\begin{figure}[!htbp]
\centering
\includegraphics[width=1\textwidth]{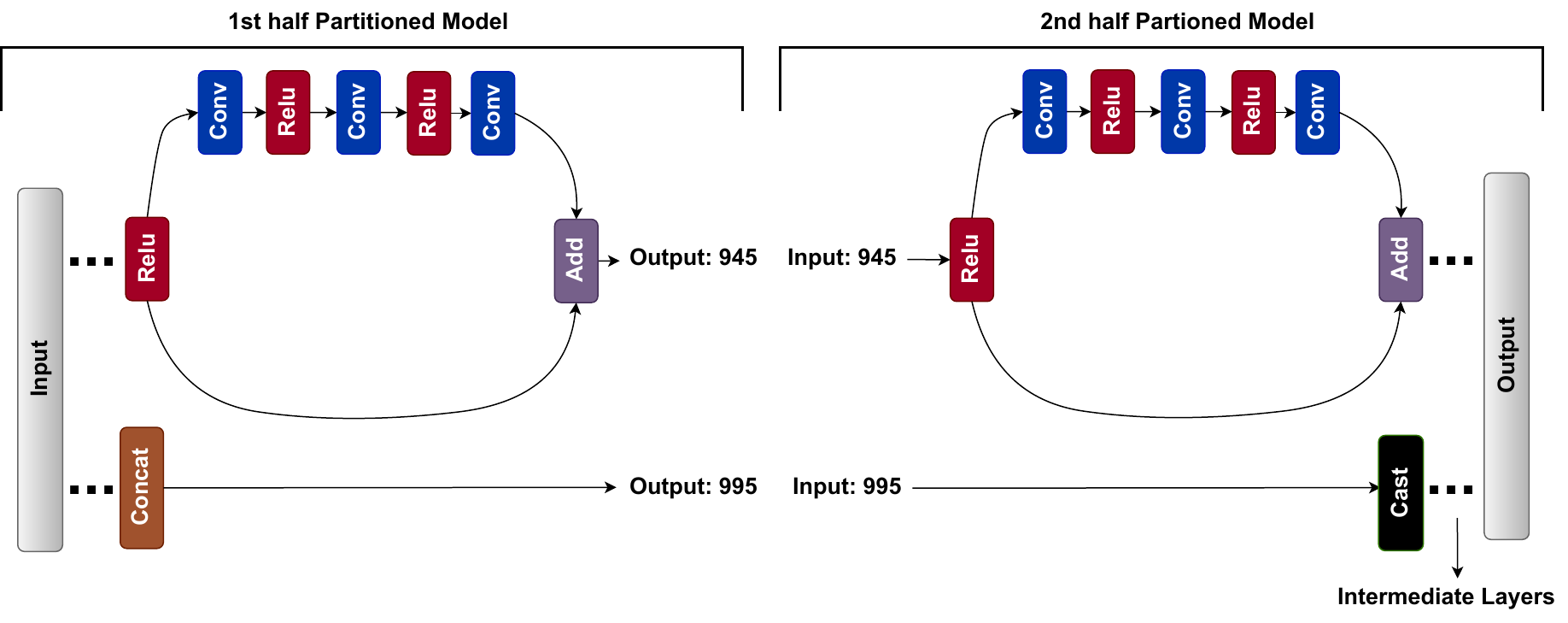}
\caption{Graphical Illustration of Model Partitioning for FCN}
\label{approach_partitioning_fcn}
\end{figure}
\begin{figure}[!htbp]
\centering
\includegraphics[width=1\textwidth]{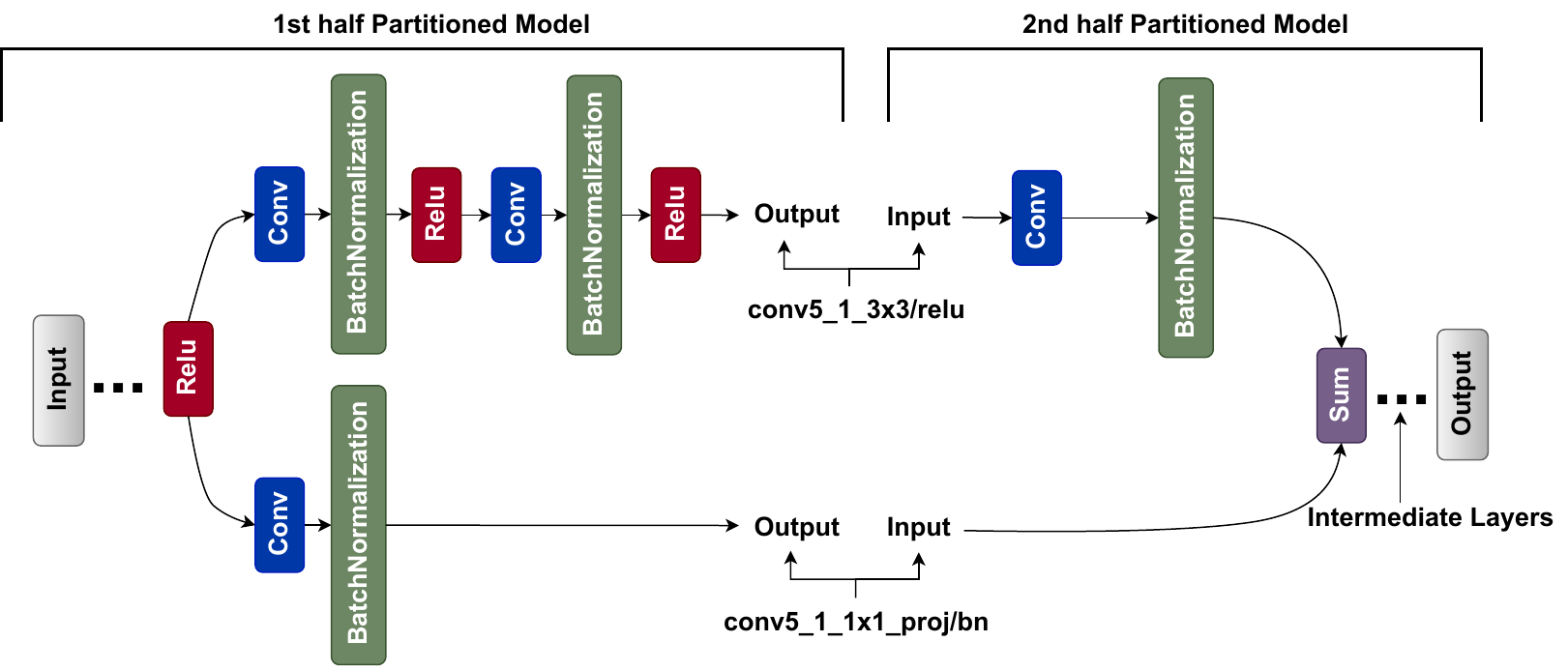}
\caption{Graphical Illustration of Model Partitioning for DUC}
\label{approach_partitioning_duc}
\end{figure}
\begin{figure}[!htbp]
\centering
\includegraphics[width=1\textwidth]{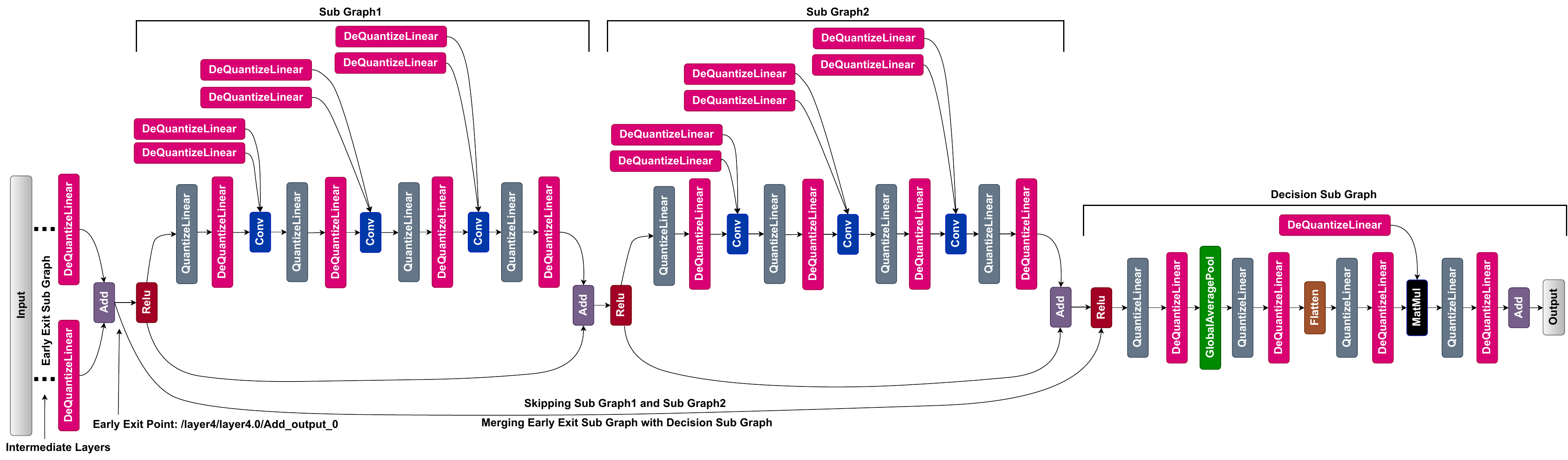}
\caption{Graphical Illustration of Quantized Early Exit for ResNet and ResNext}
\label{approach_quantized_earlyexit_resnet_resnext}
\end{figure}
\begin{figure}[!htbp]
\centering
\includegraphics[width=1\textwidth]{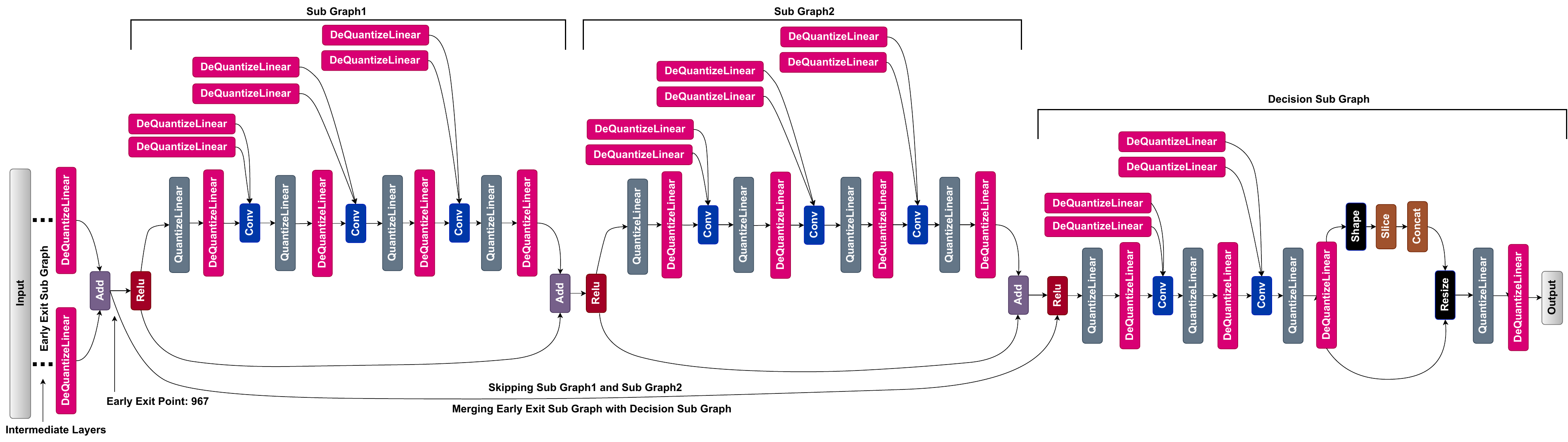}
\caption{Graphical Illustration of Quantized Early Exit for FCN}
\label{approach_quantized_earlyexit_fcn}
\end{figure}
\begin{figure}[!htbp]
\centering
\includegraphics[width=1\textwidth]{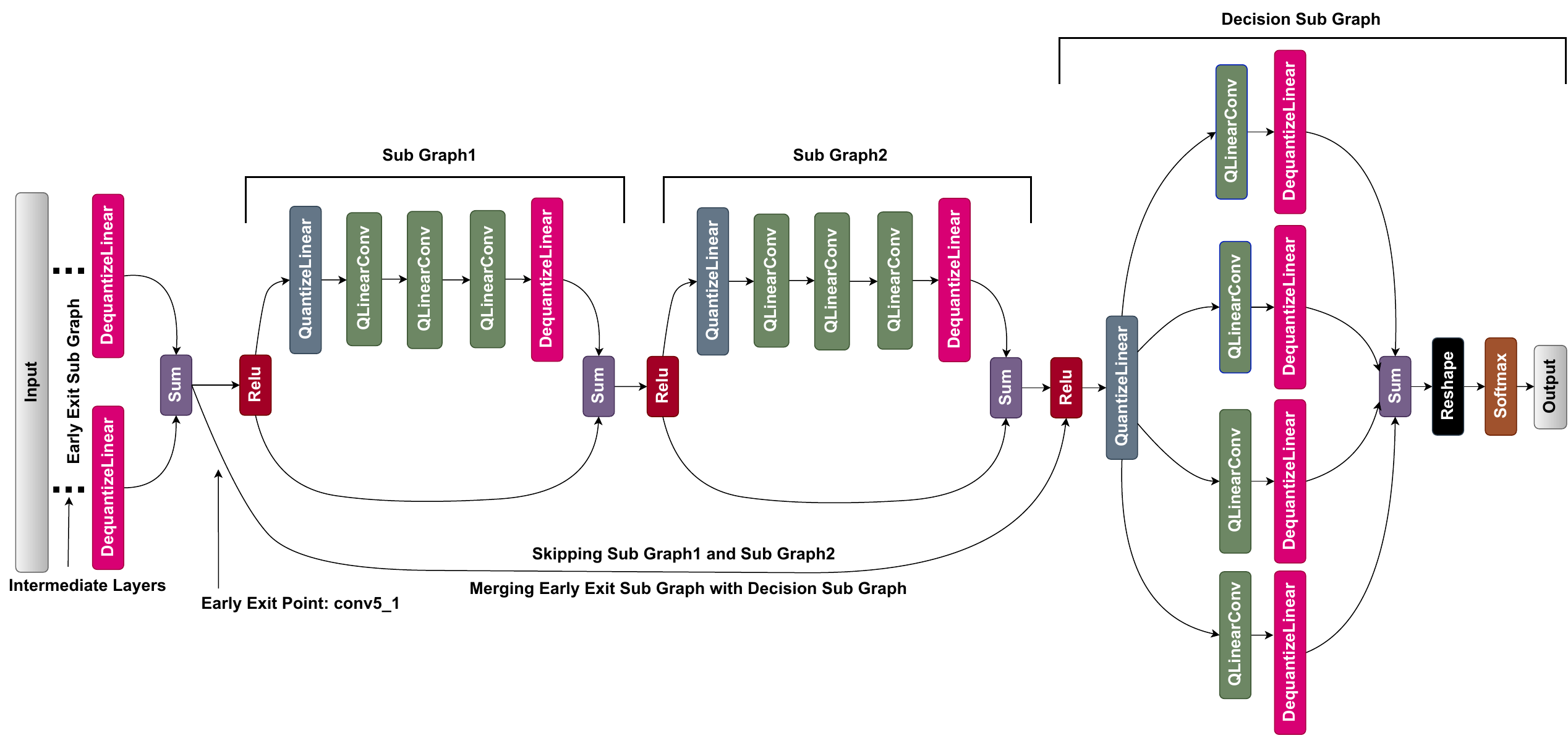}
\caption{Graphical Illustration of Quantized Early Exit for DUC}
\label{approach_quantized_earlyexit_duc}
\end{figure}
\begin{figure}[!htbp]
\centering
\includegraphics[width=1\textwidth]{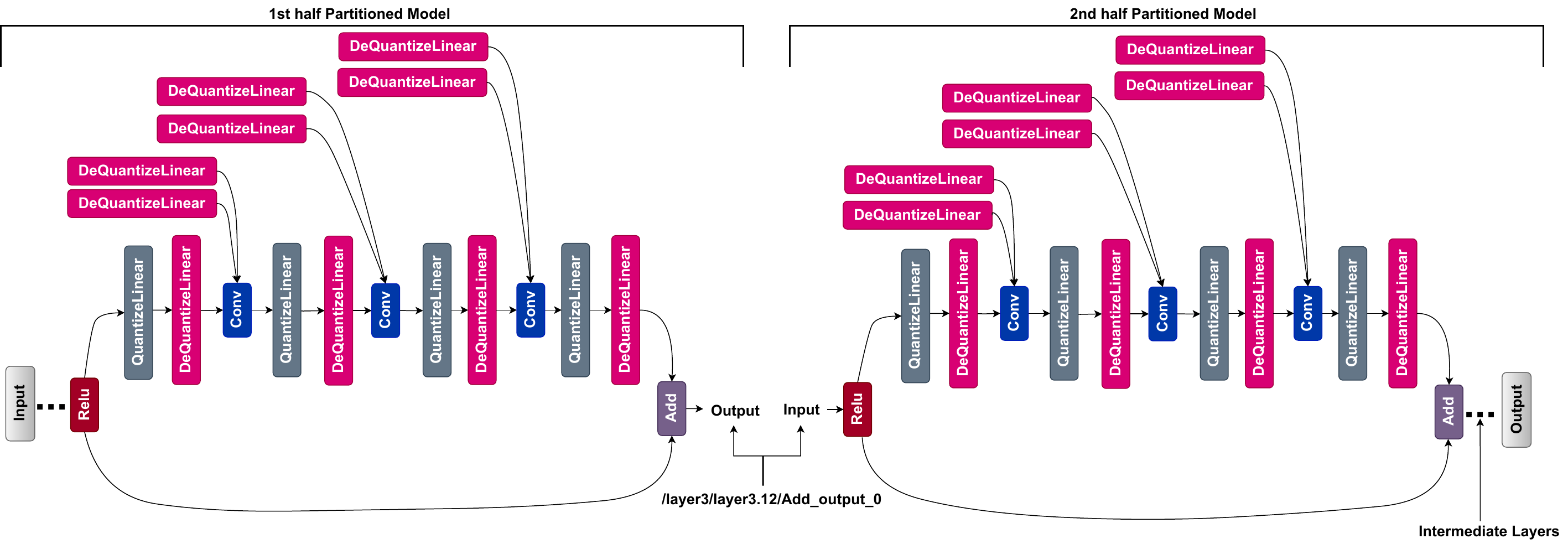}
\caption{Graphical Illustration of Quantized Early Exit Partitioning for ResNet and ResNext}
\label{approach_quantized_earlyexit_partition_resnet_resnext}
\end{figure}
\begin{figure}[!htbp]
\centering
\includegraphics[width=1\textwidth]{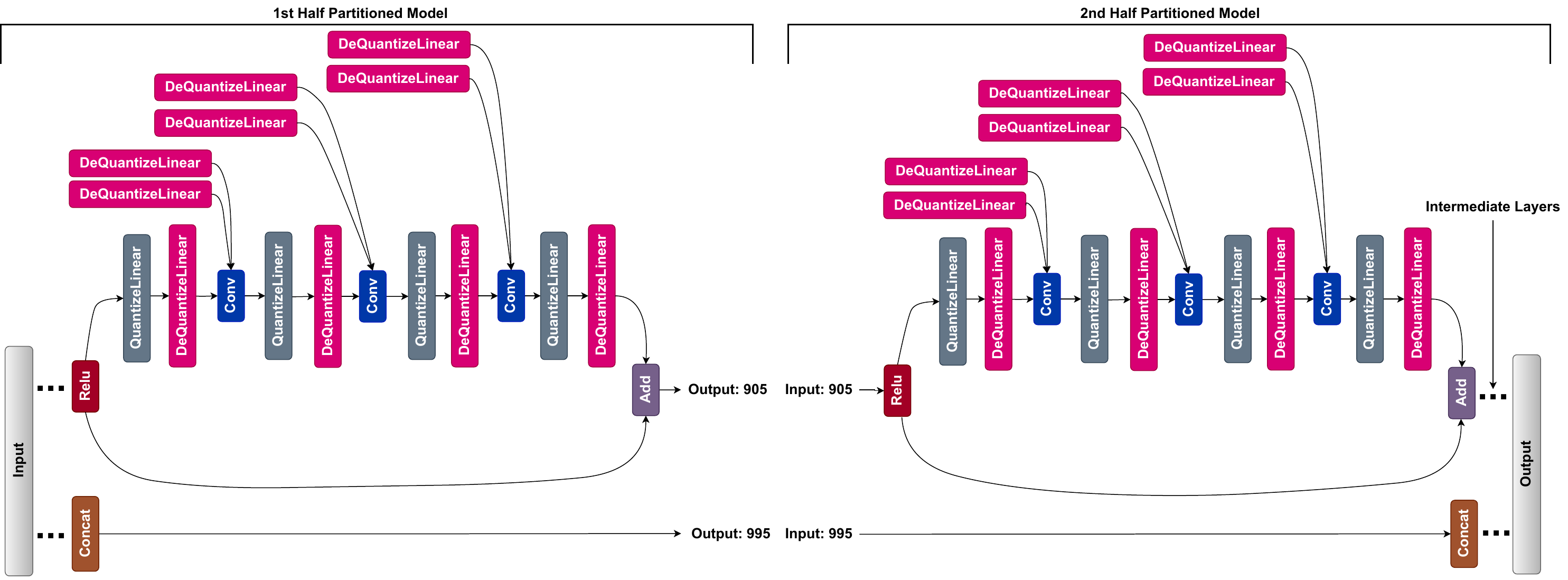}
\caption{Graphical Illustration of Quantized Early Exit Partitioning for FCN}
\label{approach_quantized_earlyexit_partition_fcn}
\end{figure}
\begin{figure}[!htbp]
\centering
\includegraphics[width=1\textwidth]{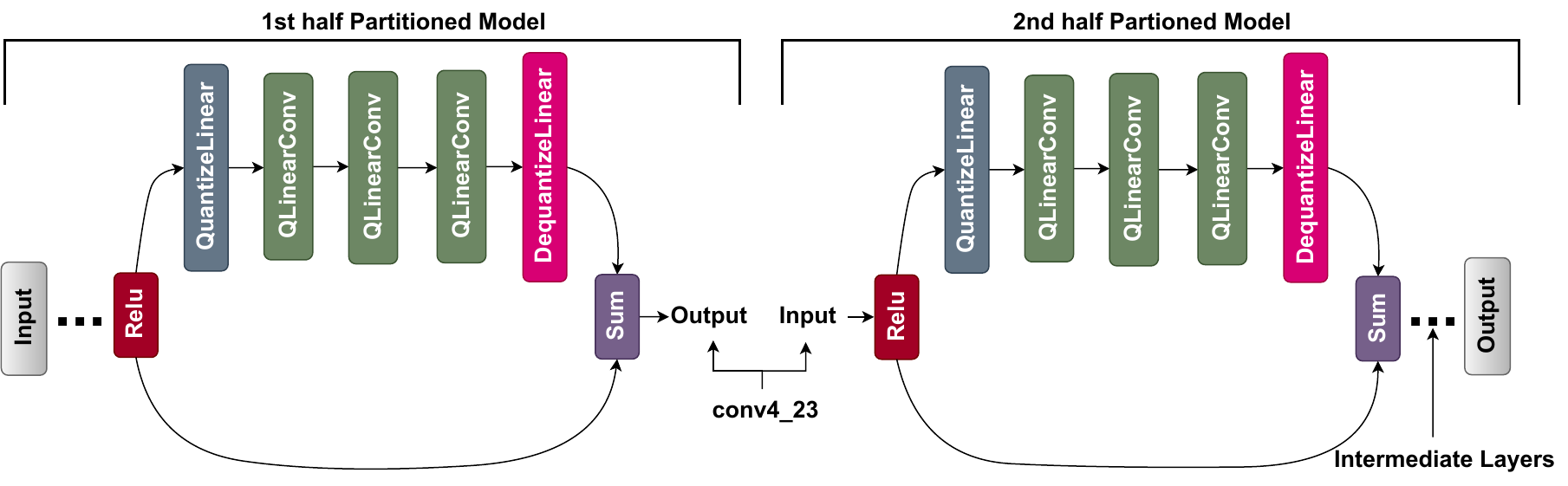}
\caption{Graphical Illustration of Quantized Early Exit Partitioning for DUC}
\label{approach_quantized_earlyexit_partition_duc}
\end{figure}
\clearpage
\section{QQ Plots}
\label{qq_plots}
\begin{figure}[htbp]
\centering
\includegraphics[width=1\textwidth]{Rq1_normality_test.png}
\caption{Graphical Illustration of QQ plots for RQ1 Deployment Strategies}
\label{normality_results_rq1}
\end{figure}
\begin{figure}[htbp]
\centering
\includegraphics[width=1\textwidth]{rq2_normailty_test.png}
\caption{Graphical Illustration of QQ plots for RQ2 Deployment Strategies}
\label{normality_results_rq2}
\end{figure}
\begin{figure}[htbp]
\centering
\includegraphics[width=1\textwidth]{rq3_normailty_test.png}
\caption{Graphical Illustration of QQ plots for RQ3 Deployment Strategies}
\label{normality_results_rq3}
\end{figure}

\begin{figure}[!htbp]
\centering
\adjustbox{max width=\textwidth, max height=0.90\textheight}{
  \includegraphics{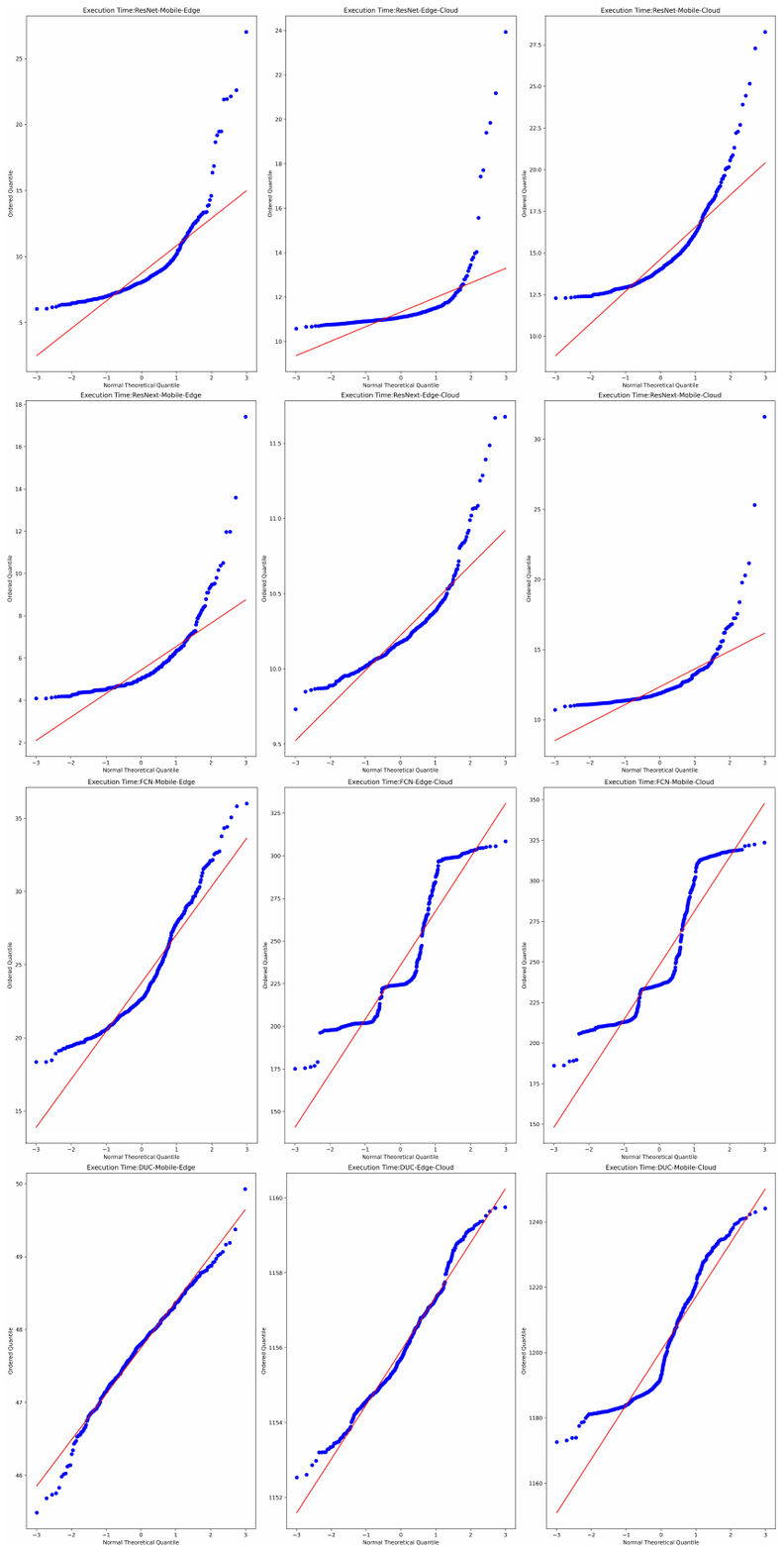}
}
\caption{Graphical Illustration of QQ plots for RQ4 Deployment Strategies}
\label{normality_results_rq4}
\end{figure}

\begin{figure}[!htbp]
\centering
\includegraphics[width=\textwidth,height=0.8\textheight,keepaspectratio]{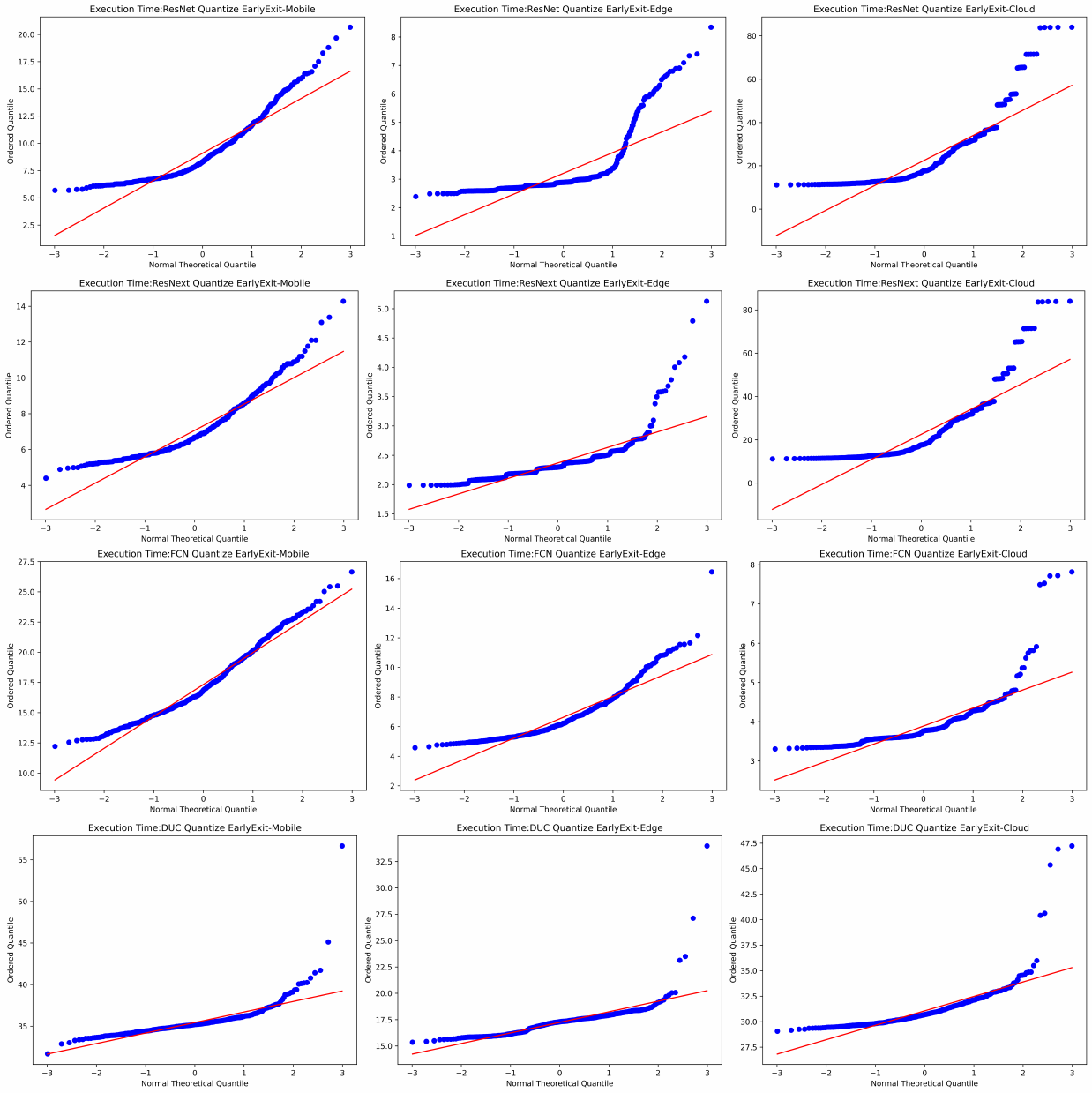}
\caption{Graphical Illustration of QQ plots for RQ5 Deployment Strategies}
\label{normality_results_rq5_1}
\end{figure}

\begin{figure}[!htbp]
\centering
\includegraphics[width=\textwidth,height=0.8\textheight,keepaspectratio]{rq5_normailty_test_2.pdf}
\caption{Graphical Illustration of QQ plots for RQ5 Deployment Strategies}
\label{normality_results_rq5_2}
\end{figure}

\begin{figure}[htbp]
\centering
\includegraphics[width=1\textwidth]{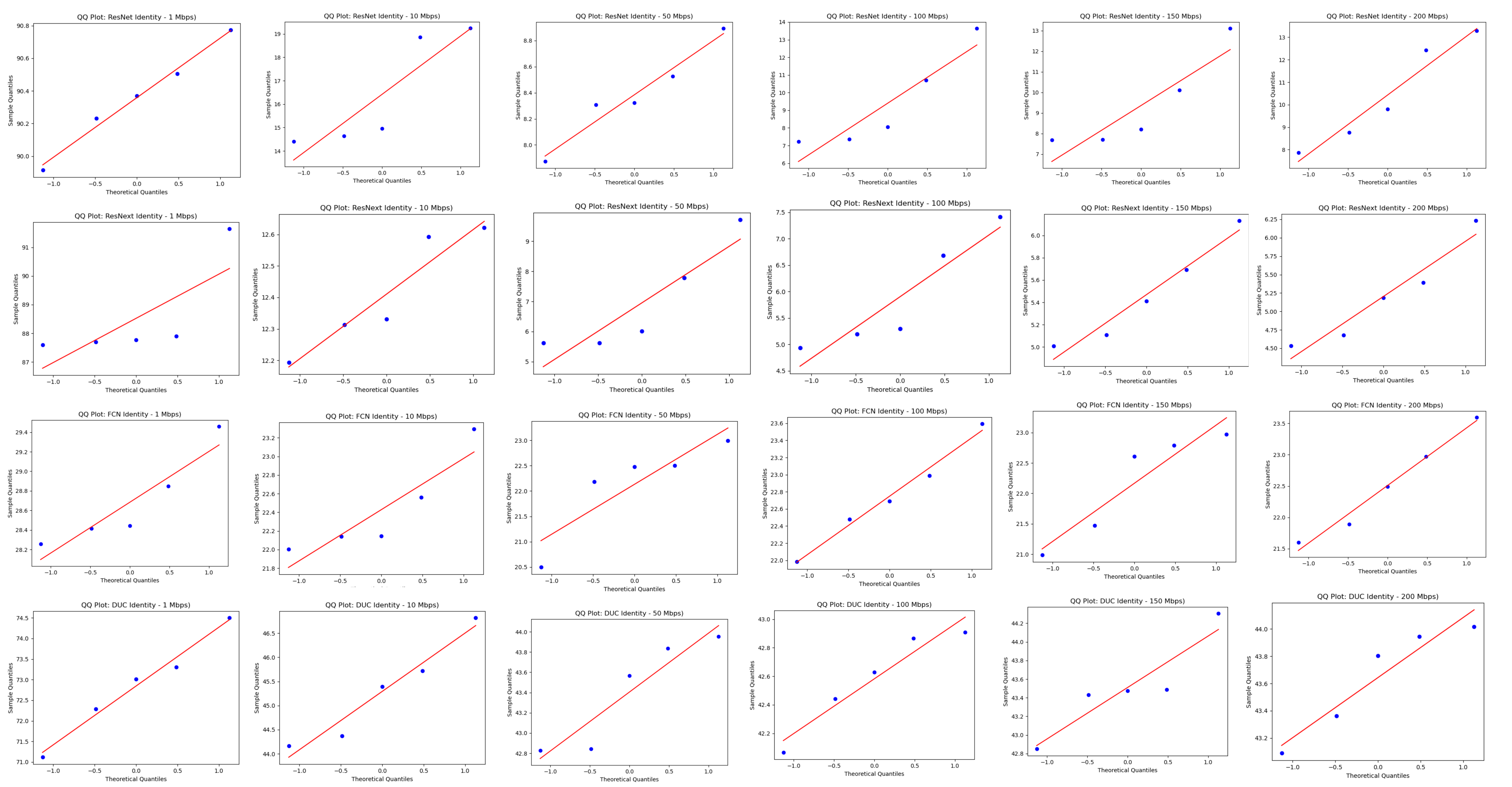}
\caption{Graphical Illustration of QQ plots for RQ6 Mobile Identity Deployment Strategies}
\label{normality_results_rq6_mobile_identity}
\end{figure}
\begin{figure}[htbp]
\centering
\includegraphics[width=1\textwidth]{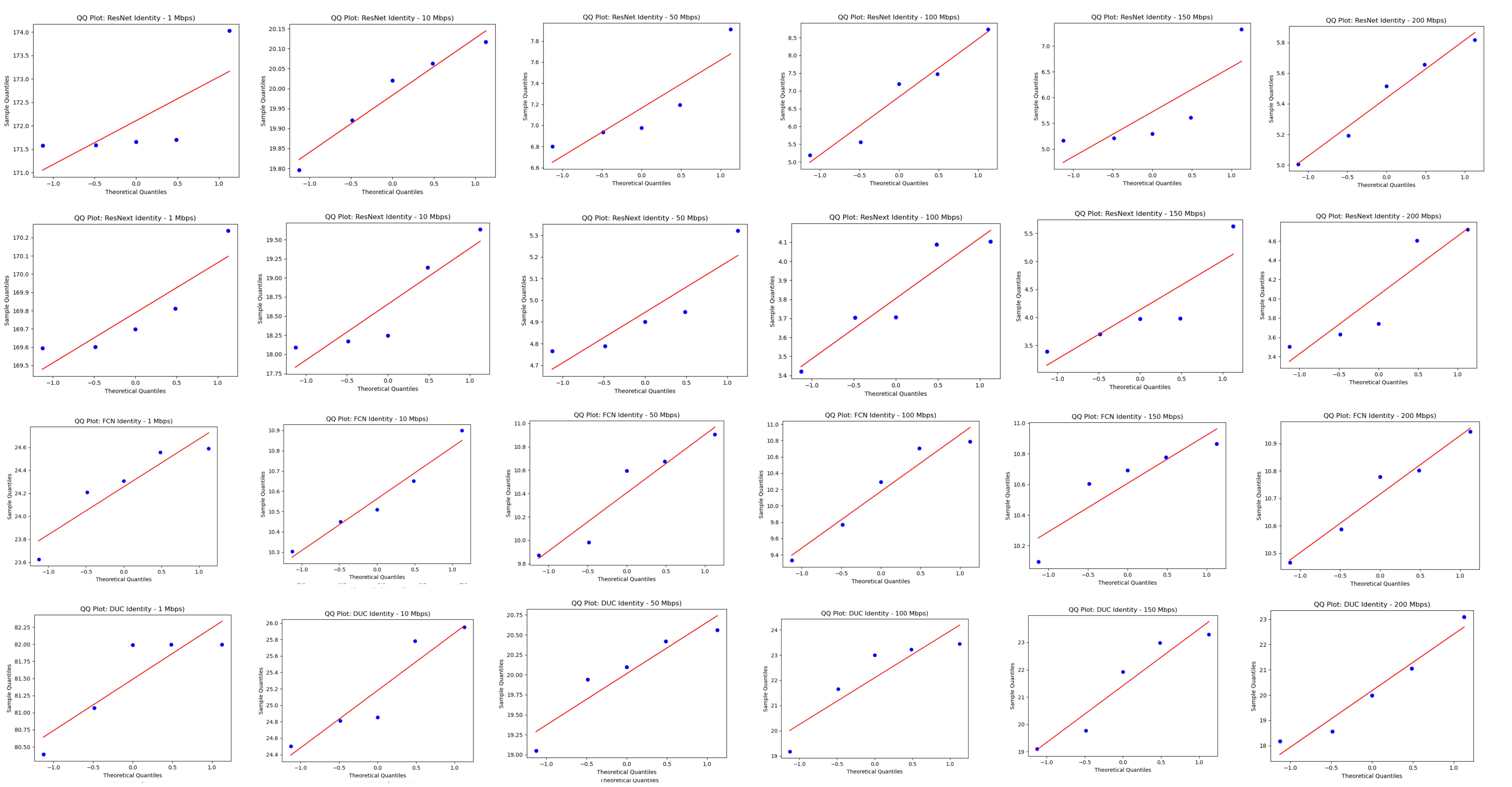}
\caption{Graphical Illustration of QQ plots for RQ6 Edge Identity Deployment Strategies}
\label{normality_results_rq6_edge_identity}
\end{figure}
\begin{figure}[htbp]
\centering
\includegraphics[width=1\textwidth]{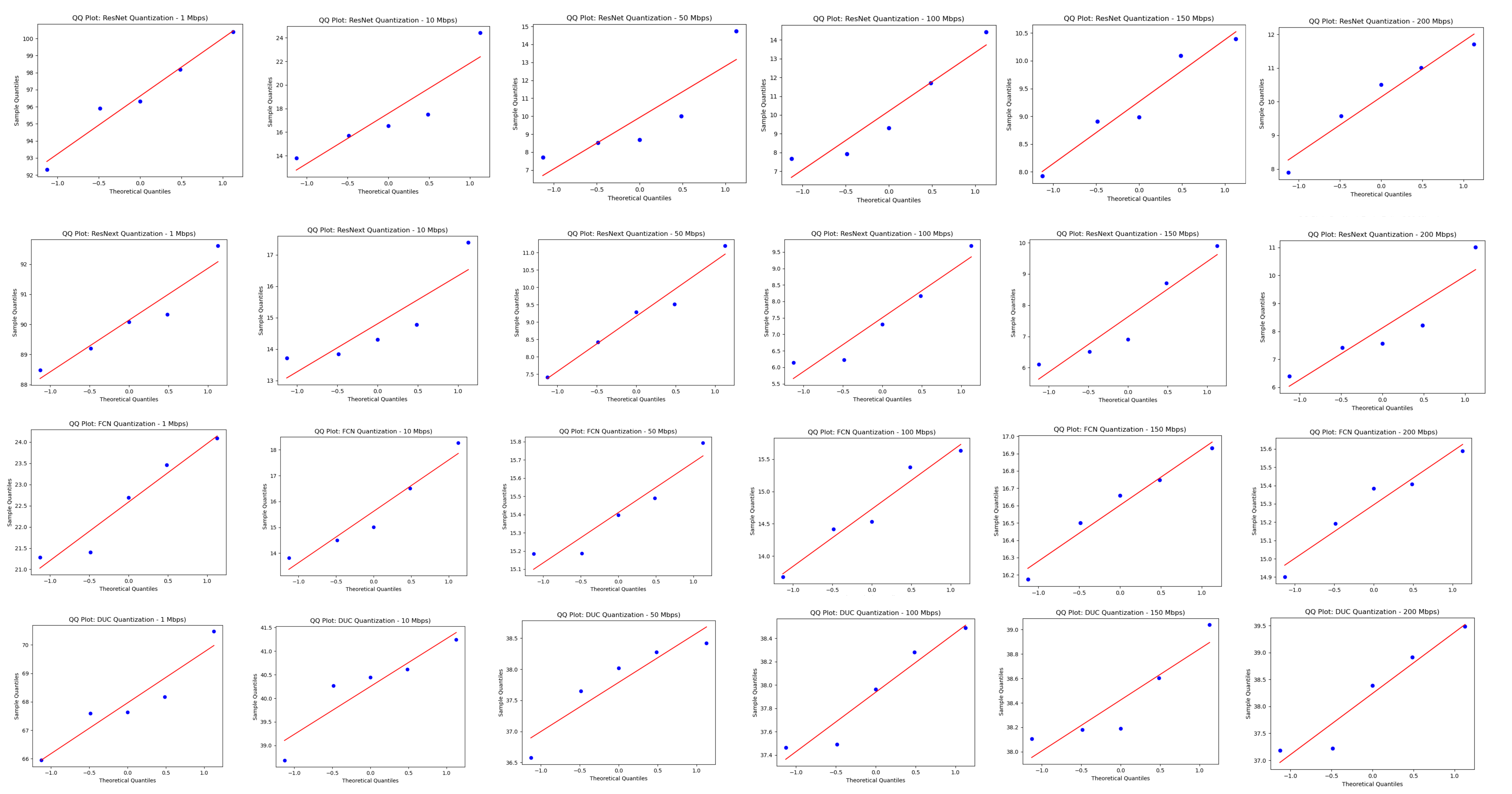}
\caption{Graphical Illustration of QQ plots for RQ6 Mobile Quantized Deployment Strategies}
\label{normality_results_rq6_mobile_quantization}
\end{figure}
\begin{figure}[htbp]
\centering
\includegraphics[width=1\textwidth]{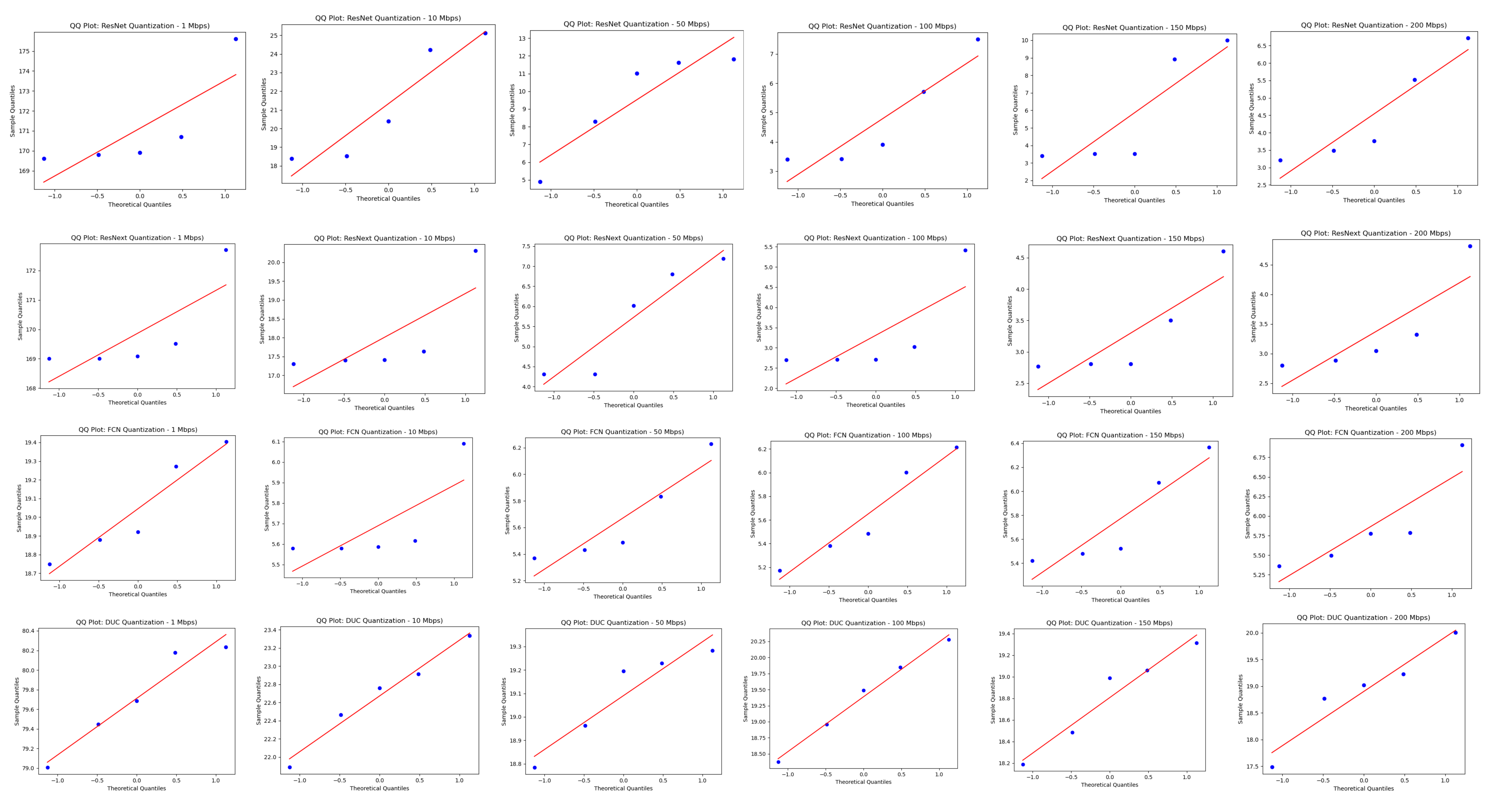}
\caption{Graphical Illustration of QQ plots for RQ6 Edge Quantized Deployment Strategies}
\label{normality_results_rq6_edge_quantization}
\end{figure}
\begin{figure}[htbp]
\centering
\includegraphics[width=1\textwidth]{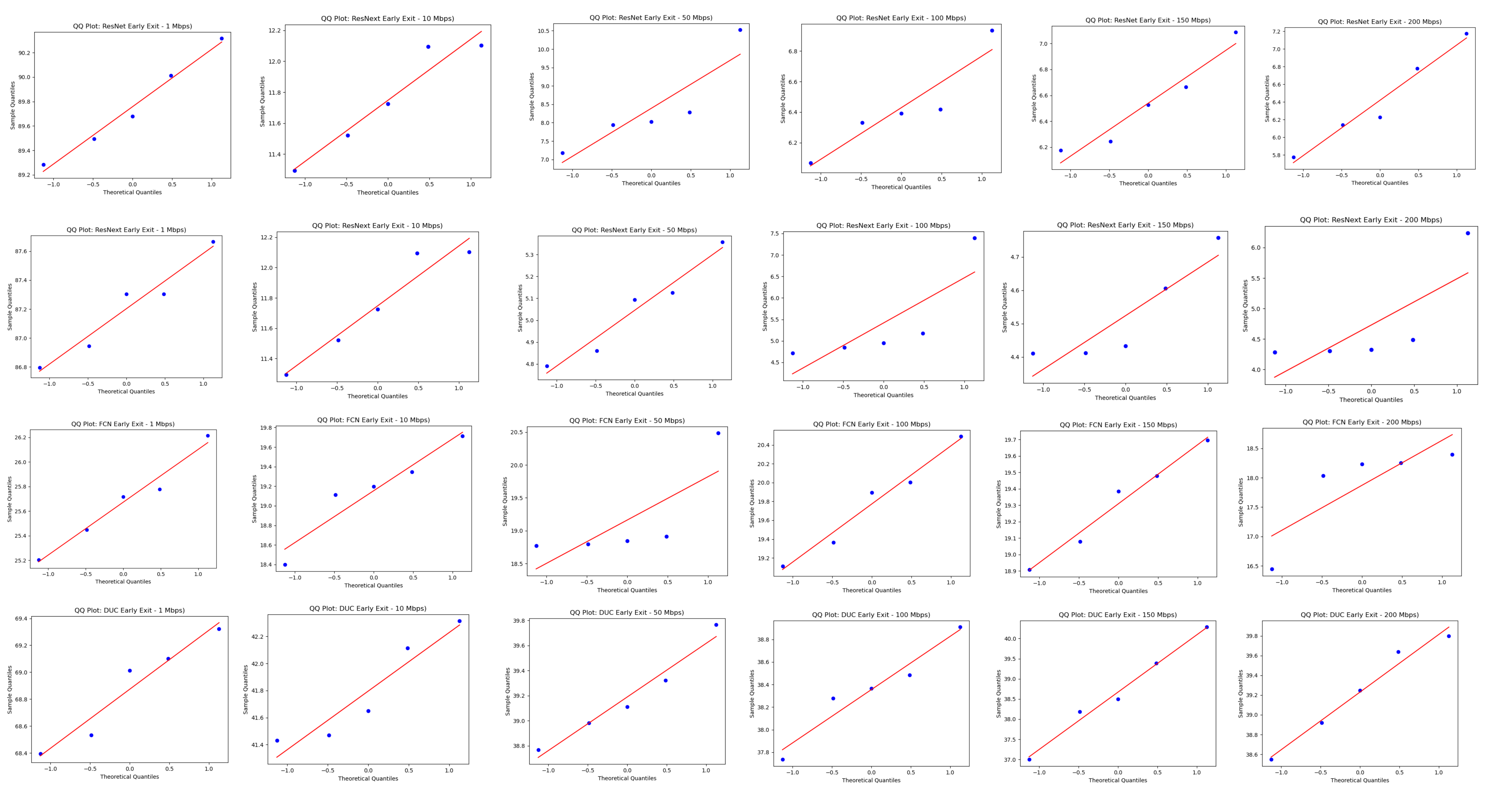}
\caption{Graphical Illustration of QQ plots for RQ6 Early Exit Mobile Deployment Strategies}
\label{normality_results_rq6_mobile_earlyexit}
\end{figure}
\begin{figure}[htbp]
\centering
\includegraphics[width=1\textwidth]{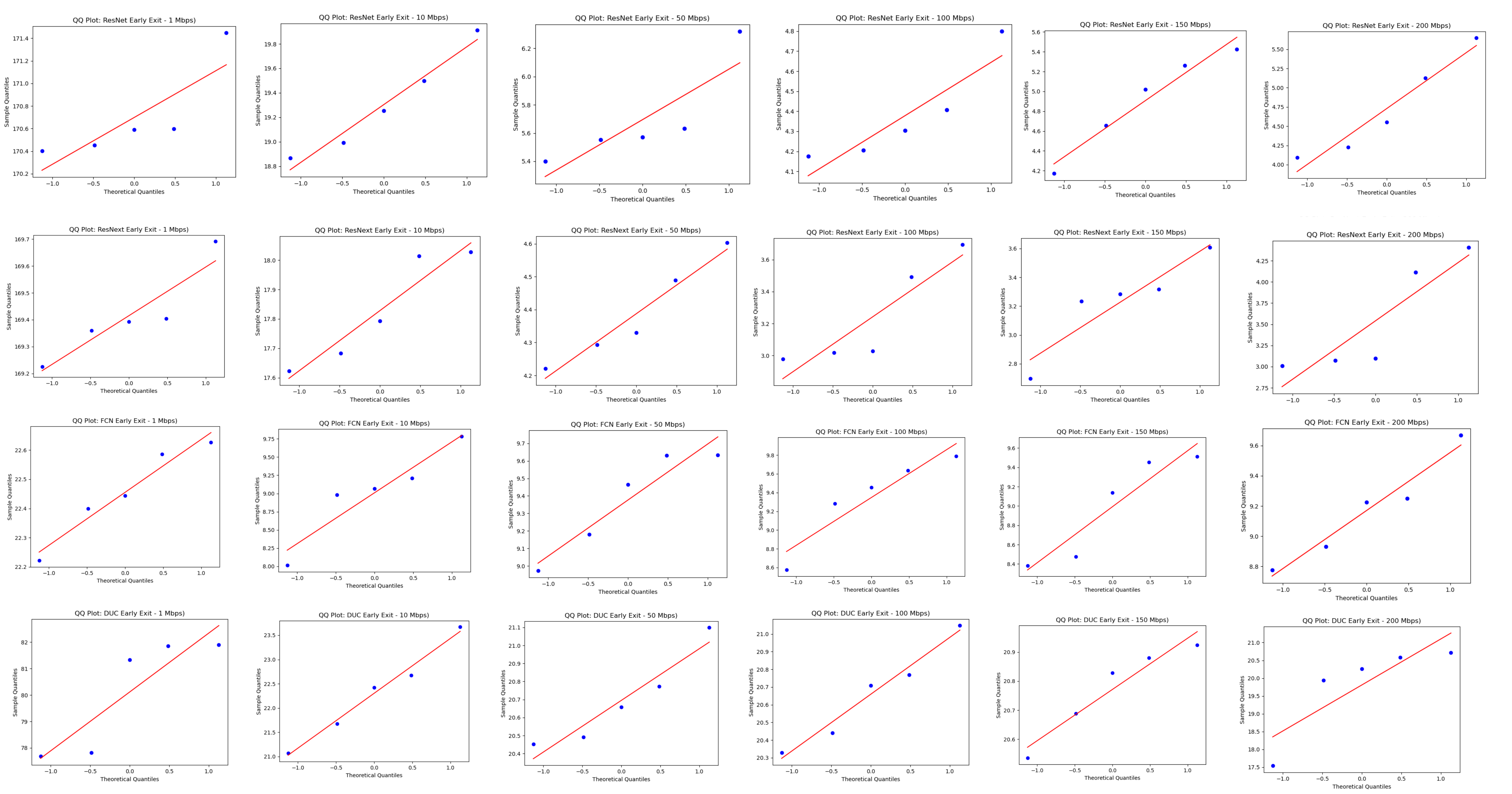}
\caption{Graphical Illustration of QQ plots for RQ6 Early Exit Edge Deployment Strategies}
\label{normality_results_rq6_edge_earlyexit}
\end{figure}
\begin{figure}[htbp]
\centering
\includegraphics[width=1\textwidth]{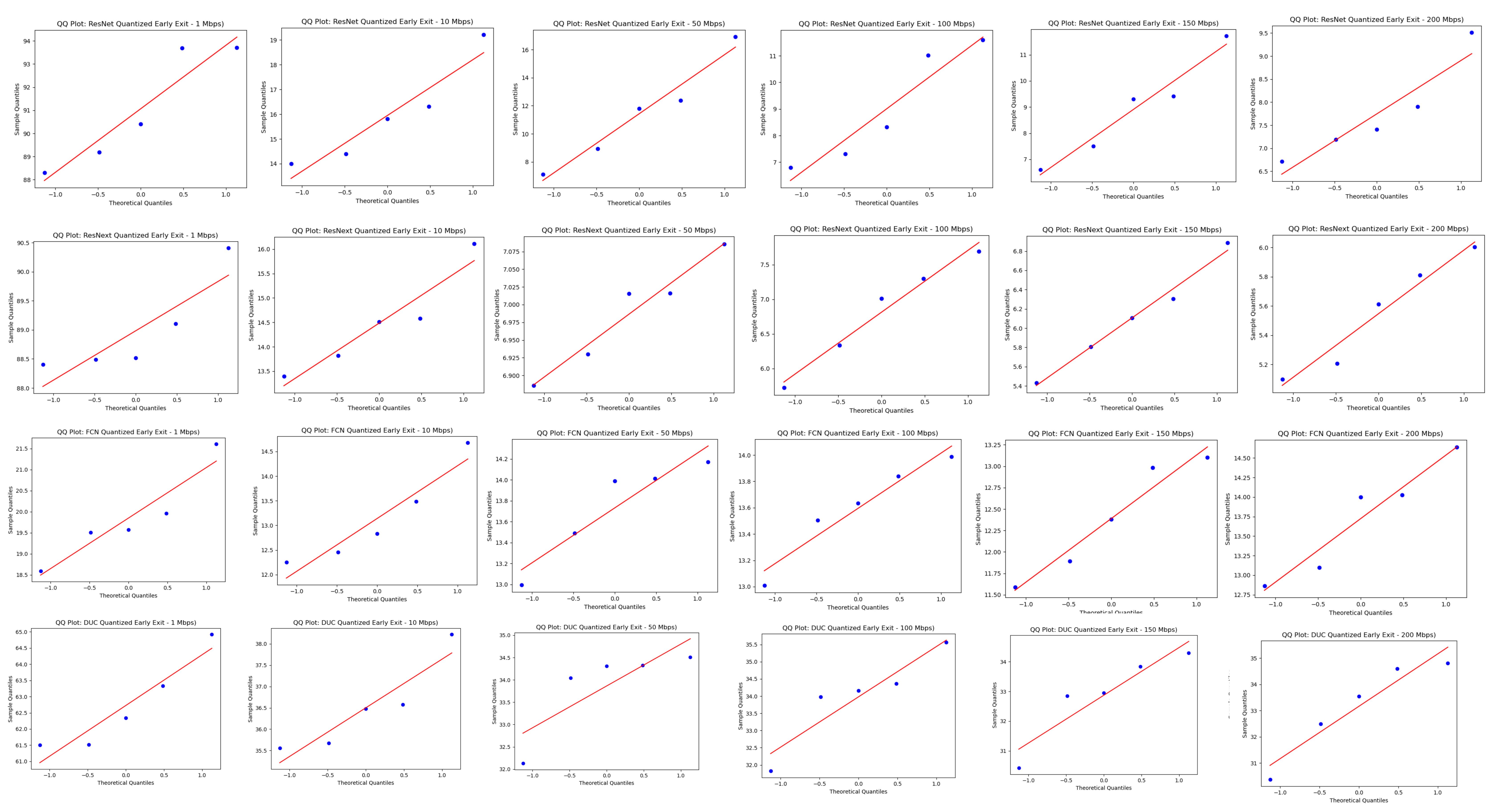}
\caption{Graphical Illustration of QQ plots for RQ6 Quantized Early Exit Mobile Deployment Strategies}
\label{normality_results_rq6_mobile_quantized_earlyexit}
\end{figure}
\begin{figure}[htbp]
\centering
\includegraphics[width=1\textwidth]{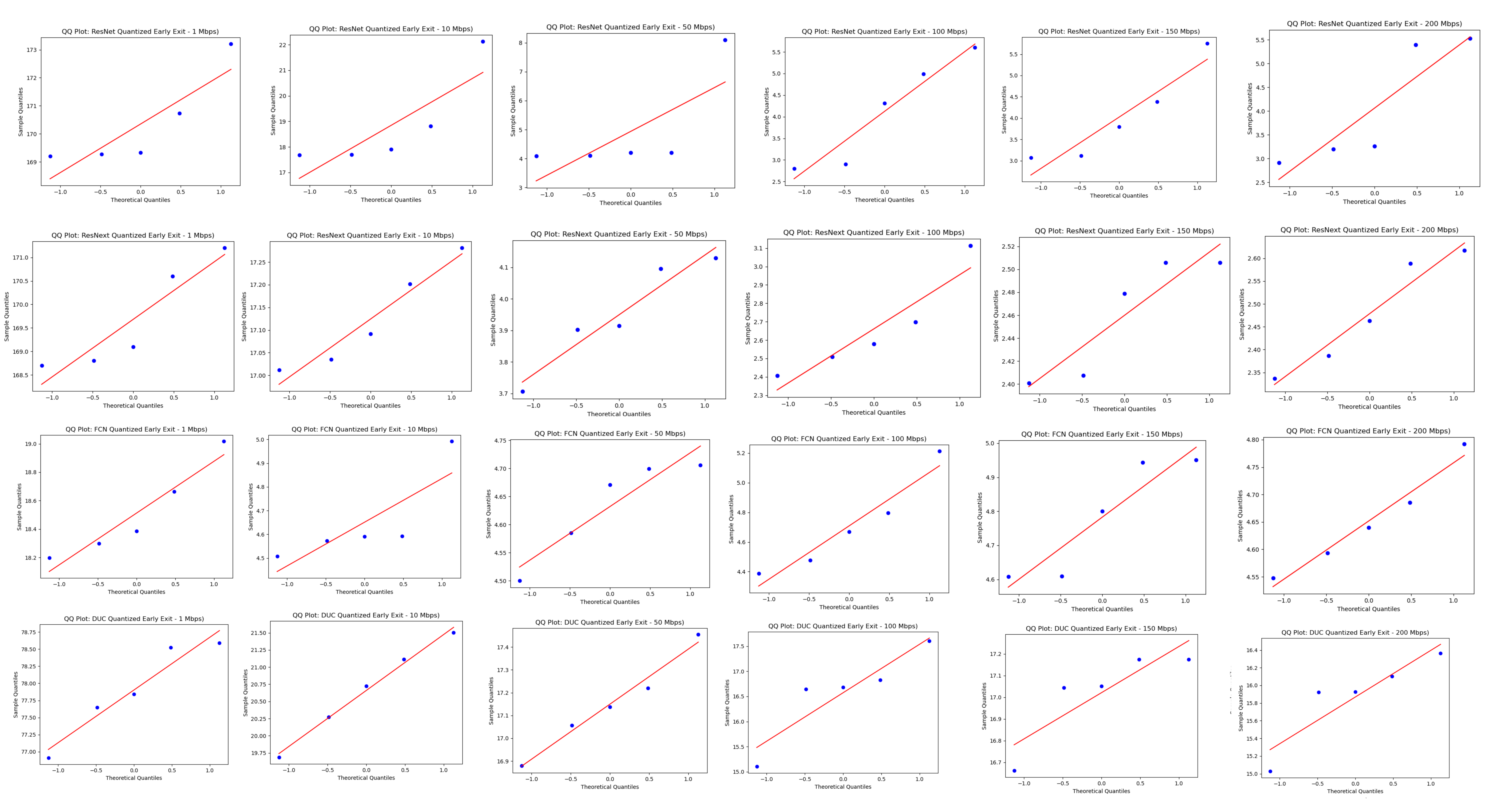}
\caption{Graphical Illustration of QQ plots for RQ6 Quantized Early Exit Edge Deployment Strategies}
\label{normality_results_rq6_edge_quantized_earlyexit}
\end{figure}
\begin{figure}[htbp]
\centering
\includegraphics[width=1\textwidth]{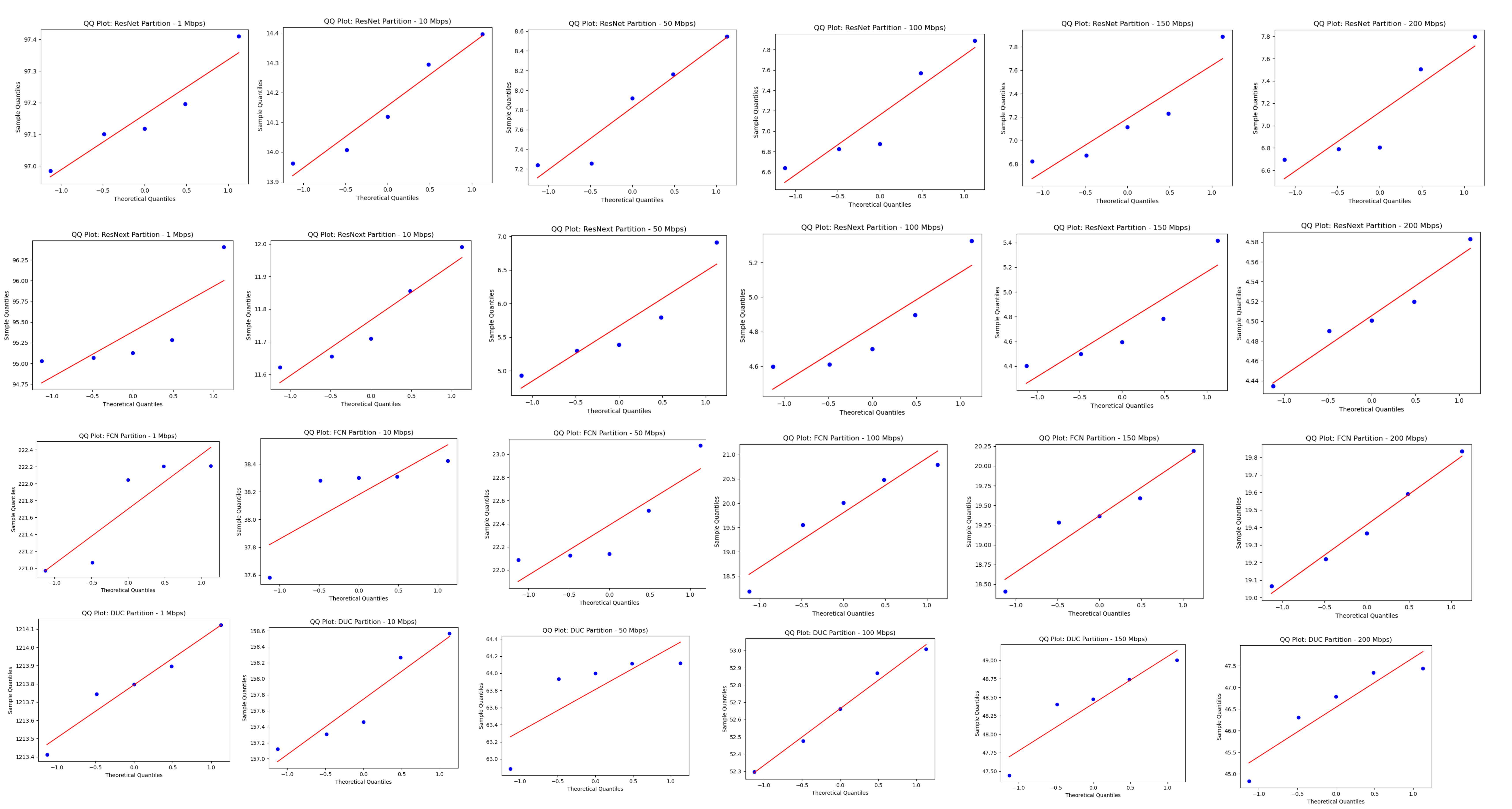}
\caption{Graphical Illustration of QQ plots for RQ6 Mobile-Edge Partition Deployment Strategies}
\label{normality_results_rq6_mobile_edge_partition}
\end{figure}
\begin{figure}[htbp]
\centering
\includegraphics[width=1\textwidth]{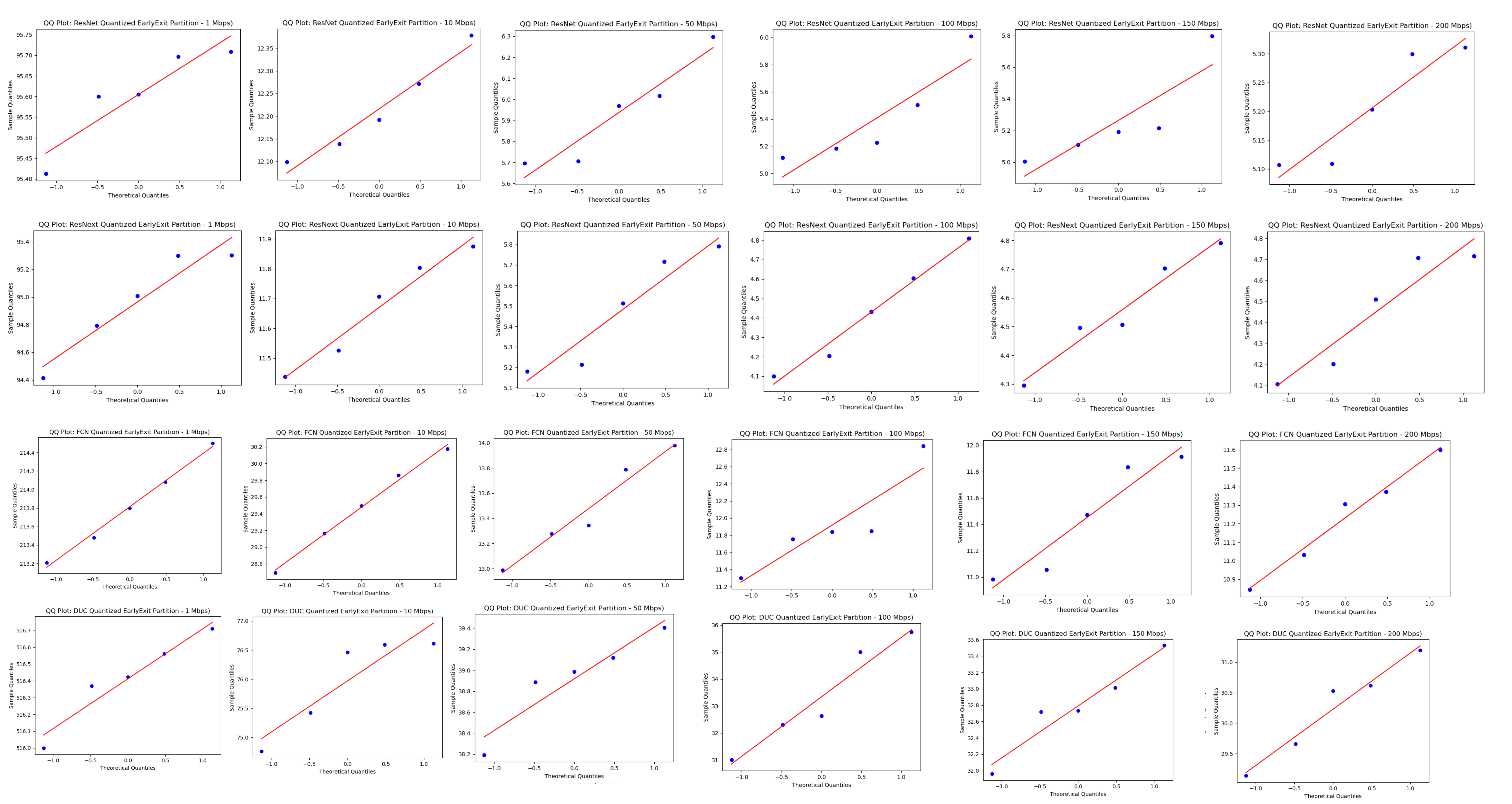}
\caption{Graphical Illustration of QQ plots for RQ6 Mobile-Edge Quantized Early Exit Partition Deployment Strategies}
\label{normality_results_rq6_mobile_edge_quantized_earlyexit_partition}
\end{figure}

\clearpage
\section{Data Analysis and Normality Assessment}
\label{data_analysis_normality_assessment}
\begin{table}[!htbp]
\caption{Descriptive statistics of the latency for RQ1, including Shapiro-Wilk p-values and normality assessment}
\centering
\resizebox{1\textwidth}{!}{%
\begin{tabular}{|l|l|l|r|r|r|r|r|c|c|}
\hline
Tier & Model & Operator & Min & Max & Median & Mean & Std & Shapiro-Wilk p & Normal? \\
\hline
Mobile & ResNet & Identity & 6.90 & 30.74 & 8.94 & 10.34 & 3.66 & $1.07 \times 10^{-27}$ & No \\
Mobile & ResNext & Identity & 4.76 & 41.58 & 6.31 & 8.27 & 5.54 & $3.66 \times 10^{-33}$ & No \\
Mobile & FCN & Identity & 21.91 & 53.66 & 27.39 & 28.78 & 4.73 & $5.30 \times 10^{-17}$ & No \\
Mobile & DUC & Identity & 42.53 & 47.89 & 45.46 & 45.48 & 0.69 & $5.21 \times 10^{-5}$ & No \\
Edge & ResNet & Identity & 4.89 & 12.90 & 5.64 & 6.10 & 1.28 & $9.59 \times 10^{-30}$ & No \\
Edge & ResNext & Identity & 3.19 & 45.67 & 3.97 & 5.56 & 4.69 & $8.59 \times 10^{-36}$ & No \\
Edge & FCN & Identity & 7.98 & 29.37 & 12.29 & 12.90 & 2.60 & $3.40 \times 10^{-22}$ & No \\
Edge & DUC & Identity & 17.93 & 25.11 & 23.65 & 23.05 & 1.50 & $9.40 \times 10^{-23}$ & No \\
Cloud & ResNet & Identity & 11.14 & 84.05 & 17.62 & 22.51 & 13.25 & $1.97 \times 10^{-26}$ & No \\
Cloud & ResNext & Identity & 11.19 & 83.89 & 17.61 & 22.49 & 13.21 & $2.04 \times 10^{-26}$ & No \\
Cloud & FCN & Identity & 3.24 & 7.83 & 3.77 & 3.89 & 0.55 & $8.64 \times 10^{-30}$ & No \\
Cloud & DUC & Identity & 27.55 & 31.34 & 28.62 & 28.85 & 0.89 & $5.86 \times 10^{-15}$ & No \\
\hline
\end{tabular}
\label{rq1_descriptive_statistics}
}
\end{table}

\begin{table}[!htbp]
\caption{Descriptive statistics of the latency for RQ2, including Shapiro-Wilk p-values and normality assessment}
\centering
\resizebox{\textwidth}{!}{%
\begin{tabular}{|l|l|l|r|r|r|r|r|c|c|}
\hline
Tier & Model & Operator & Min & Max & Median & Mean & Std & Shapiro-Wilk p & Normal? \\
\hline
Mobile & ResNet & Quantize & 6.76 & 23.09 & 10.67 & 11.37 & 3.34 & $1.88 \times 10^{-19}$ & No \\
Mobile & ResNext & Quantize & 5.22 & 15.49 & 7.29 & 7.92 & 1.89 & $5.74 \times 10^{-11}$ & No \\
Mobile & FCN & Quantize & 13.71 & 32.78 & 19.86 & 20.29 & 2.89 & $2.02 \times 10^{-11}$ & No \\
Mobile & DUC & Quantize & 36.16 & 41.47 & 38.70 & 38.81 & 1.00 & $2.32 \times 10^{-7}$ & No \\
Edge & ResNet & Quantize & 2.89 & 10.96 & 3.48 & 4.56 & 1.96 & $2.63 \times 10^{-13}$ & No \\
Edge & ResNext & Quantize & 2.19 & 7.02 & 2.67 & 2.83 & 0.66 & $1.50 \times 10^{-12}$ & No \\
Edge & FCN & Quantize & 5.47 & 16.53 & 7.49 & 7.95 & 1.78 & $6.52 \times 10^{-9}$ & No \\
Edge & DUC & Quantize & 17.15 & 21.42 & 19.59 & 19.50 & 0.65 & $1.79 \times 10^{-13}$ & No \\
Cloud & ResNet & Quantize & 11.16 & 83.96 & 17.64 & 22.49 & 13.21 & $1.98 \times 10^{-26}$ & No \\
Cloud & ResNext  & Quantize & 11.17 & 83.85 & 17.66 & 22.49 & 13.21 & $2.04 \times 10^{-26}$ & No \\
Cloud & FCN & Quantize & 3.15 & 7.73 & 3.78 & 3.90 & 0.56 & $7.97 \times 10^{-30}$ & No \\
Cloud & DUC  & Quantize & 29.60 & 36.28 & 31.00 & 31.24 & 1.00 & $2.01 \times 10^{-14}$ & No \\
\hline
\end{tabular}
\label{rq2_descriptive_statistics}
}
\end{table}

\begin{table}[!htbp]
\caption{Descriptive statistics of the latency for RQ3, including Shapiro-Wilk p-values and normality assessment}
\centering
\resizebox{\textwidth}{!}{%
\begin{tabular}{|l|l|l|r|r|r|r|r|c|c|}
\hline
Tier & Model & Operator & Min & Max & Median & Mean & Std & Shapiro-Wilk p & Normal? \\
\hline
Mobile & ResNet & Early Exit & 5.78 & 19.21 & 7.47 & 7.97 & 1.77 & $1.37 \times 10^{-12}$ & No \\
Mobile & ResNext & Early Exit & 4.00 & 22.47 & 5.02 & 5.61 & 2.00 & $1.42 \times 10^{-12}$ & No \\
Mobile & FCN & Early Exit & 17.19 & 48.63 & 26.02 & 26.74 & 4.86 & $2.22 \times 10^{-9}$ & No \\
Mobile & DUC & Early Exit & 38.88 & 46.83 & 40.90 & 40.99 & 0.72 & $7.85 \times 10^{-8}$ & No \\
Edge & ResNet & Early Exit & 3.75 & 5.79 & 4.40 & 4.44 & 0.31 & $6.44 \times 10^{-7}$ & No \\
Edge & ResNext & Early Exit & 2.46 & 4.63 & 2.99 & 3.05 & 0.31 & $1.58 \times 10^{-6}$ & No \\
Edge & FCN & Early Exit & 6.94 & 26.04 & 11.35 & 12.12 & 2.98 & $1.03 \times 10^{-11}$ & No \\
Edge & DUC & EarlyExit & 15.70 & 25.37 & 20.34 & 19.90 & 1.60 & $1.88 \times 10^{-12}$ & No \\
Cloud & ResNet & Early Exit & 11.17 & 83.95 & 17.55 & 22.51 & 13.23 & $1.95 \times 10^{-26}$ & No \\
Cloud & ResNext & Early Exit & 11.13 & 84.05 & 17.66 & 22.50 & 13.22 & $2.16 \times 10^{-26}$ & No \\
Cloud & FCN & Early Exit & 3.25 & 7.90 & 3.74 & 3.89 & 0.56 & $8.61 \times 10^{-30}$ & No \\
Cloud & DUC & Early Exit & 27.42 & 49.49 & 28.62 & 29.02 & 1.44 & $2.78 \times 10^{-9}$ & No \\
\hline
\end{tabular}
\label{rq3_descriptive_statistics}
}
\end{table}

\begin{table}[!htbp]
\caption{Descriptive statistics of the latency for RQ4, including Shapiro-Wilk p-values and normality assessment}
\centering
\resizebox{\textwidth}{!}{%
\begin{tabular}{|l|l|l|r|r|r|r|r|c|c|}
\hline
Tier & Model & Operator & Min & Max & Median & Mean & Std & Shapiro-Wilk p & Normal? \\
\hline
Mobile-Edge & ResNet & Partition & 6.03 & 27.02 & 8.06 & 8.73 & 2.46 & $3.65 \times 10^{-11}$ & No \\
Mobile-Edge & ResNext & Partition & 4.09 & 17.42 & 5.00 & 5.42 & 1.33 & $4.03 \times 10^{-9}$ & No \\
Mobile-Edge & FCN & Partition & 18.35 & 36.01 & 22.66 & 23.77 & 3.44 & $2.11 \times 10^{-10}$ & No \\
Mobile-Edge & DUC & Partition & 45.49 & 49.93 & 47.81 & 47.75 & 0.63 & $2.79 \times 10^{-5}$ & No \\
Edge-Cloud & ResNet & Partition & 10.58 & 23.93 & 11.10 & 11.33 & 1.10 & $1.02 \times 10^{-7}$ & No \\
Edge-Cloud & ResNext & Partition & 9.73 & 11.68 & 10.18 & 10.22 & 0.25 & $3.19 \times 10^{-6}$ & No \\
Edge-Cloud & FCN & Partition & 175.07 & 308.50 & 224.46 & 235.73 & 34.07 & $7.53 \times 10^{-18}$ & No \\
Edge-Cloud & DUC & Partition & 1152.53 & 1159.75 & 1155.72 & 1155.91 & 1.45 & $2.39 \times 10^{-6}$ & No \\
Mobile-Cloud & ResNet & Partition & 12.29 & 28.26 & 14.02 & 14.62 & 2.17 & $5.18 \times 10^{-10}$ & No \\
Mobile-Cloud & ResNext & Partition & 10.69 & 31.60 & 11.88 & 12.34 & 1.67 & $3.23 \times 10^{-9}$ & No \\
Mobile-Cloud & FCN & Partition & 186.09 & 323.57 & 235.80 & 247.88 & 35.81 & $1.68 \times 10^{-11}$ & No \\
Mobile-Cloud & DUC & Partition & 1172.62 & 1244.12 & 1192.89 & 1200.53 & 17.37 & $5.87 \times 10^{-6}$ & No \\
\hline
\end{tabular}
\label{rq4_descriptive_statistics}
}
\end{table}

\begin{table}[!htbp]
\caption{Descriptive statistics of the latency for RQ5, including Shapiro-Wilk p-values and normality assessment}
\centering
\resizebox{\textwidth}{!}{%
\begin{tabular}{|l|l|l|r|r|r|r|r|c|c|}
\hline
Tier & Model & Operator & Min & Max & Median & Mean & Std & Shapiro-Wilk p & Normal? \\
\hline
Mobile & ResNet & Quantize Early Exit & 5.69 & 20.66 & 8.29 & 9.09 & 2.67 & $2.12 \times 10^{-9}$ & No \\
Edge & ResNet & Quantize Early Exit & 2.39 & 8.34 & 2.89 & 3.20 & 0.94 & $2.98 \times 10^{-4}$ & No \\
Cloud & ResNet & Quantize Early Exit & 11.18 & 83.95 & 17.60 & 22.50 & 13.22 & $3.24 \times 10^{-12}$ & No \\
Mobile & ResNext & Quantize Early Exit & 4.40 & 14.28 & 6.68 & 7.06 & 1.55 & $1.02 \times 10^{-6}$ & No \\
Edge & ResNext & Quantize Early Exit & 1.99 & 5.13 & 2.30 & 2.37 & 0.32 & $2.61 \times 10^{-4}$ & No \\
Cloud & ResNext & Quantize Early Exit & 11.17 & 83.98 & 17.67 & 22.50 & 13.21 & $2.97 \times 10^{-13}$ & No \\
Mobile & FCN & Quantize Early Exit & 12.22 & 26.65 & 16.85 & 17.33 & 2.67 & $9.52 \times 10^{-7}$ & No \\
\hline
\end{tabular}
\label{rq5_descriptive_statistics}
}
\end{table}

\begin{table}[!htbp]
\caption{Descriptive statistics of the latency for RQ6 (Cloud Identity Models), including Shapiro-Wilk p-values and normality assessment (Part 1)}
\centering
\resizebox{\textwidth}{!}{%
\begin{tabular}{|c||c||c||c||c||c||c||c||c||c||c||c|}
\toprule
Tier & Model & Operator & Mobile-Edge BW & Edge-Cloud BW & Min & Max & Median & Mean & Std & Shapiro-Wilk p & Normal? \\
\midrule
Cloud & ResNet & Identity & 1 & 1 & 250.811300 & 250.973800 & 250.938000 & 250.922300 & 0.059400 & 0.140800 & Yes \\
Cloud & ResNext & Identity & 1 & 1 & 250.588000 & 250.790100 & 250.698300 & 250.693400 & 0.068000 & 0.997600 & Yes \\
Cloud & FCN & Identity & 1 & 1 & 20.544300 & 20.689900 & 20.585200 & 20.610800 & 0.056400 & 0.411200 & Yes \\
Cloud & DUC & Identity & 1 & 1 & 91.961000 & 92.196400 & 92.038100 & 92.057300 & 0.084000 & 0.726200 & Yes \\
Cloud & ResNet & Identity & 1 & 10 & 175.653100 & 175.813400 & 175.794100 & 175.755700 & 0.063400 & 0.169900 & Yes \\
Cloud & ResNext & Identity & 1 & 10 & 175.091500 & 175.209100 & 175.101900 & 175.125300 & 0.044000 & 0.048200 & No \\
Cloud & FCN & Identity & 1 & 10 & 14.890900 & 15.124900 & 14.919300 & 14.968000 & 0.087300 & 0.165300 & Yes \\
\bottomrule
\end{tabular}
\label{rq6_cloud_identity_descriptive_statistics_part1}
}
\end{table}

\begin{table}[!htbp]
\caption{Descriptive statistics of the latency for RQ6 (Cloud Identity Models), including Shapiro-Wilk p-values and normality assessment (Part 2)}
\centering
\resizebox{\textwidth}{!}{%
\begin{tabular}{|c||c||c||c||c||c||c||c||c||c||c||c|}
\toprule
Tier & Model & Operator & Mobile-Edge BW & Edge-Cloud BW & Min & Max & Median & Mean & Std & Shapiro-Wilk p & Normal? \\
\midrule
Cloud & ResNet & Identity & 100 & 1 & 84.889400 & 85.448300 & 85.044300 & 85.120300 & 0.195200 & 0.689300 & Yes \\
Cloud & ResNext & Identity & 100 & 1 & 84.101500 & 84.396900 & 84.220500 & 84.226400 & 0.099500 & 0.803300 & Yes \\
Cloud & FCN & Identity & 100 & 1 & 8.026500 & 8.216100 & 8.160500 & 8.143300 & 0.067000 & 0.573800 & Yes \\
Cloud & DUC & Identity & 100 & 1 & 32.770200 & 32.967400 & 32.891900 & 32.884200 & 0.065000 & 0.691600 & Yes \\
Cloud & ResNet & Identity & 100 & 10 & 8.324600 & 8.455700 & 8.402600 & 8.395100 & 0.043200 & 0.941600 & Yes \\
Cloud & ResNext & Identity & 100 & 10 & 8.099700 & 8.263700 & 8.215700 & 8.190300 & 0.062900 & 0.477900 & Yes \\
Cloud & FCN & Identity & 100 & 10 & 0.461600 & 0.843200 & 0.636500 & 0.647200 & 0.163500 & 0.204700 & Yes \\
\bottomrule
\end{tabular}
\label{rq6_cloud_identity_descriptive_statistics_part2}
}
\end{table}

\begin{table}[!htbp]
\caption{Descriptive statistics of the latency for RQ6 (Cloud Quantized Models), including Shapiro-Wilk p-values and normality assessment (Part 1)}
\centering
\resizebox{\textwidth}{!}{%
\begin{tabular}{|c||c||c||c||c||c||c||c||c||c||c||c|}
\toprule
Tier & Model & Operator & Mobile-Edge BW & Edge-Cloud BW & Min & Max & Median & Mean & Std & Shapiro-Wilk p & Normal? \\
\midrule
Cloud & ResNet & Quantization & 1 & 1 & 250.400700 & 251.653200 & 250.571700 & 250.943900 & 0.557000 & 0.048400 & No \\
Cloud & ResNext & Quantization & 1 & 1 & 250.425100 & 250.578600 & 250.491600 & 250.497000 & 0.051100 & 0.985900 & Yes \\
Cloud & FCN & Quantization & 1 & 1 & 20.501000 & 20.582400 & 20.552100 & 20.545700 & 0.029300 & 0.842600 & Yes \\
Cloud & DUC & Quantization & 1 & 1 & 93.663800 & 93.787000 & 93.762100 & 93.743800 & 0.044300 & 0.306500 & Yes \\
Cloud & ResNet & Quantization & 1 & 10 & 174.600800 & 174.713400 & 174.684400 & 174.672200 & 0.041300 & 0.421600 & Yes \\
Cloud & ResNext & Quantization & 1 & 10 & 174.505500 & 174.711000 & 174.567200 & 174.582000 & 0.069300 & 0.289900 & Yes \\
Cloud & FCN & Quantization & 1 & 10 & 13.887100 & 14.094800 & 13.992500 & 13.992500 & 0.089500 & 0.192200 & Yes \\
\bottomrule
\end{tabular}
\label{rq6_cloud_quantization_descriptive_statistics_part1}
}
\end{table}

\begin{table}[!htbp]
\caption{Descriptive statistics of the latency for RQ6 (Cloud Quantized Models), including Shapiro-Wilk p-values and normality assessment (Part 2)}
\centering
\resizebox{\textwidth}{!}{%
\begin{tabular}{|c||c||c||c||c||c||c||c||c||c||c||c|}
\toprule
Tier & Model & Operator & Mobile-Edge BW & Edge-Cloud BW & Min & Max & Median & Mean & Std & Shapiro-Wilk p & Normal? \\
\midrule
Cloud & ResNet & Quantization & 100 & 1 & 83.696400 & 83.836400 & 83.746600 & 83.764300 & 0.053100 & 0.574500 & Yes \\
Cloud & ResNext & Quantization & 100 & 1 & 83.590100 & 83.726400 & 83.696400 & 83.682300 & 0.047500 & 0.052400 & Yes \\
Cloud & FCN & Quantization & 100 & 1 & 7.440400 & 7.782000 & 7.651500 & 7.612800 & 0.143000 & 0.233900 & Yes \\
Cloud & DUC & Quantization & 100 & 1 & 33.015700 & 34.681600 & 33.080300 & 33.389100 & 0.647000 & 0.000600 & No \\
Cloud & ResNet & Quantization & 100 & 10 & 7.975800 & 8.249700 & 8.104500 & 8.119500 & 0.095200 & 0.936600 & Yes \\
Cloud & ResNext & Quantization & 100 & 10 & 7.991900 & 8.205400 & 7.997400 & 8.045200 & 0.081500 & 0.006700 & No \\
Cloud & FCN & Quantization & 100 & 10 & 0.377600 & 0.832600 & 0.600400 & 0.605900 & 0.145000 & 0.733100 & Yes \\
\bottomrule
\end{tabular}
\label{rq6_cloud_quantization_descriptive_statistics_part2}
}
\end{table}

\begin{table}[!htbp]
\caption{Descriptive statistics of the latency for RQ6 (Cloud Early Exit Models), including Shapiro-Wilk p-values and normality assessment (Part 1)}
\centering
\resizebox{\textwidth}{!}{%
\begin{tabular}{|c||c||c||c||c||c||c||c||c||c||c||c|}
\toprule
Tier & Model & Operator & Mobile-Edge BW & Edge-Cloud BW & Min & Max & Median & Mean & Std & Shapiro-Wilk p & Normal? \\
\midrule
Cloud & ResNet & EarlyExit & 1 & 1 & 250.631100 & 250.896600 & 250.798300 & 250.784900 & 0.086100 & 0.430800 & Yes \\
Cloud & ResNext & EarlyExit & 1 & 1 & 250.479000 & 250.769800 & 250.561200 & 250.606500 & 0.112900 & 0.429600 & Yes \\
Cloud & FCN & EarlyExit & 1 & 1 & 20.507800 & 20.638900 & 20.554400 & 20.567400 & 0.043500 & 0.852800 & Yes \\
Cloud & DUC & EarlyExit & 1 & 1 & 92.176000 & 92.219400 & 92.200700 & 92.200200 & 0.014000 & 0.563100 & Yes \\
Cloud & ResNet & EarlyExit & 1 & 10 & 175.241800 & 175.521500 & 175.409200 & 175.379600 & 0.108700 & 0.490700 & Yes \\
Cloud & ResNext & EarlyExit & 1 & 10 & 174.798400 & 175.024400 & 174.998700 & 174.944200 & 0.084400 & 0.197700 & Yes \\
Cloud & FCN & EarlyExit & 1 & 10 & 14.770600 & 14.874900 & 14.785400 & 14.805800 & 0.037700 & 0.187200 & Yes \\
\bottomrule
\end{tabular}
\label{rq6_cloud_earlyexit_descriptive_statistics_part1}
}
\end{table}

\begin{table}[!htbp]
\caption{Descriptive statistics of the latency for RQ6 (Cloud Early Exit Models), including Shapiro-Wilk p-values and normality assessment (Part 2)}
\centering
\resizebox{\textwidth}{!}{%
\begin{tabular}{|c||c||c||c||c||c||c||c||c||c||c||c|}
\toprule
Tier & Model & Operator & Mobile-Edge BW & Edge-Cloud BW & Min & Max & Median & Mean & Std & Shapiro-Wilk p & Normal? \\
\midrule
Cloud & ResNet & EarlyExit & 100 & 1 & 84.374000 & 84.566600 & 84.450600 & 84.460800 & 0.078500 & 0.344100 & Yes \\
Cloud & ResNext & EarlyExit & 100 & 1 & 83.999900 & 84.193700 & 84.113100 & 84.117300 & 0.071700 & 0.475600 & Yes \\
Cloud & FCN & EarlyExit & 100 & 1 & 7.946900 & 8.056000 & 7.993000 & 7.994000 & 0.035900 & 0.748500 & Yes \\
Cloud & DUC & EarlyExit & 100 & 1 & 32.493600 & 32.713100 & 32.518400 & 32.552000 & 0.081500 & 0.008600 & No \\
Cloud & ResNet & EarlyExit & 100 & 10 & 8.222800 & 8.377900 & 8.294100 & 8.295800 & 0.052100 & 0.993300 & Yes \\
Cloud & ResNext & EarlyExit & 100 & 10 & 8.049500 & 8.206800 & 8.164800 & 8.139700 & 0.054900 & 0.666100 & Yes \\
Cloud & FCN & EarlyExit & 100 & 10 & 0.382900 & 0.708300 & 0.657300 & 0.606700 & 0.119300 & 0.121100 & Yes \\
\bottomrule
\end{tabular}
\label{rq6_cloud_earlyexit_descriptive_statistics_part2}
}
\end{table}

\begin{table}[!htbp]
\caption{Descriptive statistics of the latency for RQ6 (Cloud Quantized Early Exit Models), including Shapiro-Wilk p-values and normality assessment (Part 1)}
\centering
\resizebox{\textwidth}{!}{%
\begin{tabular}{|c||c||c||c||c||c||c||c||c||c||c||c|}
\toprule
Tier & Model & Operator & Mobile-Edge BW & Edge-Cloud BW & Min & Max & Median & Mean & Std & Shapiro-Wilk p & Normal? \\
\midrule
Cloud & ResNet & Quantized EarlyExit & 1 & 1 & 250.453800 & 250.601900 & 250.505200 & 250.506700 & 0.054400 & 0.309100 & Yes \\
Cloud & ResNext & Quantized EarlyExit & 1 & 1 & 250.395100 & 250.589000 & 250.480500 & 250.470400 & 0.070300 & 0.396700 & Yes \\
Cloud & FCN & Quantized EarlyExit & 1 & 1 & 20.481200 & 20.594500 & 20.499800 & 20.528500 & 0.048500 & 0.101500 & Yes \\
Cloud & DUC & Quantized EarlyExit & 1 & 1 & 93.567000 & 93.859000 & 93.765000 & 93.743100 & 0.096500 & 0.398900 & Yes \\
Cloud & ResNet & Quantized EarlyExit & 1 & 10 & 174.457300 & 174.613600 & 174.599600 & 174.554900 & 0.065100 & 0.092500 & Yes \\
Cloud & ResNext & Quantized EarlyExit & 1 & 10 & 174.382300 & 174.611100 & 174.435200 & 174.468500 & 0.083600 & 0.453500 & Yes \\
Cloud & FCN & Quantized EarlyExit & 1 & 10 & 13.921600 & 14.087400 & 13.961900 & 13.982700 & 0.056700 & 0.319200 & Yes \\
\bottomrule
\end{tabular}
\label{rq6_cloud_quantized_earlyexit_descriptive_statistics_part1}
}
\end{table}

\begin{table}[!htbp]
\caption{Descriptive statistics of the latency for RQ6 (Cloud Quantized Early Exit Models), including Shapiro-Wilk p-values and normality assessment (Part 2)}
\centering
\resizebox{\textwidth}{!}{%
\begin{tabular}{|c||c||c||c||c||c||c||c||c||c||c||c|}
\toprule
Tier & Model & Operator & Mobile-Edge BW & Edge-Cloud BW & Min & Max & Median & Mean & Std & Shapiro-Wilk p & Normal? \\
\midrule
Cloud & ResNet & Quantized EarlyExit & 100 & 1 & 83.645500 & 83.721600 & 83.691200 & 83.690100 & 0.028900 & 0.538600 & Yes \\
Cloud & ResNext & Quantized EarlyExit & 100 & 1 & 83.566600 & 83.630000 & 83.602100 & 83.602800 & 0.024200 & 0.546300 & Yes \\
Cloud & FCN & Quantized EarlyExit & 100 & 1 & 7.458900 & 7.951000 & 7.589300 & 7.647700 & 0.168300 & 0.481500 & Yes \\
Cloud & DUC & Quantized EarlyExit & 100 & 1 & 32.855900 & 33.045500 & 32.878600 & 32.906100 & 0.070900 & 0.012900 & No \\
Cloud & ResNet & Quantized EarlyExit & 100 & 10 & 8.016300 & 8.116500 & 8.018100 & 8.040700 & 0.038500 & 0.005400 & No \\
Cloud & ResNext & Quantized EarlyExit & 100 & 10 & 8.012700 & 8.087900 & 8.072100 & 8.057400 & 0.028700 & 0.372200 & Yes \\
Cloud & FCN & Quantized EarlyExit & 100 & 10 & 0.331300 & 0.792300 & 0.619100 & 0.624800 & 0.168000 & 0.310300 & Yes \\
\bottomrule
\end{tabular}
\label{rq6_cloud_quantized_earlyexit_descriptive_statistics_part2}
}
\end{table}

\begin{table}[!htbp]
\caption{Descriptive statistics of the latency for RQ6 (Mobile-Cloud Partitioning Models), including Shapiro-Wilk p-values and normality assessment (Part 1)}
\centering
\resizebox{\textwidth}{!}{%
\begin{tabular}{|c||c||c||c||c||c||c||c||c||c||c||c|}
\toprule
Tier & Model & Operator & Mobile-Edge BW & Edge-Cloud BW & Min & Max & Median & Mean & Std & Shapiro-Wilk p & Normal? \\
\midrule
Mobile-Cloud & ResNet & Partition & 1 & 1 & 102.916000 & 106.020500 & 103.488400 & 103.799800 & 1.133300 & 0.018800 & No \\
Mobile-Cloud & ResNext & Partition & 1 & 1 & 101.801800 & 103.981500 & 102.142200 & 102.441900 & 0.781400 & 0.010700 & No \\
Mobile-Cloud & FCN & Partition & 1 & 1 & 414.219500 & 414.745900 & 414.467100 & 414.446800 & 0.177200 & 0.730600 & Yes \\
Mobile-Cloud & DUC & Partition & 1 & 1 & 2345.790600 & 2348.402400 & 2347.423700 & 2347.469400 & 0.950400 & 0.259300 & Yes \\
Mobile-Cloud & ResNet & Partition & 1 & 10 & 95.938800 & 97.674100 & 96.392500 & 96.576900 & 0.593800 & 0.339400 & Yes \\
Mobile-Cloud & ResNext & Partition & 1 & 10 & 94.309200 & 96.202800 & 94.839300 & 95.019500 & 0.663600 & 0.497800 & Yes \\
Mobile-Cloud & FCN & Partition & 1 & 10 & 236.611100 & 237.865300 & 236.801100 & 237.089300 & 0.499700 & 0.198900 & Yes \\
\bottomrule
\end{tabular}
\label{rq6_mobile_cloud_partitioning_descriptive_statistics_part1}
}
\end{table}

\begin{table}[!htbp]
\caption{Descriptive statistics of the latency for RQ6 (Mobile-Cloud Partitioning Models), including Shapiro-Wilk p-values and normality assessment (Part 2)}
\centering
\resizebox{\textwidth}{!}{%
\begin{tabular}{|c||c||c||c||c||c||c||c||c||c||c||c|}
\toprule
Tier & Model & Operator & Mobile-Edge BW & Edge-Cloud BW & Min & Max & Median & Mean & Std & Shapiro-Wilk p & Normal? \\
\midrule
Mobile-Cloud & ResNet & Partition & 100 & 1 & 12.988600 & 14.480300 & 13.214300 & 13.525900 & 0.584200 & 0.213600 & Yes \\
Mobile-Cloud & ResNext & Partition & 100 & 1 & 11.592000 & 12.176900 & 11.859700 & 11.895200 & 0.196500 & 0.950400 & Yes \\
Mobile-Cloud & FCN & Partition & 100 & 1 & 208.887800 & 212.363100 & 211.248900 & 211.052100 & 1.168900 & 0.310900 & Yes \\
Mobile-Cloud & DUC & Partition & 100 & 1 & 1176.858800 & 1177.779900 & 1177.172800 & 1177.233700 & 0.299700 & 0.201000 & Yes \\
Mobile-Cloud & ResNet & Partition & 100 & 10 & 4.598200 & 6.102900 & 4.965300 & 5.100800 & 0.523700 & 0.112500 & Yes \\
Mobile-Cloud & ResNext & Partition & 100 & 10 & 3.356900 & 4.015000 & 3.768100 & 3.737400 & 0.213500 & 0.443000 & Yes \\
Mobile-Cloud & FCN & Partition & 100 & 10 & 33.835100 & 34.608000 & 34.336600 & 34.287700 & 0.269400 & 0.796200 & Yes \\
\bottomrule
\end{tabular}
\label{rq6_mobile_cloud_partitioning_descriptive_statistics_part2}
}
\end{table}

\begin{table}[!htbp]
\caption{Descriptive statistics of the latency for RQ6 (Edge-Cloud Partitioning Models), including Shapiro-Wilk p-values and normality assessment}
\centering
\resizebox{\textwidth}{!}{%
\begin{tabular}{|c||c||c||c||c||c||c||c||c||c||c||c|}
\toprule
Tier & Model & Operator & Mobile-Edge BW & Edge-Cloud BW & Min & Max & Median & Mean & Std & Shapiro-Wilk p & Normal? \\
\midrule
Edge-Cloud & ResNet & Partition & 1 & 1 & 177.165600 & 177.699900 & 177.330900 & 177.401100 & 0.181600 & 0.785400 & Yes \\
Edge-Cloud & ResNext & Partition & 1 & 1 & 176.452000 & 176.801900 & 176.702600 & 176.641400 & 0.134800 & 0.429400 & Yes \\
Edge-Cloud & FCN & Partition & 1 & 1 & 215.169000 & 217.434500 & 216.748200 & 216.606800 & 0.802000 & 0.419600 & Yes \\
Edge-Cloud & DUC & Partition & 1 & 1 & 1220.037700 & 1220.476800 & 1220.454200 & 1220.303300 & 0.200400 & 0.024400 & No \\
Edge-Cloud & ResNet & Partition & 1 & 10 & 170.128000 & 170.620100 & 170.422800 & 170.400300 & 0.165200 & 0.945000 & Yes \\
Edge-Cloud & ResNext & Partition & 1 & 10 & 168.870800 & 169.067200 & 169.001200 & 168.974500 & 0.074400 & 0.568100 & Yes \\
Edge-Cloud & FCN & Partition & 1 & 10 & 38.821300 & 40.080700 & 39.887400 & 39.649200 & 0.457300 & 0.223400 & Yes \\
\bottomrule
\end{tabular}
\label{rq6_edge_cloud_partitioning_descriptive_statistics_part1}
}
\end{table}

\begin{table}[!htbp]
\caption{Descriptive statistics of the latency for RQ6 (Edge-Cloud Partitioning Models), including Shapiro-Wilk p-values and normality assessment}
\centering
\resizebox{\textwidth}{!}{%
\begin{tabular}{|c||c||c||c||c||c||c||c||c||c||c||c|}
\toprule
Tier & Model & Operator & Mobile-Edge BW & Edge-Cloud BW & Min & Max & Median & Mean & Std & Shapiro-Wilk p & Normal? \\
\midrule
Edge-Cloud & ResNet & Partition & 100 & 1 & 11.814200 & 12.277900 & 12.103400 & 12.072300 & 0.156600 & 0.927500 & Yes \\
Edge-Cloud & ResNext & Partition & 100 & 1 & 10.581200 & 10.902300 & 10.690200 & 10.692800 & 0.114100 & 0.199700 & Yes \\
Edge-Cloud & FCN & Partition & 100 & 1 & 201.670200 & 202.270100 & 201.997700 & 201.987400 & 0.192600 & 0.797700 & Yes \\
Edge-Cloud & DUC & Partition & 100 & 1 & 1149.414100 & 1150.620000 & 1150.223600 & 1150.150000 & 0.400800 & 0.360600 & Yes \\
Edge-Cloud & ResNet & Partition & 100 & 10 & 3.222000 & 3.890300 & 3.431300 & 3.520300 & 0.244000 & 0.700800 & Yes \\
Edge-Cloud & ResNext & Partition & 100 & 10 & 2.394100 & 2.964700 & 2.589000 & 2.615800 & 0.188500 & 0.224500 & Yes \\
Edge-Cloud & FCN & Partition & 100 & 10 & 24.983100 & 25.816800 & 25.419900 & 25.357500 & 0.300600 & 0.717200 & Yes \\
\bottomrule
\end{tabular}
\label{rq6_edge_cloud_partitioning_descriptive_statistics_part2}
}
\end{table}

\begin{table}[!htbp]
\caption{Descriptive statistics of the latency for RQ6 (Mobile-Cloud Quantized Early Exit Partitioning Models), including Shapiro-Wilk p-values and normality assessment (Part 1)}
\centering
\resizebox{\textwidth}{!}{%
\begin{tabular}{|c||c||c||c||c||c||c||c||c||c||c||c|}
\toprule
Tier & Model & Operator & Mobile-Edge BW & Edge-Cloud BW & Min & Max & Median & Mean & Std & Shapiro-Wilk p & Normal? \\
\midrule
Mobile-Cloud & ResNet & Quantized Early Exit Partition & 1 & 1 & 102.331500 & 103.296800 & 103.018100 & 102.977900 & 0.349100 & 0.166100 & Yes \\
Mobile-Cloud & ResNext & Quantized Early Exit Partition & 1 & 1 & 101.765400 & 102.971600 & 102.479900 & 102.443200 & 0.392500 & 0.722000 & Yes \\
Mobile-Cloud & FCN & Quantized Early Exit Partition & 1 & 1 & 407.431000 & 408.659200 & 407.804800 & 407.905500 & 0.414200 & 0.464600 & Yes \\
Mobile-Cloud & DUC & Quantized Early Exit Partition & 1 & 1 & 967.185700 & 968.940200 & 968.343300 & 968.077200 & 0.703600 & 0.322500 & Yes \\
Mobile-Cloud & ResNet & Quantized Early Exit Partition & 1 & 10 & 94.872800 & 95.313700 & 95.237400 & 95.137200 & 0.177500 & 0.203300 & Yes \\
Mobile-Cloud & ResNext & Quantized Early Exit Partition & 1 & 10 & 93.975200 & 94.600700 & 94.394500 & 94.358700 & 0.213800 & 0.541500 & Yes \\
Mobile-Cloud & FCN & Quantized Early Exit Partition & 1 & 10 & 230.248000 & 230.682800 & 230.307400 & 230.399600 & 0.162200 & 0.224400 & Yes \\
\bottomrule
\end{tabular}
\label{rq6_mobile_cloud_quantized_earlyexit_partitioning_descriptive_statistics_part1}
}
\end{table}

\begin{table}[!htbp]
\caption{Descriptive statistics of the latency for RQ6 (Mobile-Cloud Quantized Early Exit Partitioning Models), including Shapiro-Wilk p-values and normality assessment (Part2)}
\centering
\resizebox{\textwidth}{!}{%
\begin{tabular}{|c||c||c||c||c||c||c||c||c||c||c||c|}
\toprule
Tier & Model & Operator & Mobile-Edge BW & Edge-Cloud BW & Min & Max & Median & Mean & Std & Shapiro-Wilk p & Normal? \\
\midrule
Mobile-Cloud & ResNet & Quantized Early Exit Partition & 100 & 1 & 12.042500 & 13.941300 & 12.427000 & 12.700600 & 0.674700 & 0.266600 & Yes \\
Mobile-Cloud & ResNext & Quantized Early Exit Partition & 100 & 1 & 11.598300 & 12.466600 & 11.860600 & 12.026400 & 0.350300 & 0.248500 & Yes \\
Mobile-Cloud & FCN & Quantized Early Exit Partition & 100 & 1 & 203.737400 & 204.396800 & 204.058000 & 204.102600 & 0.247200 & 0.594300 & Yes \\
Mobile-Cloud & DUC & Quantized Early Exit Partition & 100 & 1 & 480.772800 & 481.749800 & 481.072300 & 481.142600 & 0.330900 & 0.400100 & Yes \\
Mobile-Cloud & ResNext & Quantized Early Exit Partition & 100 & 10 & 3.773900 & 4.276200 & 4.081100 & 4.028900 & 0.183100 & 0.773900 & Yes \\
Mobile-Cloud & FCN & Quantized Early Exit Partition & 100 & 10 & 27.148600 & 27.779600 & 27.687300 & 27.552100 & 0.234700 & 0.261200 & Yes \\
Mobile-Cloud & DUC & Quantized Early Exit Partition & 100 & 10 & 69.464200 & 72.770800 & 72.015000 & 71.667400 & 1.138800 & 0.041900 & No \\
\bottomrule
\end{tabular}
\label{rq6_mobile_cloud_quantized_earlyexit_partitioning_descriptive_statistics_part2}
}
\end{table}

\begin{table}[!htbp]
\caption{Descriptive statistics of the latency for RQ6 (Edge-Cloud Quantized Early Exit Partitioning Models), including Shapiro-Wilk p-values and normality assessment}
\centering
\resizebox{0.8\textwidth}{!}{%
\begin{tabular}{|c||c||c||c||c||c||c||c||c||c||c||c|}
\toprule
Tier & Model & Operator & Edge-Cloud BW & Mobile-Edge BW & Min & Max & Median & Mean & Std & Shapiro-Wilk p & Normal? \\
\midrule
Edge-Cloud & ResNet & Quantized Early Exit Partition & 1 & 1 & 176.072300 & 178.099300 & 176.186800 & 176.566500 & 0.772200 & 0.003500 & No \\
Edge-Cloud & ResNext & Quantized Early Exit Partition & 1 & 1 & 175.789300 & 176.090700 & 175.910300 & 175.931800 & 0.105800 & 0.942000 & Yes \\
Edge-Cloud & FCN & Quantized Early Exit Partition & 1 & 1 & 212.782700 & 213.123600 & 212.914500 & 212.961300 & 0.123000 & 0.625900 & Yes \\
Edge-Cloud & DUC & Quantized Early Exit Partition & 1 & 1 & 528.544200 & 529.855700 & 529.665400 & 529.483500 & 0.478400 & 0.020200 & No \\
Edge-Cloud & ResNet & Quantized Early Exit Partition & 1 & 10 & 168.282300 & 168.614800 & 168.496800 & 168.479300 & 0.123200 & 0.615400 & Yes \\
Edge-Cloud & ResNext & Quantized Early Exit Partition & 1 & 10 & 168.017800 & 168.524200 & 168.109700 & 168.189700 & 0.183100 & 0.236800 & Yes \\
Edge-Cloud & FCN & Quantized Early Exit Partition & 1 & 10 & 35.587000 & 36.077100 & 35.794100 & 35.789300 & 0.164000 & 0.593900 & Yes \\
\bottomrule
\end{tabular}
\label{rq6_edge_cloud_quantized_earlyexit_partitioning_descriptive_statistics}
}
\end{table}

\begin{table}[!htbp]
\caption{Descriptive statistics of the latency for RQ6 (Mobile Identity, Quantized, Early Exit, and Quantized Early Exit Models), including Shapiro-Wilk p-values and normality assessment}
\centering
\resizebox{0.8\textwidth}{!}{%
\begin{tabular}{|c||c||c||c||c||c||c||c||c||c||c|}
\toprule
Tier & Model & Operator & Mobile BW & Min & Max & Median & Mean & Std & Shapiro-Wilk p & Normal? \\
\midrule
Mobile & DUC & Identity & 1 & 71.115000 & 74.498700 & 73.010600 & 72.844300 & 1.120600 & 0.984300 & Yes \\
Mobile & FCN & Identity & 1 & 28.255700 & 29.457500 & 28.442400 & 28.684100 & 0.433500 & 0.261700 & Yes \\
Mobile & ResNext & Identity & 1 & 87.602500 & 91.638800 & 87.776300 & 88.521100 & 1.561900 & 0.000900 & No \\
Mobile & ResNet & Identity & 1 & 89.913500 & 90.772700 & 90.368500 & 90.358200 & 0.285300 & 0.996600 & Yes \\
Mobile & FCN & Identity & 10 & 22.002100 & 23.293200 & 22.143100 & 22.428100 & 0.471300 & 0.135300 & Yes \\
Mobile & DUC & Identity & 10 & 44.163700 & 46.815700 & 45.393100 & 45.293500 & 0.963000 & 0.671900 & Yes \\
Mobile & ResNet & Identity & 10 & 14.399700 & 19.240200 & 14.958800 & 16.417200 & 2.158600 & 0.045500 & No \\
\bottomrule
\end{tabular}
\label{rq6_mobile_models_descriptive_statistics}
}
\end{table}

\begin{table}[!htbp]
\caption{Descriptive statistics of the latency for RQ6 (Edge Identity, Quantized, Early Exit, and Quantized Early Exit Models), including Shapiro-Wilk p-values and normality assessment}
\centering
\resizebox{0.8\textwidth}{!}{%
\begin{tabular}{|c||c||c||c||c||c||c||c||c||c||c|}
\toprule
Tier & Model & Operator & Mobile BW & Min & Max & Median & Mean & Std & Shapiro-Wilk p & Normal? \\
\midrule
Edge & DUC & Identity & 1 & 80.387600 & 81.998600 & 81.988100 & 81.487500 & 0.656700 & 0.052400 & Yes \\
Edge & ResNet & Identity & 1 & 171.571600 & 174.023900 & 171.653300 & 172.105700 & 0.960200 & 0.000600 & No \\
Edge & FCN & Identity & 1 & 23.624200 & 24.588000 & 24.306000 & 24.256400 & 0.347500 & 0.262000 & Yes \\
Edge & ResNext & Identity & 1 & 169.594000 & 170.237400 & 169.697500 & 169.788000 & 0.238100 & 0.092600 & Yes \\
Edge & ResNet & Identity & 10 & 19.795400 & 20.116500 & 20.020600 & 19.983200 & 0.113800 & 0.727000 & Yes \\
Edge & DUC & Identity & 10 & 24.502600 & 25.953000 & 24.852500 & 25.179600 & 0.576700 & 0.243000 & Yes \\
Edge & ResNext & Identity & 10 & 18.092700 & 19.631900 & 18.243300 & 18.655300 & 0.617000 & 0.134300 & Yes \\
\bottomrule
\end{tabular}
\label{rq6_edge_models_descriptive_statistics}
}
\end{table}

\begin{table}[!htbp]
\caption{Descriptive statistics of the latency for RQ6 (Mobile-Edge Partition and Mobile-Edge Quantized Early Exit Partition Models), including Shapiro-Wilk p-values and normality assessment}
\centering
\resizebox{\textwidth}{!}{%
\begin{tabular}{|c||c||c||c||c||c||c||c||c||c||c|}
\toprule
Tier & Model & Operator & Mobile-Edge BW & Min & Max & Median & Mean & Std & Shapiro-Wilk p & Normal? \\
\midrule
Mobile-Edge & DUC & Partition & 1 & 1213.411300 & 1214.121900 & 1213.797800 & 1213.794600 & 0.231100 & 0.876800 & Yes \\
Mobile-Edge & ResNet & Partition & 1 & 96.984500 & 97.409300 & 97.117600 & 97.161500 & 0.141100 & 0.638100 & Yes \\
Mobile-Edge & FCN & Partition & 1 & 220.971700 & 222.211200 & 222.044500 & 221.699500 & 0.559600 & 0.045500 & No \\
Mobile-Edge & ResNext & Partition & 1 & 95.031800 & 96.405500 & 95.127400 & 95.382700 & 0.518500 & 0.008000 & No \\
Mobile-Edge & ResNet & Partition & 10 & 13.961100 & 14.395900 & 14.119000 & 14.155600 & 0.166200 & 0.591400 & Yes \\
Mobile-Edge & DUC & Partition & 10 & 157.121200 & 158.564800 & 157.462000 & 157.745000 & 0.566500 & 0.327100 & Yes \\
Mobile-Edge & ResNext & Partition & 10 & 11.621000 & 11.990700 & 11.708900 & 11.766000 & 0.138100 & 0.461300 & Yes \\
\bottomrule
\end{tabular}
\label{rq6_cloud_models_descriptive_statistics}
}
\end{table}

\subsection{Hypothesis Testing}
\label{hypothesis_testing}
\begin{table}[!htbp]
\centering
\caption{Pairwise Conover post-hoc p-values for RQ6 (Cloud Identity Models) with Mobile-Edge BW = 1Mbps}
\label{tab:conover_cloud_identity_me1}
\smallskip
Kruskal-Wallis p-values for each model: FCN = 0.0028, ResNet = 0.0001, ResNext = 0.0001, DUC = 0.0008.

\begin{tabular}{lcccc}
\toprule
Edge-Cloud BW Pair & FCN & ResNet & ResNext & DUC \\
\midrule
1 vs 10 & 0.001 & 0.001 & 0.001 & 0.001 \\
1 vs 50 & 0.001 & 0.001 & 0.001 & 0.001 \\
1 vs 100 & 0.001 & 0.001 & 0.001 & 0.001 \\
1 vs 150 & 0.001 & 0.001 & 0.001 & 0.001 \\
1 vs 200 & 0.001 & 0.001 & 0.001 & 0.001 \\
10 vs 50 & 0.016 & 0.001 & 0.001 & 0.001 \\
10 vs 100 & 0.497 & 0.001 & 0.001 & 0.001 \\
10 vs 150 & 0.774 & 0.001 & 0.001 & 0.001 \\
10 vs 200 & 0.275 & 0.001 & 0.001 & 0.001 \\
50 vs 100 & 0.016 & 0.001 & 0.001 & 0.378 \\
50 vs 150 & 0.378 & 0.001 & 0.001 & 0.774 \\
50 vs 200 & 0.001 & 0.001 & 0.001 & 0.924 \\
100 vs 150 & 0.497 & 0.275 & 0.924 & 0.378 \\
100 vs 200 & 0.631 & 0.631 & 0.037 & 0.275 \\
150 vs 200 & 0.378 & 0.189 & 0.005 & 0.924 \\
\bottomrule
\end{tabular}
\end{table}

\begin{table}[!htbp]
\centering
\caption{Pairwise Conover post-hoc p-values for RQ6 (Cloud Identity Models) with Mobile-Edge BW = 10Mbps}
\label{tab:conover_cloud_identity_me10}
\smallskip
Kruskal-Wallis p-values for each model: FCN = 0.0266, ResNet = 0.0001, ResNext = 0.0001, DUC = 0.0010.

\begin{tabular}{lcccc}
\toprule
Edge-Cloud BW Pair & FCN & ResNet & ResNext & DUC \\
\midrule
1 vs 10 & 0.001 & 0.001 & 0.001 & 0.001 \\
1 vs 50 & 0.001 & 0.001 & 0.001 & 0.001 \\
1 vs 100 & 0.001 & 0.001 & 0.001 & 0.001 \\
1 vs 150 & 0.001 & 0.001 & 0.001 & 0.001 \\
1 vs 200 & 0.001 & 0.001 & 0.001 & 0.001 \\
10 vs 50 & 0.924 & 0.001 & 0.001 & 0.001 \\
10 vs 100 & 0.631 & 0.001 & 0.001 & 0.001 \\
10 vs 150 & 0.774 & 0.001 & 0.001 & 0.001 \\
10 vs 200 & 0.497 & 0.001 & 0.001 & 0.001 \\
50 vs 100 & 0.774 & 0.001 & 0.001 & 0.924 \\
50 vs 150 & 0.924 & 0.001 & 0.001 & 0.378 \\
50 vs 200 & 0.631 & 0.001 & 0.001 & 0.774 \\
100 vs 150 & 0.631 & 0.497 & 0.071 & 0.924 \\
100 vs 200 & 0.924 & 0.924 & 0.122 & 0.924 \\
150 vs 200 & 0.497 & 0.378 & 0.497 & 0.497 \\
\bottomrule
\end{tabular}
\end{table}

\begin{table}[!htbp]
\centering
\caption{Pairwise Conover post-hoc p-values for RQ6 (Cloud Identity Models) with Mobile-Edge BW = 50Mbps}
\label{tab:conover_cloud_identity_me50}
\smallskip
Kruskal-Wallis p-values for each model: FCN = 0.0008, ResNet = 0.0001, ResNext = 0.0001, DUC = 0.0003.

\begin{tabular}{lcccc}
\toprule
Edge-Cloud BW Pair & FCN & ResNet & ResNext & DUC \\
\midrule
1 vs 10 & 0.001 & 0.001 & 0.001 & 0.001 \\
1 vs 50 & 0.001 & 0.001 & 0.001 & 0.001 \\
1 vs 100 & 0.001 & 0.001 & 0.001 & 0.001 \\
1 vs 150 & 0.001 & 0.001 & 0.001 & 0.001 \\
1 vs 200 & 0.001 & 0.001 & 0.001 & 0.001 \\
10 vs 50 & 0.189 & 0.001 & 0.001 & 0.001 \\
10 vs 100 & 0.016 & 0.001 & 0.001 & 0.001 \\
10 vs 150 & 0.016 & 0.001 & 0.001 & 0.001 \\
10 vs 200 & 0.275 & 0.001 & 0.001 & 0.001 \\
50 vs 100 & 0.001 & 0.001 & 0.001 & 0.122 \\
50 vs 150 & 0.189 & 0.001 & 0.001 & 0.631 \\
50 vs 200 & 0.497 & 0.001 & 0.001 & 0.189 \\
100 vs 150 & 0.497 & 0.189 & 0.497 & 0.378 \\
100 vs 200 & 0.005 & 0.189 & 0.016 & 0.016 \\
150 vs 200 & 0.071 & 0.774 & 0.924 & 0.189 \\
\bottomrule
\end{tabular}
\end{table}

\begin{table}[!htbp]
\centering
\caption{Pairwise Conover post-hoc p-values for RQ6 (Cloud Identity Models) with Mobile-Edge BW = 100Mbps}
\label{tab:conover_cloud_identity_me100}
\smallskip
Kruskal-Wallis p-values for each model: FCN = 0.0033, ResNet = 0.0001, ResNext = 0.0001, DUC = 0.0009.

\begin{tabular}{lcccc}
\toprule
Edge-Cloud BW Pair & FCN & ResNet & ResNext & DUC \\
\midrule
1 vs 10 & 0.001 & 0.001 & 0.001 & 0.001 \\
1 vs 50 & 0.001 & 0.001 & 0.001 & 0.001 \\
1 vs 100 & 0.001 & 0.001 & 0.001 & 0.001 \\
1 vs 150 & 0.001 & 0.001 & 0.001 & 0.001 \\
1 vs 200 & 0.001 & 0.001 & 0.001 & 0.001 \\
10 vs 50 & 0.037 & 0.001 & 0.001 & 0.001 \\
10 vs 100 & 0.016 & 0.001 & 0.001 & 0.001 \\
10 vs 150 & 0.275 & 0.001 & 0.001 & 0.001 \\
10 vs 200 & 0.037 & 0.001 & 0.001 & 0.001 \\
50 vs 100 & 0.924 & 0.001 & 0.001 & 0.631 \\
50 vs 150 & 0.189 & 0.001 & 0.001 & 0.924 \\
50 vs 200 & 0.497 & 0.001 & 0.001 & 0.497 \\
100 vs 150 & 0.122 & 0.924 & 0.037 & 0.631 \\
100 vs 200 & 0.631 & 0.774 & 0.774 & 0.774 \\
150 vs 200 & 0.631 & 0.774 & 0.275 & 0.275 \\
\bottomrule
\end{tabular}
\end{table}

\begin{table}[!htbp]
\centering
\caption{Pairwise Conover post-hoc p-values for RQ6 (Cloud Identity Models) with Mobile-Edge BW = 150Mbps}
\label{tab:conover_cloud_identity_me150}
\smallskip
Kruskal-Wallis p-values for each model: FCN = 0.0120, ResNet = 0.0001, ResNext = 0.0001, DUC = 0.0005.

\begin{tabular}{lcccc}
\toprule
Edge-Cloud BW Pair & FCN & ResNet & ResNext & DUC \\
\midrule
1 vs 10 & 0.001 & 0.001 & 0.001 & 0.001 \\
1 vs 50 & 0.001 & 0.001 & 0.001 & 0.001 \\
1 vs 100 & 0.001 & 0.001 & 0.001 & 0.001 \\
1 vs 150 & 0.001 & 0.001 & 0.001 & 0.001 \\
1 vs 200 & 0.001 & 0.001 & 0.001 & 0.001 \\
10 vs 50 & 0.497 & 0.001 & 0.001 & 0.001 \\
10 vs 100 & 0.378 & 0.001 & 0.001 & 0.001 \\
10 vs 150 & 0.189 & 0.001 & 0.001 & 0.001 \\
10 vs 200 & 0.071 & 0.001 & 0.001 & 0.001 \\
50 vs 100 & 0.774 & 0.001 & 0.001 & 0.122 \\
50 vs 150 & 0.497 & 0.001 & 0.001 & 0.071 \\
50 vs 200 & 0.378 & 0.001 & 0.001 & 0.275 \\
100 vs 150 & 0.774 & 0.071 & 0.774 & 0.774 \\
100 vs 200 & 0.497 & 0.037 & 0.774 & 0.774 \\
150 vs 200 & 0.631 & 0.497 & 0.631 & 0.631 \\
\bottomrule
\end{tabular}
\end{table}

\begin{table}[!htbp]
\centering
\caption{Pairwise Conover post-hoc p-values for RQ6 (Cloud Identity Models) with Mobile-Edge BW = 200Mbps}
\label{tab:conover_cloud_identity_me200}
\smallskip
Kruskal-Wallis p-values for each model: FCN = 0.0137, ResNet = 0.0001, ResNext = 0.0001, DUC = 0.0004.

\begin{tabular}{lcccc}
\toprule
Edge-Cloud BW Pair & FCN & ResNet & ResNext & DUC \\
\midrule
1 vs 10 & 0.001 & 0.001 & 0.001 & 0.001 \\
1 vs 50 & 0.001 & 0.001 & 0.001 & 0.001 \\
1 vs 100 & 0.001 & 0.001 & 0.001 & 0.001 \\
1 vs 150 & 0.001 & 0.001 & 0.001 & 0.001 \\
1 vs 200 & 0.001 & 0.001 & 0.001 & 0.001 \\
10 vs 50 & 0.631 & 0.001 & 0.001 & 0.001 \\
10 vs 100 & 0.631 & 0.001 & 0.001 & 0.001 \\
10 vs 150 & 0.378 & 0.001 & 0.001 & 0.001 \\
10 vs 200 & 0.497 & 0.001 & 0.001 & 0.001 \\
50 vs 100 & 0.497 & 0.001 & 0.001 & 0.774 \\
50 vs 150 & 0.189 & 0.001 & 0.001 & 0.631 \\
50 vs 200 & 0.774 & 0.001 & 0.001 & 0.016 \\
100 vs 150 & 0.071 & 0.037 & 0.189 & 0.631 \\
100 vs 200 & 0.497 & 0.071 & 0.275 & 0.122 \\
150 vs 200 & 0.631 & 0.378 & 0.378 & 0.189 \\
\bottomrule
\end{tabular}
\end{table}

\begin{table}[!htbp]
\centering
\caption{Pairwise Conover post-hoc p-values for RQ6 (Cloud Identity Models) with Edge-Cloud BW = 1Mbps}
\label{tab:conover_cloud_identity_ec1}
\smallskip
Kruskal-Wallis p-values for each model: FCN = 0.0057, ResNet = 0.0001, ResNext = 0.0001, DUC = 0.0008.

\begin{tabular}{lcccc}
\toprule
Mobile-Edge BW Pair & FCN & ResNet & ResNext & DUC \\
\midrule
1 vs 10 & 0.001 & 0.001 & 0.001 & 0.001 \\
1 vs 50 & 0.001 & 0.001 & 0.001 & 0.001 \\
1 vs 100 & 0.001 & 0.001 & 0.001 & 0.001 \\
1 vs 150 & 0.001 & 0.001 & 0.001 & 0.001 \\
1 vs 200 & 0.001 & 0.001 & 0.001 & 0.001 \\
10 vs 50 & 0.189 & 0.001 & 0.001 & 0.001 \\
10 vs 100 & 0.497 & 0.001 & 0.001 & 0.001 \\
10 vs 150 & 0.924 & 0.001 & 0.001 & 0.001 \\
10 vs 200 & 0.275 & 0.001 & 0.001 & 0.001 \\
50 vs 100 & 0.071 & 0.001 & 0.001 & 0.774 \\
50 vs 150 & 0.189 & 0.001 & 0.071 & 0.189 \\
50 vs 200 & 0.016 & 0.001 & 0.001 & 0.924 \\
100 vs 150 & 0.774 & 0.924 & 0.071 & 0.275 \\
100 vs 200 & 0.924 & 0.071 & 0.378 & 0.631 \\
150 vs 200 & 0.275 & 0.001 & 0.378 & 0.378 \\
\bottomrule
\end{tabular}
\end{table}

\begin{table}[!htbp]
\centering
\caption{Pairwise Conover post-hoc p-values for RQ6 (Cloud Identity Models) with Edge-Cloud BW = 10Mbps}
\label{tab:conover_cloud_identity_ec10}
\smallskip
Kruskal-Wallis p-values for each model: FCN = 0.0309, ResNet = 0.0001, ResNext = 0.0001, DUC = 0.0011.

\begin{tabular}{lcccc}
\toprule
Mobile-Edge BW Pair & FCN & ResNet & ResNext & DUC \\
\midrule
1 vs 10 & 0.001 & 0.001 & 0.001 & 0.001 \\
1 vs 50 & 0.001 & 0.001 & 0.001 & 0.001 \\
1 vs 100 & 0.001 & 0.001 & 0.001 & 0.001 \\
1 vs 150 & 0.001 & 0.001 & 0.001 & 0.001 \\
1 vs 200 & 0.001 & 0.001 & 0.001 & 0.001 \\
10 vs 50 & 0.631 & 0.001 & 0.001 & 0.001 \\
10 vs 100 & 0.631 & 0.001 & 0.001 & 0.001 \\
10 vs 150 & 0.924 & 0.001 & 0.001 & 0.001 \\
10 vs 200 & 0.774 & 0.001 & 0.001 & 0.001 \\
50 vs 100 & 0.924 & 0.001 & 0.001 & 0.924 \\
50 vs 150 & 0.924 & 0.001 & 0.001 & 0.924 \\
50 vs 200 & 0.631 & 0.001 & 0.001 & 0.497 \\
100 vs 150 & 0.924 & 0.189 & 0.631 & 0.631 \\
100 vs 200 & 0.774 & 0.037 & 0.378 & 0.924 \\
150 vs 200 & 0.924 & 0.378 & 0.497 & 0.774 \\
\bottomrule
\end{tabular}
\end{table}

\begin{table}[!htbp]
\centering
\caption{Pairwise Conover post-hoc p-values for RQ6 (Cloud Identity Models) with Edge-Cloud BW = 50Mbps}
\label{tab:conover_cloud_identity_ec50}
\smallskip
Kruskal-Wallis p-values for each model: FCN = 0.0020, ResNet = 0.0001, ResNext = 0.0001, DUC = 0.0002.

\begin{tabular}{lcccc}
\toprule
Mobile-Edge BW Pair & FCN & ResNet & ResNext & DUC \\
\midrule
1 vs 10 & 0.001 & 0.001 & 0.001 & 0.001 \\
1 vs 50 & 0.001 & 0.001 & 0.001 & 0.001 \\
1 vs 100 & 0.001 & 0.001 & 0.001 & 0.001 \\
1 vs 150 & 0.001 & 0.001 & 0.001 & 0.001 \\
1 vs 200 & 0.001 & 0.001 & 0.001 & 0.001 \\
10 vs 50 & 0.189 & 0.001 & 0.001 & 0.001 \\
10 vs 100 & 0.016 & 0.001 & 0.001 & 0.001 \\
10 vs 150 & 0.016 & 0.001 & 0.001 & 0.001 \\
10 vs 200 & 0.189 & 0.001 & 0.001 & 0.001 \\
50 vs 100 & 0.071 & 0.001 & 0.001 & 0.122 \\
50 vs 150 & 0.122 & 0.001 & 0.001 & 0.001 \\
50 vs 200 & 0.631 & 0.001 & 0.001 & 0.497 \\
100 vs 150 & 0.774 & 0.774 & 0.774 & 0.924 \\
100 vs 200 & 0.189 & 0.497 & 0.071 & 0.275 \\
150 vs 200 & 0.122 & 0.631 & 0.497 & 0.005 \\
\bottomrule
\end{tabular}
\end{table}

\begin{table}[!htbp]
\centering
\caption{Pairwise Conover post-hoc p-values for RQ6 (Cloud Identity Models) with Edge-Cloud BW = 100Mbps}
\label{tab:conover_cloud_identity_ec100}
\smallskip
Kruskal-Wallis p-values for each model: FCN = 0.0010, ResNet = 0.0001, ResNext = 0.0001, DUC = 0.0007.

\begin{tabular}{lcccc}
\toprule
Mobile-Edge BW Pair & FCN & ResNet & ResNext & DUC \\
\midrule
1 vs 10 & 0.001 & 0.001 & 0.001 & 0.001 \\
1 vs 50 & 0.001 & 0.001 & 0.001 & 0.001 \\
1 vs 100 & 0.001 & 0.001 & 0.001 & 0.001 \\
1 vs 150 & 0.001 & 0.001 & 0.001 & 0.001 \\
1 vs 200 & 0.001 & 0.001 & 0.001 & 0.001 \\
10 vs 50 & 0.016 & 0.001 & 0.001 & 0.001 \\
10 vs 100 & 0.037 & 0.001 & 0.001 & 0.001 \\
10 vs 150 & 0.122 & 0.001 & 0.001 & 0.001 \\
10 vs 200 & 0.122 & 0.001 & 0.001 & 0.001 \\
50 vs 100 & 0.378 & 0.001 & 0.001 & 0.924 \\
50 vs 150 & 0.189 & 0.001 & 0.001 & 0.378 \\
50 vs 200 & 0.005 & 0.001 & 0.001 & 0.189 \\
100 vs 150 & 0.631 & 0.189 & 0.497 & 0.631 \\
100 vs 200 & 0.016 & 0.189 & 0.071 & 0.275 \\
150 vs 200 & 0.037 & 0.631 & 0.122 & 0.497 \\
\bottomrule
\end{tabular}
\end{table}

\begin{table}[!htbp]
\centering
\caption{Pairwise Conover post-hoc p-values for RQ6 (Cloud Identity Models) with Edge-Cloud BW = 150Mbps}
\label{tab:conover_cloud_identity_ec150}
\smallskip
Kruskal-Wallis p-values for each model: FCN = 0.0039, ResNet = 0.0001, ResNext = 0.0001, DUC = 0.0003.

\begin{tabular}{lcccc}
\toprule
Mobile-Edge BW Pair & FCN & ResNet & ResNext & DUC \\
\midrule
1 vs 10 & 0.001 & 0.001 & 0.001 & 0.001 \\
1 vs 50 & 0.001 & 0.001 & 0.001 & 0.001 \\
1 vs 100 & 0.001 & 0.001 & 0.001 & 0.001 \\
1 vs 150 & 0.001 & 0.001 & 0.001 & 0.001 \\
1 vs 200 & 0.001 & 0.001 & 0.001 & 0.001 \\
10 vs 50 & 0.037 & 0.001 & 0.001 & 0.001 \\
10 vs 100 & 0.189 & 0.001 & 0.001 & 0.001 \\
10 vs 150 & 0.037 & 0.001 & 0.001 & 0.001 \\
10 vs 200 & 0.122 & 0.001 & 0.001 & 0.001 \\
50 vs 100 & 0.275 & 0.001 & 0.001 & 0.122 \\
50 vs 150 & 0.924 & 0.001 & 0.001 & 0.774 \\
50 vs 200 & 0.378 & 0.001 & 0.001 & 0.497 \\
100 vs 150 & 0.275 & 0.275 & 0.122 & 0.122 \\
100 vs 200 & 0.275 & 0.275 & 0.924 & 0.016 \\
150 vs 200 & 0.189 & 0.378 & 0.037 & 0.189 \\
\bottomrule
\end{tabular}
\end{table}

\begin{table}[!htbp]
\centering
\caption{Pairwise Conover post-hoc p-values for RQ6 (Cloud Identity Models) with Edge-Cloud BW = 200Mbps}
\label{tab:conover_cloud_identity_ec200}
\smallskip
Kruskal-Wallis p-values for each model: FCN = 0.0033, ResNet = 0.0001, ResNext = 0.0001, DUC = 0.0001.

\begin{tabular}{lcccc}
\toprule
Mobile-Edge BW Pair & FCN & ResNet & ResNext & DUC \\
\midrule
1 vs 10 & 0.001 & 0.001 & 0.001 & 0.001 \\
1 vs 50 & 0.001 & 0.001 & 0.001 & 0.001 \\
1 vs 100 & 0.001 & 0.001 & 0.001 & 0.001 \\
1 vs 150 & 0.001 & 0.001 & 0.001 & 0.001 \\
1 vs 200 & 0.001 & 0.001 & 0.001 & 0.001 \\
10 vs 50 & 0.122 & 0.001 & 0.001 & 0.001 \\
10 vs 100 & 0.071 & 0.001 & 0.001 & 0.001 \\
10 vs 150 & 0.016 & 0.001 & 0.001 & 0.001 \\
10 vs 200 & 0.122 & 0.001 & 0.001 & 0.001 \\
50 vs 100 & 0.189 & 0.001 & 0.001 & 0.001 \\
50 vs 150 & 0.189 & 0.001 & 0.001 & 0.016 \\
50 vs 200 & 0.497 & 0.001 & 0.001 & 0.924 \\
100 vs 150 & 0.378 & 0.774 & 0.631 & 0.924 \\
100 vs 200 & 0.378 & 0.189 & 0.378 & 0.001 \\
150 vs 200 & 0.275 & 0.122 & 0.774 & 0.001 \\
\bottomrule
\end{tabular}
\end{table}

\begin{table}[!htbp]
\centering
\caption{Pairwise Conover post-hoc p-values for RQ6 (Cloud Early Exit Models) with Mobile-Edge BW = 1Mbps}
\label{tab:conover_cloud_earlyexit_me1}
\smallskip
Kruskal-Wallis p-values for each model: FCN = 0.0130, ResNet = 0.0002, ResNext = 0.0001, DUC = 0.0004.

\begin{tabular}{lcccc}
\toprule
Edge-Cloud BW Pair & FCN & ResNet & ResNext & DUC \\
\midrule
1 vs 10 & 0.001 & 0.001 & 0.001 & 0.001 \\
1 vs 50 & 0.001 & 0.001 & 0.001 & 0.001 \\
1 vs 100 & 0.001 & 0.001 & 0.001 & 0.001 \\
1 vs 150 & 0.001 & 0.001 & 0.001 & 0.001 \\
1 vs 200 & 0.001 & 0.001 & 0.001 & 0.001 \\
10 vs 50 & 0.774 & 0.001 & 0.001 & 0.001 \\
10 vs 100 & 0.631 & 0.001 & 0.001 & 0.001 \\
10 vs 150 & 0.631 & 0.001 & 0.001 & 0.001 \\
10 vs 200 & 0.275 & 0.001 & 0.001 & 0.001 \\
50 vs 100 & 0.275 & 0.001 & 0.001 & 0.631 \\
50 vs 150 & 0.774 & 0.001 & 0.001 & 0.005 \\
50 vs 200 & 0.275 & 0.122 & 0.001 & 0.275 \\
100 vs 150 & 0.924 & 0.071 & 0.497 & 0.122 \\
100 vs 200 & 0.122 & 0.497 & 0.924 & 0.774 \\
150 vs 200 & 0.275 & 0.497 & 0.774 & 0.378 \\
\bottomrule
\end{tabular}
\end{table}

\begin{table}[!htbp]
\centering
\caption{Pairwise Conover post-hoc p-values for RQ6 (Cloud Early Exit Models) with Mobile-Edge BW = 10Mbps}
\label{tab:conover_cloud_earlyexit_me10}
\smallskip
Kruskal-Wallis p-values for each model: FCN = 0.0286, ResNet = 0.0001, ResNext = 0.0001, DUC = 0.0003.

\begin{tabular}{lcccc}
\toprule
Edge-Cloud BW Pair & FCN & ResNet & ResNext & DUC \\
\midrule
1 vs 10 & 0.001 & 0.001 & 0.001 & 0.001 \\
1 vs 50 & 0.001 & 0.001 & 0.001 & 0.001 \\
1 vs 100 & 0.001 & 0.001 & 0.001 & 0.001 \\
1 vs 150 & 0.001 & 0.001 & 0.001 & 0.001 \\
1 vs 200 & 0.001 & 0.001 & 0.001 & 0.001 \\
10 vs 50 & 0.774 & 0.001 & 0.001 & 0.001 \\
10 vs 100 & 0.924 & 0.001 & 0.001 & 0.001 \\
10 vs 150 & 0.631 & 0.001 & 0.001 & 0.001 \\
10 vs 200 & 0.631 & 0.001 & 0.001 & 0.001 \\
50 vs 100 & 0.774 & 0.001 & 0.001 & 0.275 \\
50 vs 150 & 0.774 & 0.001 & 0.001 & 0.275 \\
50 vs 200 & 0.378 & 0.001 & 0.001 & 0.005 \\
100 vs 150 & 0.774 & 0.774 & 0.071 & 0.924 \\
100 vs 200 & 0.774 & 0.275 & 0.275 & 0.275 \\
150 vs 200 & 0.924 & 0.189 & 0.275 & 0.037 \\
\bottomrule
\end{tabular}
\end{table}

\begin{table}[!htbp]
\centering
\caption{Pairwise Conover post-hoc p-values for RQ6 (Cloud Early Exit Models) with Mobile-Edge BW = 50Mbps}
\label{tab:conover_cloud_earlyexit_me50}
\smallskip
Kruskal-Wallis p-values for each model: FCN = 0.0018, ResNet = 0.0001, ResNext = 0.0001, DUC = 0.0002.

\begin{tabular}{lcccc}
\toprule
Edge-Cloud BW Pair & FCN & ResNet & ResNext & DUC \\
\midrule
1 vs 10 & 0.001 & 0.001 & 0.001 & 0.001 \\
1 vs 50 & 0.001 & 0.001 & 0.001 & 0.001 \\
1 vs 100 & 0.001 & 0.001 & 0.001 & 0.001 \\
1 vs 150 & 0.001 & 0.001 & 0.001 & 0.001 \\
1 vs 200 & 0.001 & 0.001 & 0.001 & 0.001 \\
10 vs 50 & 0.189 & 0.001 & 0.001 & 0.001 \\
10 vs 100 & 0.005 & 0.001 & 0.001 & 0.001 \\
10 vs 150 & 0.071 & 0.001 & 0.001 & 0.001 \\
10 vs 200 & 0.189 & 0.001 & 0.001 & 0.001 \\
50 vs 100 & 0.016 & 0.001 & 0.001 & 0.001 \\
50 vs 150 & 0.378 & 0.122 & 0.001 & 0.001 \\
50 vs 200 & 0.774 & 0.001 & 0.001 & 0.071 \\
100 vs 150 & 0.275 & 0.071 & 0.037 & 0.924 \\
100 vs 200 & 0.037 & 0.037 & 0.275 & 0.774 \\
150 vs 200 & 0.378 & 0.774 & 0.005 & 0.774 \\
\bottomrule
\end{tabular}
\end{table}

\begin{table}[!htbp]
\centering
\caption{Pairwise Conover post-hoc p-values for RQ6 (Cloud Early Exit Models) with Mobile-Edge BW = 100Mbps}
\label{tab:conover_cloud_earlyexit_me100}
\smallskip
Kruskal-Wallis p-values for each model: FCN = 0.0012, ResNet = 0.0001, ResNext = 0.0001, DUC = 0.0003.

\begin{tabular}{lcccc}
\toprule
Edge-Cloud BW Pair & FCN & ResNet & ResNext & DUC \\
\midrule
1 vs 10 & 0.001 & 0.001 & 0.001 & 0.001 \\
1 vs 50 & 0.001 & 0.001 & 0.001 & 0.001 \\
1 vs 100 & 0.001 & 0.001 & 0.001 & 0.001 \\
1 vs 150 & 0.001 & 0.001 & 0.001 & 0.001 \\
1 vs 200 & 0.001 & 0.001 & 0.001 & 0.001 \\
10 vs 50 & 0.122 & 0.001 & 0.001 & 0.001 \\
10 vs 100 & 0.071 & 0.001 & 0.001 & 0.001 \\
10 vs 150 & 0.122 & 0.001 & 0.001 & 0.001 \\
10 vs 200 & 0.071 & 0.001 & 0.001 & 0.001 \\
50 vs 100 & 0.001 & 0.001 & 0.001 & 0.016 \\
50 vs 150 & 0.016 & 0.001 & 0.001 & 0.774 \\
50 vs 200 & 0.001 & 0.001 & 0.001 & 0.631 \\
100 vs 150 & 0.497 & 0.189 & 0.275 & 0.122 \\
100 vs 200 & 0.497 & 0.497 & 0.122 & 0.001 \\
150 vs 200 & 0.774 & 0.774 & 0.774 & 0.924 \\
\bottomrule
\end{tabular}
\end{table}

\begin{table}[!htbp]
\centering
\caption{Pairwise Conover post-hoc p-values for RQ6 (Cloud Early Exit Models) with Mobile-Edge BW = 150Mbps}
\label{tab:conover_cloud_earlyexit_me150}
\smallskip
Kruskal-Wallis p-values for each model: FCN = 0.0047, ResNet = 0.0001, ResNext = 0.0001, DUC = 0.0002.

\begin{tabular}{lcccc}
\toprule
Edge-Cloud BW Pair & FCN & ResNet & ResNext & DUC \\
\midrule
1 vs 10 & 0.001 & 0.001 & 0.001 & 0.001 \\
1 vs 50 & 0.001 & 0.001 & 0.001 & 0.001 \\
1 vs 100 & 0.001 & 0.001 & 0.001 & 0.001 \\
1 vs 150 & 0.001 & 0.001 & 0.001 & 0.001 \\
1 vs 200 & 0.001 & 0.001 & 0.001 & 0.001 \\
10 vs 50 & 0.016 & 0.001 & 0.001 & 0.001 \\
10 vs 100 & 0.122 & 0.001 & 0.001 & 0.001 \\
10 vs 150 & 0.189 & 0.001 & 0.001 & 0.001 \\
10 vs 200 & 0.037 & 0.001 & 0.001 & 0.001 \\
50 vs 100 & 0.774 & 0.001 & 0.005 & 0.001 \\
50 vs 150 & 0.378 & 0.001 & 0.001 & 0.497 \\
50 vs 200 & 0.774 & 0.001 & 0.001 & 0.016 \\
100 vs 150 & 0.497 & 0.631 & 0.924 & 0.189 \\
100 vs 200 & 0.497 & 0.631 & 0.497 & 0.275 \\
150 vs 200 & 0.275 & 0.924 & 0.378 & 0.497 \\
\bottomrule
\end{tabular}
\end{table}

\begin{table}[!htbp]
\centering
\caption{Pairwise Conover post-hoc p-values for RQ6 (Cloud Early Exit Models) with Mobile-Edge BW = 200Mbps}
\label{tab:conover_cloud_earlyexit_me200}
\smallskip
Kruskal-Wallis p-values for each model: FCN = 0.0031, ResNet = 0.0001, ResNext = 0.0001, DUC = 0.0001.

\begin{tabular}{lcccc}
\toprule
Edge-Cloud BW Pair & FCN & ResNet & ResNext & DUC \\
\midrule
1 vs 10 & 0.001 & 0.001 & 0.001 & 0.001 \\
1 vs 50 & 0.001 & 0.001 & 0.001 & 0.001 \\
1 vs 100 & 0.001 & 0.001 & 0.001 & 0.001 \\
1 vs 150 & 0.001 & 0.001 & 0.001 & 0.001 \\
1 vs 200 & 0.001 & 0.001 & 0.001 & 0.001 \\
10 vs 50 & 0.037 & 0.001 & 0.001 & 0.001 \\
10 vs 100 & 0.016 & 0.001 & 0.001 & 0.001 \\
10 vs 150 & 0.037 & 0.001 & 0.001 & 0.001 \\
10 vs 200 & 0.071 & 0.001 & 0.001 & 0.001 \\
50 vs 100 & 0.189 & 0.001 & 0.001 & 0.071 \\
50 vs 150 & 0.774 & 0.001 & 0.001 & 0.001 \\
50 vs 200 & 0.631 & 0.001 & 0.001 & 0.189 \\
100 vs 150 & 0.378 & 0.774 & 0.924 & 0.001 \\
100 vs 200 & 0.275 & 0.275 & 0.631 & 0.001 \\
150 vs 200 & 0.924 & 0.037 & 0.631 & 0.005 \\
\bottomrule
\end{tabular}
\end{table}

\begin{table}[!htbp]
\centering
\caption{Pairwise Conover post-hoc p-values for RQ6 (Cloud Early Exit Models) with Edge-Cloud BW = 1Mbps}
\label{tab:conover_cloud_earlyexit_ec1}
\smallskip
Kruskal-Wallis p-values for each model: FCN = 0.0013, ResNet = 0.0001, ResNext = 0.0001, DUC = 0.0003.

\begin{tabular}{lcccc}
\toprule
Mobile-Edge BW Pair & FCN & ResNet & ResNext & DUC \\
\midrule
1 vs 10 & 0.001 & 0.001 & 0.001 & 0.001 \\
1 vs 50 & 0.001 & 0.001 & 0.001 & 0.001 \\
1 vs 100 & 0.001 & 0.001 & 0.001 & 0.001 \\
1 vs 150 & 0.001 & 0.001 & 0.001 & 0.001 \\
1 vs 200 & 0.001 & 0.001 & 0.001 & 0.001 \\
10 vs 50 & 0.122 & 0.001 & 0.001 & 0.001 \\
10 vs 100 & 0.497 & 0.001 & 0.001 & 0.001 \\
10 vs 150 & 0.378 & 0.001 & 0.001 & 0.001 \\
10 vs 200 & 0.924 & 0.001 & 0.001 & 0.001 \\
50 vs 100 & 0.016 & 0.001 & 0.001 & 0.275 \\
50 vs 150 & 0.001 & 0.001 & 0.001 & 0.497 \\
50 vs 200 & 0.005 & 0.001 & 0.001 & 0.016 \\
100 vs 150 & 0.037 & 0.071 & 0.275 & 0.275 \\
100 vs 200 & 0.497 & 0.189 & 0.275 & 0.071 \\
150 vs 200 & 0.189 & 0.122 & 0.497 & 0.037 \\
\bottomrule
\end{tabular}
\end{table}

\begin{table}[!htbp]
\centering
\caption{Pairwise Conover post-hoc p-values for RQ6 (Cloud Early Exit Models) with Edge-Cloud BW = 10Mbps}
\label{tab:conover_cloud_earlyexit_ec10}
\smallskip
Kruskal-Wallis p-values for each model: FCN = 0.0308, ResNet = 0.0001, ResNext = 0.0001, DUC = 0.0006.

\begin{tabular}{lcccc}
\toprule
Mobile-Edge BW Pair & FCN & ResNet & ResNext & DUC \\
\midrule
1 vs 10 & 0.001 & 0.001 & 0.001 & 0.001 \\
1 vs 50 & 0.001 & 0.001 & 0.001 & 0.001 \\
1 vs 100 & 0.001 & 0.001 & 0.001 & 0.001 \\
1 vs 150 & 0.001 & 0.001 & 0.001 & 0.001 \\
1 vs 200 & 0.001 & 0.001 & 0.001 & 0.001 \\
10 vs 50 & 0.774 & 0.001 & 0.001 & 0.001 \\
10 vs 100 & 0.774 & 0.001 & 0.001 & 0.001 \\
10 vs 150 & 0.774 & 0.001 & 0.001 & 0.001 \\
10 vs 200 & 0.924 & 0.001 & 0.001 & 0.001 \\
50 vs 100 & 0.774 & 0.001 & 0.001 & 0.924 \\
50 vs 150 & 0.774 & 0.001 & 0.001 & 0.189 \\
50 vs 200 & 0.774 & 0.001 & 0.001 & 0.497 \\
100 vs 150 & 0.774 & 0.924 & 0.631 & 0.497 \\
100 vs 200 & 0.924 & 0.275 & 0.275 & 0.189 \\
150 vs 200 & 0.631 & 0.275 & 0.071 & 0.189 \\
\bottomrule
\end{tabular}
\end{table}

\begin{table}[!htbp]
\centering
\caption{Pairwise Conover post-hoc p-values for RQ6 (Cloud Early Exit Models) with Edge-Cloud BW = 50Mbps}
\label{tab:conover_cloud_earlyexit_ec50}
\smallskip
Kruskal-Wallis p-values for each model: FCN = 0.0004, ResNet = 0.0001, ResNext = 0.0001, DUC = 0.0004.

\begin{tabular}{lcccc}
\toprule
Mobile-Edge BW Pair & FCN & ResNet & ResNext & DUC \\
\midrule
1 vs 10 & 0.001 & 0.001 & 0.001 & 0.001 \\
1 vs 50 & 0.001 & 0.001 & 0.001 & 0.001 \\
1 vs 100 & 0.001 & 0.001 & 0.001 & 0.001 \\
1 vs 150 & 0.001 & 0.001 & 0.001 & 0.001 \\
1 vs 200 & 0.001 & 0.001 & 0.001 & 0.001 \\
10 vs 50 & 0.071 & 0.001 & 0.001 & 0.001 \\
10 vs 100 & 0.122 & 0.001 & 0.001 & 0.001 \\
10 vs 150 & 0.005 & 0.001 & 0.001 & 0.001 \\
10 vs 200 & 0.071 & 0.001 & 0.001 & 0.001 \\
50 vs 100 & 0.016 & 0.001 & 0.001 & 0.122 \\
50 vs 150 & 0.016 & 0.001 & 0.001 & 0.497 \\
50 vs 200 & 0.774 & 0.001 & 0.001 & 0.071 \\
100 vs 150 & 0.001 & 0.924 & 0.631 & 0.275 \\
100 vs 200 & 0.016 & 0.275 & 0.631 & 0.631 \\
150 vs 200 & 0.037 & 0.016 & 0.924 & 0.189 \\
\bottomrule
\end{tabular}
\end{table}

\begin{table}[!htbp]
\centering
\caption{Pairwise Conover post-hoc p-values for RQ6 (Cloud Early Exit Models) with Edge-Cloud BW = 100Mbps}
\label{tab:conover_cloud_earlyexit_ec100}
\smallskip
Kruskal-Wallis p-values for each model: FCN = 0.0010, ResNet = 0.0001, ResNext = 0.0001, DUC = 0.0009.

\begin{tabular}{lcccc}
\toprule
Mobile-Edge BW Pair & FCN & ResNet & ResNext & DUC \\
\midrule
1 vs 10 & 0.001 & 0.001 & 0.001 & 0.001 \\
1 vs 50 & 0.001 & 0.001 & 0.001 & 0.001 \\
1 vs 100 & 0.001 & 0.001 & 0.001 & 0.001 \\
1 vs 150 & 0.001 & 0.001 & 0.001 & 0.001 \\
1 vs 200 & 0.001 & 0.001 & 0.001 & 0.001 \\
10 vs 50 & 0.005 & 0.001 & 0.001 & 0.001 \\
10 vs 100 & 0.005 & 0.001 & 0.001 & 0.001 \\
10 vs 150 & 0.001 & 0.001 & 0.001 & 0.001 \\
10 vs 200 & 0.016 & 0.001 & 0.001 & 0.001 \\
50 vs 100 & 0.774 & 0.001 & 0.001 & 0.275 \\
50 vs 150 & 0.378 & 0.001 & 0.001 & 0.497 \\
50 vs 200 & 0.071 & 0.001 & 0.001 & 0.631 \\
100 vs 150 & 0.631 & 0.497 & 0.378 & 0.774 \\
100 vs 200 & 0.189 & 0.378 & 0.037 & 0.924 \\
150 vs 200 & 0.497 & 0.275 & 0.122 & 0.631 \\
\bottomrule
\end{tabular}
\end{table}

\begin{table}[!htbp]
\centering
\caption{Pairwise Conover post-hoc p-values for RQ6 (Cloud Early Exit Models) with Edge-Cloud BW = 150Mbps}
\label{tab:conover_cloud_earlyexit_ec150}
\smallskip
Kruskal-Wallis p-values for each model: FCN = 0.0029, ResNet = 0.0001, ResNext = 0.0001, DUC = 0.0002.

\begin{tabular}{lcccc}
\toprule
Mobile-Edge BW Pair & FCN & ResNet & ResNext & DUC \\
\midrule
1 vs 10 & 0.001 & 0.001 & 0.001 & 0.001 \\
1 vs 50 & 0.001 & 0.001 & 0.001 & 0.001 \\
1 vs 100 & 0.001 & 0.001 & 0.001 & 0.001 \\
1 vs 150 & 0.001 & 0.001 & 0.001 & 0.001 \\
1 vs 200 & 0.001 & 0.001 & 0.001 & 0.001 \\
10 vs 50 & 0.037 & 0.001 & 0.001 & 0.001 \\
10 vs 100 & 0.016 & 0.001 & 0.001 & 0.001 \\
10 vs 150 & 0.016 & 0.001 & 0.001 & 0.001 \\
10 vs 200 & 0.016 & 0.001 & 0.001 & 0.001 \\
50 vs 100 & 0.497 & 0.001 & 0.001 & 0.497 \\
50 vs 150 & 0.631 & 0.001 & 0.001 & 0.497 \\
50 vs 200 & 0.497 & 0.001 & 0.001 & 0.001 \\
100 vs 150 & 0.378 & 0.924 & 0.774 & 0.924 \\
100 vs 200 & 0.631 & 0.005 & 0.037 & 0.001 \\
150 vs 200 & 0.774 & 0.037 & 0.122 & 0.071 \\
\bottomrule
\end{tabular}
\end{table}

\begin{table}[!htbp]
\centering
\caption{Pairwise Conover post-hoc p-values for RQ6 (Cloud Early Exit Models) with Edge-Cloud BW = 200Mbps}
\label{tab:conover_cloud_earlyexit_ec200}
\smallskip
Kruskal-Wallis p-values for each model: FCN = 0.0013, ResNet = 0.0001, ResNext = 0.0000, DUC = 0.0002.

\begin{tabular}{lcccc}
\toprule
Mobile-Edge BW Pair & FCN & ResNet & ResNext & DUC \\
\midrule
1 vs 10 & 0.001 & 0.001 & 0.001 & 0.001 \\
1 vs 50 & 0.001 & 0.001 & 0.001 & 0.001 \\
1 vs 100 & 0.001 & 0.001 & 0.001 & 0.001 \\
1 vs 150 & 0.001 & 0.001 & 0.001 & 0.001 \\
1 vs 200 & 0.001 & 0.001 & 0.001 & 0.001 \\
10 vs 50 & 0.037 & 0.001 & 0.001 & 0.001 \\
10 vs 100 & 0.016 & 0.001 & 0.001 & 0.001 \\
10 vs 150 & 0.001 & 0.001 & 0.001 & 0.001 \\
10 vs 200 & 0.071 & 0.001 & 0.001 & 0.001 \\
50 vs 100 & 0.275 & 0.001 & 0.001 & 0.122 \\
50 vs 150 & 0.071 & 0.001 & 0.001 & 0.924 \\
50 vs 200 & 0.774 & 0.001 & 0.001 & 0.122 \\
100 vs 150 & 0.497 & 0.774 & 0.378 & 0.275 \\
100 vs 200 & 0.497 & 0.037 & 0.005 & 0.001 \\
150 vs 200 & 0.122 & 0.001 & 0.001 & 0.005 \\
\bottomrule
\end{tabular}
\end{table}

\begin{table}[!htbp]
\centering
\caption{Pairwise Conover post-hoc p-values for RQ6 (Cloud Quantized Models) with Mobile-Edge BW = 1Mbps}
\label{tab:conover_cloud_quantized_me1}
\smallskip
Kruskal-Wallis p-values for each model: FCN = 0.0146, ResNet = 0.0001, ResNext = 0.0001, DUC = 0.0002.

\begin{tabular}{lcccc}
\toprule
Edge-Cloud BW Pair & FCN & ResNet & ResNext & DUC \\
\midrule
1 vs 10 & 0.001 & 0.001 & 0.001 & 0.001 \\
1 vs 50 & 0.001 & 0.001 & 0.001 & 0.001 \\
1 vs 100 & 0.001 & 0.001 & 0.001 & 0.001 \\
1 vs 150 & 0.001 & 0.001 & 0.001 & 0.001 \\
1 vs 200 & 0.001 & 0.001 & 0.001 & 0.001 \\
10 vs 50 & 0.631 & 0.001 & 0.001 & 0.001 \\
10 vs 100 & 0.497 & 0.001 & 0.001 & 0.001 \\
10 vs 150 & 0.631 & 0.001 & 0.001 & 0.001 \\
10 vs 200 & 0.631 & 0.001 & 0.001 & 0.001 \\
50 vs 100 & 0.189 & 0.001 & 0.001 & 0.016 \\
50 vs 150 & 0.774 & 0.001 & 0.001 & 0.497 \\
50 vs 200 & 0.189 & 0.001 & 0.001 & 0.071 \\
100 vs 150 & 0.378 & 0.378 & 0.924 & 0.005 \\
100 vs 200 & 0.631 & 0.189 & 0.378 & 0.497 \\
150 vs 200 & 0.378 & 0.071 & 0.631 & 0.016 \\
\bottomrule
\end{tabular}
\end{table}

\begin{table}[!htbp]
\centering
\caption{Pairwise Conover post-hoc p-values for RQ6 (Cloud Quantized Models) with Mobile-Edge BW = 10Mbps}
\label{tab:conover_cloud_quantized_me10}
\smallskip
Kruskal-Wallis p-values for each model: FCN = 0.0313, ResNet = 0.0001, ResNext = 0.0001, DUC = 0.0001.

\begin{tabular}{lcccc}
\toprule
Edge-Cloud BW Pair & FCN & ResNet & ResNext & DUC \\
\midrule
1 vs 10 & 0.001 & 0.001 & 0.001 & 0.001 \\
1 vs 50 & 0.001 & 0.001 & 0.001 & 0.001 \\
1 vs 100 & 0.001 & 0.001 & 0.001 & 0.001 \\
1 vs 150 & 0.001 & 0.001 & 0.001 & 0.001 \\
1 vs 200 & 0.001 & 0.001 & 0.001 & 0.001 \\
10 vs 50 & 0.774 & 0.001 & 0.001 & 0.001 \\
10 vs 100 & 0.631 & 0.001 & 0.001 & 0.001 \\
10 vs 150 & 0.924 & 0.001 & 0.001 & 0.001 \\
10 vs 200 & 0.774 & 0.001 & 0.001 & 0.001 \\
50 vs 100 & 0.924 & 0.001 & 0.001 & 0.378 \\
50 vs 150 & 0.924 & 0.001 & 0.016 & 0.071 \\
50 vs 200 & 0.774 & 0.001 & 0.005 & 0.005 \\
100 vs 150 & 0.774 & 0.631 & 0.071 & 0.497 \\
100 vs 200 & 0.774 & 0.122 & 0.275 & 0.001 \\
150 vs 200 & 0.924 & 0.122 & 0.189 & 0.001 \\
\bottomrule
\end{tabular}
\end{table}

\begin{table}[!htbp]
\centering
\caption{Pairwise Conover post-hoc p-values for RQ6 (Cloud Quantized Models) with Mobile-Edge BW = 50Mbps}
\label{tab:conover_cloud_quantized_me50}
\smallskip
Kruskal-Wallis p-values for each model: FCN = 0.0016, ResNet = 0.0000, ResNext = 0.0001, DUC = 0.0003.

%
\end{table}

\begin{table}[!htbp]
\centering
\caption{Pairwise Conover post-hoc p-values for RQ6 (Cloud Quantized Models) with Edge-Cloud BW = 1Mbps}
\label{tab:conover_cloud_quantized_ec1}
\smallskip
Kruskal-Wallis p-values for each model: FCN = 0.0253, ResNet = 0.0001, ResNext = 0.0001, DUC = 0.0006.

%
\end{table}

\begin{table}[!htbp]
\centering
\caption{Pairwise Conover post-hoc p-values for RQ6 (Cloud Quantized Models) with Edge-Cloud BW = 10Mbps}
\label{tab:conover_cloud_quantized_ec10}
\smallskip
Kruskal-Wallis p-values for each model: FCN = 0.0299, ResNet = 0.0001, ResNext = 0.0001, DUC = 0.0002.

%
\end{table}

 \clearpage
\begin{table}[!htbp]
\centering
\caption{Pairwise Conover post-hoc p-values for RQ6 (Cloud Quantized Models) with Edge-Cloud BW = 50Mbps}
\label{tab:conover_cloud_quantized_ec50}
\smallskip
Kruskal-Wallis p-values for each model: FCN = 0.0024, ResNet = 0.0001, ResNext = 0.0001, DUC = 0.0003.

%
\end{table}

\clearpage
\begin{table}[!htbp]
\centering
\caption{Pairwise Conover post-hoc p-values for RQ6 (Cloud Quantized Models) with Edge-Cloud BW = 100Mbps}
\label{tab:conover_cloud_quantized_ec100}
\smallskip
Kruskal-Wallis p-values for each model: FCN = 0.0100, ResNet = 0.0001, ResNext = 0.0001, DUC = 0.0002.

%
\end{table}

\begin{table}[!htbp]
\centering
\caption{Pairwise Conover post-hoc p-values for RQ6 (Cloud Quantized Models) with Edge-Cloud BW = 150Mbps}
\label{tab:conover_cloud_quantized_ec150}
\smallskip
Kruskal-Wallis p-values for each model: FCN = 0.0076, ResNet = 0.0001, ResNext = 0.0001, DUC = 0.0009.

%
\end{table}

\begin{table}[!htbp]
\centering
\caption{Pairwise Conover post-hoc p-values for RQ6 (Cloud Quantized Early Exit Models) with Mobile-Edge BW = 1Mbps}
\label{tab:conover_cloud_quantized_earlyexit_me1}
\smallskip
Kruskal-Wallis p-values for each model: FCN = 0.0222, ResNet = 0.0002, ResNext = 0.0001, DUC = 0.0011.

%
\end{table}

\clearpage
\begin{table}[!htbp]
\centering
\caption{Pairwise Conover post-hoc p-values for RQ6 (Cloud Quantized Early Exit Models) with Mobile-Edge BW = 10Mbps}
\label{tab:conover_cloud_quantized_earlyexit_me10}
\smallskip
Kruskal-Wallis p-values for each model: FCN = 0.0314, ResNet = 0.0001, ResNext = 0.0002, DUC = 0.0002.

%
\end{table}

\clearpage
\begin{table}[!htbp]
\centering
\caption{Pairwise Conover post-hoc p-values for RQ6 (Cloud Quantized Early Exit Models) with Mobile-Edge BW = 50Mbps}
\label{tab:conover_cloud_quantized_earlyexit_me50}
\smallskip
Kruskal-Wallis p-values for each model: FCN = 0.0010, ResNet = 0.0001, ResNext = 0.0000, DUC = 0.0001.

%
\end{table}

\begin{table}[!htbp]
\centering
\caption{Pairwise Conover post-hoc p-values for RQ6 (Cloud Quantized Early Exit Models) with Mobile-Edge BW = 100Mbps}
\label{tab:conover_cloud_quantized_earlyexit_me100}
\smallskip
Kruskal-Wallis p-values for each model: FCN = 0.0009, ResNet = 0.0001, ResNext = 0.0001, DUC = 0.0001.

%
\end{table}

\begin{table}[!htbp]
\centering
\caption{Pairwise Conover post-hoc p-values for RQ6 (Cloud Quantized Early Exit Models) with Mobile-Edge BW = 150Mbps}
\label{tab:conover_cloud_quantized_earlyexit_me150}
\smallskip
Kruskal-Wallis p-values for each model: FCN = 0.0057, ResNet = 0.0001, ResNext = 0.0001, DUC = 0.0003.

%
\end{table}

\begin{table}[!htbp]
\centering
\caption{Pairwise Conover post-hoc p-values for RQ6 (Cloud Quantized Early Exit Models) with Edge-Cloud BW = 1Mbps}
\label{tab:conover_cloud_quantized_earlyexit_ec1}
\smallskip
Kruskal-Wallis p-values for each model: FCN = 0.0293, ResNet = 0.0001, ResNext = 0.0001, DUC = 0.0010.

%
\end{table}

\begin{table}[!htbp]
\centering
\caption{Pairwise Conover post-hoc p-values for RQ6 (Cloud Quantized Early Exit Models) with Edge-Cloud BW = 10Mbps}
\label{tab:conover_cloud_quantized_earlyexit_ec10}
\smallskip
Kruskal-Wallis p-values for each model: FCN = 0.0311, ResNet = 0.0001, ResNext = 0.0001, DUC = 0.0001.

%
\end{table}

\begin{table}[!htbp]
\centering
\caption{Pairwise Conover post-hoc p-values for RQ6 (Cloud Quantized Early Exit Models) with Edge-Cloud BW = 50Mbps}
\label{tab:conover_cloud_quantized_earlyexit_ec50}
\smallskip
Kruskal-Wallis p-values for each model: FCN = 0.0010, ResNet = 0.0001, ResNext = 0.0001, DUC = 0.0001.

%
\end{table}

\begin{table}[!htbp]
\centering
\caption{Pairwise Conover post-hoc p-values for RQ6 (Mobile-Cloud Partition Models) with Mobile-Edge BW = 1Mbps}
\label{tab:conover_mobile_cloud_partition_me1}
\smallskip
Kruskal-Wallis p-values for each model: FCN = 0.0001, ResNet = 0.0180, ResNext = 0.0183, DUC = 0.0000.

%
\end{table}

\begin{table}[!htbp]
\centering
\caption{Pairwise Conover post-hoc p-values for RQ6 (Mobile-Cloud Partition Models) with Mobile-Edge BW = 10Mbps}
\label{tab:conover_mobile_cloud_partition_me10}
\smallskip
Kruskal-Wallis p-values for each model: FCN = 0.0002, ResNet = 0.0198, ResNext = 0.0064, DUC = 0.0000.

%
\end{table}

\begin{table}[!htbp]
\centering
\caption{Pairwise Conover post-hoc p-values for RQ6 (Mobile-Cloud Partition Models) with Mobile-Edge BW = 50Mbps}
\label{tab:conover_mobile_cloud_partition_me50}
\smallskip
Kruskal-Wallis p-values for each model: FCN = 0.0000, ResNet = 0.0023, ResNext = 0.0015, DUC = 0.0000.

%
\end{table}

\begin{table}[!htbp]
\centering
\caption{Pairwise Conover post-hoc p-values for RQ6 (Mobile-Cloud Partition Models) with Edge-Cloud BW = 1Mbps}
\label{tab:conover_mobile_cloud_partition_ec1}
\smallskip
Kruskal-Wallis p-values for each model: FCN = 0.0001, ResNet = 0.0006, ResNext = 0.0001, DUC = 0.0000.

%
\end{table}

\begin{table}[!htbp]
\centering
\caption{Pairwise Conover post-hoc p-values for RQ6 (Mobile-Cloud Partition Models) with Edge-Cloud BW = 10Mbps}
\label{tab:conover_mobile_cloud_partition_ec10}
\smallskip
Kruskal-Wallis p-values for each model: FCN = 0.0000, ResNet = 0.0002, ResNext = 0.0001, DUC = 0.0000.

%
\end{table}

\begin{table}[!htbp]
\centering
\caption{Pairwise Conover post-hoc p-values for RQ6 (Mobile-Cloud Partition Models) with Edge-Cloud BW = 50Mbps}
\label{tab:conover_mobile_cloud_partition_ec50}
\smallskip
Kruskal-Wallis p-values for each model: FCN = 0.0000, ResNet = 0.0001, ResNext = 0.0001, DUC = 0.0000.

%
\end{table}

\begin{table}[!htbp]
\centering
\caption{Pairwise Conover post-hoc p-values for RQ6 (Edge-Cloud Partition Models) with Mobile-Edge BW = 1Mbps}
\label{tab:conover_edge_cloud_partition_me1}
\smallskip
Kruskal-Wallis p-values for each model: FCN = 0.0002, ResNet = 0.0097, ResNext = 0.0023, DUC = 0.0001.

%
\end{table}

\begin{table}[!htbp]
\centering
\caption{Pairwise Conover post-hoc p-values for RQ6 (Edge-Cloud Partition Models) with Mobile-Edge BW = 10Mbps}
\label{tab:conover_edge_cloud_partition_me10}
\smallskip
Kruskal-Wallis p-values for each model: FCN = 0.0001, ResNet = 0.0253, ResNext = 0.0011, DUC = 0.0000.

%
\end{table}

\begin{table}[!htbp]
\centering
\caption{Pairwise Conover post-hoc p-values for RQ6 (Edge-Cloud Partition Models) with Mobile-Edge BW = 50Mbps}
\label{tab:conover_edge_cloud_partition_me50}
\smallskip
Kruskal-Wallis p-values for each model: FCN = 0.0001, ResNet = 0.0014, ResNext = 0.0024, DUC = 0.0000.

%
\end{table}

\begin{table}[!htbp]
\centering
\caption{Pairwise Conover post-hoc p-values for RQ6 (Edge-Cloud Partition Models) with Edge-Cloud BW = 1Mbps}
\label{tab:conover_edge_cloud_partition_ec1}
\smallskip
Kruskal-Wallis p-values for each model: FCN = 0.0172, ResNet = 0.0001, ResNext = 0.0001, DUC = 0.0005.

%
\end{table}

\begin{table}[!htbp]
\centering
\caption{Pairwise Conover post-hoc p-values for RQ6 (Edge-Cloud Partition Models) with Edge-Cloud BW = 10Mbps}
\label{tab:conover_edge_cloud_partition_ec10}
\smallskip
Kruskal-Wallis p-values for each model: FCN = 0.0045, ResNet = 0.0001, ResNext = 0.0001, DUC = 0.0002.

%
\end{table}

\begin{table}[!htbp]
\centering
\caption{Pairwise Conover post-hoc p-values for RQ6 (Edge-Cloud Partition Models) with Edge-Cloud BW = 50Mbps}
\label{tab:conover_edge_cloud_partition_ec50}
\smallskip
Kruskal-Wallis p-values for each model: FCN = 0.0138, ResNet = 0.0001, ResNext = 0.0000, DUC = 0.0004.

%
\end{table}

\begin{table}[!htbp]
\centering
\caption{Pairwise Conover post-hoc p-values for RQ6 (Mobile-Cloud Quantized Early Exit Partition Models) with Mobile-Edge BW = 1Mbps}
\label{tab:conover_mobile_cloud_quantized_earlyexit_partition_me1}
\smallskip
Kruskal-Wallis p-values for each model: FCN = 0.0002, ResNet = 0.0097, ResNext = 0.0023, DUC = 0.0001.

%
\end{table}

\begin{table}[!htbp]
\centering
\caption{Pairwise Conover post-hoc p-values for RQ6 (Mobile-Cloud Quantized Early Exit Partition Models) with Mobile-Edge BW = 10Mbps}
\label{tab:conover_mobile_cloud_quantized_earlyexit_partition_me10}
\smallskip
Kruskal-Wallis p-values for each model: FCN = 0.0001, ResNet = 0.0253, ResNext = 0.0011, DUC = 0.0000.

%
\end{table}

\begin{table}[!htbp]
\centering
\caption{Pairwise Conover post-hoc p-values for RQ6 (Mobile-Cloud Quantized Early Exit Partition Models) with Mobile-Edge BW = 50Mbps}
\label{tab:conover_mobile_cloud_quantized_earlyexit_partition_me50}
\smallskip
Kruskal-Wallis p-values for each model: FCN = 0.0001, ResNet = 0.0014, ResNext = 0.0024, DUC = 0.0000.

%
\end{table}

\end{appendices}

\end{document}